\documentclass[a4paper,11pt,twoside]{book}
\usepackage{caption}
\usepackage{bm}       % bold mathematic symbols
\usepackage{comment}
\usepackage{setspace}
\usepackage{graphicx}
\usepackage{amssymb}
\usepackage{amsbsy}   % ApJ recommened bold-math symbols
\usepackage{amsmath}
\usepackage{url}
\usepackage{lscape}
\usepackage{natbib}
\usepackage{aastex_hack}
\usepackage{fancyhdr}
\usepackage[avantgarde]{quotchap}
\usepackage[calcwidth]{titlesec}
\usepackage{swinburnetitle}
\usepackage{subfigure}
\usepackage{varioref} % Similar to \ref, but allows the labels to be changed 
\usepackage{lipsum}
\usepackage{pdfpages}

\usepackage{xurl}
\usepackage{rotating,tabularx,siunitx}
\usepackage{arydshln}

\usepackage[sectionbib]{chapterbib}
\usepackage{multibib}
\newcites{main}{Main References}
\newcites{other}{Bibliography}
\usepackage{cite}
\usepackage{enumerate}
\usepackage[inline]{enumitem}

\usepackage[nottoc,section]{tocbibind}

\newcommand{\fsps}{$\mathrm{fs~s^{-1}}$}
\newcommand{\msun}{\ifmmode\mbox{M}_{\odot}\else$\mbox{M}_{\odot}$\fi}
\newcommand{\Vsun}{\ifmmode\mbox{V}_{\odot}\else$\mbox{V}_{\odot}$\fi}

\newcommand{\maspy}{$\rm mas~yr^{-1}$}
\newcommand{\kmps}{$\rm km~s^{-1}$}
\newcommand{\mjypb}{$\rm mJy~beam^{-1}$}
\newcommand{\psrpi}{\ensuremath{\mathrm{PSR}\pi}}
\newcommand{\mspsrpi}{\ensuremath{\mathrm{MSPSR}\pi}}
\newcommand{\rcsl}{reduced chi-square}
\newcommand{\rcs}{$\chi_{\nu}^{2}$}
 
\newcommand{\gw}{gravitational wave}

\newcommand{\psrb}{PSR~J0030$+$0451}
\newcommand{\psrc}{PSR~J0610$-$2100}
\newcommand{\psrd}{PSR~J0621$+$1002}
\newcommand{\psrea}{PSR~J1012$+$5307}
\newcommand{\psreb}{PSR~J1024$-$0719}
\newcommand{\psrfa}{PSR~J1518$+$4904}
\newcommand{\psrfb}{PSR~J1537$+$1155}
\newcommand{\psrfbb}{PSR~B1534$+$12}
\newcommand{\psrga}{PSR~J1640$+$2224}
\newcommand{\psrgb}{PSR~J1643$-$1224}
\newcommand{\psrha}{PSR~J1721$-$2457
}
\newcommand{\psrhb}{PSR~J1730$-$2304}
\newcommand{\psri}{PSR~J1738$+$0333}
\newcommand{\psro}{PSR~J1824$-$2452A}
\newcommand{\psrka}{PSR~J1853$+$1303}
\newcommand{\psrl}{PSR~J1910$+$1256}

\newcommand{\psrmb}{PSR~J1918$-$0642}
\newcommand{\psrkb}{PSR~J1939$+$2134}

\newcommand{\Psrb}{J0030$+$0451}
\newcommand{\Psrc}{J0610$-$2100}
\newcommand{\Psrd}{J0621$+$1002}
\newcommand{\Psrea}{J1012$+$5307}
\newcommand{\Psreb}{J1024$-$0719}
\newcommand{\Psrfa}{J1518$+$4904}
\newcommand{\Psrfb}{J1537$+$1155}
\newcommand{\Psrga}{J1640$+$2224}
\newcommand{\Psrgb}{J1643$-$1224}
\newcommand{\Psrha}{J1721$-$2457
}
\newcommand{\Psrhb}{J1730$-$2304}
\newcommand{\Psri}{J1738$+$0333}
\newcommand{\Psro}{J1824$-$2452A}
\newcommand{\Psrka}{J1853$+$1303}
\newcommand{\Psrl}{J1910$+$1256}
\newcommand{\Psrma}{J1911$-$1114}
\newcommand{\Psrmb}{J1918$-$0642}
\newcommand{\Psrkb}{J1939$+$2134}
\newcommand{\Psrna}{J0437$-$4715}
\newcommand{\Psrnb}{J1713$+$0747}

\newcommand{\multilinecomment}[1]{}

\newcommand{\aql}{Aql~X$-$1}
\newcommand{\xte}{XTE~J1810$-$197}
\newcommand{\swift}{Swift~J1818.0$-$1607}

\newcommand{\gx}{GX~17$+$2}
\newcommand{\cyg}{Cyg~X$-$2}
\newcommand{\cen}{Cen~X$-$4}
\newcommand{\uua}{4U~0919$-$54}
\newcommand{\xb}{XB~2129$+$47}
\newcommand{\sax}{SAX~J1808.4$-$3658}

\def\tabnotefont{\footnotesize}
\newcommand{\tabnote}[1]{\par\vskip1pt{\raggedright\tabnotefont #1\par}}

\newcommand{\pasa}{PASA}
\usepackage[bookmarks=true,breaklinks=true,hypertexnames=false,bookmarksnumbered=true]{hyperref}
\hypersetup{
pdfauthor = { YourName },
pdftitle = { YourThesisTitle },
pdfsubject = { Astronomy },
pdfkeywords = { -- },
pdfcreator = {LaTeX with hyperref package},
pdfproducer = {dvips + ps2pdf}}

\usepackage{array}
\renewcommand{\arraystretch}{1.2}

\fancypagestyle{plain}{
\fancyhf{}
\fancyfoot[CO]{\thepage}

}

\makeatletter
\def\cleardoublepage{\clearpage\if@twoside \ifodd\c@page\else
	\hbox{}
	\vspace*{\fill}
	\thispagestyle{empty}
	\newpage
	\if@twocolumn\hbox{}\newpage\fi\fi\fi}
\makeatother

\renewcommand\chapterheadstartvskip{\vspace*{-5\baselineskip}}

\titleformat{\section}[hang]{\sffamily\bfseries}
{\Large\thesection}{12pt}{\Large}[{\titlerule[0.5pt]}]

\labelformat{chapter}{Ch.~#1}
\labelformat{section}{Sec.~#1}
\labelformat{appendix}{App.~#1}
\labelformat{subsection}{Sec.~#1}
\labelformat{subsubsection}{Sec.~#1}
\labelformat{figure}{Fig.~#1}
\labelformat{subfigure}{Fig.~\thefigure #1}
\labelformat{table}{Tab.~#1}
\labelformat{equation}{Eq.~(#1)}

\DeclareFontFamily{OT1}{pzc}{} 
\DeclareFontShape{OT1}{pzc}{m}{it}{<->[1.5] pzcmi8t}{}
\DeclareMathAlphabet{\mathpzc}{OT1}{pzc}{m}{it}

\begin{document}
%\pagestyle{headings}

% Titles, acknowledgements and declarations
\frontmatter
\author{Hao Ding}
\title{Enhancing the use of Galactic neutron stars as physical laboratories with precise astrometry}
% Warning: This field cannot be empty or you will get terrible terrible errors!
\date{ October 2022 }
\maketitle

\addcontentsline{toc}{chapter}{Abstract}
\Huge \noindent Abstract
\normalsize
\\

Neutron stars (NSs) are extreme stellar remnants, whose central density is greater than that of nuclear matter and is exceeded only by gravitational singularities (black holes).
Their combination of extreme magnetisation, rotation, and density produces effects that are observable across the electromagnetic spectrum, and with gravitational-wave (GW) detectors.
NS observations not only offer the unique key to understanding the uncertain NS interior composition, studying high-energy activities involving NSs, and probing NS formation channels, but also supercharge the study of ionized interstellar medium (IISM), and the test of gravitational theories.
However, many of these scientific motivations require precise astrometric information of NSs, especially NS distances and proper motions.
To address this need, this thesis focuses on high-precision astrometry of three distinct NS divisions --- magnetars, NS X-ray binaries (XRBs) and millisecond pulsars (MSPs).
The thesis encompasses the whole \mspsrpi\ project (including the re-analysis of a published MSP) --- so far the largest astrometric survey of MSPs, and 2 of the 5 constant radio magnetars, and all (4) type I X-ray bursters with significant Gaia parallaxes. 

Methodologically, magnetars and MSPs were astrometrically measured using the Very Long Baseline Array (VLBA), while a specific NS XRB sub-group known as type I X-ray bursters (bursters) were astrometrically studied with the data collected from the Gaia space telescope (Gaia).
On the VLBI (very long baseline interferometry) side, numerous methodological advances were developed and implemented in this thesis to maximise the astrometric precision.  In-beam calibration (which eliminates temporal interpolation) was available for all sources, but even better results were seen for some sources via the use of phase solution interpolation.
VLBI  data were reduced with the {\tt psrvlbireduce} package, while the astrometric parameters for the \mspsrpi\ pulsars were derived using the new {\tt sterne} Bayesian astrometric inference package. 
On the Gaia side, a novel method was employed to estimate the parallax zero-point and its uncertainty for each burster using the background quasars.

Scientifically, the thesis obtained significant VLBI parallaxes for 15 MSPs and (for the first time) a magnetar. The median VLBI parallax uncertainty of all the 18 \mspsrpi\ pulsars is 0.09\,mas, which is improved to 0.07\,mas when incorporating pulsar timing results.
Based on the precise astrometric results of 16 MSPs, a multi-modal distribution of MSP transverse space velocities $v_\perp$ was revealed. Though overall smaller than the previous studies, the new $v_\perp$ distribution can likely address both Galactic centre $\gamma$-ray excess and the MSP retention problem in globular clusters. The precise astrometric parameters determined for the two radio magnetars can refine the poorly constrained magnetar $v_\perp$ distribution, which, as proposed in this thesis, can potentially probe the formation channels of magnetars. In addition, incorporating precise pulsar timing results, the new parallax-based distances refined {\bfseries 1)} the orbital-decay test of general relativity (GR) using the double neutron star (DNS) \psrfb, and {\bfseries 2)} the constraints on alternative theories of gravity with two white-dwarf-pulsars --- \psrea\ and \psri. Among the other major scientific outputs, the thesis provides the road map for probing the simplistic model of photospheric radius expansion (PRE) bursts with the Gaia parallaxes of the bursters. The simplistic model stands the preliminary test from the Gaia Early Data Release 3, but may be falsified in the future as Gaia parallax precision improves.

The rich scientific outputs achieved in this thesis will motivate future astrometric survey of magnetars, MSPs and type I X-ray bursters. Particularly, VLBI astrometry of new MSPs can play an important part in enhancing the sensitivity of pulsar timing array that will likely directly detect the GW background in few years.
Since these new MSPs will typically be fainter and more distant, the advanced calibration techniques detailed in this thesis will be crucial for studying these new sources.

%\textit{ quote --- }

\hfill
\addcontentsline{toc}{chapter}{Acknowledgements}
\Huge \noindent Acknowledgements
\normalsize
\\

This thesis cannot be possible without the continuous mentoring and supports from my star supervisors --- Adam Deller, Ryan Shannon and Matthew Bailes, who have shown me the integrity, academic excellence and communication skills that a great astronomer should possess. 
My biggest and deepest thank belongs to my principal supervisor Adam, who brought me to the \mspsrpi\ project he had been leading, and has been prioritizing the mentoring over other work. 
It is probably much easier to measure the parallax of a neutron star at the remote edge of the Galaxy than my gratitude to Adam, to whom I must have thanked hundreds of times throughout the PhD program, for the cheer-ups at my lows, for the countless productive discussions, for the tons of useful advice, etc.
%I hope one day I can become a supervisor as good as you

During the PhD program, it is a privilege to work with and get to know so many friendly, encouraging and helpful colleagues (in addition to my supervisors).
The papers presented in this thesis involve the efforts from all the collaborators (see the {\it Declaration}), from whom I have really learnt a lot.
The progress review panel, which consists of Virginia Kilborn, Ivo Labbe, Alister Graham and Simon Stevenson, has reviewed this PhD program at four different stages, and provided helpful advice.
Karl Glazebrook and Michael Murphy have offered valuable supports as the CAS guide and the PhD student coordinator, respectively.
I appreciate the great administrative supports from Lisa and Erin, and the enlightening discussions with Chris Flynn, Marcus, Xingjiang, Harry, Poojan, Hannah, Rahul, Debatri, Ayushi, Cherie and Jacob.

The PhD journey would not be enjoyable without the wonderful persons I have encountered in Australia. Nick and Usman, I feel blessed to have you as my house mates and friends, especially during the difficult time of Covid lockdowns. I have really enjoyed our chats of broad topics, the countless foosball matches, the Xbox FIFA rivalries, the movie watching, the food sharing, and the state-wide and inter-state road trips. 
Mohsen, you are my longest office mate. The trip we made together in Tasmania adds to my best memories in Australia! Vivek, I will remember our fun bike rides to different venues around Melbourne, as well as the casual australian pool games. 
Jielei, you always have an upbeat vibe, and it is great joy to chat with you. 
Pravir and Wahl, I enjoyed the chats we had and the road trips we shared. The stargazing at Wilsons Promontory is truly magnificent! 
To the badminton/basketball friends, including Jacob, Morrison, Paul, William, thanks for the relaxing games in the lovely days!
%All of the above have played a part in helping me survive the stress and lows during the PhD program.

My final rounds of thanks go to my biggest supporters --- my family. Wei and Weilan, thanks for understanding and fully supporting my pursuit of PhD research overseas.
Life in your vicinity is more meaningful and hence more satisfying. I will try to make up for my absence in the past few years. 
Huge thanks to my parents-in-law, Mum and Dad for the generous help while I was away.

This PhD program was financially supported by the ACAMAR scholarship (November 2018 -- November 2021) and the OzGrav scholarship (December 2021 -- June 2022).
\hfill
\addcontentsline{toc}{chapter}{Declaration}
\Huge \noindent Declaration
\normalsize
\\

\noindent
The work presented in this thesis has been carried out in the Centre for
Astrophysics \& Supercomputing at Swinburne University of Technology between
November 2018 and July 2022. This thesis contains no material that has been accepted for the
award of any other degree or diploma. To the best of my knowledge, this thesis
contains no material previously published or written by another author, except
where due reference is made in the text of the thesis. The content of the
chapters listed below has appeared in refereed journals. Minor alterations have
been made to the published papers in order to maintain argument continuity and
consistency of spelling and style.
The major contribution of each author for a given paper is also mentioned below. The author contribution percentages, evaluated by nominal working hours needed to fulfill one's contribution, are summarized in the authorship indication forms that are attached to the thesis appendix (as requested).

\begin{itemize}

\item Chapter 3 has been published as Ding H., et~al., 2020, MNRAS, 498, 3736 with the title ``A magnetar parallax''. The paper is authored by Hao Ding (HD), Adam T. Deller (ATD), Marcus E. Lower (MEL), Chris Flynn (CF), Shami Chatterjee (SC), Walter Brisken (WB), Natasha Hurley-Walker (NH), Fernando Camilo (FC), John Sarkissian (JS) and Vivek Gupta (VG). The publication is based on 14 new Very Long Baseline Array (VLBA) observations, and 2 archival ones. The initial 11 new VLBA observations and the 3 extended ones were led by ATD and HD, respectively. MEL, CF, SC and WB also contributed to the VLBA proposal drafting. The pulsar gating of the VLBA data makes use of the pulse ephemerides produced by MEL based on timing observations at the Parkes and the Molonglo telescopes. The Parkes-based observations were led by FC and supported by JS, while the Molonglo-based observations were assisted by CF and VG. HD carried out VLBA data reduction (which includes the implementation of 1D interpolation) and the follow-up analysis.
All authors contributed to the paper preparation. Particularly, HD finished the first draft, and contributed to about 95\% of the final draft. During the draft iterations, ATD provided most suggestions and comments.
NH wrote the third paragraph for \ref{subsec:J1810_snr_association} that discusses the postulated magnetar-SNR association. 

\item Chapter 4 has been accepted for publication in {\it Proceedings of the International Astronomical Union} as ``Probing magnetar formation channels with
high-precision astrometry: The progress of
VLBA astrometry of the fastest-spinning
magnetar Swift~J1818.0$-$1607''. The conference paper, authored by HD, ATD, MEL and Ryan Shannon (RS), is the written submission for the IAU Symposium 363 --- ``Neutron Star Astrophysics at the Crossroads: Magnetars and the Multimessenger Revolution''. The astrometry of \swift\ is based on the VLBA observations led predominantly by HD. All the 4 authors contributed to the VLBA proposal drafting. MEL provided the pulse ephemerides that were used for pulsar gating of the VLBA data. HD carried out VLBA data reduction and the follow-up analysis. All authors contributed to the conference paper preparation. 
In particular, HD completed the first draft, and contributed to about 95\% of the accepted draft.

\item Chapter 5 has been published as Ding H., et al., 2021, PASA, 38, e048. The paper, entitled ``Gaia EDR3 parallaxes of
type I X-ray bursters and their implications on the models of type I X-ray bursts: A
generic approach to the Gaia parallax zero point and its uncertainty'', is authored by HD, ATD and James C. A. Miller-Jones (JCAM). 
HD made data analysis, and wrote the first paper draft. ATD and JCAM offered equal amount of valuable comments during the draft iterations.

\item Chapter 6 involves the publication Ding H., et al., 2020, ApJ, 896, 85 as updated by the erratum Ding et al. 2020, ApJ, 900, 89. The paper, entitled ``Very Long Baseline Astrometry of \psrea\ and its Implications on
Alternative Theories of Gravity'', is authored by HD, ATD, Paulo C. C. Freire (PCCF), David L. Kaplan (DLK), T. Joseph W. Lazio (TJWL), RS and Benjamin W. Stappers (BWS).
The publication is based on the VLBA data of the \mspsrpi\ project led by ATD. All authors except for HD contributed to the observation proposal drafting. The VLBA observations were led predominantly by ATD. BWS provided the pulse ephemerides for the pulsar gating of the VLBA data.
HD reduced the VLBA data. HD carried out the follow-up data analysis. 
All authors contributed to the paper preparation. In particular, HD finished the first paper draft, and addressed the comments from other authors during the draft iterations. 
ATD offered most comments on the paper, followed by PCCF. DLK added \ref{subsec:J1012_Gal_path} to the manuscript. 

\item Chapter 7 has been published as Ding H., et~al., ApJ, 921, L19 with the title ``The Orbital-decay Test of General Relativity to the 2\% Level with 6 yr VLBA Astrometry of the Double Neutron
Star PSR J1537+1155''. The authors of the paper are HD, ATD, Emmanuel Fonseca (EF), Ingrid H. Stairs (IHS), BWS and Andrew Lyne (AL). The paper is mainly based on the VLBA observations proposed by ATD ($\sim$60\%) and HD ($\sim$40\%). All authors contributed to observation proposal drafting.
EF, BWS and AL provided the pulse ephemerides for the pulsar gating of the VLBA data. 
All authors contributed to the paper preparation. In particular, HD wrote the first draft, and addressed comments from other authors during draft iterations. ATD offered most comments on the manuscript. EF made \ref{fig:B1534_mass_mass_diagram} based on the new intrinsic orbital decay estimate.

\item Chapter 8 has been submitted to {\it Monthly Notices of Royal Astronomical Society} as ``The \mspsrpi\ catalogue: VLBA
astrometry of 18 millisecond pulsars''. The paper is authored by HD, ATD, BWS, TJWL, DLK, SC, WB, James Cordes (JC), PCCF, EF, IHS, Lucas Guillemot (LG), AL, Isma\"el Cognard (IC), Daniel J. Reardon (DJR) and Gilles Theureau (GT).
The paper is mainly based on the VLBA data of the \mspsrpi\ project proposed by ATD. 
All authors except for HD, DJR and GT contributed to the main VLBA proposal for the \mspsrpi\ project.
The VLBA scheduling was predominantly led by ATD. 
To enable the pulsar gating of the VLBA data, BWS, EF, LG, AL provided pulse ephemerides based on different pulsar timing facilities. Pulsar timing observations at the Nan\c{c}ay Radio Observatory were assisted by IC and GT.
HD reduced the VLBA data, years after the epoch-to-epoch preliminary data examination by ATD. 
HD made the analysis based on the results of VLBA data reduction. 
All authors contributed to the paper preparation. In particular, HD wrote the first paper draft, and addressed the comments from other authors during the draft iterations.
ATD offered most comments on the manuscript, and made small additions and/or re-wordings of the draft text (phrases to sentences long) during the draft iterations. Accordingly, ATD contributed to $\sim$10\% of the paper preparation. 
TJWL analyzed the discrepancy between the parallax-based distance and the dispersion-measure-based distance of \psrc, and wrote \ref{subsubsec:mspsrpi_J0610_DM}.
DJR estimated the temporal scattering of \psrgb\ and made 
\ref{fig:mspsrpi_tau_sc}.

\end{itemize}

%My contribution to these papers was ..., accounting for X percent of the final
%manuscripts. My co-authors contributed by ..., accounting for Y percent of the final manuscripts.

\vspace{1.5cm}
\begin{flushright}
\textit{ Hao Ding}

\textit{14 October 2022}
\end{flushright}

\multilinecomment{
\vspace{1.0cm}
\begin{flushright}
\textit{ Co-author 1 }

\textit{year--}
\end{flushright}

\vspace{1.0cm}
\begin{flushright}
\textit{ Co-author 2 }

\textit{year--}
\end{flushright}
}
\hfill
%\addcontentsline{toc}{chapter}{Thesis structure and norms}
%\include{prefix}
%\hfill
% Including a dedication is entirely optional
\newpage
\vspace*{3.5in}
\begin{center}\parbox{11cm}{\begin{center}
\textit{Dedicated to my daughter Weilan}
\end{center} } \end{center}

%\addcontentsline{toc}{chapter}{Acronyms, Abbreviations and Conventions}
%\include{acronyms/acronyms}
%\hfill

\tableofcontents
\addcontentsline{toc}{chapter}{List of Figures}
\listoffigures
\addcontentsline{toc}{chapter}{List of Tables}
\listoftables

\mainmatter

\chapter[Introduction]{Introduction}
\label{ch:Introduction}

% This can be used to include a quotation if desired.
%\begin{center}\parbox{11cm}{\begin{center}
%\textit{To be, or not to be, that is the question.}
%
%---\href{http://en.wikipedia.org/wiki/Hamlet}{Hamlet}
%\end{center} } \end{center}

%For example \citep{YorkEtal2000}.

%\section{Purpose of the Thesis}

%\lipsum[1-9]

%\section*{Chapter Outline}
%\label{sec:outline_intro}

This chapter introduces the broad context (of this thesis) that interlinks the paper chapters, and sets up the stages for the upcoming chapters.
\ref{sec:NSs} and \ref{sec:astrometry101} explain the basics and the scope of this thesis. \ref{sec:NSs} concisely introduces neutron stars from the perspectives of their discovery, formation, composition, observational properties and power sources (of electromagnetic radiation). \ref{sec:astrometry101} summarizes the approaches to achieve high-resolution astrometry, and points out the two pathways taken in this thesis; additionally, the \mspsrpi\ project is highlighted following a review of previous large pulsar astrometry programs.
Sec.~1.3 -- Sec.~1.6 unfold the main scientific themes around high-resolution astrometry of Galactic neutron stars.
\ref{sec:thesis_norms} draws an end to this chapter with the structure and norms followed by this thesis.

\section{A brief introduction to neutron stars}
\label{sec:NSs}

\subsection{The discovery}
\label{subsec:NS_discovery}
 
Neutron stars (NSs) were first proposed by \citet{Baade34} to exist as the remaining cores of supernovae. 
In the following 3 decades, almost no attempt was made to find these hypothetical compact stars, as they were believed too small to be observed. The advancement of radio and space technology achieved during the second world war and the cold war opened up new windows of electromagnetic observations, including radio, X-ray and $\gamma$-ray. 
These new observational windows enabled the discovery of NSs.

In hindsight, the first NS was observed in the form of an X-ray binary (XRB) later known as Sco~X$-$1 \citep{Giacconi62}, though at that time the link between Sco~X$-$1 and XRB was not clear, let alone the nature of the XRB. 
The first proposal that solitary NSs can be directly observed was made by \citet{Chiu64}: it is suggested the thermal radiation from NSs are observable at $10^6$\,K surface temperature.
\citet{Pacini67} put forward that a rapidly rotating neutron star inside the Crab Nebula has energized the environment and accelerated the expansion of the nebula.
Not long after this accurate prediction, NSs made their debut, though at an unexpected wavelength --- radio.

In 1967, a train of radio pulses with 1.3\,s periodicity was distinguished by Jocelyn Bell from radio frequency interference (RFI), which led to the first discovery of a pulsar (i.e., PSR~B1919$+$21) \citep{Hewish69}. Following this discovery, more pulsars were quickly identified \citep[e.g.][]{Lyne68}, including the famous 89\,ms Vela Pulsar \citep{Large68} and the 33\,ms Crab Pulsar \citep{Staelin68}. 
The short periods of pulsars must come from a relatively small emission region in either a rotating or an oscillating style. The steady changes in the pulse periods of pulsars \citep[e.g.][]{Richards69} favoured the rotation scenario over the oscillation one \citep{Gold68,Pacini68}. 
For a fast-spinning degenerate star (i.e., NS or a white dwarf), its self-gravity must be large enough to bind the star together against the centrifugal force. Namely,
\begin{equation}
\frac{GM}{R^2} > \left( \frac{2\pi}{P}\right)^2 R \,,
\end{equation}
where $M$, $R$, $P$ and $G$ stand for NS mass, radius, spin period and the Newton's gravitational constant, respectively. Given the average density
\begin{equation}
\rho = \frac{M}{\frac{4}{3}\pi R^3} \,,
\end{equation}
\begin{equation}
\label{eq:rho_P_relation}
\rho > \frac{3\pi}{G P^2} = 1.412\times10^{12} \left( \frac{P}{10\,\mathrm{ms}}\right)^{-2}\,\mathrm{g~cm^{-3}} \,.
\end{equation}
The discovery of the Crab pulsar renders $\rho>1.3\times10^{11}\,\mathrm{g~cm^{-3}}$, which is 4 orders higher than the typical white dwarf (WD) density \citep{Chandrasekhar31}. Hence, WDs were ruled out as the pulsar counterparts, and the connection between NSs and pulsars had become solid by 1970. So far, the fastest-spinning pulsar has a spin period of only 1.4\,ms \citep{Hessels06}, which corresponds to $\rho>7\times10^{13}\,\mathrm{g~cm^{-3}}$ according to \ref{eq:rho_P_relation}, comparable to the nuclear density.

\subsection{The mainstream formation scenario}
\label{subsec:NS_formation}
It is widely believed that NSs are the end products of massive ($>8$\,\msun) stars \citep[e.g.][]{Lattimer04}.
When such a massive star runs out of hydrogen and helium fuel towards the end of its life, it starts to fuse heavier elements \citep{Hoyle46} and segregates into different layers in the inner region. The deeper inside the star, the heavier elements are formed at higher temperatures \citep{Woosley05}.
When an iron core is eventually formed at the centre, the fusion of stable iron nuclei into even heavier ones would consume more energy than what the fusion releases, causing a deficit of light pressure to counter-balance the gravity of the star. As a result, the core collapse begins. The release of massive neutrinos effectively \citep{Hirata88,Bionta87} disperses huge gravitational energy and enables a quick collapse.
When the infalling materials reach the nuclear density of about $2.3\times10^{14}\,\mathrm{g~cm^{-3}}$, the contraction is halted abruptly by neutron degeneracy pressure, leading to a shock wave that triggers the violent type II supernova \citep{Janka96}.

After a type II supernova, the majority of the progenitor star is thrown into space, which would be seen as a diffuse supernova remnant (SNR). At the centre of this SNR, the compressed core forms a NS. As the iron core contracts from $\sim10^6$\,km radius to only $\sim10$\,km, the magnetic field strength is intensified by $\sim10^{10}$ times due to conservation of magnetic fluxes; likewise, the spin frequency would also be increased by a factor of $\sim10^{10}$ owing to the conversion of angular momentum \citep{Condon16}. Should the progenitor star be more massive than 30\,\msun, the immense gravity would overcome the neutron degeneracy pressure, compressing the star into a black hole (BH). Apart from the main formation scenario described above, NSs may be born in other channels, which will be discussed in \ref{sec:NS_formation} and \ref{ch:J1818}.

\subsection{The constraints on the neutron star composition}
\label{subsec:NS_composition}

%Their immense self-gravity is counter-balanced with degeneracy pressure (and centrifugal force for fast-spinning NSs).
NSs are the densest state of matter apart from BHs, with typically 1--2\,\msun\ mass \citep{Shao20} and only $\sim10$\,km radius \citep[e.g.][]{Miller19,Capano20}. For a typical 1.4\,\msun\ NS, its Schwarzschild radius is about 4\,km, which is only a factor of 2.5 smaller than the nominal NS radius (i.e., $\sim10$\,km). Namely, NSs are close to collapsing into singularities under their own gravity. 
%baryon-made celestial bodies, 
The interior of NSs is not homogeneous. A sudden increase of the NS spin frequency, aka. pulsar glitch, followed by slow recovery to the pre-glitch level, is a common phenomenon that has been observed from more than 200 pulsars (see the Jodrell bank glitch catalogue\footnote{\url{https://www.jb.man.ac.uk/pulsar/glitches/gTable.html}}, \citealp{Espinoza11,Basu22}). The long relaxation after a pulsar glitch hints the existence of the neutron superfluid, while the pulsar glitches are interpreted as the unpinning and re-pinning of the superfluid to the NS crust, which is known as the two-component model \citep{Baym69,Anderson75}. 
5 decades after the first pulsar glitch is recorded \citep{Radhakrishnan69}, the interior structure of NSs remains poorly understood \citep[e.g.][]{Andersson12}. Pulsar glitches continue to serve as an essential probe of the superfluidity inside NSs \citep[e.g.][]{Link92,Gugercinoglu20}, alongside other proposals aiming at the same goal (e.g., using NS precession, \citealp{Link03}).

Ultimately, the composition of NSs depends on the relation between pressure and density inside NSs \citep{Oppenheimer39}. This relation is known as the equation of state (EoS), which remains highly uncertain due to the inadequacy of both experimental techniques and nuclear-theory tools. 
Given a NS EoS, the NS mass, radius and moment of inertia $I$ can be derived. Conversely, the determination of at least two of the mass, radius and $I$ can in principle constrain the EoS of NSs \citep[e.g.][]{Lattimer01}.
%characterized by the NS equation of state (EoS)
While NS masses can be precisely determined for some binary pulsars using pulsar timing or spectroscopic observations of optically bright companions (see \ref{subsec:pulsar_timing}), estimating NS radius or $I$ in conjunction with NS mass is difficult, even for the most precisely timed pulsar systems \citep{Kramer21a}. This difficulty starts to be overcome recently for few nearby NSs using the Neutron Star Interior Composition ExploreR (NICER, \citealp{Gendreau12}) --- an X-ray timing and spectroscopic device aboard the International Space Station. 
NICER observations can be used to model the rotation of the hot spots on the NS surface, which is complicated by the gravitational redshift and the light bending caused by the immense gravity of the NS. Nevertheless, the two post-Newtonian gravitational effects enable the determination of the NS mass and radius, as the two  effects would change with the NS mass and radius \citep{Miller19,Bogdanov19a}.
In parallel to the NICER breakthrough, gravitational-wave observations of NS-NS mergers and NS-BH mergers can reveal both the NS mass and its tidal deformability (i.e., the degree that the NS is tidally deformed), which can effectively truncate the phase space of the NS EoS \citep{Read09,Annala18,Tan20}. In light of these two recent developments, the observational constrains on the NS EoS will be substantially tightened in few years; a golden era of the NS EoS study may be right on the horizon.

%The largest mass that a non-rotating NS can undertake is about 1.5--3\,\msun \citep{Kalogera96}, which depends on the their unclear equation of state (EoS) \citep{Oppenheimer39}. The composition of NS cores, formulated by the equation of state, remains highly uncertain, which hinges on the uncertain distributions of NS mass and radius. NSs are the remnants of the violent supernova explosions \citep{Pacini68}, as supergiant stars between 10 and 25\,\msun\ reach their end of life.

\subsection{Observational manifestations}
\label{subsec:NS_observations}
%NSs have been observed in different ways. 
The observations of solitary neutron stars have covered an unparalleled wide range of the electromagnetic spectrum, from low radio  ($\sim10$\,MHz, \citealp{Bridle70}) to high $\gamma$-ray frequencies ($\gtrsim100$\,GeV, \citealp{VERITAS-collaboration11}). %The first discovered neutron star, PSR~B1919$+$21 \citep{Hewish69}, manifests itself as a stellar object periodically pulsating at radio frequencies. 
%Such pulsating stellar objects are later known as pulsars. After the discovery of the first pulsar, it soon became clear that pulsars are neutron stars that constantly release electromagnetic radiation from the emitting region in the magnetosphere \citep[e.g.][]{Pacini68}. 
As the first manifestation, pulsars can sometimes also be identified at optical \citep[e.g.][]{Davidsen72,Wallace77,Ambrosino17,Vigeland18}, X-ray \citep[e.g.][]{Fritz69,Fahlman81} and $\gamma$-ray \citep[e.g.][]{Abdo09,Abdo13} wavelengths. The spin period of active pulsars ranges from 1.4\,ms \citep{Hessels06} to 18.18\,min \citep{Hurley-Walker22}.
Despite the beaming effect and limited observing sensitivity, $\sim3000$ pulsars have been detected so far (see the ATNF Pulsar Catalogue\footnote{\label{footnote:psrcat}\url{https://www.atnf.csiro.au/research/pulsar/psrcat}}, \citealp{Manchester05}).
Thanks to the large moment of inertia ($\sim10^{45}\,\mathrm{g~cm^2}$) of a neutron star, the pulses from a pulsar normally display high temporal stability, which makes it possible to carry out the so-called pulsar timing technique based on observed pulse time-of-arrivals (ToAs). 
Pulsar timing compares the observed ToAs against model predictions, and obtains estimates for a large number of model parameters (e.g., spin period, spin phase etc.) that minimize the residual ToAs (i.e., the observed ToAs subtracted by the model prediction) \citep[e.g.][]{Detweiler79,Helfand80}. 
The list of model parameters should cover all physical effects that lead to observable changes in ToAs. Such model parameters would include basic astrometric parameters (reference position, proper motion and parallax), orbital parameters (for pulsars in binary systems) and post-Newtonian ones.

Many Galactic NSs are discovered in orbit with a companion, which can be either a main-sequence star or another degenerate star (i.e., white dwarf or NS). For a neutron star having a main-sequence-star companion, the outer layer of the companion could be accreted onto the surface of the NS.
%, if the companion is close enough to the NS. 
In this process, X-ray and $\gamma$-ray radiations are emitted when the accretion is congested, or when fast nuclear burning takes place on the NS surface. The latter case is also known as type I X-ray bursts, where the light pressure induced by fast nuclear burning can sometimes exceeds the gravity on the nuclear ``fuel'' (accumulated on the NS surface), causing the so-called photospheric expansion (PRE) bursts (see \ref{subsec:PRE_bursts} and \ref{sec:PRE_intro} for more explanations). 
Given the high-energy emissions, the binary systems in which a NS accretes from its ``donor'' star are called NS X-ray binaries (XRBs). Though BH XRBs bear observational similarities with NS XRBs, BHs do not host fast nuclear burning on the surface of their event horizons. Thus, fast nuclear burning serves as one way to distinguish NS XRBs from BH XRBs. To date, hundreds of NSs are found in XRB systems \citep{Liu06,Liu07}, which include some of the confirmed transitional pulsars \citep{Deller12,Ambrosino17}. 

NSs accrete from their donor stars in different ways, which depend on the donor mass. Low-mass ($\lesssim1.5$\,\msun) donors would gradually evolve and fill its Roche Lobe \citep{Savonije78}, due to the tidal force from the NS. Consequently, materials are accreted from these donors via an accretion disk. In contrast, high-mass ($\gtrsim10$\,\msun) main-sequence companion would blow out their outer materials in the form of stellar winds. Eventually, some of the blown-away materials are captured by the NS.
While the NS in an XRB accretes from its donor star via an accretion disk, part of the orbital angular momentum would be converted to the spin angular momentum of the NS, causing the NS to spin up. At the end of this accretion process, a fully spun-up NS may likely reach a spin period as short as $\sim1.4$\,ms \citep{Hessels06}. 
If the electromagnetic beam of such a spun-up NS happens to sweep across the Earth-to-NS sightline, we would see a fast-rotating pulsar, known as a millisecond pulsar (MSP) or recycled pulsars.
Most NSs studied in this thesis are MSPs. To be explicit, MSPs in this thesis refer to spun-up pulsars with $\lesssim40$\,ms spin periods and $\lesssim10^{10}$\,G surface magnetic strength. This criterion would capture most pulsars that have undergone any recycling process.

The standard formation scenario (of MSPs) described above \citep{Alpar82} is supported by the discovery of transitional pulsars (e.g. PSR~J1023$+$0038, \citealp{Archibald09}) in the X-ray binary systems. 
However, a number of MSPs are found to be solitary (i.e., not in orbit with a companion), such as PSR~J0030$+$0451, PSR~J1730$-$2304 and PSR~B1937$-$21. This absence of companion is likely caused by the disintegration of the companion by the NS, or the disturbance by a third object.
The discovery of the black widow pulsars burning away their companions (e.g. PSR~J1959$+$2048, \citealp{Fruchter88}) reinforces the former explanation, while lone pulsars found in globular clusters are probably formed in the latter pathway (e.g., PSR~J1824$-$2452A, \citealp{Lyne87a}). 
%MSPs are also known as recycled or spun-up pulsars, which used to be low-mass X-ray binaries (LMXBs) accreting effectively from a donor star.
As an interesting exception, PSR~J1024$-$0719 in an extremely wide ($\gtrsim2$\,kyr) orbit is found in a sparse stellar neighbourhood, which is also postulated to have been ejected from a dense region \citealp{Bassa16,Kaplan16}).
Last but not least, despite the success of the standard MSP formation scenario that can explain the properties of almost all MSPs found in our vicinity, a small proportion of MSPs are hypothesized to follow different formation pathways \citep[e.g.][]{Michel87,Bailyn90,Freire13}. This argument will be elaborated in light of the neutron star retention problems in \ref{subsec:NS_kinematics}.

NSs possess the strongest magnetic strength in the universe. At the most magnetized end of the NS family are magnetars. Without exception, magnetars display energetic emissions at X-rays and $\gamma$-rays. Their link to fast radio bursts (FRBs) has been strengthened with the detection of FRB-like bursts from the Galactic magnetar SGR~J1935$+$2154 \citep{Andersen20,Bochenek20}. So far, about 30 magnetars have been discovered (see the McGill Online Magnetar Catalogue\footnote{\label{footnote:magnetar_catalogue}\url{http://www.physics.mcgill.ca/~pulsar/magnetar/main.html}}, \citealp{Olausen14}), but only 6 of them are seen at radio frequencies. Among the 6 radio magnetars, SGR~J1935$+$2154 is a transient radio burster, from which only sparse and random-like radio bursts have been recorded; the other five are constant radio magnetars behaving similar to pulsars.
Though radio magnetars can be considered a special kind of radio pulsars, their observational characteristics are alien to other radio pulsars.
Compared to normal radio pulsars that display steep radio spectra, radio magnetars normally show flat radio spectra, which allows them to be observed at higher radio frequencies.
The formation of magnetars is still poorly understood, and requires more observational constraints \citep[e.g.][]{Ding23a}. No companion star has been found for any magnetar, which implies either a magnetar was formed in a violent binary merger, or the companion was disrupted at the birth of the magnetar.

Due to observing sensitivity limitations, most neutron stars are found in the Milky Way; a few are discovered inside the Large/Small Magellanic Clouds \citep{Kaspi94a,Majid04,Ridley13,Titus20}. However, at $\gtrsim1$\,Mpc distances, violent NS-NS mergers have been identified with gravitational-wave (GW) detectors (see \ref{subsec:PTAs} for more description of GW events),  accompanied by electromagnetic flares \citep[e.g.][]{Abbott17,Abbott17a,Goldstein17,Mooley18,Pozanenko19}. In addition, there are strong indications that the observed short $\gamma$-ray bursts (SGRBs) at further distances are also originated from DNS mergers \citep[e.g.][]{Coward12}, which is strongly supported by the GW170817 event \citep{Goldstein17}. 
%To wrap up, NSs can be observed as pulsars, X-ray binaries, gravitational-wave events and SGRBs. 
Finally, the birth of a NS through core collapse is associated with neutrino events \citep{Woosley05} that have been observed with neutrino observatories \citep{Hirata88,Bionta87}.
Among the five aforementioned observational manifestations of NSs (i.e., pulsars, X-ray binaries, gravitational-wave events, SGRBs and neutrino events), pulsars and NS XRBs in the Galaxy will be the main focus of this thesis. Within the vast pulsar population, this thesis only covers two distinct (and probably the most interesting) pulsar divisions --- millisecond pulsars and magnetars. 
%More introduction to MSPs can be found in the introductory sections of \ref{ch:J1012}, \ref{ch:B1534} and \ref{ch:mspsrpi}.

\subsection{Two sources of neutron star electromagnetic emission}
\label{subsec:NS_emission_sources}

%The observations of solitary neutron stars have covered an unparalleled wide range of the electromagnetic spectrum, from low radio  ($\sim10$\,MHz, \citealp{Bridle70}) to high $\gamma$-ray frequencies ($\gtrsim100$\,GeV, \citealp{VERITAS-collaboration11}). 
The radiation mechanism of NSs is not yet fully understood. As a general picture, 
%Due to the enormous electric field strength generated from the NS interior, electron-positron pairs arising from the vacuum outside the NS depart each other and migrate along the electric field lines. This migration subsequently develops into a plasma cloud surrounding the NS, which is known as the magnetosphere. For non-rotating NSs, 
the electromagnetic radiation (EMR) at different wavelengths is mostly generated from distinct regions of the NS magnetosphere (i.e., the NS surrounding filled with plasma). As NSs rotate, electrons and positrons in the magnetosphere are accelerated to relativistic velocities, then give away their momentum to photons. 
The consensus is that optical to $\gamma$-ray photons are generated by curvature and synchrotron radiation from an outer region of the magnetosphere (aka. the outer gap), while radio photons are emitted from an inner region of magnetosphere over the magnetic pole (aka. the polar gap) \citep{Lyne12}.  
In addition to the magnetosphere-born EMR, thermal radiation from the NS surface can be observed at X-rays \citep{Chiu64,Bogdanov19,Riley19}.

The ultimate sources of NS electromagnetic emissions include the rotational kinematic energy and the electromagnetic energy reservoir stored in the magnetosphere. 
%The latter source is the main contributor of the EMR from magnetars \citep{Duncan92,Woods06}. 
Provided the moment of inertia $I$ of a NS, the NS rotational kinematic energy is
\begin{equation}
\label{eq:E_rot}
    E_\mathrm{rot} = \frac{1}{2} I \left(\frac{2\pi}{P}\right)^2 \,.
\end{equation}
 Hence, the upper limit of the EMR luminosity equals to
\begin{equation}
\label{eq:EMR_budget}
    L_\mathrm{EMR} = -\dot{E}_\mathrm{rot} - \dot{E}_\mathrm{mag} = 4 \pi^2 I \frac{\dot{P}}{P^3} - \dot{E}_\mathrm{mag} \,,
\end{equation}
where $\dot{E}_\mathrm{mag}$ represents the time derivative of the stored electromagnetic energy, and $-\dot{E}_\mathrm{rot}$ is known as the spin-down power.
Recycled MSPs are all rotation-powered. Due to their fast spins, the spin-down power of MSPs are generally larger than other rotation-powered pulsars, and their magnetosphere inside the light cylinder is denser. As a result, MSPs amount to about half of the $\gamma$-ray pulsar population. 
Standing in contrast to the rotation-powered pulsars are magnetars. These exotic pulsars release
EMR much higher than their $-\dot{E}_\mathrm{rot}$ \citep{Duncan92,Woods06}. Namely, $\dot{E}_\mathrm{rot} \ll \dot{E}_\mathrm{mag}$ for magnetars.

In the $\dot{E}_\mathrm{rot} \gg \dot{E}_\mathrm{mag}$ regime, the ensemble of EMR emitted from different regions of the magnetosphere can be approximated with a rotating magnetic dipole calculated from an evenly magnetized sphere, which renders
\begin{equation}
\label{eq:dipole_radiation}
L_\mathrm{EMR} \approx -\dot{E}_\mathrm{rot} = \frac{2}{3c^3} \left(BR^3 \sin{\alpha}\right)^2 \left( \frac{2\pi}{P}\right)^4 
\end{equation}
(Eq.~6.12 of \citealp{Condon16}), 
where $R$ and $B$ denote the sphere radius and surface magnetic field strength, respectively; $\alpha$ is the included angle between the magnetic dipole moment and the rotation axis. Incorporating \ref{eq:dipole_radiation} with \ref{eq:EMR_budget}, we reach 
\begin{equation}
    \label{eq:dipole_EMR}
    \frac{2}{3c^3} \left(BR^3 \sin{\alpha}\right)^2 \left( \frac{2\pi}{P}\right)^4 = 4 \pi^2 I \frac{\dot{P}}{P^3} \,.
\end{equation}
Solving \ref{eq:dipole_EMR} gives the surface field strength 
%and the surface magnetic field strength
\begin{equation}
\label{eq:B_surf}
   B > 3.2 \times 10^{19} \left(\frac{R}{10\,\mathrm{km}} \right)^{-3} \left( \frac{I}{10^{45}\,\mathrm{g~cm^2}}\right)^{1/2} \left(\frac{P \dot{P}}{1\,\text{s}} \right)^{1/2}\,\text{G} \,;
\end{equation}
and the NS age can be approximated with the characteristic age
%The diverse NS zoo can be categorized with two basic observables --- the spin period $P$ (of a NS) and its derivative $\dot{P}$. Provided $P$ and $\dot{P}$, one can calculate both the characteristic age 
\begin{equation}
\label{eq:tau_c}
    \tau_\mathrm{c} \equiv \frac{P}{2 \dot{P}} = 15.8 \left( \frac{P}{1\,\text{s}}\right) \left( \frac{\dot{P}}{10^{-15}}\right)^{-1}\,\text{Myr} \,.
\end{equation}
Both indicative $B$ and $\tau_c$ are formulated as functions of $P$ and $\dot{P}$. Hence, in a $P-\dot{P}$ diagram of NSs (e.g. Fig.~1 of \citealp{Caleb22}), one can not only distinguish MSPs from other pulsars, but also separate highly magnetic magnetars from other NSs. It can be seen from a typical $P-\dot{P}$ diagram (e.g. Fig.~6.3 of \citealp{Condon16}) that binary pulsars show larger variety of spin periods, which agrees with the scenario where NSs in binaries are spun up during the accretion from the companions. 

\multilinecomment{
\begin{figure*}
    \centering
	\includegraphics[width=13cm]{Figures/P_Pdot_diagram.pdf}
    \caption{The $P-\dot{P}$ diagram (spin period versus its derivative) taken from Page~219 of \citet{Condon16}. At the lower-left corner of the diagram sit millisecond pulsars, while the opposite corner of the diagram is occupied by the slow-rotating magnetars.
    }    
    \label{fig:P_Pdot}
\end{figure*}
}

%On the other hand, in the $\dot{E}_\mathrm{rot} \ll \dot{E}_\mathrm{mag}$ regime, approximating $L_\mathrm{EMR}$ with a rotating magnetic dipole may be no longer valid. Otherwise, $B$ of magnetars would be much larger than the estimate derived from \ref{eq:B_surf}.

%For rotation-powered pulsars, strong correlation has been found between $L_\mathrm{sd}$ and the X-ray luminosity.
%The assumption that pulsars are rotation-powered is generally true, as correlation between X-ray flux density and $L_\mathrm{sd}$ has been found.

%high-energy energy in the form of Soft $\gamma$-ray repeaters (SGRs) and anomalous X-ray pulsars (AXPs) (see the McGill Online Magnetar Catalogue\footnote{\url{http://www.physics.mcgill.ca/~pulsar/magnetar/main.html}}).  As the observed electromagnetic radiation from magnetars is larger than their $L_\mathrm{sd}$, .

%As we will mention in \ref{section:pulsar_astrometry}, Among all the aforementioned manifestations of neutron stars, pulsars in the binary (or multiple-stellar) system provide the best test of gravitational theories, thanks to the realization and development of the pulsar timing technique.

\section{Astrometry of Galactic neutron stars}
\label{sec:astrometry101}

\subsection{Overview of high-spatial-resolution astrometry}
\label{subsec:overview_astrometry}

Broadly speaking, astrometry is a sub-field of astronomy where astrometric parameters are determined with various methods (which include the pulsar timing technique). 
%The astrometric parameters determined in this way ar
The astrometric parameters normally include (but are not limited to) reference position, proper motion, parallax, where parallax refers to the magnitude of the annual position oscillation (of the stellar object) due to the changing Earth-to-Sun vector. 
Based purely on geometry, astrometric parameters can be derived by modelling a time series of high-spatial-resolution sky locations (of a target stellar object). This theory-independent (or model-independent) approach is referred to as high-spatial-resolution (or simply high-resolution) astrometry.
%Nevertheless, this thesis focuses on the astrometry purely based on geometry. The resultant astrometric parameters determined with geometry-based astrometry are considered theory-independent or model-independent.
%An astrometric model is purely based on geometry. Hence, astrometric parameter
In this thesis, the spatial resolution at the 10\,mas level is considered high spatial (or angular) resolution, which corresponds to $\sim0.5$\,mas positional precision for a compact source detected at 10\,$\sigma$ significance.

Facilities achieving high-spatial-resolution observations can be categorized into two groups --- single telescopes and synthesized telescopes.
The former group consists of infrared/optical space-based telescopes, such as the Hubble Space Telescope (HST), the Hipparcos space telescope and the Gaia space telescope. 
Both the Hipparcos space telescope \citep{van-Leeuwen97} and the Gaia space telescope \citep{Gaia-Collaboration16} are dedicated to astrometry. As the successor of Hipparcos, Gaia operates at optical and infrared bands, and can achieve higher spatial resolution than Hipparcos \citep{Gaia-Collaboration16}. On the other hand, the versatile HST remains highly useful for astrometry of individual sources \citep[e.g.][]{Deutsch99,Lyman22}.

In the other facility group, synthesized telescopes refer to largely ground-based interferometers formed with an array of telescopes at a distance to each other.
%Example facilities of the former means include . 
In the order of operating frequencies (low to high), famous high-resolution interferometers include the VLBA (Very Long Baseline Array), the EVN (European VLBI Network), the LBA (Long Baseline Array), the EHT (Event Horizon Telescope), the VLTI (Very Large Telescope Interferometer) and the CHARA (Center for High Angular Resolution Astronomy) array.  
Among these interferometers, the VLTI and the CHARA array are the only devices operating outside radio frequencies. 
Owing to the technical challenges to perform interferometry at infrared/optical bands, the separation between the constituent mirrors of the VLTI or the CHARA array is limited. Despite this, the VLTI and the CHARA array can achieve, respectively, 2\,mas and sub-mas angular resolution thanks to the high operating frequencies.
To help understand how other interferometers work at radio frequencies, the radio interferometry technique is concisely summarized in \ref{sec:VLBI_in_a_nutshell}.

When positional precision allows, finer astrometry such as orbital motion inference becomes possible. This has been achieved at cm wavelengths using VLBA for pulsars at $\lesssim300$\,pc distances \citep{Deller13,Deller16,Guo21}. A better demonstration (of the power of fine astrometry) has been made for a close (merely 45\,pc away) quadruple system HD~98800  \citep{Zuniga-Fernandez21} with the VLTI. Incorporating radial-velocities (measured with spectroscopic observations), the authors are close to inferring the full orbital parameter space of the complicated quadruple system.

Depending on the type of neutron stars, observing setup for the astrometry of a neutron star differs. For example, radio pulsars are normally observed at low radio frequencies (around 1.5\,GHz, e.g. \citealp{Brisken02,Chatterjee09,Deller19,Ding21a}), as they typically show steep radio spectra. In comparison, radio magnetars displaying flat radio spectra are normally astrometrically observed at higher radio frequencies (4\,GHz to 15\,GHz, e.g. \citealp{Deller12a,Bower14,Bower15,Ding20c,Ding23a}), in order to increase angular resolution and reduce propagation effects due to interstellar medium (see \ref{subsec:scattering_screen}) and Earth atmosphere. 
Additionally, transient radio emissions can be occasionally observed from NS XRBs during their X-ray outbursts \citep[e.g.][]{Tudose09a,Miller-Jones10}.
At optical/infrared frequencies, non-thermal emissions have been observed from magnetars and other pulsars, which would potentially allow dedicated astrometric campaign with the HST \citep{Tendulkar12a,Lyman22}.
In addition, some NSs relatively close to the Earth have optically bright companions (including white dwarfs and non-degenerate stars) that can be astrometrically measured with Gaia \citep{Jennings18,Antoniadis21}. 
%Likewise, local ($\lesssim10$\,kpc) X-ray binaries and NS- can potentially be astrometrically measured with Gaia. 

This thesis only involves two specific astrometry facilities --- VLBA and Gaia. As mentioned in \ref{sec:NSs}, three distinct types of NSs are studied in this thesis: millisecond pulsars, magnetars and NS X-ray binaries.
The studies around MSPs and magnetars are carried out with VLBA at different radio frequencies (see \ref{ch:J1810}, \ref{ch:J1818}, \ref{ch:J1012}, \ref{ch:B1534} and \ref{ch:mspsrpi}), while NS XRBs are investigated with Gaia data (see \ref{ch:PRE}).
%-- millisecond pulsars (see \ref{section:pulsar_astrometry}), magnetars (\ref{sec:magnetar_astrometry}) and X-ray binaries (\ref{sec:XRB_astrometry}).

%In the following sections of this chapter, the scientific theme for each NS type is laid out. These contexts are complemented by the introductory sections of the ``paper'' chapters.

\subsection{Pulsar astrometry with radio interferometers}
\label{subsec:VLBI_astrometry}

Though the first pulsar parallax \citep{Bailes90} was measured with pulsar timing, most pulsar parallaxes are determined with the very long baseline interferometry (VLBI) technique at radio frequencies (see the pulsar parallax catalogue\footnote{\url{http://hosting.astro.cornell.edu/research/parallax/}}). 
The majority of VLBI parallaxes were measured with the Very Long Baseline Array (VLBA), followed by the Australian Long Baseline Array (LBA) and the European VLBI Network (EVN).
Previous large VLBI pulsar astrometry programs are summarized in \ref{tab:astrometry_programs}. 
As pulsars are generally steep-spectrum radio sources, all the large programs are conducted at low frequencies (around 1.6\,GHz). Benefiting from the increasingly large bandwidth of telescope receivers, later VLBI programs can astrometrically measure fainter (and potentially further) pulsars.

\begingroup
\renewcommand{\arraystretch}{1.4} % Default value: 1

\begin{table*}
\raggedright
\caption{Large VLBI pulsar astrometry programs}
\label{tab:astrometry_programs}
%\begin{tabular}{@{}l@{\:}l@{\:}l@{}} % manual @ spacing to prevent this being too wide for a page
\resizebox{\textwidth}{!}{
\begin{tabular}{ccccccccc}
\hline
\hline
Program & Obs. time & Pulsar & Parallax & Obs. freq & bandwidth & VLBI & Polarization &  Reference    \\
name &   & number  & number & (GHz) & (MHz) & array &  &   \\
\hline
--- & 10/1999--10/2000 & 9 & 9 & 1.6 & 64 & VLBA & single & \citet{Brisken02}  \\
--- & 06/2002--03/2005 & 14 & 14 & 1.5 & 32 & VLBA & dual & \citet{Chatterjee09} \\
--- & 08/2006--02/2008 & 8 & 7 & 1.6 & 64 & LBA & single  & \citet{Deller09a}\\
\psrpi\ & 2011--2013 & 60 & 57 & 1.66 & 64 & VLBA & dual  & \citet{Deller19}\\
\mspsrpi\ & 06/2015--2018 & 18 & 15 & 1.55 & 128 & VLBA & dual  & \citet{Ding23} \\

\hline

%\multicolumn{6}{l}{$\bullet$ $m_\mathrm{p}$, $m_\mathrm{c}$ and $q$ stand for, respectively, pulsar mass, companion mass and}\\

\end{tabular}
}
\end{table*}
\endgroup

\citet{Chatterjee09} was the first pulsar astrometry program that systematically searched for and used in-beam phase calibrators (see \ref{subsec:ch2_in_beam_astrometry} for explanation). The strategy to search for in-beam calibrators was optimized and streamlined for the \psrpi\ project --- the largest pulsar astrometry program \citep{Deller19}. This strategy was shared by the mJIVE-20 project dedicated to finding phase calibrators at 20\,cm wavelength \citep{Deller13a}.
The \psrpi\ project capitalized on the improved VLBA capacity, and nearly tripled the number of pulsars with precise parallaxes.

\subsection{The \mspsrpi\ project}
\label{subsec:mspsrpi}

To follow on from the success of the \psrpi\ project \citep{Deller19} and to achieve the scientific goals to be outlined in coming sections, the \mspsrpi\ program was conceived by the \psrpi\ collaboration to systematically improve the astrometric precision of MSPs. Compared to the time of the \psrpi\ program, the maximum recording bandwidth of VLBA had been further improved by the beginning of the \mspsrpi\ program, allowing $\sqrt{2}$ times as high image sensitivity to be achieved. This sensitivity enhancement paves the way for VLBA astrometry of MSPs, as MSPs are generally fainter than canonical pulsars.  
The \mspsrpi\ program contains 18 MSPs (including 2 DNSs) with the highest scientific values. More descriptions of the program can be found in \ref{subsec:mspsrpi_mspsrpi}. The works detailed in \ref{ch:J1012}, \ref{ch:B1534} and \ref{ch:mspsrpi} are based on the \mspsrpi\ datasets.

%\subsection{NS astrometry with Gaia}
%\label{subsec:Gaia_astrometry}

\section{Probing neutron star formation theories with high-resolution astrometry}
\label{sec:NS_formation}

\subsection{Neutron star kinematics and the retention problems}
\label{subsec:NS_kinematics}
Due to the supernova asymmetry, a neutron star would receive a natal kick at its birth \citep[e.g.][]{Bailes89}, which likely results in a high space velocity (i.e., the NS velocity with respect to its stellar neighbourhood). 
\citet{Hobbs05} compiled 233 pulsars with proper motion measurements, and found that pulsar velocity distribution is well described by a Maxwellian distribution. \citet{Hobbs05} estimates a 3-D birth velocity of $400\pm40$\,\kmps\ for all pulsars. The 2-D velocity for canonical pulsars are $246\pm22$\,\kmps, while recycled pulsars display lower ($87\pm13$\,\kmps) 2-D velocity \citep{Hobbs05}.
Though the overall pulsar velocity distribution by \citet{Hobbs05} is very useful, it runs into difficulties, at the low-velocity end, to address the globular-cluster retention problem \citep[e.g.][]{Pfahl02}. 
In specific, more pulsars are found in globular clusters than expected with the globular cluster escape velocities (of only $\sim50$\,\kmps) and the pulsar velocity distribution by \citet{Hobbs05}. 

In addition, the MSP transverse (or ``2D'') space velocity distribution, estimated by \citet{Hobbs05,Gonzalez11} to be around $\sim100$\,\kmps, is challenged by another retention problem.  
MSPs are suspected to contribute to the excessive $\gamma$-ray emissions from the Galactic Centre \citep[e.g.][]{Abazajian12}. However, \citet{Boodram22} suggests that the space velocities of MSPs have to be very small to explain the Galactic Centre $\gamma$-ray excess. Given that the transverse space velocities of MSPs are $\sim100$\,\kmps\ \citep{Hobbs05,Gonzalez11}, unless the velocity distribution of MSPs is bi-modal (or multi-modal) with one mode at near zero (see \ref{subsec:mspsrpi_v_t}), the Galactic Centre $\gamma$-ray excess might be attributed to exotic sources, such as annihilating dark matter particles \citep[e.g.][]{Hooper18}. 

Conceivably, it is hard to render a bi-modal MSP velocity distribution with the combination of the standard MSP formation scenario (see \ref{subsec:NS_observations}) and the mainstream NS formation mechanism \ref{subsec:NS_formation}. In other words, the bi-modality implies another co-existing MSP formation channel and/or alternative NS formation channel that would lead to small kick velocities. One prime candidate for the additional MSP formation channel is the accretion-induced collapse (AIC) channel, where a super-Chandrasekhar WD (overfed by its non-degenerate companion) turns into an MSP via type Ia supernova \citep{Bailyn90}. Alternatively, WD-WD mergers (WDM) may also give birth to MSPs \citep{Michel87}. In either way, the kick velocity would be significantly smaller than the standard core-collapse supernovae (CCSN) scenario \citep{Tauris13}. 

In addition, NSs born from 
%ultra-stripped supernovae (USSNe) \citep[e.g.][]{Tauris13a} and 
electron-capture supernovae (ECSNe) of burnt-out transitional-mass ($\sim8$--10\,\msun) stars \citep{Miyaji80} would also have small kick velocities \citep{Gessner18}. 
Such NSs may likely eventually turn into MSPs, following the standard pathway of MSP formation.
%For the USSN-born NSs that must originally have degenerate companions, this evolution pathway requires change of companion star, which is not unlikely in the dense stellar regions such as globular clusters or the Galactic Centre.
Therefore, a postulated MSP population born from the above alternative channels in the globular clusters \citep{Bailyn90} and Galactic Centre can likely explain both aforementioned retention problems. 
This explanation is supported by a recent binary population synthesis study: \citet{Gautam22} shows that AIC-born MSPs can reproduce the Galactic Centre $\gamma$-ray excess signals.
To sum up, despite few direct indications from pulsar timing \citep{Freire13}, the retention problems in the globular cluster and the Galactic centre indirectly suggest an MSP sub-population born from alternative NS and/or MSP formation channels, which is partly supported by recent binary population synthesis \citep{Gautam22}.

Another potential retention problem is related to double neutron stars (DNSs). DNSs are among the prime sources of r-process elements \citep{Eichler89,Korobkin12}. To account for the observed abundance of r-process elements in the ultra-dwarf galaxies (UDGs) that have escape velocities of only $\sim15$\,\kmps, a considerable fraction of DNSs should have space velocities of $\lesssim15$\,\kmps.
%Unlike the globular-cluster retention problem and the Galactic centre $\gamma$-ray excess issue, the tension 
Due to the small sample size, the velocity distribution for DNSs is yet not well defined, hence the UDG retention problem is only considered ``potential''.

%On the low-velocity end, most more to constrain the fraction of pulsars with low kick velocities is essential for addressing various NS retention problems. 

\subsection{Probing magentar formation channels with VLBI astrometry}
\label{subsec:probe_magnetar_formation}

Similar to DNSs, few magnetars have been precisely measured astrometrically, owing to the small number of magnetars sufficiently bright at optical/infrared or radio (see \ref{subsec:overview_astrometry} or \ref{sec:J1818_intro}). 
%the other NS division -- magnetar, has even less constrained space velocity distribution, due to a smaller sample (\ref{subsec:probe_magnetar_formation} and \ref{ch:J1818}). 
So far, proper motions have been determined for nine magnetars \citep{Tendulkar13,Deller12a,Bower15,Ding20c,Ding23a,Lyman22}, while only one has precise model-independent distance (see \ref{ch:J1810} or \citealp{Ding20c}). Despite the small sample size, no magnetar displays a space velocity higher than normal pulsars, which disfavors the scenario where magnetars have to be born with $\gtrsim1000$\,\kmps\ kick velocities \citep[e.g.][]{Duncan92}. %This scenario is disfavored with astrometry of few magnetars \citep{Tendulkar13,Deller12a,Ding20c,Lyman22} showing magnetar space velocities not inconsistent with normal pulsars.

As mentioned in \ref{sec:NSs}, the formation of magnetars is poorly constrained. Abundant formation channels have been proposed for magnetars, including the core-collapse supernovae (CCSN) channel \citep{Schneider19}, the AIC channel \citep{Duncan92}, the WDM channel \citep{Levan06}, the NS-WD merger channel \citep{Zhong20} and the DNS merger channel \citep{Giacomazzo13,Xue19}. It is likely that more than one formation channel contributes to the magnetar population. 
%In other words, magnetars might be born in various channels. 
The DNS merger channel is supported by light-curve analysis of extragalatic short $\gamma$-ray bursts \citep{Xue19}. However, the low Galactic latitudes of all Galactic magnetars disfavor the DNS scenario that expects a $\approx10$\,kpc vertical distance from the Galactic plane (see \ref{sec:J1818_intro} or \citealp{Ding23a}).
On the other hand, the $\approx10$ associations between magnetars and supernova remnants (SNRs) strongly reinforces the CCSN channel as a major formation channel for Galactic magnetars \citep{Olausen14}. 
To probe the formation mechanism of Galactic magnetars without solid SNR associations, magnetar astrometry can be used, as each channel would correspond to a different velocity distribution (see \ref{sec:J1818_intro} or \citealp{Ding23a}). Specifically, the AIC and WDM channels may lead to a group of magnetars with $\sim10$\,\kmps\ kick velocities, similar to the discussion in \ref{subsec:NS_kinematics}.
%As pointed out by \citet{Ding23a},

Up till now, the median tangential velocity of magnetars is estimated to be around 150\,\kmps\ \citep{Lyman22} on the smaller side of the 2-D velocity of canonical pulsars ($246\pm22$\,\kmps, \citealp{Hobbs05}). Future magnetar astrometry based on a larger sample would determine the tangential space velocity distribution of Galactic magnetars, hence shedding light on their formation mechanism. 
This thesis involves VLBI astrometry of 2 (out of 5) constant radio magnetars.
In \ref{ch:J1810}, VLBI astrometry of the first discovered radio magnetar XTE~J1810$+$197 is detailed; the work leads to the first and the hitherto only significant parallax measurement for a magnetar.
In \ref{sec:J1818_astrometry_progress}, the progress of VLBI astrometry of \swift, the hitherto fastest-spinning magnetar, is provided.

\section{Studying ionized interstellar media with VLBI}
\label{sec:IISM}

Cold ionized interstellar media (IISM) would slow electromagnetic waves (EMWs) at frequency $\nu$ to the group velocity
\begin{equation}
    \label{eq:group_velocity}
    v_\mathrm{g} \approx c \left(1 - \frac{\nu_\mathrm{p}^2}{2 \nu^2}\right) \,,
\end{equation}
with the plasma frequency
\begin{equation}
    \label{eq:plasma_frequency}
    \nu_\mathrm{p} = \left( \frac{e^2 n_\mathrm{e}}{\pi m_\mathrm{e}} \right)^{1/2} 
\end{equation}
\citep[e.g.][]{Condon16},
where $c$, $e$, $m_\mathrm{e}$ and $n_\mathrm{e}$ stand for the speed of light, the electron charge, mass, and number density, respectively. 
The delay $\Delta t$ corresponding to $\nu_\mathrm{g}$ is
\begin{equation}
    \label{eq:dispersion_delay}
    \Delta t = \int_{0}^{D} \frac{\mathrm{d}r}{v_\mathrm{g}} - \frac{D}{c} = \left( \frac{e^2}{2 \pi c m_\mathrm{e}} \right) \nu^{-2} \int_{0}^{D} n_\mathrm{e}(r) \mathrm{d}r
\end{equation}
for a source at distance $D$ \citep[e.g.][]{Lorimer12}. The propagation effect described by \ref{eq:dispersion_delay} is known as dispersion, as EMWs with different wavelengths would arrive asynchronously. Accordingly, $\Delta t$ is also called dispersion delay. Though in principle other cold plasma can cause dispersion, cold free electrons are the dominant contributor, as they have the highest $e^2/m_\mathrm{e}$ (hence the largest $\Delta t$). 
%Besides, the number of relativistic electrons along the line of sight is normally negligible compared to that of cold free electrons.
\ref{eq:dispersion_delay} can be simplified to
\begin{equation}
    \label{eq:Delta_t_simp}
    \left( \frac{\Delta t}{\mathrm{1\,\mu s}} \right) \approx 4.149\times10^3 \left( \frac{\mathrm{DM}}{\mathrm{1\,pc~cm^{-3}}} \right) \left( \frac{\nu}{\mathrm{1\,GHz}} \right)^{-2} \,,
\end{equation}
where
\begin{equation}
    \label{eq:DM}
    \mathrm{DM} \equiv \int^{D}_{0} n_\mathrm{e}(r) \mathrm{d}r
\end{equation}
is the abbreviation for the dispersion measure, which denotes the column density integrated over the line of sight \citep[e.g.][]{Condon16}.

As $\Delta t \propto \nu^{-2}$, the dispersion effect is commonly observed at cm wavelengths. The effect is typically seen from sources releasing short and powerful radio emissions, such as radio pulsars and fast radio bursts (FRBs). If a bright pulse is recorded across a wide radio band, the DM can be fitted with \ref{eq:Delta_t_simp}.  Hence, almost all radio pulsars and wide-band FRBs have precisely determined DMs (see the ATNF Pulsar Catalogue\textsuperscript{\ref{footnote:psrcat}}, \citealp{Manchester05}, and the FRB Catalogue\footnote{\label{footnote:frbcat}\url{https://www.frbcat.org/}}, \citealp{Petroff16}).
%The process of fitting DM and removing the effect of dispersion is called de-dispersion.
%is typically observed for sources releasing brief and powerful 

\subsection{Refining Galactic free-electron distribution models with pulsar astrometry}
\label{subsec:refine_n_e}

The knowledge of free-electron number density distribution $n_\mathrm{e}(\Vec{x})$ in the Galaxy is essential for several reasons. Firstly, such an $n_\mathrm{e}(\Vec{x})$ model can be used to roughly estimate the distance to a new Galactic pulsar with its DM \citep{Taylor93,Cordes02,Yao17}. Secondly, FRBs can be used as probes of intergalactic interstellar medium on a cosmological scale \citep{Macquart20,Mannings21}. This use demands the deduction of Galactic DM contribution from the total observed DM (of an FRB), which, again, requires a reasonable $n_\mathrm{e}(\Vec{x})$ model.

Most constraints on $n_\mathrm{e}(\Vec{x})$ are obtained with pulsar observations \citep{Taylor93,Cordes02,Yao17}. Using pulsar timing, DM of a pulsar can be precisely determined. Given an independent distance $D$ estimated for the pulsar, $n_\mathrm{e}(\Vec{x})$ along the Earth-to-pulsar sightline (up to the distance of the pulsar) can be constrained using \ref{eq:DM}. Provided a large number of \{DM, $D$\} pairs for pulsars across the sky, an $n_\mathrm{e}(\Vec{x})$ model can be established. The quality of an $n_\mathrm{e}(\Vec{x})$ model depends on the availability of pulsars with precise distance measurements. This availability changes with sky regions. As most pulsars sit around the Galactic plane \citep{Manchester05}, the high-Galactic-latitude areas are relatively sparsely sampled with pulsars, leading to worse $n_\mathrm{e}(\Vec{x})$ constraints (in these sky regions).

In addition, most pulsars with precise distances are in the vicinity of the solar system \citep{Yao17,Deller19}. 
Hence, at larger distance, $n_\mathrm{e}(\Vec{x})$ extrapolated from the measurements of nearby pulsars is increasingly affected by local $n_\mathrm{e}(\Vec{x})$ inhomogeneities. 
The vast majority ($>90$\%) of pulsars do not have precise $D$, which is the main limiting factor against the establishment of a precise $n_\mathrm{e}(\Vec{x})$ model. Hence, high-precision pulsar astrometry, in particular VLBI astrometry (see \ref{subsec:VLBI_astrometry}), plays a leading role in refining the $n_\mathrm{e}(\Vec{x})$ model. 
%Compared to pulsar timing, VLBI astrometry takes much shorter time to obtain a parallax at the same significance \citep{Chatterjee09,Deller19}, therefore serving as the prime method of pulsar astrometry. 
In \ref{subsubsec:mspsrpi_DM_distances}, the two prevailing $n_\mathrm{e}(\Vec{x})$ models \citep{Cordes02,Yao17} are tested with parallax-based distances of 15 MSPs.
Further improvement on the $n_\mathrm{e}(\Vec{x})$ model can be made with the independent distances of 15 MSPs (\ref{ch:mspsrpi}) and a magnetar (\ref{ch:J1810}) provided in this thesis.

\subsection{Constraining scattering screens with pulsar angular broadening}
\label{subsec:scattering_screen}

In parallel to dispersion, other propagation effects are also induced by IISM, which include the scattering of pulsar radio emissions due to the inhomogeneity of IISM \citep[e.g.][]{Lyne12}. The scattering can be observed in two aspects --- scintillation and multi-path propagation. Scintillation refers to the time variation of pulsar brightness \citep[e.g.][]{Romani86,Narayan92,Mall22}, due to the changing Earth-to-pulsar sightline that passes through different parts of the inhomogeneous IISM. 
On the other hand, the multi-path propagation would lead to two observational phenomena --- temporal pulse broadening and the image-domain angular broadening of apparent pulsar size \citep[e.g.][]{Lyne12}. 
The former phenomenon is pronounced with a broadened pulse profile obtained with pulsar timing, while the latter one can only be detected with high-resolution VLBI observations \citep[e.g.][]{Bower14}.
Similar to scintillation, both temporal and angular broadening are found to fluctuate with time \citep[e.g.][]{Brisken09,Lentati17}.
In addition, both phenomena are frequency-dependent: they become more prominent at lower radio frequencies. 
Since both phenomena are caused by multi-path propagation, they can offer joint constraints on the geometry of scattering screens \citep[e.g.][]{Brisken09,Bower14}.
In \ref{subsec:mspsrpi_J1643}, the mean angular broadening of \psrgb\ is measured with VLBA at 1.55\,GHz, and is used to test the scintillation model \citep{Mall22} of the pulsar in conjunction with a temporal broadening estimate.

%diffractive scintillation and refractive scintillation due to the inhomogeneity of IISM \citep[e.g.][]{Romani86,Narayan92}. Relatedly,  

\section{Enhancing studies of gravity with VLBI astrometry of millisecond pulsars}
\label{sec:pulsar_astrometry}
\subsection{Probing gravitational theories using pulsar timing of millisecond pulsars}
\label{subsec:pulsar_timing}
%A pulsar is a neutron star that constantly beams electromagnetic radiation from the emitting region in the magnetosphere. 

As a joint result of faster and more stable spin \citep{Hobbs10} compared to other pulsars, MSPs are considered ideal testbeds for probing physical effects that would affect the pulse ToAs, particularly in the gravity-related aspects.
%No explicit upper limit for the spin period has been defined for MSPs. 
As mentioned in \ref{sec:NSs}, MSPs in this thesis refer to pulsars having spin periods of $\lesssim40$\,ms, which would include some intermediately spun-up pulsars like double neutron stars \citep[e.g.][]{Hulse75,Wolszczan91}. 
Compared to other MSPs, pulsars in double neutron stars (DNSs) normally spin one order slower in more compact orbits. Due to much deeper gravitational potentials than other MSPs, DNSs offer one of the best tests of gravitational theories in the strong-field regime \citep[e.g.][]{Fonseca14,Weisberg16,Kramer21a}.

The theory of general relativity (GR) \citep{Einstein16} is derived from the theory of special relativity and the principle of equivalence. %Reshaping our understanding of gravity, 
The theory deepens the understanding of gravity by interpreting gravity as the curvature of spacetime, and serves as a generalization of the Newtonian formalism of gravity.
However, GR is not the only plausible post-Newtonian gravitational theory. Instead, GR takes the simplest form among a group of candidate post-Newtonian gravitational theories. Some GR alternatives suggest temporal variation of the Newton's gravitational constant, which is usually associated with dipole \gw\ radiation \citep{Will93}. 

In testing GR and other alternative gravitational theories, MSPs have been playing a central role.
In general, tests of gravitational theories are made with MSPs in orbit with either another NS or a white dwarf (WD).
The first indirect evidence of gravitational-wave emissions came from the pulsar timing of the first discovered DNS system \citep{Hulse75}. The most precise GR test is achieved with the hitherto only double pulsar system \citep{Lyne04,Kramer21a}. Additionally, the most stringent constraint on dipole gravitational radiation (predicted by some alternative gravitational theories) is obtained with four pulsar-WD systems \citep{Deller08,Freire12,Zhu19,Ding20}. As another example, the sharpest test of the strong equivalence principle is provided by timing of a pulsar (i.e., PSR~J0337$+$1715) in a triple star system \citep{Archibald18}.

To quantify post-Newtonian gravitational effects, the so-called post-Keplerian (PK) parameters are introduced, which include (but are not limited to) the intrinsic orbital decay (the intrinsic time derivative of orbital period), the advance of periastron longitude, the Doppler coefficient (related to the gravitational redshift) and the ``range'' and ``shape'' of the Shapiro delay effect \citep{Damour92,Stairs03}. 
In the context of testing gravitational theories with an MSP in a binary system (hereafter referred to as a binary MSP), each of the PK parameters is a theory-dependent function of the masses of the NS and its companion.
This statement has two indications. Firstly, mass determination is the crux of testing gravitational theories with a binary MSP. Secondly, the masses of the NS and its companion can be inferred based on two PK parameters measured with pulsar timing \citep[e.g.][]{Damour91,Kramer06}. Since the PK parameters are theory-dependent, the masses also rely on the underlying gravitational theory.
Nevertheless, this underlying theory can be tested with more than two PK parameters.
More explicitly, provided $N$ ($N>2$) known PK parameters, $N-2$ tests of the underlying gravitational theory can be made. 
Such PK-parameter-based tests of GR have been made with multiple DNSs \citep{Fonseca14,Weisberg16,Kramer21a}, where GR has largely passed all of the tests.
%as DNSs generally exhibit much stronger post-Newtonian gravitational effects than pulsar-WD systems.

Pulsar timing is mostly self-sufficient for PK parameter estimation. As an exception, the determination of the intrinsic orbital decay requires precise distance to the pulsar, which is normally acquired elsewhere (though a small number of MSPs can achieve precise parallax with pulsar timing).
The radial acceleration $a_\mathrm{r}$ of a pulsar would give rise to an extrinsic orbital decay
\begin{equation}
\label{eq:Pbdot_ex}
    \dot{P}_\mathrm{b}^\mathrm{ex}=\frac{a_\mathrm{r}}{c} P_\mathrm{b} \,,
\end{equation}
where $P_\mathrm{b}$ stands for the orbital period. The radial acceleration equals to
\begin{equation}
\label{eq:A_r}
    a_\mathrm{r}= -e_\mathrm{r} \cdot \triangledown \varphi + \mu^2 D \,,
\end{equation}
where $\varphi$, $\mu$ and $D$ refers to the Galactic potential, proper motion and distance, respectively. Namely, the first term of \ref{eq:A_r} represents the radial acceleration due to Galactic potential gradient; the second term stands for the the radial acceleration owing to tangential motion \citep{Shklovskii70}.
Further explanations of orbital period decay contributions can be found in \ref{subsec:J1012_alternative_gravity}, \ref{sec:B1534_Pbdot_test} and \ref{sec:mspsrpi_orbital_decay_tests}.

Due to the much weaker constraints on the PK parameters compared to DNSs, the mass determination for the NS and its companion in a pulsar-WD system normally takes a different pathway. For a pulsar-WD system $\lesssim1$\,kpc away, the WD can be potentially observed at optical wavelengths. Hence, spectroscopic observations of the WD can reveal its radial velocity curve, which, combining the orbital period and the pulsar projected semi-major axis obtained with pulsar timing, can lead to a precise mass ratio determination \citep[e.g.][]{Antoniadis12}.
By comparing the observed spectral energy distribution (SED) against the theoretical one, the ratio between the WD radius and its distance from the Earth can be derived \citep[e.g.][]{Mata-Sanchez20}. Given a precise distance to the WD, the WD radius can be calculated. 
Once a WD radius is acquired, the WD mass can be inferred with WD mass-radius relations \citep[e.g.][]{Nauenberg72,Suh00,Althaus05} or WD evolutionary models \citep[e.g.][]{Istrate16}.
Finally, combining the inferred WD mass and the mass ratio (between the NS and the WD), the NS mass can be obtained.
Compared to the PK-parameter-based mass determination, the spectroscopy-based method does not depend on gravitational theories, but is more reliant on measurements extrinsic to the binary system, such as the distance to the binary. 
%The paper Chapters~3, 4 and 5 of this thesis  will demonstrate how theories of gravity can be tested with pulsar-WD systems and DNSs.

%To test GR or its alternatives with the timing of an MSP, the crux is to determine the masses for the NS and its companion. Mass constraints can be acquired in mainly two ways: pulsar timing and spectroscopy (of the WD companion).

%There are several ways to test GR and constrain alternative theories of gravity with MSPs. Highly relativistic DNS systems \citep[e.g.][]{Damour92,Burgay03} and pulsars in orbit with white dwarf (WD) companions \citep[e.g.][]{Lazaridis09,Freire12,Antoniadis13,Zhu15} have probed different regions of phase space for deviations from the predictions of general relativity. 

\subsection{Searching for gravitational-wave background with pulsar timing arrays}
\label{subsec:PTAs}

Gravitational waves (GWs), also termed Einstein–Rosen cylindrical waves, are a  prediction of GR \citep{Einstein37}. 
GWs are produced when the fabric of spacetime is disturbed by a mass quadrapole moment. While rippling through spacetime at the speed of light, GWs carry away energy (and information) from their emitter.
As mentioned in \ref{subsec:pulsar_timing}, the first strong confirmation of GWs came from the orbital decay \citep{Taylor82} of PSR~B1913$+$16, the first discovered pulsar in a DNS system \citep{Hulse75}.
At the advent of mature laser interferometer techniques, the first direct GW event, indexed GW150914, was detected in 2015 \citep{Abbott16} with the two detectors of the Laser Interferometer Gravitational-wave Observatory (LIGO). 
Modeling of GW150914 reveals 5\% of the initial mass budget, equivalent to 3\,\msun\ worth of energy, is lost to GWs during the brief BH-BH coalescence \citep{Abbott16}.
At the time of writing, $\sim100$ individual GW events\footnote{\url{https://www.ligo.org/detections.php}} have been detected (with GW observatories), the majority of which are originated from BH-BH mergers. Electromagnetic counterparts %\citep{Drout17,Mooley18,Pozanenko19} 
have only been identified conclusively for the GW170817 event associated with a NS-NS merger \citep{Abbott17a,Drout17,Mooley18}, and arguably for the GW190425 event \citep{Abbott20,Pozanenko19}.

Similar to the cosmic microwave background (CMB) observed at radio frequencies, a stochastic gravitational-wave background (GWB) that is anisotropic by a small fraction is expected, but has not yet been detected at any GW frequency band. The GWB is a superposition of historically emitted GWs. The detection of the first individual GW event \citep{Abbott16} indicates that individual astrophysical events must at least partially contribute to the GWB.
Apart from this contribution, it is postulated that GWB is partly of cosmological origin, as the so-called relic GWB is predicted with the inflation theory \citep{Guth81,Turner97}. Other theoretical possibilities, such as cosmic strings \citep{Kibble76}, may also contribute to the GWB \citep{Damour01,Kuroyanagi12}. 

Across the electromagnetic spectrum, the furthest and earliest universe that mankind can observe is the CMB, which was released at the recombination epoch. 
Compared to the CMB emitted roughly 380\,kyr after the Big Bang, the GWB can be used to probe the universe that was only $\lesssim10^{-33}$\,s old \citep{Guth81}.
Though the imprint left by the relic GWB into the CMB might be potentially observed with the B-mode of the CMB \citep{Pagano16,Calabrese20}, such an effort is limited predominantly by the contamination due to foreground Galactic dust \citep{Ade16}. As a result, no evidence of GWB has been confirmed from the CMB observations.

Besides the CMB approach, other methods have been employed to search for the GWB at different GW frequencies (see \citealp{Lasky16} for a review). 
Among them, pulsar timing is thought to be sensitive for GWs at nano-Hz frequencies \citep{Sazhin78,Detweiler79}. When the light emitted from a pulsar is intercepted by a passing ultra-long-period GW before reaching the Earth, it becomes Doppler-shifted \citep{Estabrook75}, leading to delays (or advances) in the pulse ToAs. As the GW moves, the delay of the ToAs changes accordingly, leaving a GW signature into the residual ToAs.
Likewise, the GWB would also be imprinted in the residual ToAs.
At $\sim1$\,nHz, the GWB is probably dominated by the GWs generated by inspiraling supermassive black hole (SMBH) binaries \citep{Sesana08}. 
Therefore, pulsar timing can be used to potentially investigate the demography of SMBH binaries in the local universe \citep{Jaffe03}, and put upper limits to the strength of relic GWB at $\sim1$\,nHz.
%The largely isotropic GWB would lead to a GW signature in the residual ToAs in 
To directly extract the GWB signal (from the residual ToAs) in the time domain is difficult, as the tiny GW signals would be buried by other noises.
Hence, one normally converts the residual ToAs into its power spectrum $S(\nu)$
(in the frequency domain) with Fourier transform. 
%The power spectrum is usually fitted to the formalism 
%\begin{equation}
%    S(\nu) = A^2 \nu^{-\gamma},
%\end{equation}
%where $A$ is an achromatic amplitude, $\nu$ stands for frequency, and $\gamma$ represents the spectral index.

Assuming the GWB at $1\sim$\,nHz is generated by inspiraling SMBH binaries, the power spectrum of the GWB should obey $S \propto \nu^{13/3}$ \citep{Phinney01}. Hence, comparing this scaling relation to the observed power spectrum can test the origin of GWB at $\sim1$\,nHz.
%To extract the tiny GW signature from the residual ToAs, an array of MSPs is necessitated. 
%observed ToAs need to be well modeled
%Gravitational waves (GWs) are predicted by Einstein's general theory of relativity to be a quadrupole radiation of gravity that travels at the speed of light. A gravitational wave background (GWB), composed of primordial GWs and GWs induced by later astrophysical events \citep{Carr80}, is widely accepted by astronomers. In the range of $10^{-9}\,Hz-0.1\,Hz$, supermassive black hole binaries are among the primary sources of the GWB \citep{Sesana08}. The pulse ToA measurements of pulsars are believed to be sensitive to nano-hertz GWs \citep{Detweiler79}, which makes pulsar timing a good way to detect the GWB in the \lf\ range and study supermassive black hole binaries in galaxy merger. When \lf\ GWs pass through the line of sight from the Earth to the pulsar, the pulse train from the pulsar would be Doppler-shifted \citep{Estabrook75}. An additional residual is expected to change slowly in time in accordance with the \lf\ GW.
However, pulsar timing involves a wealth of model parameters, thus subject to complicated error estimation \citep{Goncharov21a}. Accordingly, searching for GWB with one pulsar is unreliable and impractical. In contrast, using a group of pulsars can increase redundancy and sensitivity of the GWB search.
Furthermore, the GWB is not the only source that can cause secular changes of ToAs at nHz. To mitigate other factors such as clock errors, interstellar medium (ISM) variations \citep{Lyne68} and solar-system ephemeris (SSE) errors \citep[e.g.][]{Champion10,Tiburzi16}, a pulsar timing array (PTA) is proposed to capture spatially correlated GWB signals in the residual ToAs \citep[see][and references therein]{Foster90} with an array of well timed MSPs scattered across the sky. It is timely to note that a PTA is only used to detect the GWs passing through the Earth. Otherwise, the GW signals in the residual ToAs cannot be spatially correlated.
Since 2005 or so, three major regional PTA consortia have been running, including the NANOGrav \citep{Alam20,Faisal-Alam20}, the EPTA \citep{Desvignes16} and the PPTA \citep{Kerr20}.
%To better distinguish the ToA residuals caused by the GWB from that of another origin such as planetary ephemeris error, clock error and interstellar medium (ISM) \citep{Lyne68}, a pulsar timing array (PTA) scattered across the sky \citep[see][for requirement of spatial distribution]{Roebber19} is suggested that includes a set of MSPs widely separated on the sky. By observing a set of MSPs and measuring the correlation of their ToAs, it is possible to discriminate between different potential signals by corresponding angular signatures \citep{Foster90}. Among the signals, a GW passing through the Earth is expected to exhibit a quadrupole signature. There are currently three operating PTA projects: the Parkes Pulsar Timing Array (PPTA), the European Pulsar Timing Array (EPTA) and the North American Nanohertz Observatory for Gravitational Waves (NANOGrav). Additionally, an organization called the International Pulsar Timing Array (IPTA) is built to facilitate collaboration between the three PTAs in order to reach the common goals \citep{Manchester13}.

%Recent results from NANOGrav \citep{Arzoumanian20} and PPTA \citep{Goncharov21} have both detected common-spectrum time-correlated signals. However, they cannot confirm (or rule out) if the signals come from the gravitational-wave background. Such a confirmation requires higher sensitivities of PTAs.

For the purpose of constraining the GWB, the cross power spectral density for any two different pulsars of a PTA is usually fitted to the GWB formalism \citep{Phinney01}
\begin{equation}
\label{eq:S_ab}
    S_\mathrm{ab} 
    = \Gamma(\theta_\mathrm{ab}) \frac{h_\mathrm{c}^2(\nu)}{12\pi^2 \nu^3}
    = \Gamma(\theta_\mathrm{ab}) \frac{1}{12\pi^2 \nu^3} \left[ A \left( \frac{\nu}{1\,\mathrm{yr}^{-1}} \right)^{\alpha} \right]^2
    = \Gamma(\theta_\mathrm{ab}) \frac{A^2}{12 \pi^2} \left( \frac{\nu}{1\,\mathrm{yr}^{-1}}\right)^{-\gamma}\,\mathrm{yr}^3 \,,
\end{equation}
where $\Gamma(\theta_\mathrm{ab})$ (known as the overlap reduction function) stands for spatial correlation factor provided the angular separation $\theta_\mathrm{ab}$ between the two pulsars; $A$ and $\gamma$ represents the achromatic amplitude and spectral index, respectively; $h_\mathrm{c}(\nu) \propto A \nu^\alpha$ refers to the characteristic GW strain. It is easy to see that $\gamma=3-2\alpha$.
Assuming the GWB (at nHz) is isotropic, the $\Gamma(\theta_\mathrm{ab})$ should follow the relation 
\begin{equation}
    \Gamma_\mathrm{GWB}(\theta_\mathrm{ab}) = \frac{1}{2} - \frac{1}{4}\left( \frac{1-\cos{\theta_\mathrm{ab}}}{2}\right) + \frac{3}{2} \left(\frac{1-\cos{\theta_\mathrm{ab}}}{2}\right) \ln{\left(\frac{1-\cos{\theta_\mathrm{ab}}}{2}\right)}
\end{equation}
\citep{Hellings83}.
To sum up, to detect the GWB at nHz with a PTA requires the following conditions to be met:
\begin{itemize}
    \item the power spectral densities of all pulsars of a PTA have to follow the same scaling relation $S \propto \nu^{-\gamma}$;
    \item $\gamma$ should be largely consistent with 13/3 \citep{Phinney01,Jaffe03};
    \item $\Gamma(\theta_\mathrm{ab})$ should generally agree with $\Gamma_\mathrm{GWB}(\theta_\mathrm{ab})$. 
\end{itemize}

So far, the first two conditions have been met, with a ``common process'' (i.e., common-$\gamma$ steep-spectrum signals, or ``red noise'', in the residual ToAs) identified by all major PTAs \citep{Arzoumanian20,Goncharov21,Chen21,Antoniadis22}. 
However, neither $\Gamma(\theta_\mathrm{ab}) = \Gamma_\mathrm{GWB}(\theta_\mathrm{ab})$ nor $\Gamma(\theta_\mathrm{ab}) \neq \Gamma_\mathrm{GWB}(\theta_\mathrm{ab})$ is suggested by any PTA analysis. In other words, a GWB detection with PTAs is not reached or ruled out.
%Assuming $\gamma=13/3$, 

\subsection{Enhancing pulsar timing array sensitivities with VLBI astrometry of millisecond pulsars}
\label{subsec:VLBI_PTA}

Looking into the future, the prospects of detecting the stochastic GWB at nHz are bright. \citet{Pol21} projects that, in 2--5 years time, sufficient timing data might have been accumulated to claim the first detection of the GWB. The S/N of the GWB detection would then improve slowly with $t^{1/2}$ \citep{Siemens13}. To accelerate the PTA sensitivity enhancement, the best strategy is to keep adding newly discovered MSPs to the PTA \citep{Siemens13}.
However, this strategy has a weak point.
For an MSP with a short (i.e., $\lesssim3000$ days) time baseline, current timing fitting technique might be incapable of stopping the red noise (i.e., steep-spectrum timing noise) from being absorbed into the timing model \citep{Madison13}, which would not only bias the timing model, but also weaken the GWB signals residing in the red timing noise.
One way to mitigate this unwanted absorption is to incorporate independent astrometric estimates (including proper motion and parallax) into timing fitting \citep{Madison13}.
%In timing analysis, despite the efforts to perform parameter fitting in the presence of the red noise \citep{Coles11}, better fitting results can be achieved if we have prior knowledge of astrometric parameters, especially for MSPs with $<$3000\,day timing data; this would significantly improve the sensitivity of PTAs \citep{Madison13}.

Very long baseline interferometry (VLBI) can achieve mas level resolution at cm band and $\mu $as level at sub-mm band. By sampling sky positions of an MSP and modeling its position variations, model-independent astrometric measurements (including reference position, proper motion and parallax) can be made. This method is known as VLBI astrometry.
It takes a relatively short time ($\approx$2 years) to acquire accurate VLBI astrometric parameters for Galactic pulsars \citep{Chatterjee09,Deller19}. In comparison, pulsar timing normally takes $\sim10$\,yr to acquire parallax as precise as that obtained with VLBI astrometry across 2 years. 
Hence, the proper motion and parallax determined for a new MSP with VLBI astrometry can be used in the timing analysis of the MSP, which can prevent its red noise from being absorbed into its timing model \citep{Madison13}, and accelerate the GWB S/N improvement \citep{Siemens13}.
In light of the high-sensitivity radio telescopes that are either recently commissioned (e.g., the MeerKat telescope, \citealp{Bailes20}, and the FAST telescope, \citealp{Nan11}) or planned (e.g. the SKA, \citealp{Dewdney09}), more relatively faint MSPs will be discovered in the coming years, which will supercharge the PTA research. 
Accordingly, VLBI astrometry of these new  MSPs will play a more important role in a PTA campaign based on a high-sensitivity telescope \citep[e.g.][]{Bailes20}. 
%By achieving high-precision parallaxes and proper motions of MSPs with VLBI astrometry, various tests for theories of gravity have been done \citep[e.g.][]{Deller08,Deller09,Deller18}.

\subsection{Testing Solar system ephemerides with VLBI astrometry of millisecond pulsars}
\label{subsec:SSE_VLBI}
In order to incorporate VLBI astrometric results into timing analysis, the offset between the reference frames used by pulsar timing and VLBI needs to be taken into account \citep{Chatterjee09,Madison13}. The radio telescopes on Earth are essentially in a non-inertial frame reference frame due to orbital motions around the Solar System Barycenter (SSB). To connect ToAs observed at different time, the ToAs need to be converted to the SSB frame. This conversion is made, to the first order, by calculating and correcting the Roemer delay (i.e., the light travel time difference between pulsar-Earth and pulsar-SSB). Unlike pulsar timing carried out in the SSB frame, VLBI astrometry is performed with respect to quasi-static quasars located in the International Celestial Reference Frame (ICRF, \citealp{Charlot20}) or the Radio Fundamental Catalogue\footnote{\url{http://astrogeo.org/}}, both of which can be considered inertial reference frames.

Presumably, the offset between the SSB frame and the ICRF frame (or alternatively the RFC frame) is reasonably small. Therefore, VLBI astrometric results have been directly incorporated into timing analysis \citep{Guo21,Kramer21a}, and vice versa \citep[e.g.][]{Ding21a} for the long-timed MSPs. 
However, %the other reason that the frame rotation (a more technical way of saying frame offset) is not yet taken into account in published pulsar astrometry work 
to have not considered the frame offset (in published pulsar astrometry work) is partly because the frame offset is still poorly constrained due to the less precise absolute VLBI position of MSPs \citep{Wang17}. 
A good understanding of the frame offset, or frame rotation, would be essential for quantifying and removing the error introduced by directly applying VLBI astrometric results to timing analysis (or vice versa). 

In a similar vein, the conversion of ground-based ToAs to the SSB requires a solar-system ephemeris (SSE) to inform where the Earth is located with respect to the SSB. 
Various realizations of SSEs have been presented by different manufacturers, which normally give consistent relative positions. It is found that SSB-to-Earth positions predicted by recent SSEs only differ by $\sim100$\,m \citep{Arzoumanian18}, which corresponds to $0.3\,\mu$s time delay.
However, the PTA has become sensitive enough to not tolerate the $0.3\,\mu$s time difference: it would significantly affect the constraints on the GWB \citep{Arzoumanian18}.
At present, SSE is one of the limiting factors against the GWB search with PTAs \citep{Tiburzi16,Vallisneri20}.
The median timing residual achieved for the 10 most prioritized NANOGrav MSPs is $0.2\,\mu$s (see Table~4 of \citealp{Arzoumanian20}), which implies that SSE-induced errors have to be suppressed to the $\lesssim0.1\,\mu$s level for the detection of a GWB signature.

Recent NANOGrav works have largely taken into account the error in SSE with the pioneering {\tt BAYESEPHEM} framework \citep{Arzoumanian18}, though unavoidably leading to larger parameter uncertainties \citep{Arzoumanian20}.
Given the new framework, independent checks on various SSEs would provide the Bayesian analysis with extra information, thus reducing the timing parameter uncertainties. 
Such a check can be made with VLBI astrometry of MSPs, as the absolute pulsar positions (as well as other astrometric parameters) of MSPs can be compared to the timing counterparts based on a specific SSE, and judge the accuracy of the SSE.
%Unfortunately, the frame rotation investigation based on the \mspsrpi\ catalogue (see \ref{subsec:mspsrpi} and \ref{subsec:mspsrpi_MSP_VLBI_astrometry}) is outside the scope of this thesis, but will be carried out in the near future.
The \mspsrpi\ results presented in \ref{ch:mspsrpi} lay the foundation for an accurate frame rotation investigation in the near future, once the MSP absolute positions have been refined (see \ref{sec:future_prospects}).

%This research eyes extremely precise VLBI  astrometry, and studies the effect of precisely measured astrometric parameters on reducing the red timing noise of MSPs and improve the sensitivity of PTAs. Besides this ultimate goal, we would also use improved results on individual MSPs to make constraints on gravitational science. Furthermore, astrometric results obtained for MSPs with VLBI can be compared to timing counterparts, thus providing clues for SSE elimination and frame tie between the two different reference frams used by VLBI and pulsar timing. At the moment, SSE is one of the limiting factors on searching for GWB with PTAs \citep{Vallisneri20}. An improvement of SSE facilitated by VLBI would directly lead to sharper PTA sensitivities.

%As new high-sensitivity radio telescopes are deployed (FAST, MeerKAT) or under development (SKA-low), more dimmer pulsars are expected to be found; new MSPs will join the PTAs, that would accelerate the discovery of \lf\ GW. 

\section{Advancing high-energy studies with neutron star astrometry}
\label{sec:XRB_astrometry}

As is mentioned in \ref{sec:NSs}, NSs are active X-ray and $\gamma$-ray sources. Precise astrometric information is important for high-energy studies of NSs in two aspects. Firstly, a precise distance is needed for calculating the luminosity at any frequency band or the fluence of a radio burst. Secondly, to estimate spin-down power (see \ref{eq:E_rot}) of a pulsar requires the determination of the intrinsic spin period decay $\dot{P}$, which demands the removal of the extrinsic spin period decay $\dot{P}^\mathrm{ext}$ that can be calculated with 
\begin{equation}
\label{Pdot_ext}
    \dot{P}^\mathrm{ext} = \frac{a_\mathrm{r}}{c} P
\end{equation}
and \ref{eq:A_r}. 
%Though NS astrometry has been made at X-ray \citep[e.g.][]{Kaplan08}, it is far from precise.
Hence, in conjunction with $\gamma$-ray observations, precise astrometry of NSs (made at radio or optical/infrared) can explore the lower limit of spin-down power that enables $\gamma$-ray radiation, which is known as the NS high-energy ``death-line''. Examples of the two aspects can be found in \ref{subsec:mspsrpi_J0610} and \ref{subsec:mspsrpi_J1730}. An extension of the first scientific aspect is the study of photospheric radius expansion bursts, which is outlined as follows.

\subsection{Probing models of photospheric radius expansion bursts with model-independent astrometry}
\label{subsec:PRE_bursts}
Low-mass X-ray binaries (LMXBs) consist of a compact object accreting material from a donor star (generally $\lesssim$1.5\,\msun) via an accretion disk. Black hole LMXBs (BH LMXBs, also known as microquasars) can be considered as scaled-down versions of Active Galactic Nuclei (AGNs) that evolve on much shorter timescales, thus providing essential insights into the accretion physics. Neutron star LMXBs (NS LMXBs), though not as akin to AGNs as BH LMXBs, are also of fundamental importance. By comparing the observational behaviors of BH LMXBs and NS LMXBs, one can potentially discriminate accretor-related phenomena from non-accretor-related ones. 

In the NS-LMXB family, there are hundreds of members \citep{Liu07}, which are sub-divided into two groups --- atolls and Z sources, according to their trajectories in the diagnostic color-color diagram (CCD) over the course of an outburst. Compared to Z sources persistently emitting at around the Eddington luminosity, atoll sources have lower accretion rates. Radio observations of LMXBs provide knowledge of disc-jet coupling, jet morphology and astrometry \citep[e.g.][]{Miller-Jones10}. The latter two motivations can be realized with VLBI observations. Extended radio features have only been confirmed in two of the bright Z sources, Sco~X$-$1 and Cir~X$-$1 \citep{Fomalont01,Fender98}, possibly due to their higher accretion rates and more powerful jets. Only one classical NS-LMXB, Sco~X$-$1, has a published trigonometric parallax \citep{Bradshaw99}. 
%while the transitional millisecond pulsar PSR~J1023+0038 (that accretes at much lower rates, when accreting at all) also has a trigonometric parallax \citep{Deller12}. 

Though various methods  have been employed to infer the distances to NS-LMXBs (e.g., using main-sequence companion-star spectral type and luminosity, \citealp{Chevalier99}), most NS-LMXB distances are estimated by using PRE bursts (see \ref{sec:NSs}) as standard candles. PRE bursts are a subset of type I X-ray bursts originated typically from NS X-ray binaries. The occurrence of PRE bursts requires matter from the outer layer of the donor star to be accreted onto the surface of the NS. When the base of the accumulated nuclear fuel is sufficiently hot, rapid nuclear burning is triggered. The local Eddington limit is met when the local radiation pressure outweighs gravity, leading to photospheric radius expansion (and contraction that follows). In the soft X-ray band, a PRE burst typically manifests itself as a double-peaked light curve \citep{Lewin93}. During the process of photospheric radius expansion and contraction, it is believed the luminosity stays roughly constant at the Eddington luminosity, thus making PRE bursts a useful tool for distance measurement. Not every NS-LMXB exhibits PRE bursts, such as Sco~X$-$1 \citep{Galloway08}. To date, PRE-based distances have been estimated for 73 out of the total 115 PRE bursters (see the MINBAR source catalog\footnote{\url{https://burst.sci.monash.edu/sources}}, \citealp{Galloway20}).

The precision of PRE-based distances is mainly limited by two factors. Firstly, \citet{Galloway08} suggests a 5-10\% variation in PRE luminosities in any individual source. Secondly, the Eddington luminosity varies with the composition of the ``nuclear fuel'' accumulated on the NS surface; it is a factor of 1.7 higher for pure helium than for the solar composition. This translates to a factor of 1.3 difference in distance. So the two uncertainty sources would jointly lead to $\approx$40\% distance uncertainty.
Therefore, a precise model-independent distance of a PRE burster holds the key to constraining the composition of accumulated nuclear fuel on the NS surface and checking the validity of the PRE-burst standard candle, as is investigated in \ref{sec:PRE_discussions}
%Once the composition of accreted materials is ascertained, the residual offset between a PRE distance and a geometric distance of a PRE burster can be used to refine the oversimplified model of PRE bursts, which assumes the PRE burst emissions are spherically symmetric.
with the Gaia Early Data Release 3 (EDR3) data.  
%Another project is long-term VLBA astrometry of \aql, a well studied X-ray burster. Though using different facilities, the ultimate goal of this branch is the same - to calibrate the widely used model of PRE bursts that is probably oversimplified.

\section{Thesis structure and norms}
\label{sec:thesis_norms}

\subsection{Thesis structure}
\label{subsec:thesis_structure}
This thesis entails 6 paper chapters submitted to journals at different time of the PhD program. These 6 chapters are grouped by the neutron star divisions into three --- magnetars, neutron star X-ray binaries and MSPs. The three groups are arranged in the order of typical NS age.
The first group (i.e., \ref{ch:J1810} and \ref{ch:J1818}) presents VLBI astrometry of two radio magnetars.
The second group (i.e., \ref{ch:PRE}) delivers the study of PRE bursters using Gaia astrometry.
The third group consists of three chapters (i.e., \ref{ch:J1012}, \ref{ch:B1534} and \ref{ch:mspsrpi}) related to MSPs, based on the results from the \mspsrpi\ project.
The paper chapters follow this ``Introduction'' chapter and the ``Methodology'' chapter (see \ref{ch:methodology}), and precede the ``Conclusion'' chapter (see \ref{ch:conclusion}). 
While both \ref{ch:Introduction} and \ref{ch:methodology} essentially serve as introduction to the paper chapters, \ref{ch:Introduction} provides the interconnected scientific context, and \ref{ch:methodology} describes the technical foundation of this thesis.

\subsection{Thesis norms}
\label{subsec:thesis_norms}
\begin{itemize}
    \item As this thesis consists of paper chapters written at different stages of the PhD program, the mathematical symbols in the paper chapters are not unified. Hence, each chapter follows its own mathematical symbol system. In other words, no mathematical symbol is inherited from preceding chapters. The same rule applies to abbreviations.
    \item To convenience readers, each chapter has its own list of references.
    \item To better convey the logical flow between topics, a generalized title different from the original paper name is given to each paper chapter,  which is clarified in the beginning (of each paper chapter).
    %To better convey the logical flow between topics, some chapter titles have been given a different name to that of the journal article which form the basis of that chapter.  In these cases, a clarification is provided at the beginning.
\end{itemize}

\bibliographystyle{mnras}
\bibliography{haoding}

\chapter[Methodology]{Methodology}
\label{ch:methodology}

% This can be used to include a quotation if desired.
%\begin{center}\parbox{11cm}{\begin{center}
%\textit{To be, or not to be, that is the question.}
%
%---\href{http://en.wikipedia.org/wiki/Hamlet}{Hamlet}
%\end{center} } \end{center}

%For example \citep{YorkEtal2000}.

%\section{Purpose of the Thesis}

%\lipsum[1-9]

%\section*{A roadmap to precise astrometry}
%\label{sec:roadmap_to_astrometry}

This chapter expands upon \ref{sec:astrometry101}, and briefly explains the procedure through which precise astrometry is realized. Arranged in the order of the astrometric workflow, the chapter involves both VLBI (very long baseline interferometry) and Gaia astrometry. The latter is not covered until \ref{sec:astrometric_validation}, as Gaia astrometric parameters are directly offered in Gaia data releases. 
VLBI is a vast topic. Hence, only a small fraction of VLBI techniques directly relevant to this thesis is introduced in this chapter. 

\section{Radio interferometry}
\label{sec:VLBI_in_a_nutshell}

Radio interferometry is a technique to combine an array of radio telescopes into a synthesized one with a large radius \citep[e.g.][]{Ryle52}, which is a development of the Michelson experiment at optical wavelengths \citep{Michelson21}. 
Thanks to the large radius of the synthesized telescope, the spatial resolution of ground-based very long baseline interferometry (VLBI) observations can reach $\sim25\,\mu$as at mm-band \citep{Akiyama19}, where a baseline refers to the distance between two constituent telescopes of the VLBI array.
The main challenge of VLBI synthesis is to align the wave fronts of electromagnetic signals received at different telescopes (which is a procedure technically referred to as phase calibration). Better alignment would achieve lower noise level and less distortion of the obtained synthesized image. 

In theory, the wavefront alignment is achieved by estimating and compensating for the inter-telescope time delay of electromagnetic plane waves from a remote source.
The time delay $\tau$ (between different telescopes) depends on the sky position $\vec{r} = \vec{r}_0 + \vec{\sigma}$, following the relation
\begin{equation}
    \label{eq:ch2_tau_g}
    \tau =  \frac{\Vec{D} \cdot ({\Vec{r}}_0 + \Vec{\sigma} )}{c} + \tau_\epsilon \,,
\end{equation}
where the first term can be understood as the light-travel time delay due to geometric
path-length difference;  $\Vec{D}$ and ${\Vec{r}}_0$ stand for the inter-telescope displacement and the unit vector pointing to a chosen reference sky position, respectively; 
$\Vec{\sigma}$ represents a small angular displacement from $\Vec{r}_0$;
$\tau_\epsilon$ accounts for additional error sources such as atmospheric propagation effects and clock errors. 
More specifically, $\tau$ can be considered a function of $\vec{\sigma}$ and time $t$. 
%is time-varying, as both $\vec{D}$ and $\tau_\epsilon$ are expected to change with time.
As the only unknown parameter of $\tau$ that needs to be determined with phase calibration,
$\tau_\epsilon$ fluctuates with time due to atmospheric turbulence and clock instabilities, and changes with direction due to anisotropic atmospheric effects (see \ref{subsec:ch2_interpolation}). 
Apart from $\tau_\epsilon$ being time-varying, $\vec{D}$ also changes with time as the Earth rotates, but in a predictable way.

Before the calibration, the first step is to measure the coherence between the radiation received at different telescopes using a correlator such as {\tt DiFX} \citep{Deller11a} and {\tt SFXC} \citep{Keimpema15}, which is briefly explained as follows.   
%To measure the coherence between the radiation received at different telescopes, from which important source parameters such as the position and flux density can be extracted, requires the use of a correlator
For quasi-monochromatic electromagnetic plane waves from a remote radio source,
the electric field component of the plane waves can be formulated as 
\begin{equation}
  \label{eq:ch2_monochromatic_plane_wave}
    E(t) = \Re \left[ B(t) \mathrm{e}^{i2\pi \nu_0 t}\right] \,,
\end{equation}
where $\Re$ denotes the real part of a complex function;
$\nu_0$ represents the frequency of the electromagnetic plane waves; $B(t)$ is a complex random process \citep{Romney95}. Hence, the cross-correlation between $E_{1}(t)$ and $E_{2}(t)$ received at two telescopes equals to 
\begin{equation}
    \label{eq:ch2_correlation_Em_En}
    \begin{split}
    \Gamma(\tau) &= \Re \left[\int_{-\infty}^{\infty}E^{*}_\mathrm{1}(t) E_\mathrm{2}(t+\tau) \text{d}t \right]\\
    &=\Re \left[\mathrm{e}^{i2\pi \nu_0 \tau} \int^{\infty}_{-\infty} B^{*}(t) B(t+\tau) \text{d}t \right] \,,
    \end{split}
\end{equation}
where $B^{*}(t)$ and $E_1^{*}(t)$ denote the complex conjugates of $B(t)$ and $E_1(t)$, respectively. 
The $\mathrm{e}^{i2\pi \nu_0 \tau}$ is known as the fringe term of the correlation function, which is proportional to the complex form of $\Gamma(\tau)$.
%and is modulated by the amplitude and phase of $\int^{\infty}_{-\infty} B^{*}(t) B(t+\tau) \text{d}t$.
The $\tau$ of the fringe term can be interpreted as the time delay between two telescopes, which is formulated as \ref{eq:ch2_tau_g}.
%The term ``fringe'' of radio interferometry is adopted from the double-slit experiment. Just as the fringe pattern on the screen reveals the information of the light source in the double-slit experiment, the fringes obtained during the phase calibration can be used to infer the visibility phase.

On the other hand, the power received from the radio sky is proportional to $A(\Vec{\sigma}) I(\Vec{\sigma}) \text{d}\nu \text{d}\Omega$, where $A(\Vec{\sigma})$ and $I(\Vec{\sigma})$ stand for effective antenna collecting area and sky brightness distribution, respectively.
As $\Gamma(\tau)$ is expected to be proportional to both the received power and the fringe term \citep{Thompson17},
\begin{equation}
    \label{eq:ch2_r_I_relation}
    \begin{split}
    \Gamma &\propto \Re \left[ \int  A(\Vec{\sigma}) I(\Vec{\sigma}) \mathrm{e}^{i2\pi \Vec{D}_\lambda \cdot  (\Vec{r}_0 + \Vec{\sigma}) + i2\pi \nu_0 \tau_\epsilon } \text{d}\Omega \right] \\
    &= \Re \left[  \mathrm{e}^{i2\pi (\Vec{D}_\lambda \cdot \Vec{r}_0 + \nu_0 \tau_\epsilon )} \int A(\Vec{\sigma}) I(\Vec{\sigma}) \mathrm{e}^{i2\pi \Vec{D}_\lambda \cdot \Vec{\sigma}} \text{d}\Omega \right] \,,
    \end{split}
\end{equation}
where $\Vec{D}_\lambda = \Vec{D} \nu_0/c$; the operation of integration assumes that $A(\vec{\sigma}) I(\vec{\sigma})$ is not spatially coherent (which is a condition satisfied for almost all radio astronomy sources).
In the calculation, $\tau_\epsilon$ is considered direction-independent, as the VLBI field of view is extremely small.
%which is a common practice in phase calibration; when practicing phase referencing (see \ref{sec:ch2_VLBI_observing_strategies}), the assumption becomes increasingly valid as the calibrator-target separation gets smaller. Nevertheless, strictly speaking, $\tau_\epsilon$ is direction-dependent due to anisotropic atmospheric effects, therefore meriting the use of the interpolation phase calibration strategy (see \ref{subsec:ch2_interpolation} and \ref{subsec:J1810_dualphscal}).
In \ref{eq:ch2_r_I_relation}, the $\mathrm{e}^{i2\pi \Vec{D}_\lambda \cdot \Vec{r}_0}$ term represents the model geometric light-travel delay (towards the reference sky position $\vec{r}_0$) that changes as the Earth rotates; the $\mathrm{e}^{i2\pi \nu_0 \tau_\epsilon}$ term stands for the additional time delay to be estimated and corrected in the post-correlation calibration; the other term, normally known as the visibility function, contains the sky brightness information.
By defining the visibility as
\begin{equation}
    \label{eq:ch2_visibility}
    V = |V| \mathrm{e}^{i\phi} \propto \int A(\Vec{\sigma}) I(\Vec{\sigma}) \mathrm{e}^{-i2\pi \Vec{D}_\lambda \cdot \Vec{\sigma}} \text{d}\Omega
\end{equation}
\citep{Thompson17}, one obtains
\begin{equation}
    \label{ch2_r_V_relation}
    \Gamma \propto \Re \left[ |V| \mathrm{e}^{i(2\pi \Vec{D}_\lambda \cdot \Vec{r}_0 + 2 \pi \nu_0 \tau_\epsilon - \phi)} \right] = |V| \cos{(2\pi \Vec{D}_\lambda \cdot \Vec{r}_0 + \phi_\epsilon - \phi)} \,,
\end{equation}
where $\phi$ and $|V|$ are, respectively, the phase and amplitude of the visibility function; $\phi_\epsilon=2 \pi \nu_0 \tau_\epsilon$.

%After $\Gamma$ is derived, the pre-determined geometric phase difference between the telescopes (i.e., $2\pi \Vec{D}_\lambda \cdot \Vec{r}_0$) is calculated and removed from $\Gamma$ with a correlator. Namely, the visibility output of a correlator is proportional to $|V|\cos{(\phi_\epsilon - \phi})$.
While calculating $\Gamma$, a correlator usually simultaneously performs high speed digital signal processing to remove the changing geometric delay (i.e., $2\pi \Vec{D}_\lambda \cdot \Vec{r}_0$), hence providing an output proportional to $|V|\cos{(\phi_\epsilon - \phi})$.
As can be seen from \ref{eq:ch2_visibility}, the visibility function is the Fourier transform of $A(\Vec{\sigma})I(\Vec{\sigma})$.
%, hence being equivalent to $A(\Vec{\sigma})I(\Vec{\sigma})$. 
%VLBI data are normally recorded and reduced in the visibility format. 
Hence, $A(\Vec{\sigma})I(\Vec{\sigma})$ can be recovered via an inverse Fourier transform from the  visibility data after this has been generated (with a correlator) and calibrated.
Accordingly, the phase calibration discussed in this thesis refers to the calibration of the visibility phase (hereafter also simply referred to as phase).
%the visibility output of a VLBI correlator can be saved and processed prior to the conversion to a VLBI image.
To make VLBI images $I(\Vec{\sigma})$, a deconvolution process is required to account for the limited sampling of the visibilities (see Lectures 7 and 8 of \citealp{Taylor99} for a detailed description of VLBI imaging).

Apart from the basic functions, a correlator can potentially fulfill other tasks including pulsar gating \citep[e.g.][]{Benson95,Brisken10,Keimpema15,Smits17}. Based on the pulse ephemeris of a pulsar, VLBI data for the pulsar can be folded using a correlator equipped with the pulsar gating feature. Subsequently, the off-pulse part (or sometimes the on-pulse part, depending on the scientific goal) of the VLBI data can be discarded, in order to increase the S/N of the pulsar (or to investigate the properties of the off-pulse emission, if present).

Following VLBI correlation, data reduction related to phase calibration estimates and removes $\phi_\epsilon$ from the correlator output in an incremental manner using tools like {\tt AIPS} \citep{Greisen03} and {\tt CASA} \citep{McMullin07}, so that the corruption of the sky brightness information by $\phi_\epsilon$ can be minimized.
The data reduction, though having a largely standard procedure, can be flexible to meet various observational needs, as long as it is in line with the observation setup.
For high-precision VLBI astrometry, the calibration of visibility phase is the most crucial part of the data reduction. Therefore, this chapter would only focus on phase calibration. 
%Due to the astrometric nature of this thesis, this chapter only covers the phase calibration strategies of VLBI astrometry. 
Other VLBI calibration steps adopted in this thesis are the same as \citet{Deller19}, hence not reiterated any more.

\section{Phase calibration strategies of VLBI astrometry}
\label{sec:ch2_VLBI_observing_strategies}

As aforementioned, phase calibration is a procedure to approach and remove phase offset $\phi_\epsilon$ incrementally, leaving only $\phi$ (in the visibility data) that is related to the sky brightness distribution.
The total phase offset equals to 
\begin{equation}
\label{eq:phase_offset}
\phi_\epsilon = \sum_{j} \phi_{\epsilon j}\ \ \ \ \  (j=1,2,...)\,,
\end{equation}
where $\phi_{\epsilon j}$ is the phase offset increment corrected at the $j$-th phase calibration step.
In data reduction, the solutions of $\phi_{\epsilon j}$, its time derivative (aka. rate) $\partial{\phi_{\epsilon j}}/\partial{t}$ and frequency derivative (aka. delay) $\partial{\phi_{\epsilon j}}/\partial{\nu}$ can be derived for bright sources using the global fringe fitting technique \citep{Schwab83}.
%, which works in a way analogous to correlators (see \ref{sec:VLBI_in_a_nutshell}). 
For a target that is not  bright enough for global fringe fitting, the so-called phase referencing technique \citep[e.g.][]{Lestrade90,Beasley95} can be used to improve the image S/N of the target. In the phase referencing configuration, a bright phase calibrator close to the target (on the sky) is observed alongside the target. At data reduction, the phase solutions obtained with the calibrator is applied to the target. As the calibrator is close to the target, the $\phi_{\epsilon j}$ solutions of the calibrator serve as good approximation to the $\phi_{\epsilon j}$ solutions of the target, hence improving the imaging quality of the target. %Accordingly, the imaging quality would improve with smaller angular separation between the target and the reference source, in that the phase solutions of the target are better approximated.

Meanwhile, as the solutions are applied to both the calibrator and the target, the target becomes tied to the calibrator. 
%Specifically, For a target source, the offset sigma is the quantity of interest to be determined, and hence tau_e must be derived, in the process called calibration.  This can be performed with a "calibrator" source of known position, for which sigma is \sim zero, and hence a measured delay can be used to calculate tau_e.
Consequently, the target position (measured in this way) would reflect the relative position between the two sources (i.e., changing the reference position of the calibrator would shift the target position accordingly), which forms the basis of relative VLBI astrometry.
Differently put, the target is phase-referenced (or simply referenced) to the phase calibrator, as sky position is equivalent to visibility phase.
In relative astrometry, adopting different reference positions for a reference source would only affect the reference position of the target, but not other astrometric parameters such as proper motion and parallax.
All the high-precision VLBI astrometric campaigns included in this thesis are variations of relative astromtry. 
It is noteworthy that a reference source is needed to carry out relative astrometry even when the target is already bright enough for global fringe fitting. However, the reference source would not be mandatory for the imaging of the bright target.
%In other words, the target is phase-referenced (or simply referenced) to the reference source.
%After correlation, the phase errors of VLBI data are dominated by propagation effects due to atmospheric turbulence.
Typical relative astrometry uses a remote quasar (whose position can be assumed to be constant on the sky, although breakdowns in this assumption are discussed below and explored in \ref{sec:mspsrpi_inference_with_priors}) as the reference source, while the target can be either an extragalactic source (such as a quasar, e.g. \citealp{Hada11}, or a fast radio burst, e.g. \citealp{Marcote17}) or a Galactic radio source (as is the focus of this thesis).

Phase referencing enables precise determination of the astrometric parameters for a Galactic source. However, the technique is not perfect: the astrometric precision is limited by two factors. Firstly, the astrometric uncertainties are dominated by systematic errors due to anisotropic propagation effects between the calibrator and the target. Accordingly, using a calibrator closer to the target (on the sky) can reduce the astrometric uncertainties \citep[e.g.][]{Chatterjee04,Kirsten15,Deller19}. However, such a close and bright reference source may not be always available, especially when observing in scatter-broadened (see \ref{subsec:scattering_screen} for explanation) sky regions at relatively low frequencies ($\lesssim4$\,GHz). Secondly, the phase calibrator needs to be compact and stable over the course of the astrometric campaign. Otherwise, additional astrometric uncertainties might be introduced into the astrometric parameters (see \ref{subsec:ch2_VLBI_check_src}).
These two limiting factors of astrometric precision motivate the advanced phase referencing strategies described as follows.

%High-precision VLBI astrometry is realized in the manner of relative astrometry,  
%The simplest and most widely used tactic of VLBI relative astrometry is phase referencing \citep[e.g.][]{Lestrade90,Beasley95}, where a relatively weak target is phase-referenced to a bright source (that is close to the target on the sky). 
%The solutions are subsequently passed to the target.  

%Ultimately, phase referencing is an imaging strategy that can improve the image S/N of the weak source, as the phase solution becomes more accurate. 

%\subsection{Phase referencing}

\subsection{Phase calibration relay and in-beam astrometry}
\label{subsec:ch2_in_beam_astrometry}

In a typical phase referencing experiment, the telescopes of the VLBI array observe the target and the phase calibrator in an alternating fashion.
To further reduce the propagation-related systematic uncertainties due to relatively large angular separation between the reference source and the target, 
the strategy of phase calibration relay (or relayed phase calibration) can be adopted, where
a relatively faint secondary phase calibrator that is closer to the target (compared to the primary calibrator) is used as the ultimate phase reference source. 
In the phase calibration relay, the residual phase $\phi_{\epsilon j}$, $\partial{\phi_{\epsilon j}}/\partial{t}$ and $\partial{\phi_{\epsilon j}}/\partial{\nu}$ are first obtained with the primary phase calibrator using global fringe fitting. The solutions are then applied to both the primary and the secondary phase calibrators.
As the delay/rate difference between the primary and secondary phase calibrators is generally negligible, the secondary phase calibrator is normally used to only search for frequency-independent residual phase $\phi_{\epsilon k}$ (where $k=2,3,...$ and $k>j$) using the self-calibration technique (also known as the hybrid mapping technique, e.g. \citealp{Readhead78}).
Compared to the global fringe fitting that fits $\phi_{\epsilon j}$ for each baseline separately, the self-calibration technique incorporates the data of all baselines for the $\phi_{\epsilon k}$ search. As a result, secondary calibrator fainter than allowed by the global fringe fitting technique can be used, which is the crux of the phase calibration relay strategy.

When the target and the reference source are close enough to be contained in the primary beam of each constituent telescope of a VLBI array, the two sources can be observed simultaneously, which saves the overhead time required for slewing. 
%These in-beam secondary calibrators are simply known as in-beam calibrators.
%but also substantially reduces the propagation-related systematic uncertainties thanks to relatively small angular separation (between the target and the reference source). 
This special observing tactic of phase referencing is known as in-beam phase referencing. Accordingly, astrometry carried out with in-beam phase referencing is referred to as in-beam astrometry.
The main challenge of in-beam astrometry is to find suitable in-beam calibrators (IBCs).
At the observing frequency of around 1.5\,GHz, IBCs have been systematically searched and identified in previous pulsar astrometric surveys (\citealp{Chatterjee09,Deller19}, also see \ref{subsec:VLBI_astrometry}), and for individual targets \citep[e.g.][]{Li18,Kramer21a}. \citet{Deller19} found that IBCs can be found for a vast majority of pulsars (or targets of any kind)  at 1.5\,GHz using the Very Long Baseline Array (VLBA) formed with 10 equally sized (25-m) radio telescopes. Suitable IBCs, identified in pilot observations using the multi-field observing tactic developed by \citet{Deller13a,Deller19}, were found for all \mspsrpi\ pulsars presented in this thesis.

Despite the great success of the 1.5\,GHz VLBA in-beam astrometric surveys, the chance of finding suitable IBCs gets smaller at higher observing frequencies, as the telescope primary beams shrink. Even at low frequencies, in-beam astrometry becomes difficult when a large telescope or phased array (that has a relatively tiny field of view) joins the VLBI array, or when the sky region of interest is highly scatter-broadened (and hence all potential IBCs are angular-broadened and become fainter, see \ref{subsec:scattering_screen}). In these challenging situations, alternatives to the in-beam phase-referencing strategy shall be considered.

\subsection{Inverse phase referencing}
\label{subsec:ch2_inverse_phase_ref}

%In the scenario where the target is brighter than the reference source, the normal phase referencing tactic is probably not optimal. Instead, inverse phase referencing should be considered.
Inverse phase referencing is a widely used  strategy of relative astrometry \citep[e.g.][]{Imai12,Yang16,Deller19}, which is applicable when the target is brighter (and ideally more compact) than the reference source.
Unlike normal phase referencing (that uses the reference source to search for phase solutions), inverse phase referencing uses the bright target to fit for phase solutions, which allows the reference source to be much fainter (compared to normal phase referencing). %Inverse phase referencing is a widely used technique
Hence, the chance of finding faint in-beam reference source is much larger than normal in-beam phase referencing \citep[e.g.][]{Li18}. 

When the target is bright enough for self-calibration but not sufficiently bright for global fringe fitting, relayed phase calibration has to be applied, where the target serves as the second phase calibrator.
When the phase solutions obtained with the target is applied to both the target and the reference source, the target is moved to the position of its assumed model, and the reference source position, again, reflects the relative position between the two sources. As a result, the reference source, which is normally a quasar, absorbs the unmodelled sky motion of the target. Assuming the reference source is static on the sky and does not have model evolution, the apparent sky motion of the reference source relative to the target can be fully attributed to the unmodelled target motion, which is the inverse of the apparent motion of the reference source. An example of relayed inverse phase referencing is the VLBA astrometry of \psrkb\ detailed in \ref{subsec:mspsrpi_sophisticated_data_reduction}.
In the scenario where all the aforementioned advanced phase referencing strategies are not applicable, the generally more resource-consuming 1D/2D interpolation strategy is the final resort of high-precision relative astrometry.

\subsection{1D/2D interpolation}
\label{subsec:ch2_interpolation}

1D/2D interpolation (hereafter referred to as interpolation) is an advanced observing and data reduction strategy of phase referencing that has only been realized in a few occasions (e.g. \citealp{Fomalont03,Doi06,Rioja17,Ding20c,Ding23a,Hyland22}, also see \ref{subsec:mspsrpi_dualphscal}). Unlike the aforementioned advanced phase referencing strategies that can only phase-reference the target to one reference source each time, interpolation phase-reference the target to multiple phase calibrators. 
In its most basic and to-date used form, interpolation assumes that, in the sky region containing the target and the phase calibrators (used by interpolation), the residual phase solutions $\phi_{\epsilon k}$ changes linearly with sky position, i.e., $\nabla \phi_{\epsilon k}$ does not change with sky position (see \ref{subsubsec:mspsrpi_implications_for_1D_interpolation} for the test of this assumption). Therefore, the phase solution at the rough position of the target can be interpolated from the phase solutions of at least 3 phase calibrators around the target, or from the phase solutions of 2 calibrators that are quasi-colinear with the target. The former and latter kind of interpolation is known as 2D and 1D interpolation, respectively.
The mathematical formalism for 1D interpolation is detailed in \ref{subsec:J1810_dualphscal}. 2D interpolation can be considered two operations of 1D interpolation. In this regard, the mathematical formalism of interpolation is not reiterated.

Without exception, all previous realizations of interpolation were made by observing the phase calibrators and the target alternatively, which limits the observing time on each source, not to mention that a lot of time is used to switch sources. Hence, interpolation is traditionally the most observing-time-consuming phase-referencing strategy.
Furthermore, a full target-calibrators cycle has to be completed within the coherence time limited by atmospheric turbulence evolution \citep[e.g.][]{Marti-Vidal10a}, in order to make sure the phase solutions obtained with the calibrators are timely enough to approximate the target phase. This constraint limits the use of interpolation at $\gtrsim15$\,GHz where the coherence time is typically $\lesssim2$\,min; in contrast, the coherence time at $\lesssim8.4$\,GHz is relaxed to around 4\,min \citep[e.g.][]{Marti-Vidal10}. 
%coherence time \citep{Marti-Vidal10}.
To overcome the observing time shortage, the Multi-View observing strategy has been proposed \citep{Rioja17} for interpolation to observe different sources simultaneously with different subsets of a VLBI array. Alternatively, arrays with beam-forming capability (such as the SKA-low, \citealp{Dewdney09}) can potentially serve as ideal Multi-View facilities, though demanding much higher computational costs.

%\section{Data reduction of VLBI astrometry}

\section{Astrometric inference}
\label{sec:ch2_astrometry_inference}

%After VLBI data calibration, the image of a point-like target can be fitted to a 2-D Gaussian model, from which the target position and its statistical uncertainty (as opposed to systematic uncertainty, see \ref{sec:mspsrpi_data_reduction}) are extracted (using, for example, the {\tt JMFIT} task of the {\tt AIPS} package, \citealp{Greisen03}).
%the end product is a target position as well its uncertainty for each epoch.
After VLBI data reduction, the target position and its statistical uncertainty can be extracted from a model fitting in the image domain (using, for example, the {\tt JMFIT} task of the {\tt AIPS} package, \citealp{Greisen03}) or in the visibility domain (e.g. \citealp{Marti-Vidal14}). 
Compared to the fitting in the image domain, visibility-domain fitting offers theoretical advantages (since there is no need to approximate the interferometer point spread function with a 2D Gaussian distribution), but its uncertainty estimation depends sensitively on the data weights, which are difficult to calibrate accurately for VLBI data (leading to image plane fitting to be often preferred).
From an astrometric campaign across $\gtrsim2$ years, a series of $\sim10$ target positions would be collected, which reflect the sky position evolution due to proper motion, parallax, etc.
Astrometric inference is a procedure to determine astrometric parameters (including reference position at a given reference epoch, proper motion and parallax) from the position series.
Three inference methods are used in this thesis --- least-square fitting, bootstrap and Bayesian inference, which are thoroughly introduced and discussed in \ref{sec:mspsrpi_parameter_inference}.

In brief, 
%least-square fitting can barely recover the ``true'' error level if the given positional uncertainties are misestimated, while the two other methods can both potentially accommodate improper positional uncertainties and render astrometric parameters with reasonable uncertainties.  
the accuracy of parameter uncertainties inferred with least squares fitting is strongly dependent on the input positional uncertainties, while the other two methods can both potentially accommodate improper positional uncertainties and render astrometric parameters with reasonable uncertainties.
Between bootstrap and Bayesian method, the latter shows better reliability in a statistical study (see \ref{subsec:mspsrpi_Bayesian_as_major}), and is better at incorporating and utilizing complex information to constrain additional parameters beyond the ``canonical'' astrometric parameters. %(though being more resource-consuming).
Particularly, the Bayesian method is the best way (among the three inference methods) to constrain orbital motion for applicable targets \citep[e.g.][]{Deller16,Guo21}.
In comparison, the astrometric inference of Gaia position series for a specific star used to not take into account the effect of orbital motion \citep{Lindegren18,Lindegren21a}, which has led to non-optimal astrometric fitting performance for Gaia sources with observable orbital motion. However, this issue starts to be addressed for the recent Gaia Data Release 3, as orbital parameters have been added to the astrometric solutions for many applicable sources \citep{Halbwachs22}.
%Compared to the astrometric inference of VLBI to add and talk about Gaia inference as well.

\section{Astrometric validation and correction using check sources}
\label{sec:astrometric_validation}

%add VLBI also into the shell.
\subsection{Gaia parallax zero point correction}
\label{subsec:ch2_px_zero_point}

Since the commissioning of the Gaia space telescope in July 2014 \citep{Gaia-Collaboration16}, Gaia has provided trigonometric parallaxes for almost 1478 million sources \citep{Brown21}. However, the parallax measurements made by Gaia are not perfect. Specifically, the variations in the ``basic angle'' between the pointing of the two constituent telescopes of Gaia cannot be singled out from the parallax determination \citep{Butkevich17}, which contributes to systematic errors in parallaxes. As a result of the imperfect instrument setup, the Gaia parallaxes for optical sources at infinity distance shift away from zero. In this thesis, we refer to such Gaia parallaxes (of sources at infinity) as parallax zero-points.

Parallax zero-points were first systematically studied with Gaia Data Release 2 (DR2) by \citet{Lindegren18}. The authors determined the all-sky median parallax zero-point of DR2 to be $-29\,\mu$as, and proposed that parallax zero-point depends on the three parameters -- sky position, source magnitude and source color. The later Gaia Early Data Release 3 (EDR3) has linked the position dependence of parallax zero-point to dependence of parallax zero-point on the ecliptic latitude \citep{Lindegren21}. Furthermore, \citet{Lindegren21} offered an empirical solution for determining parallax zero-points, while humbly describing the derived parallax zero-point merely ``indicative''. Though this empirical solution of parallax zero-points has been largely supported from independent distance measurements \citep{Huang21,Ren21}, there are implications that the empirical solutions are inaccurate around the Galactic plane \citep{Ren21}.
Moreover, the uncertainty of the empirically derived parallax zero-point is still not available and hard to estimate. In \ref{subsec:PRE_s_pi0}, a novel method based solely on the nearby Gaia quasars (on the sky) will be introduced to not only determine the parallax zero-point, but also its uncertainty.

\subsection{Probing reference source structure evolution with check sources}
\label{subsec:ch2_VLBI_check_src}

Unlike Gaia parallaxes, there is no indication that VLBI parallax (or proper motion) measurements are systematically biased. 
However, there are strong evidences suggesting secular radio structure evolution in a fraction of the quasar population \citep[e.g.][]{Kovalev17,Perger18}. 
If a target happens to be phase-referenced to a quasar undergoing significant structure evolution, the resultant proper motion and parallax would likely be biased \citep[e.g.][]{Deller13}.
Such structure evolution in reference source can be astrometrically investigated with check sources, which is analogous to the Gaia parallax zero-point determination.
To conduct the investigation, the target, the main reference source and the check sources should be ideally close to each other on the sky. Otherwise, the large propagation-related systematic errors would preclude the detection of any astrometric effect due to structure evolution.
If multiple check sources show consistent (and significant) proper motions with respect to the main reference source, it would serve as a good indication for the structure evolution in the main reference source.
%This indication would become stronger with more check sources
Such an indication would become increasingly conclusive with larger number of check sources.
In \ref{sec:mspsrpi_inference_with_priors}, the structure evolution in the main reference sources of the \mspsrpi\ pulsars is probed with respective check sources (when available).

\bibliographystyle{mnras}
\bibliography{mybibliography,haoding}

\chapter[The First Magnetar Parallax]{The First Magnetar Parallax}
\label{ch:J1810}

\newcommand{\snra}{G11.0$-$0.0}

This chapter is adapted from \citet{Ding20c} entitled ``A magnetar parallax'', where the studied magnetar is \xte. 
Along with the astrometry of another radio magnetar \swift\ outlined in \ref{sec:J1818_astrometry_progress}, this chapter enriches the small sample of magnetar space velocities, which is essential for probing magnetar formation channels (see \ref{subsec:probe_magnetar_formation} and \ref{sec:J1818_intro}). 
In addition, \ref{subsec:J1810_dualphscal} outlines the 1D interpolation technique and lays out its mathematical formalism, which paves the way for other chapters using 1D interpolation (see \ref{sec:J1818_astrometry_progress} and \ref{subsec:mspsrpi_dualphscal}).
In response to the comments given by the thesis examiners, minor corrections are made to the text.

\section{Abstract}

\xte\ (J1810) was the first magnetar identified to emit radio pulses, and has been extensively studied during a radio-bright phase in 2003--2008. 
It is estimated to be relatively nearby compared to other Galactic magnetars, and provides a useful prototype for the physics of high magnetic fields, magnetar velocities, and the plausible connection to extragalactic fast radio bursts.
Upon the re-brightening of the magnetar at radio wavelengths in late 2018, we resumed an astrometric campaign on J1810 with the \textit{Very Long Baseline Array}, and sampled 14 new positions of J1810 over 1.3 years. 
The phase calibration for the new observations was performed with two phase calibrators that are quasi-colinear  on the sky with J1810, enabling substantial improvement of the resultant astrometric precision.
Combining our new observations with two archival observations from 2006, we have refined the proper motion and reference position of the magnetar and have measured its annual geometric parallax, the first such measurement for a  magnetar. The parallax of $0.40\pm0.05$~mas corresponds to a most probable distance $2.5^{\,+0.4}_{\,-0.3}$\,kpc for J1810.
Our new astrometric results confirm an unremarkable transverse peculiar velocity of $\approx200$\,\kmps\ for J1810, which is only at the average level among the pulsar population. The magnetar proper motion vector points back to the central region of a supernova remnant (SNR) at a compatible distance at $\approx70$\,kyr ago, but a direct association is disfavored by the estimated SNR age of $\sim3$\,kyr.

\section{Introduction}
\label{sec:J1810_intro}

Magnetars are a class of highly magnetized, slowly rotating neutron stars (NSs) with surface magnetic field strengths typically inferred in the range $10^{14}$--$10^{15}$\,G, making them the most magnetic objects in the known universe.
They have been observed to emit high energy electromagnetic radiation, and to undergo powerful X-ray and gamma-ray outbursts. The high energy emission from these objects is thought to be powered by the decay of their magnetic fields \citep{Thompson95} as opposed to dipole radiation for classical pulsars. 
To date, 29 magnetars and 6 magnetar candidates have been discovered \citep{Olausen14}\footnote{Catalogue: \url{http://www.physics.mcgill.ca/~pulsar/magnetar/main.html}}; however, only 6 magnetars have ever been observed to emit radio pulsations, partly due to a small birthrate for the class \citep{Gill07}.
SGR~J1935$+$2154 recently joined the other 5 magnetars that have been observed to emit radio pulses. Its radio emission was detected in the form of an unprecedented radio pulse with a fluence of $1.5\pm0.3$\,MJy~ms \citep[][]{Andersen20,Bochenek20}. That burst is the highest-fluence radio pulse ever recorded from the Galaxy and confirms magnetars are plausible sources of extragalactic Fast Radio Bursts (FRBs). 
However, the mechanism by which such strong radio pulses are produced from magnetars is poorly understood \citep[e.g.][]{Margalit20}, as is the birth mechanism of magnetars. 

Multi-wavelength observations of Galactic magnetars, including long-term timing and \textit{Neutron Star Interior Composition Explorer} (NICER) observations, allow us to study the morphology and evolution of their magnetic fields, and potentially probe their internal structure \citep{Kaspi17}.
However, such studies are usually limited by the uncertainties in the underlying magnetar distances (and uncertain proper motions as well, in some cases).
For instance, the X-ray spectrum fitting technique that has recently been made possible by observations with NICER requires a well-constrained, pre-determined distance to the target (which can be a magnetar) in order to infer its radius along with its mass \citep{Bogdanov19}. 
Besides, owing to the enormous instability in spin-down rates (period derivative $\dot{P}$) of magnetars \citep[e.g.][]{Camilo07,Archibald15,Scholz17}, measuring the proper motion (not to mention parallax) via timing is difficult for magnetars; as such, using an accurate, {\em a priori} proper motion and parallax in the timing analysis of a magnetar can improve the reliability of the timing model, thus facilitating the study of long-term $\dot{P}$ evolution.
Furthermore, an accurate distance would enable unbiased estimation of the absolute flux of X-ray flares or the absolute fluence of giant radio pulses. 
On top of the studies focusing on the magnetars, accurate distance and proper motion for a magnetar also enables constraints to be placed on the distance to the dominant foreground (scattering) interstellar-medium (ISM) screen (\citealp{Putney06}; see \citealp{Bower14,Bower15} for an example).

Proper motion measurements for magnetars are significant in their own right.
Both the space velocities of neutron stars and their surface magnetic field strengths have been connected to the progenitor stellar masses and the processes of core-collapse supernovae. 
\citet{Duncan92} suggested that the high magnetic fields of magnetars could be associated with very high space velocities of $\rm 10^3\,km~s^{-1}$ through e.g., asymmetric mass loss during core collapse or in the form of an anisotropic magnetized wind, or through a neutrino and/or photon rocket effect. Such processes would be ineffective for ``ordinary'' neutron star field strengths ($\lesssim 10^{13}$~G), leading to great interest in magnetar velocity estimates as diagnostics of their natal processes.
So far the transverse velocity measurements that have been made (modulo large uncertainties) do not support a higher-than-average kick velocity for magnetars, with transverse velocities around the range of 200\,\kmps\ inferred for \xte, PSR~J1550$-$5418, and PSR~J1745$-$2900 \citep[][respectively]{Helfand07,Deller12,Bower15}.
Additionally, the proper motion of a magnetar could provide a crucial test of its association with nearby supernova remnants (SNR), especially when the magnetar is outside the SNR. More importantly, the proper motion enables us to infer the kinematic age of the magnetar as well as the SNR from the underlying association, which is more reliable than the characteristic age of either.

Both distance and proper motion for a magnetar can be geometrically measured with VLBI (very long baseline interferometry) astrometry at radio wavelengths.
Observations of magnetar radio pulses reveal that they are quite distinct from the radio emission seen in pulsars -- most of them have flat radio spectra and their pulse profiles are highly variable on timescales ranging from seconds to years. Radio emission from magnetars is also a relatively short-lived phenomenon, generally starting out bright after an outburst, then fading over the following months to years. After radio emission ceases, they then spend long periods of time in a radio-silent, quiescent state before the next outburst.
With a current sample size of only 3 magnetars with precise VLBI proper motions, the (re-)appearance of a radio-emitting magnetar offers a valuable opportunity, particularly for pulsar timing and astrometry, both of which can be performed much more precisely with radio observations than with X-rays.

\xte\ (hereafter J1810) was discovered in 2003 due to an outburst at X-ray wavelengths \citep{Ibrahim04}, and was subsequently seen to be pulsating at radio wavelengths with a period of 5.4 seconds \citep{Camilo06}, the first time radio pulsations had been detected for a magnetar. During its brief period of radio brightness over a decade ago, two VLBA observations separated by 106 days were made, allowing the measurement of a proper motion of $13.5 \pm 1.0$\,\maspy\ -- also the first for a magnetar \citep{Helfand07}. The implied transverse velocity was $212 \pm 35$\,km\,s$^{-1}$ (assuming a distance of $3.5 \pm 0.5$\,kpc) and was the first indication that magnetar velocities might be much lower than the expectations outlined earlier in this section.
J1810 subsequently faded into a 10 year quiescence at radio wavelengths until 2018 December 8, when it was found to be radio bright (and pulsating) again at Jodrell Bank \citep{Lyne18}, Molonglo Radio Observatory \citep{Lower18} and Effelsberg \citep{Desvignes18}. The flux density of the source ranged between 9 and 20~mJy from 835\,MHz to 8\,GHz, showing a flat spectrum. 

Here we present new VLBI astrometric results for J1810 in \ref{sec:J1810_results}, and lay out the direct indications of the results in \ref{sec:J1810_discussion}. We also describe dual-calibrator phase calibration technique (also known as 1-D interpolation \citealp{Fomalont03}) and detail the relevant data analysis in \ref{sec:J1810_data_reduction} and \ref{sec:J1810_results}.
Throughout this paper, the uncertainties are provided at 68\% confidence level unless otherwise stated.

\section{Observations and correlation}
\label{sec:J1810_observations}

After the re-activation of J1810 at radio wavelengths in December 2018, we observed the magnetar with the \textit{Very Long Baseline Array} (VLBA) from January 2019 to November 2019 on a monthly basis (project code BD223).
We re-visited J1810 with three consecutive VLBA observations in the same observing setup on 28 March, 6 April and 13 April 2020 (when J1810 was at its parallax maximum; project code BD231). Altogether there are 14 new VLBA observation epochs.

\begin{figure}
    \centering
	\includegraphics[width=14cm]{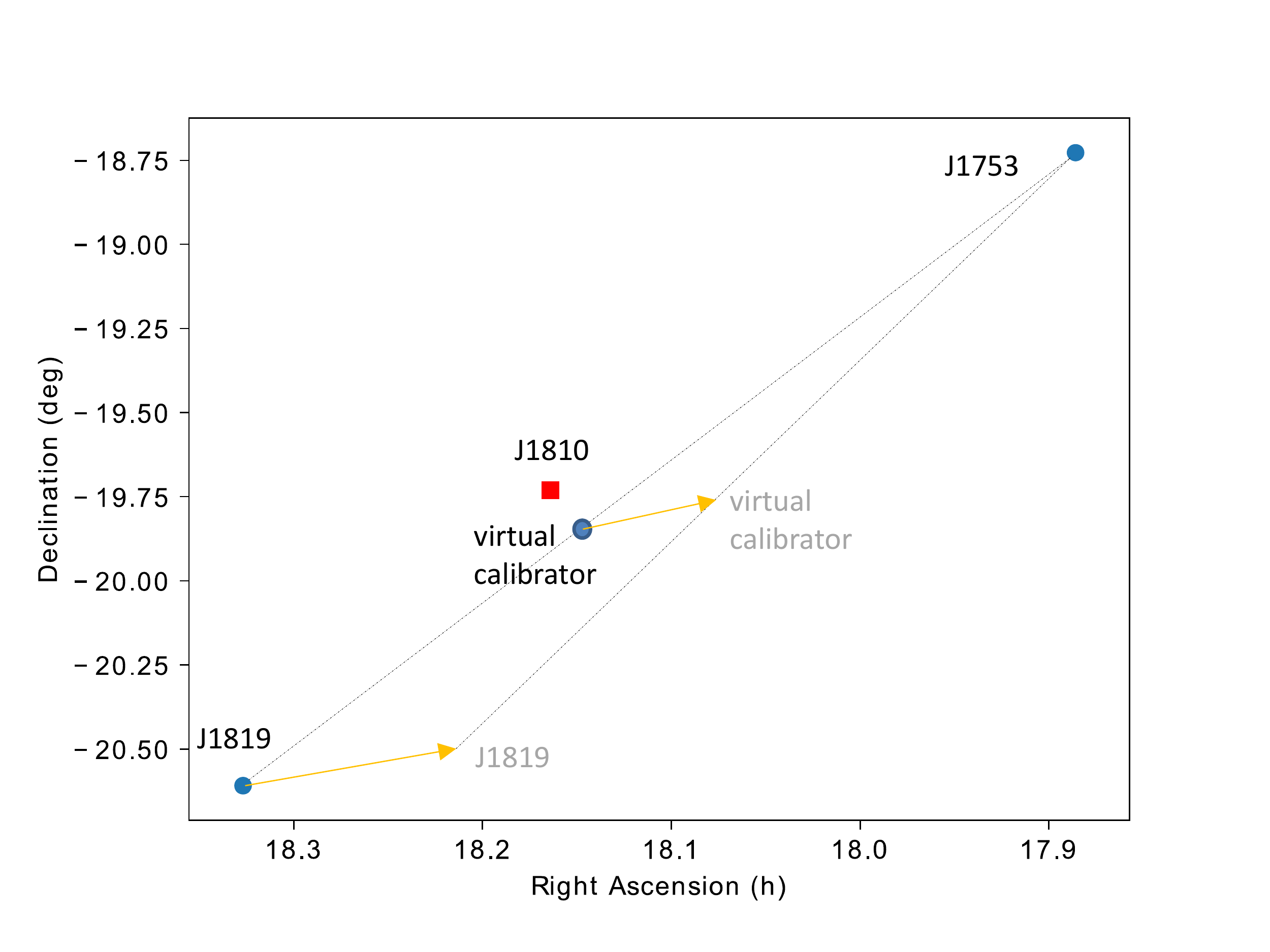}
    \caption{
    {\bf a)} The diagram shows the positions of the magnetar (marked with red rectangle), the primary phase calibrator J1753 and secondary phase calibrator J1819. The virtual calibrator is chosen as the closest point (12\farcm5 away) to J1810 on the J1753-to-J1819 geodesic line, which is at 62.43\% of the distance to J1819 from J1753. {\bf b)} Yellow arrows are overlaid on the calibrator plan to illustrate the discussion in \ref{subsec:J1810_systematics} and \ref{subsec:J1810_absolute_position}: if the position for J1819 was in error (greatly exaggerated here for visual effect -- any position offset is in reality only at the $\sim1$\,mas level), the position of the virtual calibrator would change in the same direction.  The degree of the position shift of the virtual calibrator would be 62.4\% that of J1819. Any position error for J1753 would likewise affect the position of the virtual calibrator.
    }
    \label{fig:calibrator_plan}
\end{figure}

All the observations were carried out at around 5.7~GHz in astrometric phase referencing mode (where the pointing of the array alternates between the target and phase calibrator throughout the observation). Unlike typical astrometric observations, J1810 was phase-referenced to two phase calibrators: ICRF~J175309.0-184338 (4\fdg1 away from J1810, hereafter J1753) and VCS6~J1819-2036 (2\fdg5 away from J1810, hereafter J1819), which are almost colinear with J1810 (see \ref{fig:calibrator_plan}). Accordingly, the two phase calibrators and J1810 were observed in turn in cycles; in each cycle that is typically 5-minute-long, nearly 3\,min are spent on J1810. The purpose of the non-standard astrometric setup is explained in \ref{subsec:J1810_dualphscal}.
ICRF~J173302.7$-$130449 was observed as the fringe finder.

The data were correlated with the {\tt DiFX} software correlator \citep{Deller11a} in two passes, standard (ungated) and gated. For the gated pass, the off-pulse durations were excluded from the correlation in order to improve the signal-to-noise ratio (S/N) on the magnetar. For all the 14 observations, pulsar gating was applied based on the pulsar ephemerides obtained with our monitoring observations of the magnetar at the Parkes and Molonglo telescopes.
The timing observations at Parkes were carried out under the project code P885; the relevant observing setup and data reduction are described in Lower et~al. (in preparation).
The timing observations at Molonglo were fulfilled as part of the UTMOST project \citep{Jankowski19,Lower20}.

\begin{figure}
    \centering
	\includegraphics[width=15cm]{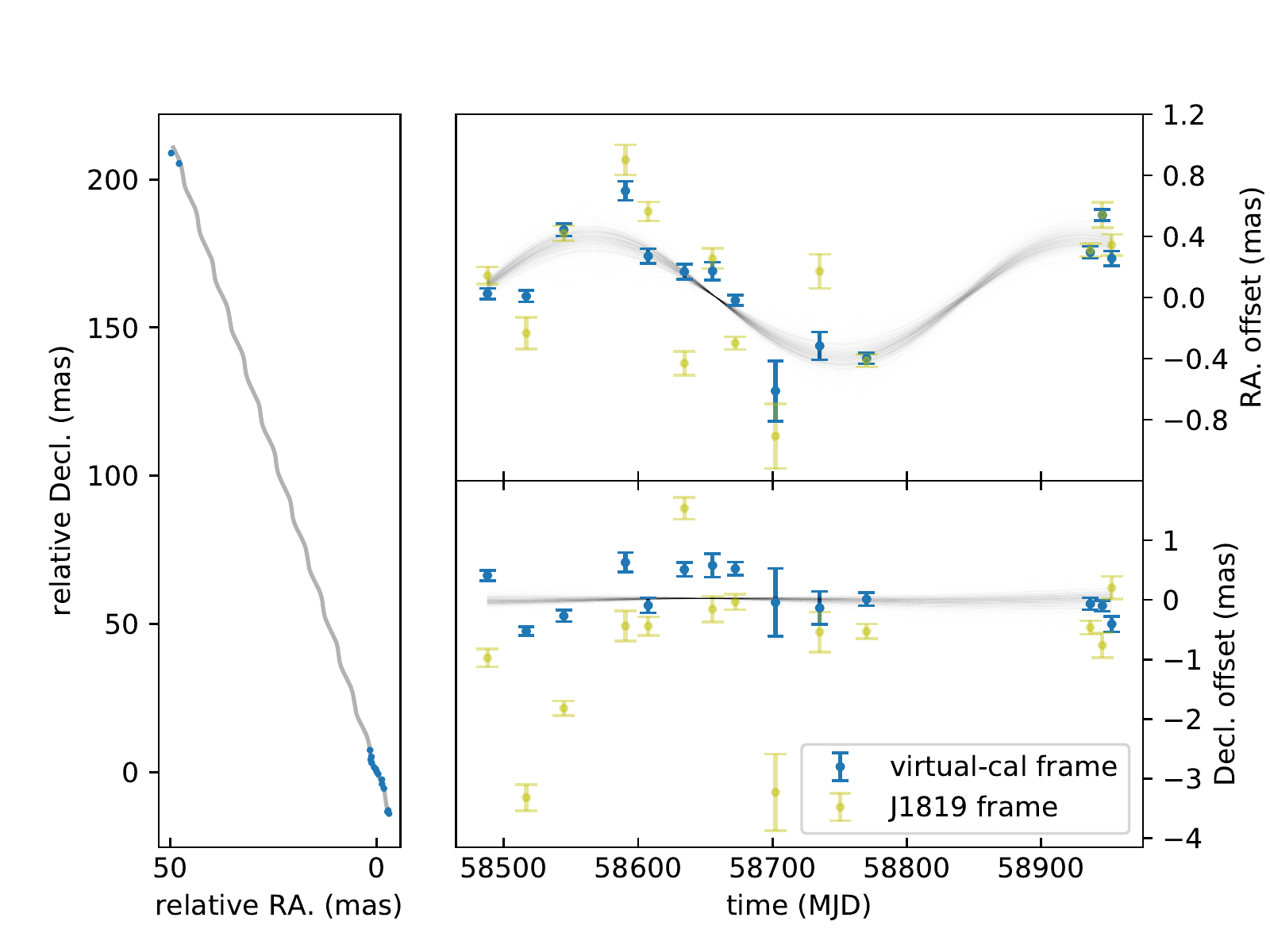}
    \caption{
    {\bf Left panel:} position evolution of J1810 relative to $18^{\rm h}09^{\rm m}51\fs 08333, -19\degr43'52\farcs1398$
    from 2006 to 2020. 
    {\bf Right panel:} recent positions of J1810 with proper motion subtracted measured in the virtual-calibrator frame (blue) and J1819 frame (yellow; see \ref{subsec:J1810_reference_frames} for explanations of different reference frames); to convenience visual comparison, the two sets of positions adopt the same systematic uncertainties at each epoch as obtained in the virtual-calibrator frame.
    Overlaid are the best-fit models for 500 bootstrap draws from the positions measured in the virtual-calibrator frame.
    Here, the systematic uncertainties for the positions measured in the virtual-calibrator frame are still under-estimated, which is addressed in \ref{subsec:J1810_pi_and_mu} using a bootstrap technique. 
    Obviously, the positions measured in the virtual-calibrator frame provide tighter constraints to the model.}
    \label{fig:positions_and_model}
\end{figure}

Apart from the 14 new VLBA observations, we re-visited two archival VLBA observations of J1810 taken in 2006 under the project code BH142 and BH145A \citep{Helfand07}.
An overview of observation dates is provided in \ref{fig:positions_and_model}. The two observations in 2006 were carried out at both 5\,GHz and 8.4\,GHz, using J1753 as the only phase calibrator. More details of the observing setup for the two observations in 2006 can be found in \citet{Helfand07}.
Hereafter, where unambiguous, positions obtained from the two epochs in 2006 and the 14 new epochs are equally referred to, respectively, as the ``year-2006 positions" and the ``recent positions".

\section{Data reduction}
\label{sec:J1810_data_reduction}
All VLBI data were reduced with the {\tt psrvlbireduce} (\url{https://github.com/dingswin/psrvlbireduce}) pipeline written in python-based {\tt ParselTongue} \citep{Kettenis06}. {\tt ParselTongue} serves as an interface to interact with {\tt AIPS} \citep{Greisen03} and {\tt DIFMAP} \citep{Shepherd94}. 
The pipeline has been incrementally developed for VLBI pulsar observations over the last decade, notably for the large VLBA programs \psrpi\ \citep{Deller19} and \mspsrpi\ \citep[e.g.][]{Ding20}.

For the two year-2006 epochs, only data taken at 5\,GHz were reduced and analyzed, in order to avoid potential position uncertainties due to any  frequency-dependent core shift in the phase calibrators \citep[e.g.][]{Bartel86,Lobanov98}.
All positions of J1810 were obtained from the gated J1810 data, the S/N of which exceed the ungated J1810 data by 40\% on average.

Due to multi-path scattering caused by the turbulent ISM along the line of sight to J1810 or J1819, the deconvolved angular size of J1810 is mildly broadened by $0.7\pm0.4$\,mas (detailed in the Appendix), and J1810 (as well as J1819) is fainter at longer baselines. In addition, J1819 exhibits intrinsic source structure, with a jet-like feature extending over $\sim$10~mas.  Accordingly, J1819 is heavily resolved by the longest baselines of the VLBA, with a flux density of at most a couple of mJy on baselines longer than 50~M$\lambda$ (mega-wavelength).
The three stations furthest from the geographic centre of VLBA are MK (Mauna Kea), SC (St. Croix) and HN (Hancock), each of which only has 0--1 baselines shorter than 50~M$\lambda$.  Unsurprisingly, we found that while valid solutions could still often be found, the phase solutions for HN, MK, and SC are much noisier than for the remainder of the array.  
On top of this, atmospheric fluctuation causes larger phase variation at longer baselines, which, combined with the worse phase solutions, makes phase wraps (to be explained in \ref{subsec:J1810_dualphscal}) at MK, SC and HN hard to determine.
We found that the image S/N for J1810 generally improves when MK, SC and HN are excluded from the target field data, and as a result, taking the three stations out of the J1810 data leads to a statistical positional error comparable to that obtained when using the full array.
Therefore, we consistently flagged the three stations from the final J1810 data for 14 recent epochs. However, we did not remove the three stations from any calibration steps, because we found the participation of the three stations allows better performance of self-calibration on J1819 at other stations.

The applied calibration steps in this work are largely the same as the \psrpi\ project \citep{Deller19} except for the phase calibration.
For the two year-2006 epochs, the phase solutions obtained with J1753 were directly transferred to J1810; whereas, for the recent 14 epochs, phase solutions were corrected based on the positions of J1753 and J1819, before being applied to J1810. Such a technique, though previously adopted by other researchers \citep[e.g.][]{Fomalont03}, is applied to a pulsar for the first time.

\subsection{Dual-calibrator phase calibration}
\label{subsec:J1810_dualphscal}
During the phase calibration of VLBI data, the calibration step $k~(k=2, 3, ...)$ provides an increment of the phase difference $\Delta\phi^{(k)}_{n}(t)$ between the station $n$ and the reference station in the VLBI array at a given time $t$ (for simplicity, frequency-dependency is not accounted for), which is then added to the previous sum of the phase difference $\phi^{(k-1)}_{n}(t)$ when the new solutions are applied.
The phase calibration of the primary phase calibrator (J1753 for this work) is followed by the self-calibration of the secondary phase calibrator (J1819 for this work), the phase solutions of which are predominantly limited by anisotropic atmospheric (including ionosphere and troposphere) propagation effect. 
Accordingly, the position-dependent phase difference should be formulated as $\Delta\phi_{n}(\vec{x},t)$, where $\vec{x}$ represents the 2-D sky position. 
In the normal phase calibration, the solutions $\Delta\phi_{n}(\vec{x}_\mathcal{S},t)$ obtained with the self-calibration of the secondary phase calibrator at $\vec{x}_\mathcal{S}$ are directly given to the target. The closer $\vec{x}_\mathcal{S}$ is to the target field, the better $\Delta\phi_{n}(\vec{x}_\mathcal{S},t)$ can approximate the unknown $\Delta\phi_{n}(\vec{x}_\mathcal{T},t)$ at the target field. 
In cases where $\vec{x}_\mathcal{S}$ is $\gtrsim1$\degr\ away from the target field, considerable offsets are expected between $\Delta\phi_{n}(\vec{x}_\mathcal{S},t)$ and $\Delta\phi_{n}(\vec{x}_\mathcal{T},t)$ (especially at lower frequencies), which are commonly treated as systematic errors when using the normal phase calibration. 
However, if the primary phase calibrator, secondary phase calibrator and the target happen to be quasi-colinear on the sky, $\Delta\phi_{n}(\vec{x}_\mathcal{T},t)$ can be well approximated by phase solutions corrected from $\Delta\phi_{n}(\vec{x}_\mathcal{S},t)$ \citep{Fomalont03}.

It is easy to prove that for three arbitrary different co-linear positions $\vec{x}$, $\vec{x}_1$ and $\vec{x}_2$,
\begin{equation}
    \phi_{n}(\vec{x},t)=\frac{\vec{x}_1-\vec{x}}{\vec{x}_1-\vec{x}_2}\phi_{n}(\vec{x}_2,t)+\frac{\vec{x}-\vec{x}_2}{\vec{x}_1-\vec{x}_2}\phi_{n}(\vec{x}_1,t),
	\label{eq:taylor_expansion}
\end{equation}
assuming higher-than-first-order terms are negligible (as supported by \citealp{Chatterjee04,Kirsten15}). Specific to the self-calibration of the secondary phase calibrator, we have $\Delta\phi_{n}(\vec{x},t)=\Delta\phi_{n}(\vec{x}_\mathcal{S},t)\cdot (\vec{x}_\mathcal{P}-\vec{x})/(\vec{x}_\mathcal{P}-\vec{x}_\mathcal{S})$, where $\vec{x}_\mathcal{P}$ and $\vec{x}_\mathcal{S}$ refer to the position of the primary phase calibrator and secondary phase calibrator, respectively; this relation allows us to extrapolate to $\Delta\phi_{n}(\vec{x},t)$ at any position $\vec{x}$ colinear with J1753 and J1819 based on $\Delta\phi_{n}(\vec{x}_\mathrm{J1819},t)$. 

We calculated the closest position to J1810 on the J1753-to-J1819 geodesic line (an arc), where it can best approximate $\Delta\phi_{n}(\vec{x}_\mathrm{J1810},t)$.
Since it acts like a phase calibrator for J1810, hereafter we term it the ``virtual calibrator'' for J1810.
The position of the virtual calibrator is $\vec{x}_\mathrm{v}=0.38\cdot \vec{x}_\mathrm{J1753}+0.62\cdot \vec{x}_\mathrm{J1819}$, which is 12\farcm5 away from J1810 (see \ref{fig:calibrator_plan}). Accordingly, the phase solutions extrapolated to the virtual calibrator are $\Delta\phi_{n}(\vec{x}_\mathrm{v},t)=0.62\cdot \Delta\phi_{n}(\vec{x}_\mathrm{J1819},t)$;
using this relation, we corrected the phase solutions of J1819 obtained with its self-calibration. The correction was implemented with a dedicated module called {\tt calibrate\_target\_phase\_with\_two\_colinear\_phscals} that was newly added to the {\tt vlbatasks.py}, as a part of the {\tt psrvlbireduce} package. 
One of the functions of the module is to solve the phase ambiguity of $\Delta\phi_{n}(\vec{x}_\mathrm{J1819},t)$ prior to multiplying $\Delta\phi_{n}(\vec{x}_\mathrm{J1819},t)$ by 0.62.

The biggest challenge of the dual-calibrator phase calibration is the phase ambiguity of $\Delta\phi_{n}(\vec{x}_\mathrm{J1819},t)$, which can be equivalently expressed as $\Delta\phi_{n}(\vec{x}_\mathrm{J1819},t)\pm2i\pi~(i=0,1,2,...)$. This phase ambiguity will not change the quality of solutions for the normal phase calibration, but will cause trouble for the dual-calibrator phase calibration, as the periodicity of the phase is broken when multiplied by a factor. 
The degree of phase ambiguity depends on observing frequency and angular distance between the main and secondary phase calibrator;
for this work (5.7\,GHz, 6.5\degr), we run into mild phase ambiguity, mainly at the longest baselines (that we do not use anyway as mentioned earlier in this section).
Among the recent epochs, the smallest size of the synthesized beam excluding MK, SC and HN is $2.9\times8.2$\,mas, more than two times larger than the level of the systematic uncertainties dominated by propagation effect (see \ref{fig:positions_and_model}).
Therefore, for this work, we consider $\Delta\phi_{n}(\vec{x}_\mathrm{J1819},t)$ less likely to turn more than one wrap, and impossible to turn more than two wraps (i.e. $|i|\leq2$).

We resolved the phase ambiguity of $\Delta\phi_{n}(\vec{x}_\mathrm{J1819},t)$ in a semi-automatic and iterative manner. The pipeline would go though all values of $\Delta\phi_{n}(\vec{x}_\mathrm{J1819},t)$ at each station. If $|\Delta\phi_{n}(\vec{x}_\mathrm{J1819},t)|<\pi/2$ holds true throughout the observation, then the solutions $\Delta\phi_{n}(\vec{x}_\mathrm{J1819},t)$ are deemed phase-unambiguous, and no human intervention is needed for the station $n$. Otherwise, $\Delta\phi_{n}(\vec{x}_\mathrm{J1819},t)$ solutions are plotted out for inspection and interactive correction. 
In most cases, no interactive correction is necessary after the inspection of the $\Delta\phi_{n}(\vec{x}_\mathrm{J1819},t)$ plot, as the solutions look continuous, oscillating around 0 within a reasonable range (e.g. between $\pm2\pi/3$). 
In the few cases where interactive corrections are needed, solutions are ambiguous in phase for at most 1 or 2 stations per observation. 
This enabled us to take a simple, brute-force approach to trialing the plausible possibilities for phase wraps (adding or subtracting an integer multiple by $2\pi$ radians to the solution) with interactive correction. For every possibility, we implemented the dual-calibrator phase calibration using the resultant solution and ran through the complete (data-reduction and imaging) pipeline. The correct $\Delta\phi_{n}(\vec{x}_\mathrm{v},t)$ should outperform other possibilities in terms of the image S/N for J1810 (and the S/N difference is normally significant), as it better approximates the $\Delta\phi_{n}(\vec{x}_\mathrm{J1810},t)$. This S/N criterium helps us find the ``real'' $\Delta\phi_{n}(\vec{x}_\mathrm{J1819},t)$, thus the right $\Delta\phi_{n}(\vec{x}_\mathrm{v},t)$.
The obtained solutions $\Delta\phi_{n}(\vec{x}_\mathrm{v},t)$ were then transferred to J1810.

More generally, if no phase-calibrator pair quasi-colinear with the target is found, one can extrapolate to $\Delta\phi_{n}(\vec{x},t)$ at any position $\vec{x}$ with three non-colinear phase calibrators \citep[also known as 2-D interpolation,][]{Fomalont03,Rioja17}.
Despite the longer observing cycle and hence sparser time-domain sampling \citep[unless using the multi-view observing setup,][]{Rioja17}, the tri-calibrator phase calibration can in principle remove all the first-order position-dependent systematics.

\section{Systematic Errors and Astrometric Fits}
\label{sec:J1810_results}

\subsection{Reference frames in relative VLBI astrometry}
\label{subsec:J1810_reference_frames}

Similar to the way a reference frame is normally defined in non-relativistic (Cartesian) contexts, a reference frame in the context of relative VLBI astrometry (hereafter reference frame or frame) generally refers to a system of an infinite amount of sky positions that are tied to a phase calibrator (not necessarily a real one), in which positions are measured relative to the phase calibrator.
In this work there are three different reference frames where we can measure the positions of J1810: the J1753 frame, the J1819 frame, and the virtual-calibrator frame.
To be more specific, in the J1753/J1819 frame, the positions are measured relative to the brightest spot of the model image for J1753/J1819 respectively. By applying an identical model of J1753/J1819\footnote{available at \url{https://data-portal.hpc.swin.edu.au/dataset/calibrator-models-used-for-vlba-astrometry-of-xte-j1810-197}} during the fringe fitting and self-calibration steps, the J1753/J1819 images at different epochs are aligned, respectively,
to $17^{\rm h}53^{\rm m}09\fs 0886, -18\degr43'38\farcs520$
and $18^{\rm h}19^{\rm m}36\fs 8955, -20\degr36'31\farcs573$;
the virtual-calibrator frame is thus anchored to $18^{\rm h}09^{\rm m}35\fs 9437, -19\degr55'49\farcs656$,
determined by the relation $\vec{x}_\mathrm{v}=0.38\cdot \vec{x}_\mathrm{J1753}+0.62\cdot \vec{x}_\mathrm{J1819}$.
For the 14 recent epochs, the final positions of J1810 were measured in the virtual-calibrator frame, though the positions of J1810 were also measured in the other two frames for various purposes (see \ref{subsec:J1810_systematics} and \ref{fig:positions_and_model}).
The two year-2006 positions were merely measured in the J1753 frame, as J1753 is the only available phase calibrator for these observations.

\subsection{Systematic errors and frame transformation}
\label{subsec:J1810_systematics}
Similar to the way in which systematic positional errors for pulsars in the \psrpi\ project were evaluated using Eqn~1 of \citet{Deller19} to account for both differential ionospheric propagation effects and thermal noise at secondary phase calibrators,
the estimation of systematic uncertainty for the measured positions of J1810 is based on the mathematical formalism
\begin{equation}
\label{eq:empirical_sys_error}
{\Delta_\mathrm{sys}}^2=\left( A \cdot \frac{s}{1\,{\rm arcmin}} \cdot \overline{\csc{\epsilon}}\right)^2 + \left(B/S\right)^2\,,
\end{equation}
where $\Delta_\mathrm{sys}$ is the ratio of the systematic error to the synthesized beam size, $\epsilon$ stands for elevation angle, $\overline{\csc{\epsilon}}$ is the average $\csc{\epsilon}$ for a given observation (over time and antennas), $s$ is the angular separation between J1810 and the calibrator of the frame, $S$ represents the image S/N of the calibrator of the frame, and $A$ and $B$ are coefficients to be determined.
However, unlike Eqn~1 of \citet{Deller19}, in \ref{eq:empirical_sys_error} the two contributing terms on the right side are added in quadrature.  Given that they should be uncorrelated, this is more appropriate than the linear summation in Eqn~1 of \citet{Deller19}.

For this work, the second term of \ref{eq:empirical_sys_error} is negligible for any of the three frames, as both J1753 and J1819 are strong sources, with a brightness $\geq$24\,\mjypb\ at our typical resolution (after MK, SC and HN have been removed from the array). 
In order to find a reasonable estimate of $A$ for this work, we measured the positions of J1810 consistently in the J1753 frame for all 16 epochs, and determined the value of $A$ that renders an astrometric fit (see ``direct fitting'' in \ref{subsec:J1810_pi_and_mu}) with unity \rcsl, or \rcs~$=1$. We obtained $A=3\times10^{-4}$.
We note that, in principle, $A$ is invariant with respect to different reference frames. 
As is mentioned in \ref{subsec:J1810_reference_frames}, the final positions of J1810 were measured in the virtual-calibrator frame for the 14 recent epochs and in the J1753 frame for the two year-2006 epochs.
Using $A=3\times10^{-4}$ and \ref{eq:empirical_sys_error} without the second term, we acquired systematic errors for each epoch, which was then added in quadrature to the random errors.

After the determination of the systematic errors, the next step is to transform positions into the same reference frame. 
Since J1753 and J1819 are remote quasars almost static in the sky, the frame transformation is simply translational.
We translated the two year-2006 positions measured from the J1753 to the virtual-calibrator frame. The translation is equivalent, but in the reverse direction, to translate from the virtual-calibrator frame to the J1753 frame, which is easier to comprehend. In order to translate the virtual-calibrator frame to the J1753 frame, the position of the virtual calibrator needs to be measured in the J1753 frame, which can be accomplished by measuring the position of J1819 in the J1753 frame. The method to estimate the position of J1819 $\vec{x}'_\mathrm{J1819}$ and its uncertainty $\vec{\sigma}'_\mathrm{J1819}$ in the J1753 frame is detailed in Section~3.2 of \citet{Ding20}.

As is shown in \ref{fig:calibrator_plan}, once $\vec{x}'_\mathrm{J1819}$ is measured in the J1753 frame, the new position of the virtual calibrator $\vec{x}'_\mathrm{v}$ in the J1753 frame is also determined, the uncertainty $\vec{\sigma}'_\mathrm{v}$ of which is 0.62 times $\vec{\sigma}'_\mathrm{J1819}$. The difference between $\vec{x}'_\mathrm{v}$ and $\vec{x}_\mathrm{v}$ (or 0.62 times the difference between $\vec{x}'_\mathrm{J1819}$ and $\vec{x}_\mathrm{J1819}$) was used to translate the two year-2006 positions of J1810 from the J1753 frame to the virtual-calibrator frame. The $\vec{\sigma}'_\mathrm{v}$ was added in quadrature to the error budget (already including systematic and random errors) of the two year-2006 positions.

\subsection{Proper motion, parallax and distance}
\label{subsec:J1810_pi_and_mu}
After including the systematic errors and unifying to the virtual-calibrator frame, the 16 positions of J1810 can be used for astrometric fitting. 
Astrometric fitting was performed using {\tt pmpar}\footnote{\url{https://github.com/walterfb/pmpar}}. The median among the 16 epochs, MJD~58645, was adopted as the reference epoch. The results out of direct fitting are reproduced in \ref{tab:mu_and_pi}, the \rcs\ of which is 10.6. The large \rcs\ suggests the systematic errors for the recent 14 positions are probably under-estimated, and the actual uncertainty for either parallax or proper motion is about 3 times larger than the uncertainty from direct fitting. 

\begin{figure}
    \centering
	\includegraphics[width=15cm]{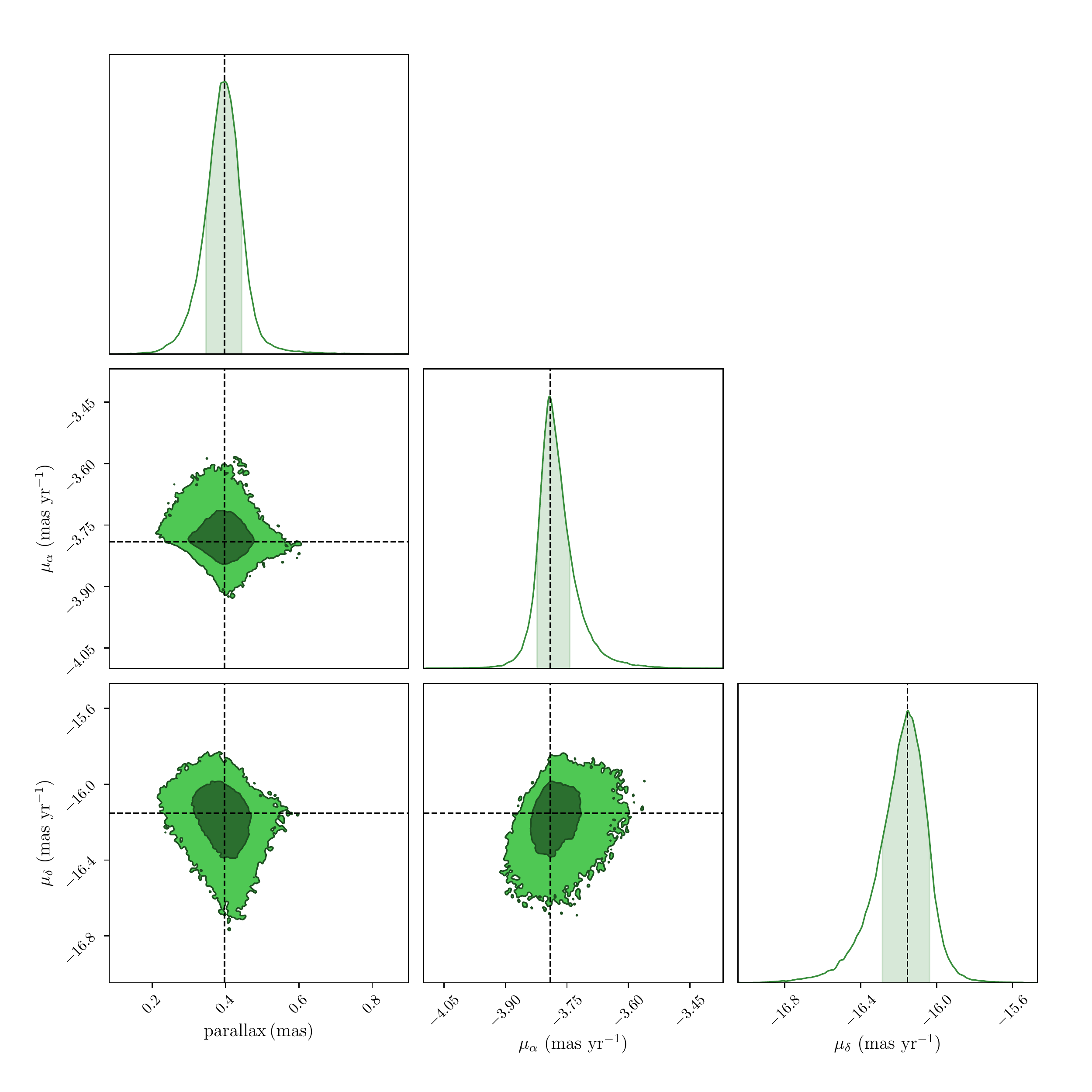}
    \caption{Error ``ellipses'' and marginalized histograms for parallax and proper motion.
    In each histogram, the dashed line marks the measured value; the shade stands for the 68\% confidence interval.
    In each error ``ellipse'', the dark and bright contour enclose, respectively, 68\% and 95\% of the bootstrapped data points.}
    \label{fig:covariance_pi_and_mu}
\end{figure}

Applying a bootstrap technique to astrometry can generally provide more conservative uncertainties, compared to direct fitting \citep[e.g.][]{Deller19}. 
In the same way as is described in Section~3.1 on \citet{Ding20}, we bootstrapped 100000 times, from which we assembled 100000 fitted parallaxes, proper motions and reference positions for J1810. The marginalized histograms for parallax and proper motion as well as their paired error ``ellipses'' are displayed in \ref{fig:covariance_pi_and_mu}. 
We reported the most probable value at the peak of each histogram as the measured value; the most compact interval containing 68\% of the sample was taken as the 68\% uncertainty range of the measured value (see \ref{fig:covariance_pi_and_mu}). The parallax and proper motion estimated with bootstrap are listed in \ref{tab:mu_and_pi}, which are highly consistent with direct fitting while over 3 times more conservative (as is expected from the \rcs\ of direct fitting). Thus, the precision achieved for parallax and proper motion gauged with bootstrap can be deemed reasonable.
The parallax corresponds to the distance $2.5^{\,+0.4}_{\,-0.3}$\,kpc. 
Compared to the final 8\,$\sigma$ parallax, normal phase calibration (in the J1819 frame) would render a consistent parallax with only $\lesssim5\,\sigma$ significance (see \ref{fig:positions_and_model}).
%Such a precision in distance would not be achieved with normal phase calibration using the same data (see \ref{fig:positions_and_model}).

\begin{table}
	\centering
	\caption{Proper motion and distance measurements for J1810}
	\label{tab:mu_and_pi}
	\resizebox{\textwidth}{!}{
	\begin{tabular}{cccccc} % four columns, alignment for each
		\hline
	method & $\mu_\alpha \equiv \dot{\alpha}\cos{\delta}$ & $\mu_\delta$ & $\varpi$ & $D$ & References \\
	    & (\maspy) & (\maspy) & (mas) & (kpc) & \\
		\hline
	direct fitting	& $-3.78\pm0.01$ & $-16.18\pm0.03$ & $0.39\pm0.01$ & $2.5\pm0.1$ & this work\\
	bootstrap	& $-3.79^{\,+0.05}_{\,-0.03}$ & $-16.2\pm0.1$ & $0.40\pm0.05$ & $2.5^{\,+0.4}_{\,-0.3}$  & this work\\
	\\
	Previous VLBI astrometry & $-6.60\pm0.06$ & $-11.72\pm1.03$ & $-$ & $-$ & \citet{Helfand07} \\
	red clump stars & $-$ & $-$ & $-$ & $3.1\pm0.5$ & \citet{Durant06} \\
	neutral hydrogen absorption & $-$ & $-$ & $-$ & 3.1$-$4.0 & \citet{Minter08} \\
		\hline
	\end{tabular}}
\end{table}

\subsection{Absolute position}
\label{subsec:J1810_absolute_position}
Along with proper motion and parallax, a reference position $18^{\rm h}09^{\rm m}51\fs 083326 \pm 0.03\,{\rm mas}$, $-19\degr43'52\farcs1398 \pm 0.1$\,mas at the reference epoch MJD~58645 was also obtained for J1810 with bootstrap.
We note again the reference position was measured in the virtual-calibrator frame. According to the relation $\vec{x}_\mathrm{v}=0.38\cdot \vec{x}_\mathrm{J1753}+0.62\cdot \vec{x}_\mathrm{J1819}$, change in $\vec{x}_\mathrm{J1753}$ or $\vec{x}_\mathrm{J1819}$ would cause the position shift of the virtual calibrator $\Delta \vec{x}_\mathrm{v}$ (hence the position of J1810), following the relation 
\begin{equation}
\label{eq:position_shift_of_virtual_cal}
\Delta \vec{x}_\mathrm{v}=0.38\cdot \Delta \vec{x}_\mathrm{J1753}+0.62\cdot \Delta \vec{x}_\mathrm{J1819}\,.  
\end{equation}
Using \ref{eq:position_shift_of_virtual_cal} and the method outlined in Section~3.2 on \citet{Ding20}, the reference position was shifted to align with the latest positions of J1753 and J1819\footnote{\url{http://astrogeo.org/vlbi/solutions/rfc_2020b/rfc_2020b_cat.html}}. 
The shifted reference position $18^{\rm h}09^{\rm m}51\fs 08333 \pm 0.3\,{\rm mas}, -19\degr43'52\farcs1418 \pm~0.5$\,mas is the absolute position of J1810 at MJD~58645, where the uncertainties of the J1753 and J1819 positions have been propagated onto the uncertainty budget.
At 5.7\,GHz, the effect of frequency-dependent core shift \citep[e.g.][]{Bartel86,Lobanov98} is at the $\lesssim0.1$\,mas level in each direction \citep{Sokolovsky11}, which makes unnoticeable difference to the uncertainty of the absolute position.

\section{Discussion}
\label{sec:J1810_discussion}
As is shown in \ref{tab:mu_and_pi}, our new proper motion significantly improves on the previous value inferred from the two year-2006 positions;
the new distance $D=2.5^{\,+0.4}_{\,-0.3}$\,kpc is consistent with $3.1\pm0.5$\,kpc estimated using red clump stars \citep{Durant06}, while in mild tension with 3.1$-$4.0\,kpc constrained with neutral-hydrogen absorption \citep{Minter08}, suggesting the distance to the neutral-hydrogen screen was over-estimated.

In models of NS kicks from the electromagnetic rocket effect \citep{Harrison75} one might expect magnetars to have higher velocities \citep{Duncan92}.
Our new parallax and proper motion corresponds to the transverse velocity $v_t=198^{\,+29}_{\,-23}$\,\kmps. Using the Galactic geometric parameters provided by \citet{Reid19} and assuming a flat rotation curve between J1810 and the Sun, the peculiar velocity (with respect to the neighbourhood of J1810) perpendicular to the line of sight was calculated to be $v_{b}=-54\pm8$\,\kmps\ and $v_{l}=-175\pm26$\,\kmps. Our refined astrometric results consolidate the conclusion by \citet{Helfand07} that J1810 has a peculiar velocity typically seen in ``normal'' pulsars, unless its radial velocity is several times larger than the transverse velocity.

\subsection{SNR Association}
\label{subsec:J1810_snr_association}
The closest cataloged SNR to J1810 is \snra\ \citep{Green19}\footnote{\url{http://www.mrao.cam.ac.uk/surveys/snrs/}}, a partial-shell SNR $9'\times11'$ in size \citep{Brogan04,Brogan06}. The position of its geometric center is $18^{\rm h}10^{\rm m}04^{\rm s}, -19\degr25'$, $19'$ away from J1810. 
The latest distance estimate of \snra\ by \citet{Shan18} is $2.4\pm0.7$\,kpc, consistent with our new distance of J1810. Using our astrometric results, we find that the projected position at 70\,kyr ago is $18^{\rm h}10^{\rm m}09\fs 8, -19\degr25'00''$, about $1'$ east to the geometric center of \snra. For the above geometric reasons, it is possible that J1810 is associated with \snra. 

The plausibility of this potential association can be tested by considering the ages of both J1810 and \snra.
The spin-down rate $\dot{P}$ is erratic for J1810 \citep{Camilo07} as well as other magnetars \citep{Archibald15,Scholz17}, making the characteristic age $\tau_\mathrm{c}$ an unreliable estimate of the true age for J1810. Over the course of a decade, the changing value of $\dot{P}$  for J1810 has led to the $\tau_\mathrm{c}$ ($\propto 1/\dot{P}$) increasing from 11\,kyr \citep{Camilo07} to 31\,kyr \citep{Pintore18}. 
While the characteristic age is currently less than the tentative kinematic age $\tau^{*}_\mathrm{k}$ that the tentative association would imply, the unreliability of the $\tau_\mathrm{c}$ estimator in the case of magnetars suggests that the association cannot be ruled out on this basis.

From the perspective of \snra, the compactness of the SNR (see Figure~1 of \citealp{Castelletti16}) suggests that it is probably in the Sedov-Taylor stage. In this stage, the relation between the SNR radius $R_\mathrm{SNR}$ and its age $\tau_\mathrm{SNR}$ can be rewritten from \citet{Sedov59} as
\begin{equation}
R_\mathrm{SNR}\approx5\left(\frac{E}{10^{51}\,\mathrm{erg}}\right)^{1/5}\left(\frac{n}{30\,\mathrm{cm}^{-3}}\right)^{-1/5}\left(\frac{\tau_\mathrm{SNR}}{1\,\mathrm{kyr}}\right)^{2/5}\,\mathrm{pc}\,,
\label{eq:snr_r2}
\end{equation}
where the injected energy $E$ is expressed in a value typical of spherical SNRs expanding into the Galactic ISM, and the ambient ISM density $n\sim30\,\mathrm{cm}^{-3}$ for the $\gamma$-ray-emitting region including \snra\ was required to power the observed $\gamma$-ray emission above 1\,TeV at a distance of 2.4\,kpc \citep{Castelletti16}.
At an SNR distance $D_\mathrm{SNR}=2.4$\,kpc \citep{Shan18}, $R_\mathrm{SNR}\leq$3.8\,pc (corresponding to the angular size along the long axis), which yields $\tau_\mathrm{SNR}\lesssim3$\,kyr using \ref{eq:snr_r2}, consistent with an SNR in the early part of the Sedov-Taylor phase. 
A $\tau_\mathrm{SNR}$ of 70\,kyr can be made possible with an injected energy 500 times smaller than the typically-assumed value of $10^{51}$\,erg, which is extremely unlikely \citep{Leahy17}. 
Therefore, we conclude \snra\ is not directly associated with J1810.
This is not too surprising, as less than half of the known magnetar population has a potential SNR association \citep{Olausen14}. Additionally, it has been proposed a strong post-birth magnetar wind can accelerate the dissipation of the SNR \citep{Duncan92}.

Though \snra\ is not directly associated with J1810, the 3-D geometric alignment might not be a coincidence. One explanation for the geometric alignment is: the progenitor star of J1810 has been in orbit with another supergiant companion;
and \snra\ is the SNR for the ``divorced'' companion of J1810.
In such a scenario, the progenitor of J1810 underwent a supernova explosion $\approx70$\,kyr ago and became unbound from its original companion.  The companion star continued evolving in isolation before itself undergoing a supernova explosion at $\lesssim3$\,kyr ago.
In this ``companion SNR'' scenario, assuming the components of the stellar system were formed at approximately the same time, the progenitor of J1810 should be slightly more massive than its companion (and hence evolve faster). 
However, given that this scenario would require that the companion underwent a supernova explosion only $\approx67$\,kyr (compared to the typical supergiant age of $\gtrsim1$\,Myr) after the first supernova, the mass difference of the two progenitor stars would have to be small. 

Assuming no peculiar velocity of the progenitor binary system (with regard to its neighbourhood mean) as well as a flat rotation curve between the Sun and J1810, the expected proper motion of the barycentre of the supergiant binary as observed from the Earth is only 1.2\,\maspy. The additional proper motion of the companion due to the orbital motion at the moment of unbinding is even smaller for supergiant binaries. Thus, the accumulated position shift of the companion star after the unbinding is at the $1.3'$ level across 67\,kyr, which does not violate the premise of the ``companion SNR'' scenario.

In principle, given the large age scatter of supergiants, the ``companion SNR'' scenario allows J1810 to be indirectly associated with an SNR further away. For example, J1810 can be traced back to 17' west to the centre of SNR~G11.4$-$0.1 (which is though far from the boundary of the SNR) at $\approx144$\,kyr ago. However, the relatively small characteristic age $\tau_\mathrm{c}=11$--31\,kyr favors the closer indirect association (or no association) with \snra.
Despite the ``companion SNR'' scenario, we note that it is highly possible that J1810 does not come from the \snra\ region; instead, it comes from an already dissipated SNR between J1810 and \snra\ (as supported by \citealp{Duncan92}).  
Longer-term $\tau_\mathrm{c}$ monitoring with timing observations on J1810 will offer a more credible range of $\tau_\mathrm{c}$ to be compared with the tentative kinematic age $\tau^{*}_\mathrm{k}\approx70$\,kyr suggested by the possible ``indirect'' association between \snra\ and J1810. Besides, a deeper search for SNR in the narrow region between J1810 and \snra\ might provide an alternative candidate for the SNR associated with J1810.

\section*{Acknowledgements}

We thank Marten van Kerkwijk for his in-depth review and helpful comments on this paper.
H.D. is supported by the ACAMAR (Australia-ChinA ConsortiuM for Astrophysical Research) scholarship, which is partly funded by the China Scholarship Council (CSC).
A.T.D is the recipient of an ARC Future Fellowship (FT150100415).
S.C. acknowledges support from the National Science Foundation (AAG~1815242).
Parts of this research were conducted by the Australian Research Council Centre of Excellence for Gravitational Wave Discovery (OzGrav), through project number CE170100004.
This work is based on observations with the Very Long Baseline Array (VLBA), which is operated by the National Radio Astronomy Observatory (NRAO). The NRAO is a facility of the National Science Foundation operated under cooperative agreement by Associated Universities, Inc.
Data reduction and analysis was performed on OzSTAR, the Swinburne-based supercomputer.
This work made use of the Swinburne University of Technology software correlator, developed as part of the Australian Major National Research Facilities Programme and operated under license. 

\section*{Data and code availability}
\label{sec:J1810_data_availability}
The pipeline for data reduction is available at \url{https://github.com/dingswin/psrvlbireduce}.\\
All VLBA data used in this work can be found at \url{https://archive.nrao.edu/archive/advquery.jsp} under the project codes bd223, bd231, bh142 and bh145a.
The calibrator models for J1753 and J1819 can be downloaded from \\ \url{https://data-portal.hpc.swin.edu.au/dataset/calibrator-models-used-for-vlba-astrometry-of-xte-j1810-197}.

\section*{Appendix: Measuring scatter-broadened size of XTE~J1810--197}
\label{sec:J1810_scatter_broadening}
The angular size of a radio source can be measured from its image deconvolved by the synthesized beam.
As magnetars are point-like radio sources, a non-zero deconvolved angular size of J1810 can be attributed to scatter-broadening effect caused by ISM.
For each epoch, the deconvolved image of J1810 is obtained as an elliptical gaussian component; the mean of its major- and minor-axis lengths is used as the scatter-broadened size of J1810. Assuming the degree of scatter-broadening did not vary across the 14 recent epochs, the 14 measurements of scatter-broadened sizes yield a scatter-broadened size of $0.7\pm0.4$\,mas for J1810.

\bibliographystyle{mnras}
\bibliography{haoding}
\chapter[Probing Magnetar Formation Channels with
High-precision Astrometry]{Probing magnetar formation channels with
high-precision astrometry}
\label{ch:J1818}

This chapter is converted from \citet{Ding23a} entitled ``Probing magnetar formation channels with
high-precision astrometry: The progress of
VLBA astrometry of the fastest-spinning
magnetar Swift~J1818.0$-$1607'', which is the written submission for the International Astronomical Union Symposium 363 with the theme ``Neutron Star Astrophysics at the Crossroads: Magnetars and the Multimessenger Revolution''. 
%At the time of thesis submission, the conference paper is still in press by the Proceedings of the International Astronomical Union.
\ref{sec:J1818_intro} lays out the main scientific motivation of high-precision magnetar astrometry for this thesis, which is to probe magnetar formation channels (also see \ref{subsec:probe_magnetar_formation}). 
%The small sample of magnetar space velocities is enriched by 
The new astrometric results measured for \swift\ (see \ref{sec:J1818_astrometry_progress}) and \xte\ (see \ref{ch:J1818}) mark two steps closer to fulfilling the scientific motivation. 
As an interesting analogy, the space velocity distribution of millisecond pulsars (MSPs) can likely be used to probe the formation channels of MSPs (see \ref{subsec:mspsrpi_v_t} and explanations in \ref{subsec:NS_kinematics}).

\section{Abstract}
Boasting supreme magnetic strengths, magnetars are among the prime candidates to generate fast radio bursts. 
Several theories have been proposed for the formation mechanism of magnetars, but have not yet been fully tested. 
As different magnetar formation theories expect distinct magnetar space velocity distributions, high-precision astrometry of Galactic magnetars can serve as a probe for the formation theories. In addition, magnetar astrometry can refine the understanding of the distribution of Galactic magnetars. This distribution can be compared against fast radio bursts (FRBs) localized in spiral galaxies, in order to test the link between FRBs and magnetars.
%To date, only one magnetar has a model-independent 3D position in the Galaxy and transverse velocity.
\swift\ is the hitherto fastest-spinning magnetar and the fifth discovered radio magnetar. 
In an ongoing astrometric campaign, we have observed \swift\ for one year using the Very Long Baseline Array, and have determined a precise proper motion as well as a tentative parallax for the magnetar.
%\keywords{radio continuum: stars, pulsars: individual (\swift), stars: neutron, techniques: interferometric, techniques: high angular resolution}
%\end{abstract}

\section{Introduction}
\label{sec:J1818_intro}
As the most magnetized objects in the universe, magnetars may account for at least 12\% of the neutron star population \citep{Beniamini19}. However, only roughly 30 magnetars have been identified \citep{Olausen14} in our Galaxy or in the Magellanic Clouds. This discrepancy can be explained by short-lived energetic electromagnetic activities of magnetars. 

Magnetars sit at the intersection of multiple research topics. Their postulated link to fast radio bursts (FRBs) has been strengthened by the FRB-like bursts recently observed from a Galactic magnetar \citep{Andersen20,Bochenek20}. But it remains unclear whether all FRBs are originated from magnetars.
Magnetars are also strongly connected to $\gamma$-ray bursts (GRBs) through the detection of giant magnetar flares from nearby galaxies (e.g. the one in NGC~253, \citealp{Roberts21,Svinkin21}), and newborn magnetars from double neutron star mergers (e.g. \citealp{Sarin21}).
On the other hand, the formation mechanism of magnetars is yet not well understood. A few distinct magnetar formation theories have been proposed, including normal core-collapse supernovae (CCSN) of magnetic massive stars \citep{Schneider19}, accretion-induced collapse (AIC) of white dwarfs \citep{Duncan92} and double neutron star mergers \citep{Giacomazzo13,Xue19}. Most magnetar formation channels require magnetars to be born with millisecond spin periods. The only exception is the normal CCSN formation channel, where a new-born magnetar simply inherits the magnetic fields from its magnetic progenitor star \citep{Schneider19}.

While almost all magnetars are identified with their soft $\gamma$-ray and X-ray activities, $\approx40$\% of them are also visible at optical/infrared or radio frequencies \citep{Olausen14}. This visibility allows precise astrometry for the magnetars. Historically, 7 magnetars have been precisely measured astrometrically, including 4 infrared-bright magnetars \citep{Tendulkar13} and 3 radio magnetars \citep{Deller12a,Bower15,Ding20c}.
The primary motivations of previous astrometric campaigns of magnetars are to {\bf 1)} establish supernova remnant associations and {\bf 2)} test whether magnetars receive extraordinary kick velocities ($\gtrsim1000$\,\kmps) as predicted by \citet{Duncan92}. With growing numbers of Galactic magnetars precisely measured astrometrically, new research opportunities start to emerge.

\begin{figure}
%\floatbox[{\capbeside\thisfloatsetup{capbesideposition={right,center},capbesidewidth=3.5cm}}]{figure}[\FBwidth]
{\caption{Galactic latitudes of Galactic magnetars calculated from their equatorial positions provided by \citet{Olausen14}, where \swift\ is highlighted in red.
%(see \url{http://www.physics.mcgill.ca/~pulsar/magnetar/main.html}). 
%Note that SGR~0526$-$66 and CXOU~J010043.1$-$721134 reside in the Magellanic Clouds. 
%All Galactic magnetars stay close to the Galactic plane.
}\label{fig:gal_lat}}
{\includegraphics[width=5.5in]{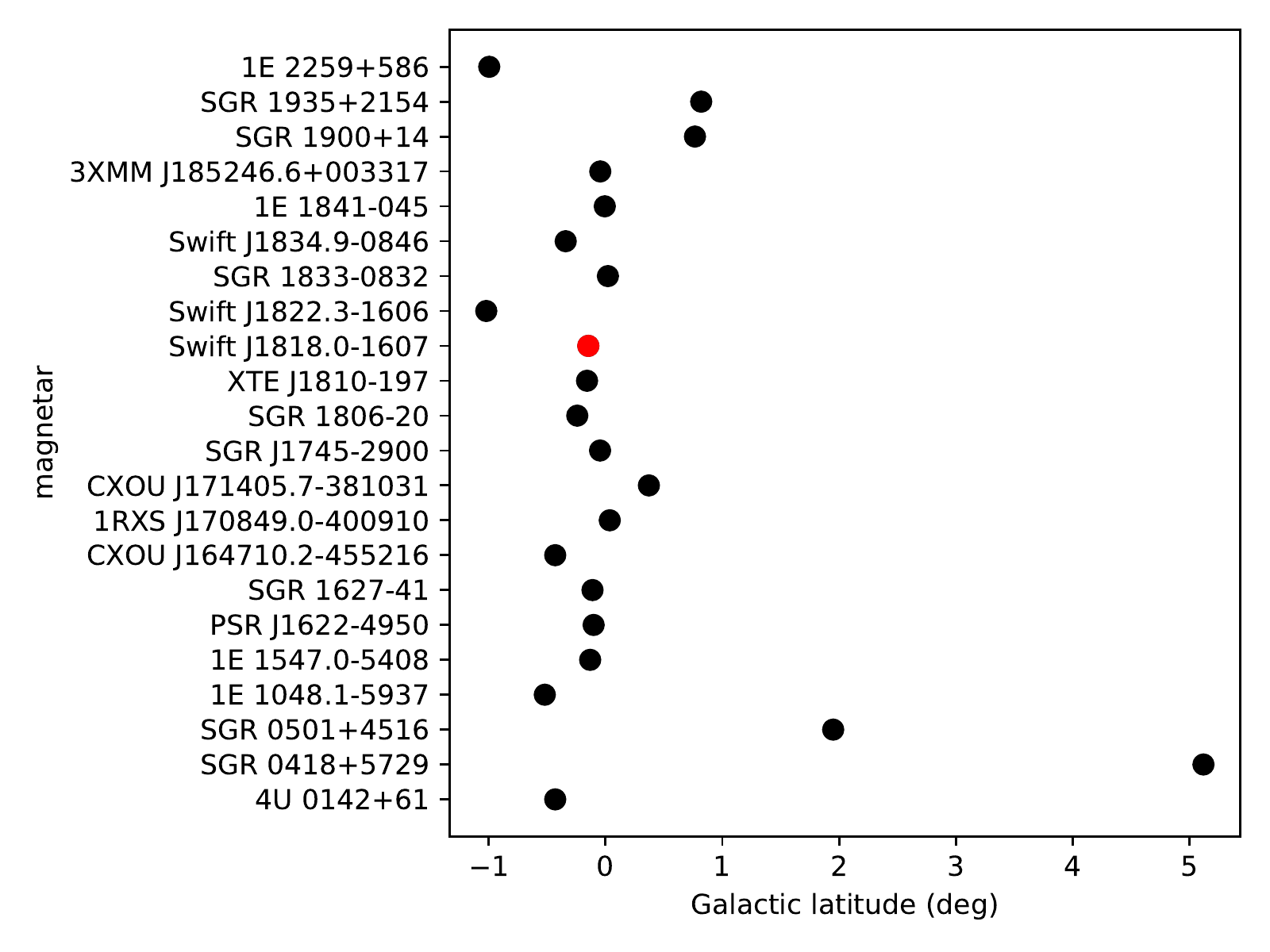}}
\end{figure}

The small sample studied by \citet{Tendulkar13}, entailing 6 astrometrically measured magnetars, offers no indication that magnetar space velocities (magnetar velocity with respect to the neighbourhood of the magnetar) follow a distribution different from that of normal pulsars \citep{Hobbs05}. This rough consistency may imply that most magnetars in spiral galaxies are born in normal CCSN similar to the ones that create typical neutron stars. The CCSN origin of magnetars is also supported by few SNR associations of magnetars (e.g. \citealp{Borkowski17}, GCN circular 16533). On the other hand, the DNS merger origin is not favored by the Galactic magnetar sample, as all Galactic magnetars are found near the Galactic plane (see \ref{fig:gal_lat}). 
%The typical merger time of double neutron stars is 
For other formation channels (e.g., the AIC channel), it remains an open question if they contribute to the formation of the Galactic magnetar population, or more generally magnetars in spiral galaxies.
This question can be approached with a refined magnetar space velocity distribution, to be established with $\gtrsim$10 Galactic magnetars precisely measured astrometrically. As different formation channels may lead to distinct magnetar space velocity distributions (e.g., the AIC channel would probably give rise to relatively small magnetar space velocities), the averaged magnetar space velocity distribution could turn out to be bi-modal or even multi-modal (if there are more than one formation channel of magnetars).

Astrometry of Galactic magnetars would also play a role in the FRB study. On one hand, FRBs have been localized to specific environments (i.e., spiral arms) of spiral galaxies \citep{Mannings21}. 
On the other hand, magnetar astrometry could potentially pinpoint the 3-D magnetar location \citep{Ding20c} in the Galaxy. 
Hence, comparing the Galactic magnetar distribution against FRB sites (localized to spiral galaxies) can test the link between FRBs and magnetars.
However, Galactic magnetars precisely measured astrometrically are usually limited to the vicinity of the Solar system. 
To better infer magnetar distributions in spiral galaxies, one needs the knowledge of contributing formation channels of magnetars in the spiral galaxies (which, again, can be approached by magnetar astrometry). This is because magnetars formed in one channel are expected to follow a specific spatial distribution (e.g., magnetars born via the CCSN and DNS-merger channel are expected to be associated with, respectively, star-forming regions and stellar mass of a galaxy). 
%Once again, the magnetar formation channels can be refined by astrometry of Galactic magnetars.
%, would improve the understanding of magnetar spatial distributions (in spiral galaxies), which can then be compared against the FRB sites (localized to spiral galaxies), in order to test the link between FRBs and magnetars.

\section{The progress of astrometry of \swift}
\label{sec:J1818_astrometry_progress}
\swift\ is the fifth discovered radio-bright \citep{Karuppusamy20} magnetar (GCN circular 27373), which is also the hitherto fastest-spinning magnetar with a spin period of 1.4\,s \citep{Enoto20}. 
%The small characteristic age ($\lesssim300$\,yr, \citealp{Champion20,Champion20a,Hu20}) of \swift\ implies its great youth.
Its short spin period and high spin-down rate correspond to a characteristic age of around 500\,yr \citep{champion20a}, implying its great youth.
Right after the radio detection of \swift, we launched an astrometric campaign of the magnetar using the Very Long Baseline Array (VLBA). The first VLBA observation was made at 1.6\,GHz on 20 April 2020, which did not lead to a detection \citep{Ding20b}. In response to the spectral flattening (of \swift) first noticed in July 2020 \citep[e.g.][]{Majid20}, we raised the observing frequency to 8.8\,GHz, and managed to detect \swift\ with VLBA on 19 August 2020 \citep{Ding20b}.

\begin{figure}
%\floatbox[{\capbeside\thisfloatsetup{capbesideposition={right,center},capbesidewidth=2.6cm}}]{figure}[\FBwidth]
{\caption{Sky positions of \swift\ relative to the reference position $18^{\rm h}18^{\rm m}00\fs 19327$, $-16\degr07'53\farcs0095$. The positions are labeled with the observing dates in MJD. We note that the positional uncertainties presented here only reflect random errors caused by image noises. The blue curve represents the best-fit astrometric model.}\label{fig:ra_dec_model}}
{\includegraphics[width=5.5in]{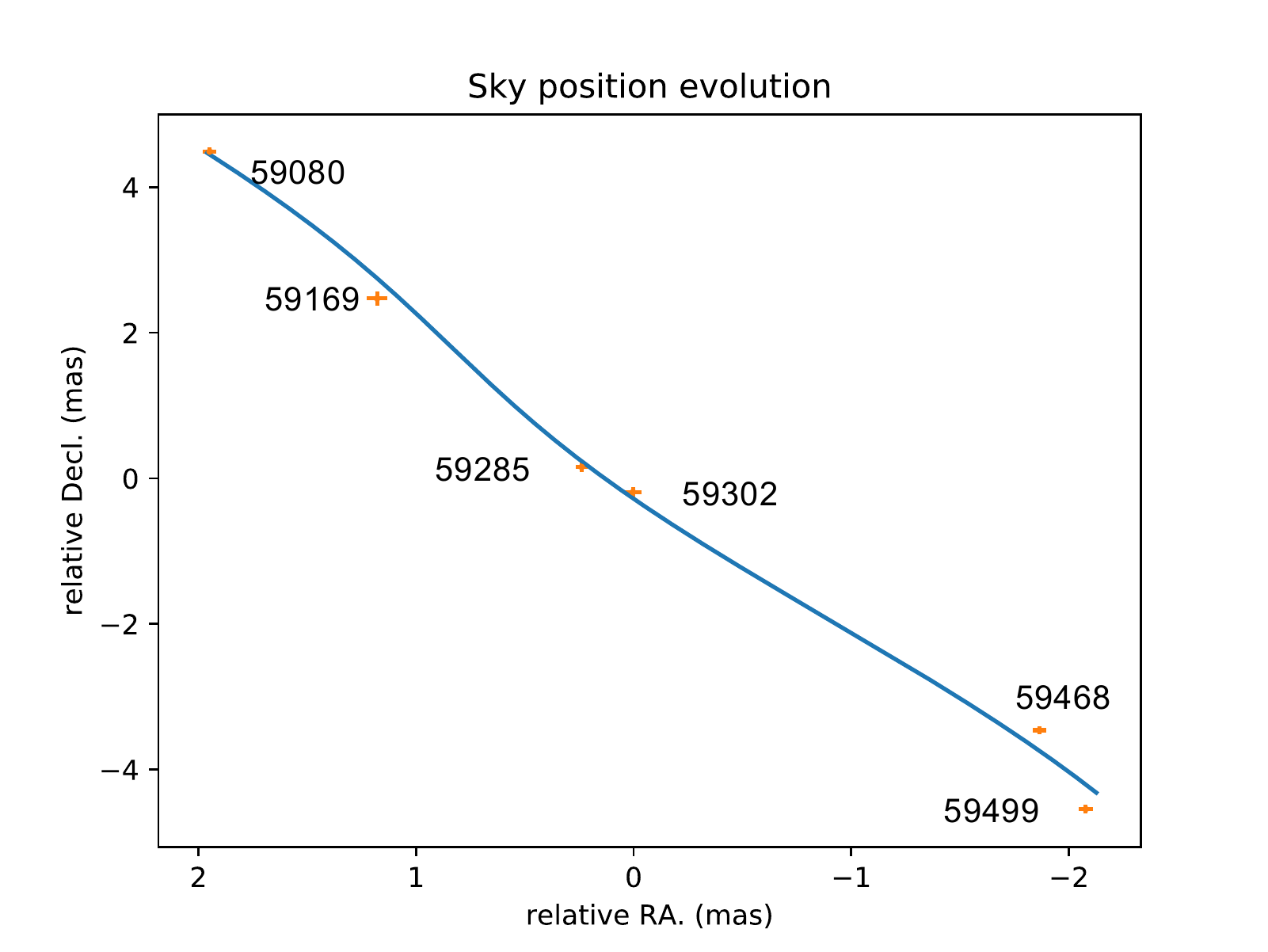}}
\end{figure}

At the time of writing, more than a year has passed since the first VLBA detection. During this period, 5 more VLBA observations have been made, all resulting in detections at sub-mas positional precision. 
%For all the 6 VLBA observations, we consistently adopt the 1-D interpolation tactic \citep[e.g.][]{Fomalont03,Ding20c}. 
We applied pulsar gating to improve image S/N, where the pulse ephemerides were generated from ongoing Parkes monitoring of \swift\ \citep{Lower20c}.
To enhance astrometric accuracy of our observations, we have employed the 1-D interpolation tactic \citep[e.g.][]{Fomalont03,Ding20c} for all the 6 VLBA observations.
After data reduction and analysis, we obtained a compelling proper motion $\mu_{\alpha}=-3.54\pm0.05$\,\maspy, $\mu_{\delta}=-7.65\pm0.09$\,\maspy, alongside a tentative (5\,$\sigma$) parallax (see \ref{fig:ra_dec_model}). Here, in order to roughly account for the systematic errors caused by atmospheric propagation effects, the uncertainty of the proper motion (as well as parallax) has been scaled by $1/\sqrt{\chi^2_{\nu}}$, where $\chi^2_{\nu}$ stands for reduced chi-square of direct parallax fitting. The final astrometric results with thorough error estimation will be obtained and discussed in a future publication, following the completion of the whole astrometric campaign that spans at least 2 years.

\section{Future prospects}
\label{sec:J1818_future prospects}
For \swift, with 6 more VLBA observations to be made in the following year, the parallax measurement is likely to become substantial ($>7\,\sigma$), which would simplify the conversion of parallax to distance (without needing to take into account the Lutz-Kelker effect, \citealp{Lutz73}). 
With regard to the whole magnetar category, to establish the magnetar space velocity distribution is mainly limited by the small sample of ($\approx11$, \citealp{Olausen14}) radio or optical/infrared magnetars. 
To expand this sample and hence accelerate the establishment of the magnetar space velocity distribution, new candidates of radio or optical/infrared magnetars, such as the newly discovered ultra-long-period sources \citep[e.g.][]{Hurley-Walker22}, are desired.

\bibliographystyle{mnras}
\bibliography{haoding}
%\setlength{\itemsep}{0mm}
%\small

%\end{document}
\chapter[Testing the Models of Photospheric Radius Expansion Bursts with Gaia Parallaxes]{Testing the models of photospheric radius expansion bursts with Gaia parallaxes}
\label{ch:PRE}

This chapter is transformed from \citet{Ding21} entitled ``Gaia EDR3 parallaxes of type I X-ray bursters and their implications
on the models of type I X-ray bursts: A generic approach to the Gaia
parallax zero point and its uncertainty''.
The type I X-ray bursters (bursters) studied in this chapter are a subset of the neutron star (NS) X-ray binary (XRB) population, and will eventually evolve into recycled pulsars (which are the focus of \ref{ch:J1012}, \ref{ch:B1534} and \ref{ch:mspsrpi}) after the phase of active accretion (see \ref{subsec:NS_observations}). 

This is the only paper chapter based on Gaia data (see \ref{subsec:overview_astrometry} for a summary of methods of high-precision astrometry). The Gaia sources (see \ref{tab:Gaia_counterparts}) identified for the 4 bursters are the optically bright donor stars of the XRB systems (see \ref{subsec:NS_observations}). 
Among the 4 bursters, radio emissions have only been reported for \cyg\ \citep[e.g.][]{Hjellming90,Spencer13}. Hence, Gaia astrometry serves as the major pathway to fulfilling the scientific goal of this chapter.
Provided the relatively large distances to the 4 bursters, the position shift due to orbital motion is too small to affect the Gaia astrometric fitting performance (see \ref{sec:ch2_astrometry_inference}). %as their distances are relatively large, and the XRB orbits are expected to be compact. 

\section{Abstract}
Light curves of photospheric radius expansion (PRE) bursts, a subset of type I X-ray bursts, have been used as standard candles to estimate the ``nominal PRE distances'' for 63\% of PRE bursters (bursters), assuming PRE burst emission is spherically symmetric.
Model-independent geometric parallaxes of bursters provide a valuable chance to test models of PRE bursts (PRE models), and can be provided in some cases by Gaia astrometry of the donor stars in bursters.
We searched for counterparts to 115 known bursters in the Gaia Early Data Release 3, and confirmed 4 bursters with Gaia counterparts that have detected ($>3\,\sigma$, prior to zero-point correction) parallaxes. 
We describe a generic approach to the Gaia parallax zero point as well as its uncertainty 
using an ensemble of Gaia quasars individually determined for each target. 
Assuming the spherically symmetric PRE model is correct, we refined the resultant nominal PRE distances of three bursters (i.e. \cen, \cyg\ and \uua), and put constraints on their compositions of the nuclear fuel powering the bursts. Finally, we describe a method for testing the correctness of the spherically symmetric PRE model using parallax measurements, and provide preliminary results.

\section{Introduction}
\label{sec:PRE_intro}

A type I X-ray burst is rapid nuclear burning on the surface of a neutron star (NS) \citep[e.g.][]{Lewin93}. 
Type I X-ray bursts happen when the accumulated ``nuclear fuel'' accreted from the donor star onto the surface of the NS is sufficiently hot. 
In some cases, the radiation pressure becomes large enough to overcome the gravity of the NS, causing the photosphere to expand. This is followed by fallback of the photosphere after extending to its outermost point. This process of expansion and contraction of the photosphere is well characterised by double-peaked soft X-ray light curves. These particular type I X-ray bursts are called photospheric radius expansion (PRE) bursts.

According to \citet{Galloway20}, there are hitherto 115 recognized PRE bursters (see \url{https://burst.sci.monash.edu/sources}), which are all low mass X-ray binaries (LMXBs).
PRE bursts can serve as standard candles \citep{Basinska84} based on the assumption that the luminosity of PRE bursts stays at the Eddington limit throughout the photospheric expansion and contraction. We refer to the distances estimated in this way as PRE distances. 
Using the standard assumption of spherically symmetric PRE burst emission, \citet{Galloway20} constrained PRE distances for 73 PRE bursters (23 of which only have upper limits of PRE distances, see Table~8 of \citealp{Galloway20}) based on the theoretical Eddington luminosity calculated by \citet{Lewin93}. 
Hereafter, we refer to this spherically symmetric model as the simplistic PRE model, and the simplistic-PRE-model-based PRE distances as nominal PRE distances.
Besides PRE models, a Bayesian framework has been recently developed to infer parameters, including the distance and the composition of nuclear fuel on the NS surface, for the type I X-ray burster \sax\ \citep{Goodwin19}, by matching the burst observables (such as burst flux and recurrence time) of non-PRE type I X-ray bursts with the prediction from a burst ignition model \citep{Cumming00}. 

At least 43\% of PRE bursters registered nominal PRE distances as their best constrained distances.
The great usefulness of the simplistic PRE model calls for careful examination of its validity. There are several uncertainties in the simplistic PRE model. Firstly, according to \citet{Lewin93}, the Eddington luminosity measured by an observer at infinity
\begin{equation}
\label{eq:Eddington_luminosity}
L_{\mathrm{Edd},\infty} \propto 
\frac{M_\mathrm{NS}}{1+z(R_\mathrm{P})} \frac{1}{1+X}
\end{equation}
depends on the NS mass $M_\mathrm{NS}$ and the gravitational redshift $z(R_\mathrm{P})=(1-2G_\mathrm{N} M_\mathrm{NS}/c^2 R_\mathrm{P})^{-1/2}-1$ (where $G_\mathrm{N}$ and $c$ stand for Newton's gravitational constant and speed of light in vacuum, respectively) at the photosphere radius $R_\mathrm{P}$. To a greater degree, $L_{\mathrm{Edd},\infty}$ hinges on the hydrogen mass fraction $X$ of the nuclear fuel on the NS surface at the time of the PRE burst: hydrogen-free nuclear fuel corresponds to a $L_{\mathrm{Edd},\infty}$ 1.7 times higher than nuclear fuel of cosmic abundances \citep{Lewin93}. Secondly, the method assumes spherically symmetric emission. 
On one hand, PRE bursts per se are not necessarily spherically symmetric given that the NS in an LMXB is spinning and accretes via an accretion disk.
On the other hand, even if PRE bursts are initially isotropic (assuming the nuclear fuel has spread evenly on the NS surface), propagation effects (such as the reflection from the surrounding accretion disk) would still potentially lead to anisotropy of PRE burst emission.
Thirdly, for each PRE burster, the peak fluxes of its recognized PRE bursts vary, typically by 13\% \citep{Galloway08}. 

Provided the uncertainties of the simplistic PRE model as well as the resultant nominal PRE distances, independent distance measurements for PRE bursters hold the key to testing the simplistic PRE model.
By 2003, 12 PRE bursters in 12 globular clusters (GCs) had distances determined from RR Lyrae stars residing in the GCs; incorporating the distances with respective peak bolometric fluxes derived from X-ray observations of PRE bursts, \citet{Kuulkers03} measured 12 observed Eddington luminosities, 9 of which are consistent with the theoretical value by \citet{Lewin93}.
The independent check by \citet{Kuulkers03} largely confirms the validity of the simplistic PRE model. The confirmation, however, is not conclusive due to the modest number of PRE bursters residing in GCs. Besides, this GC-only sample of PRE bursters might lead to some systematic offsets of the measured Eddington luminosities, as a result of potential systematic offsets of distances determined with RR Lyrae stars or X-ray flux decay that becomes more prominent in the dense regions of GCs.

Alternatively, geometric parallaxes of PRE bursters provided by the Gaia mission \citep{Gaia-Collaboration16} can also be used to test the simplistic PRE model.
This effort normally requires determination of Gaia parallax zero point $\pi_0$ for each PRE burster, as $\pi_0$ is found to be offset from 0 \citep{Lindegren18,Lindegren21}.
Recently, \citet{Arnason21} (A21) has identified Gaia Data Release 2 (DR2) counterparts for 10 PRE bursters with Gaia parallaxes. After applying the Gaia DR2 global $\pi_0$ \citep{Lindegren18}, A21 converted these parallaxes into distances, incorporating the prior information described in \citet{Bailer-Jones18}. A21 noted that the inferred distances of the 10 PRE bursters are systematically smaller than the nominal PRE distances estimated before 2008 (this discrepancy is relieved with the latest nominal PRE distances by \citealp{Galloway20}, see the discussion in \ref{subsubsec:PRE_bayes_factor_about_X}).  
However, due to the predominance of low-significance parallax constraints, this discrepancy is strongly dependent on the underlying Galactic distribution of PRE bursters (as is shown in Figure~3 of A21), which is far from well constrained. 
Moreover, adopting the Gaia DR2 global $\pi_0$, instead of one individually estimated for each source, would introduce an extra systematic error.

This work furthers the effort of A21 to test the simplistic PRE model with Gaia parallaxes of PRE bursters, making use of the latest Gaia Early Data Release 3 (EDR3, \citealp{Brown21}). 
Unlike A21, this work only focuses on the PRE bursters with relatively significant Gaia parallaxes; we apply a locally-determined $\pi_0$ for each PRE burster, and probe the simplistic PRE model as well as the composition of nuclear fuel.

In addition to the above-mentioned motivation, Gaia astrometry of PRE bursters (hereafter simplified as bursters, when unambiguous) can provide more than geometric parallaxes (and hence distances). A Gaia counterpart of a burster also brings a reference position precise to sub-mas level, and possibly a detected proper motion as well.
The latter can be combined with the parallax-based or model-dependent distance estimates to yield the burster's transverse space velocity (the transverse velocity with respect to its Galactic neighbourhood).
The space velocity distribution of PRE bursters (or LMXBs) is important for testing binary evolution theories (see \citealp{Tauris17} as an analogy).
Both reference position and proper motion can be used to confirm or rule out candidate counterparts at other wavelengths, especially when the immediate neighbourhood of a burster is crowded on a $\sim$1'' scale.

In this paper, all uncertainties are quoted at the $1\,\sigma$ confidence level unless otherwise stated; all sky positions are J2000 positions.
As we mainly deal with Gaia photometric passbands \citep{Jordi10} in this work, a conventional optical passband is referred to in the form of $\mathrm{Y}^*$, while its magnitude is denoted by $Y^*$; by contrast, a Gaia passband is referred to in the form of Y band, and its magnitude is denoted by $m_\mathrm{Y}$.

\section{Gaia counterparts of PRE bursters with detected parallaxes}
\label{sec:PRE_identification}
To be able to effectively refine nominal PRE distances, constrain the composition of nuclear fuel on the NS surface and test the simplistic PRE model, we only search for Gaia counterparts with detected parallaxes $\pi_1$ ($>3\,\sigma$, a criterion we apply throughout this paper). This cutoff leads to a smaller sample of PRE bursters compared to A21, which included many marginal parallax detections. 
Our search is based on the positions of 115 PRE bursters compiled in Table~1 of \citet{Galloway20}. 
For each PRE burster, we found its closest Gaia EDR3 source within 10\farcs0 using {\tt TOPCAT}\footnote{\url{http://www.star.bris.ac.uk/~mbt/topcat/}}. 
From the resultant 110 candidates, we shortlisted 16 Gaia counterpart candidates with detected $\pi_1$. 
Among the 16 Gaia counterpart candidates, the Gaia counterparts for \cyg, \cen\ and \uua\ have been recognized by VizieR \citep{Ochsenbein00} in an automatic manner, simply based on the $\approx1$\,mas angular distance (at the reference epoch year 2015.5 disregarding proper motion) that is comparable to the Gaia positional uncertainties (see \ref{tab:Gaia_counterparts}); the Gaia counterpart for XB~2129$+$47 has been identified by A21.

To identify Gaia counterparts of PRE bursters from the remaining 12 candidates, we adopted a more complicated cross-match criterion,
which requires the identification of an optical counterpart (either confirmed or potential) for the PRE burster. For an optical source $\mathcal{S}$ that is a confirmed or potential counterpart, we consider it associated with the Gaia source $\mathcal{G}$ if all the following conditions are met:
\begin{enumerate}[label=(\roman*)]
    \item $\mathcal{S}$ is sufficiently bright ($3\leq G^*\leq 21$\footnote{\url{https://www.cosmos.esa.int/web/gaia/earlydr3}}) for Gaia detection; the conventional apparent magnitude measured closest to $\mathrm{G}^*$ band is looked at if $G^*$ is not available (as compiled in \ref{tab:Gaia_counterparts});
    \item $\mathcal{G}$ falls into the 1-$\sigma_\mathcal{S}$-radius circle around $\mathcal{S}$ (where the position of a confirmed radio or infrared counterpart of $\mathcal{S}$ would be adopted if it is more precise than the optical position);
    \item $\mathcal{G}$ is the only Gaia source within the 5-$\sigma_\mathcal{S}$-radius circle around $\mathcal{S}$.
\end{enumerate}
Here, to account for the effect of proper motion (and the smaller contribution of parallax), 
the position of the candidate has been extrapolated to 
the epoch at which the position of $\mathcal{S}$ was measured, with the astrometric parameters of $\mathcal{G}$ using the ``predictor mode'' of {\tt pmpar} (available at \url{https://github.com/walterfb/pmpar}). This extrapolation allows the evaluation of the same-epoch $\Delta_\mathcal{S-G}$, the separation between $\mathcal{S}$ and $\mathcal{G}$, which is free from proper motion and parallax effects. 

In principle, the apparent magnitude of $\mathcal{G}$ should match that of the optical counterpart of $\mathcal{S}$. However, making such a comparison is complicated by {\bf 1)} the wide photometric passbands designed for the in-flight Gaia \citep{Jordi10} and {\bf 2)} magnitude variability of stars. 

On top of the 4 Gaia counterparts of PRE bursters (\cen, \cyg, \uua\ and \xb) already recognized by VizieR and A21, we identified one Gaia counterpart with detected $\pi_1$ (from the  remaining 12 objects), which was NP~Ser (see \ref{tab:Gaia_counterparts}). However, as we will discuss in \ref{subsec:PRE_NpSer_neq_Gx17}, NP~Ser is not the optical counterpart of \gx, meaning that our final sample consists of 5 Gaia sources with detected $\pi_1$, of which 4 are PRE bursters.

Hereafter, the 5 Gaia sources are sometimes referred to as ``targets''.
The (Gaia) B band covers the conventional $\mathrm{G}^*$, $\mathrm{V}^*$, $\mathrm{B}^*$ bands and part of the $\mathrm{R}^*$ band. According to \ref{tab:Gaia_counterparts}, the magnitude $V^*$ of NP~Ser measured at conventional $\mathrm{V}^*$ band generally agrees with $m_\mathrm{B}$ of its Gaia counterpart.
The 5 astrometric parameters of the Gaia counterparts for the 4 PRE bursters and NP~Ser are summarized in \ref{tab:before_calibration}.

\begin{table*}
\caption{Gaia EDR3 counterparts with detected parallaxes $\pi_1$ for 4 PRE bursters and NP~Ser}
\label{tab:Gaia_counterparts}
\centering
\resizebox{\textwidth}{!}{
\begin{tabular}{@{}cccccccc@{}}
\hline\hline
source name & $m^*$ $^{a_1}$ & $\sigma_\mathcal{S}$ $^{a_2}$ & Gaia EDR3 & $m_\mathrm{B}$ $^{a_3}$ & $\Delta_\mathcal{S-G}$ $^{a_4}$ & $R_1$ $^{a_5}$ & associated\\
& (mag) & ('') & counterpart & (mag) & ('') & ('') & by? \\
\hline%
NP~Ser & $V^*\!=\!17.42 ^{e_1}$ & 0.5$^{e_2}$ & 4146621775597340544 & 17.72 & 0.3 & 12 & this work\\
\cyg\ & $-$ & $-$ & 1952859683185470208 & $-$ & $-$ & $-$ & VizieR$^i$\\
\cen\ & $-$ & $-$ & 6205715168442046592 & $-$ & $-$ & $-$ & VizieR$^i$\\
\uua\ & $-$ & $-$ & 5310395631798303104 & $-$ & $-$ & $-$ & VizieR$^i$\\
\xb\ & $-$ & $-$ & 1978241050130301312 & $-$ & $-$ & $-$ & A21$^k$\\
\hline\hline
\end{tabular}
}

\tabnote{$^{a_1}$Apparent magnitude at a conventional passband;
$^{a_2}$positional uncertainty (adding in quadrature uncertainties in both directions) of the source (see \ref{sec:PRE_identification}); 
$^{a_3}$apparent magnitude at the Gaia B band \citep{Jordi10}, which covers the conventional $\mathrm{G^*}$, $\mathrm{V^*}$, $\mathrm{B^*}$ bands and part of the $\mathrm{R^*}$ band;
$^{a_4}$angular separation between the source and its Gaia counterpart, where $\alpha_\mathcal{G}$ and $\delta_\mathcal{G}$ were extrapolated to the respective epoch of $\alpha_\mathcal{S}$ (and $\delta_\mathcal{S}$) using the astrometric parameters in \ref{tab:before_calibration} (see \ref{sec:PRE_identification} for more explanation); 
$^{a_5}$maximum search radius that contains only one Gaia source.}

\tabnote{$^{e_1}$\citet{Deutsch96};
$^{e_2}$\citet{Deutsch99}, where $\alpha_\mathcal{S}=18^{\rm h}16^{\rm m}01\fs$380, $\delta_\mathcal{S}=-$14\degr02'11\farcs34 out of R-band CCD observation, with uncertainty radius of 0\farcs5 at 90\% confidence, measured at MJD~50674.}

\tabnote{
$^i$\citet{Ochsenbein00}; $^k$\citet{Arnason21}.}
\end{table*}

\begin{table*}
\caption{Five astrometric parameters from Gaia EDR3 counterparts at year 2016.0 prior to calibration of Gaia parallaxes}
\label{tab:before_calibration}
\centering
\resizebox{\textwidth}{!}{
\begin{tabular}{@{}cccccc@{}}
\hline\hline
source name & $\alpha_\mathcal{G} \pm \sigma_\alpha$ & $\delta_\mathcal{G}$ & $\pi_1$ & $\mu_\alpha \equiv \dot{\alpha} \cos\delta$ & $\mu_\delta$ \\
& ($\sigma_\alpha$ in mas) &  & (mas) & (\maspy) & (\maspy)\\
\hline%
NP~Ser & $18^{\rm h}16^{\rm m}01\fs 392863 \pm 0.07$ & $-14\degr02'11\farcs 75580(6)$ & 0.69(8) & 3.35(8) & -6.97(6) \\
\cyg\ & $21^{\rm h}44^{\rm m}41\fs 152001 \pm 0.01$ & $38\degr19'17\farcs 06138(1)$ & 0.068(19) & -1.79(2) & -0.32(2) \\
\cen\ & $14^{\rm h}58^{\rm m}21\fs 93584 \pm 0.1$ & $-31\degr40'08\farcs 40635(9)$ & 0.53(13) & 0.84(15) & -55.68(13) \\
\uua\ & $09^{\rm h}20^{\rm m}26\fs 471033 \pm 0.05$ & $-55\degr12'24\farcs 47694(6)$ & 0.24(6) & -5.77(7) & 2.24(8) \\
\xb\ & $21^{\rm h}31^{\rm m}26\fs 209631 \pm 0.06$ & $47\degr17'24\farcs 44432(7)$ & 0.50(8) & -2.34(8) & -4.23(8) \\
\hline\hline
\end{tabular}
}
\end{table*}

\section{Calibration of Gaia parallaxes}
\label{sec:PRE_parallax_calibration}
Before being applied for scientific purposes, each of the Gaia parallaxes $\pi_1$ needs to be calibrated by determining its (Gaia) parallax zero point $\pi_0$, as $\pi_0$ has been shown to be systematically offset from zero \citep[with an all-sky median of $-0.02$\,mas for EDR3;][]{Lindegren21}.
Previous studies have revealed the dependence of $\pi_0$ on sky position, apparent magnitude and color \citep{Lindegren18,Chen18,Huang21}. Furthermore, \citet{Lindegren21} proposed that the position dependence of $\pi_0$ can be mostly attributed to the evolution of $\pi_0$ with respect to ecliptic latitude $\beta$, and derived an empirical global solution for five-parameter (see \citealp{Lindegren21a} for explanation) Gaia EDR3 sources and another such solution for six-parameter sources.
While this approach is convenient to use, the empirical solutions do not provide an estimate of the uncertainty of $\pi_0$, and are only considered indicative \citep{Lindegren21}. 

In this work, we attempted a new pathway of generic $\pi_0$ determination. Unlike the global empirical solutions, the new $\pi_0$ determination technique offers locally acquired $\pi_0$, and provides the uncertainty of the determined $\pi_0$.
We estimated $\pi_0$ for the 5 targets using a subset of the {\tt agn\_cross\_id} table publicized along with EDR3 (Klioner et al. in preparation).
We selected 1,592,629 quasar-like sources (out of the 1,614,173 sources in the {\tt agn\_cross\_id} table), which {\bf a)} have $m_\mathrm{B-R}$ (denoted as ``{\tt bp\_rp}'' in EDR3) values and {\bf b)} show no detected ($>3\,\sigma$ level) proper motion in either direction. Hereafter, we refer to this subset of 1,592,629 sources as the quasar catalog, and its sources as background quasars or quasars (when unambiguous).

On the sky, spatial patterns of $\pi_0$ on angular scales of up to $\sim10$\degr\ are seen in Figure~2 of  \citet{Lindegren21}.
We attempted different radius cuts (around each of the 5 targets) of up to 40\degr, but found no clear evidence showing quasar parallaxes representative of the target $\pi_0$ beyond a 10\degr\ radius cut.
In most cases, a 10\degr\ radius cut provides a sufficiently large sample of quasars for $\pi_0$ estimation. Hence, we only present analysis of quasars within 10\degr\ around each target.

\subsection{A new pathway to parallax zero points}
\label{subsec:PRE_s_pi0}

Our $\pi_0$ determination is solely based on the quasar catalog (of 1,592,629 sources).
The vital step of (Gaia) parallax calibration (or $\pi_0$ determination) is to select appropriate quasars in the same sky region with similar G-band apparent magnitudes $m_\mathrm{G}$ and colors $m_\mathrm{B-R}$.
We implemented this selection for each of the 5 targets by applying 4 filters to the quasar catalog: the angular-distance filter, the $\beta$ filter, the $m_\mathrm{G}$ filter and the $m_\mathrm{B-R}$ filter. 
Once appropriate background quasars are chosen for each target, one can calculate $\pi_0$ as the weighted mean parallax of this ``sub-sample'', and the formal uncertainty of $\pi_0$
\begin{equation}
    \tilde{\sigma}_{\pi_0}=\left({\sum\limits_{i}\frac{1}{{\sigma_i}^2}}\right)^{-\frac{1}{2}}
	\label{eq:parallax_zero_point_uncertainty}
\end{equation}
(Equation~4.19 in \citealp{Bevington03}), where $\sigma_i$ is the (Gaia) parallax error of each background quasar in the sub-sample. 

We parameterized {\bf 1)} the angular-distance filter using the search radius $r$ around the target, {\bf 2)} the $\beta$ filter using $\Delta \sin{\beta}$, the half width of the $\sin{\beta}$ filter centred about $\sin{\beta^\mathrm{target}}$, {\bf 3)} the $m_\mathrm{G}$ filter using $\underline{\Delta} m_\mathrm{G}$, which stands for the relative half width (or $\Delta m_\mathrm{G}/m_\mathrm{G}^\mathrm{target}$) of the $m_\mathrm{G}$ filter around $m_\mathrm{G}^\mathrm{target}$, and {\bf 4)} the $m_\mathrm{B-R}$ filter using $\Delta m_\mathrm{B-R}$, the half width of the $m_\mathrm{B-R}$ filter centred about $m_\mathrm{B-R}^\mathrm{target}$.

\begin{figure}
    \centering
	\includegraphics[width=12cm]{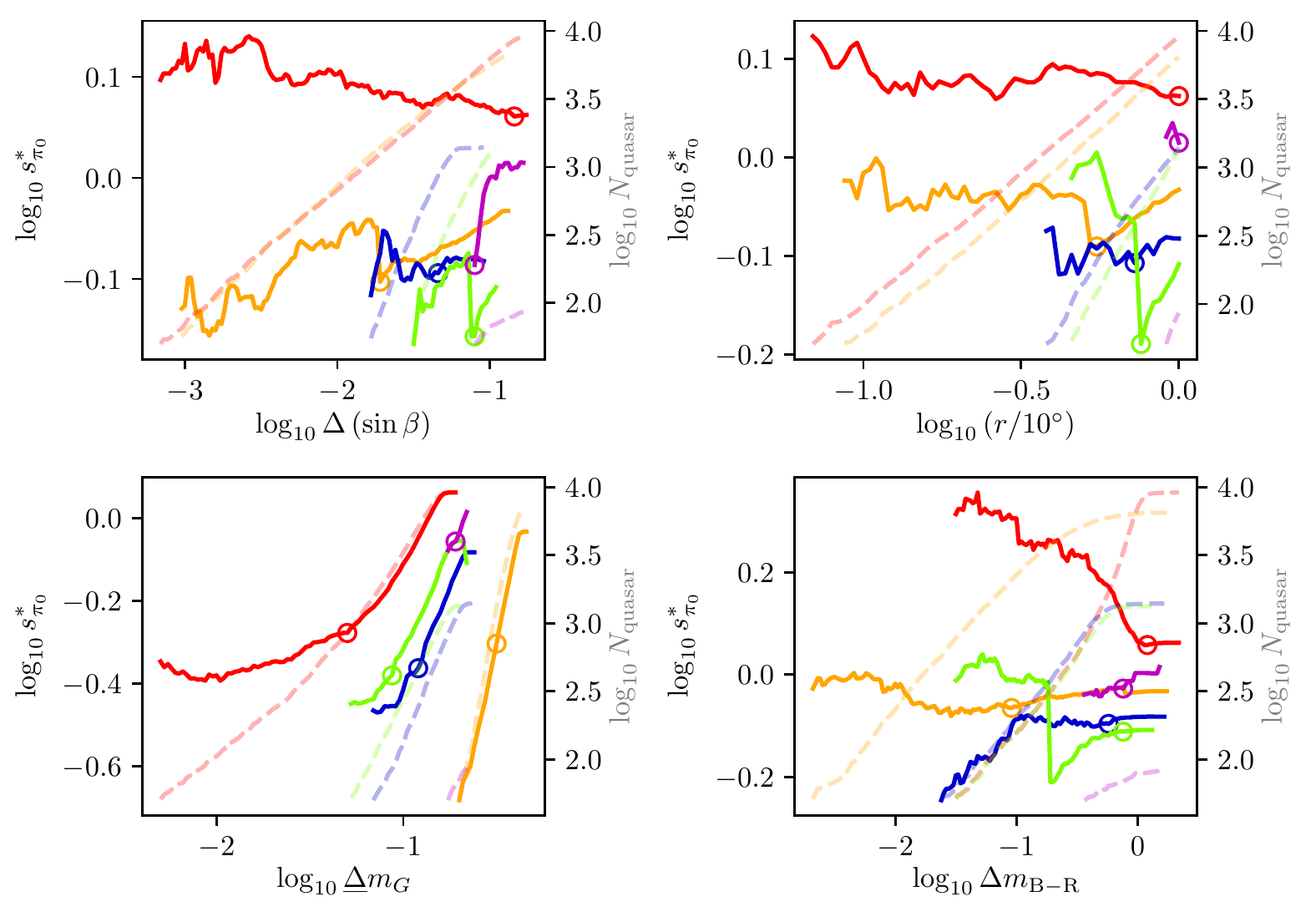}
	\includegraphics[width=12cm]{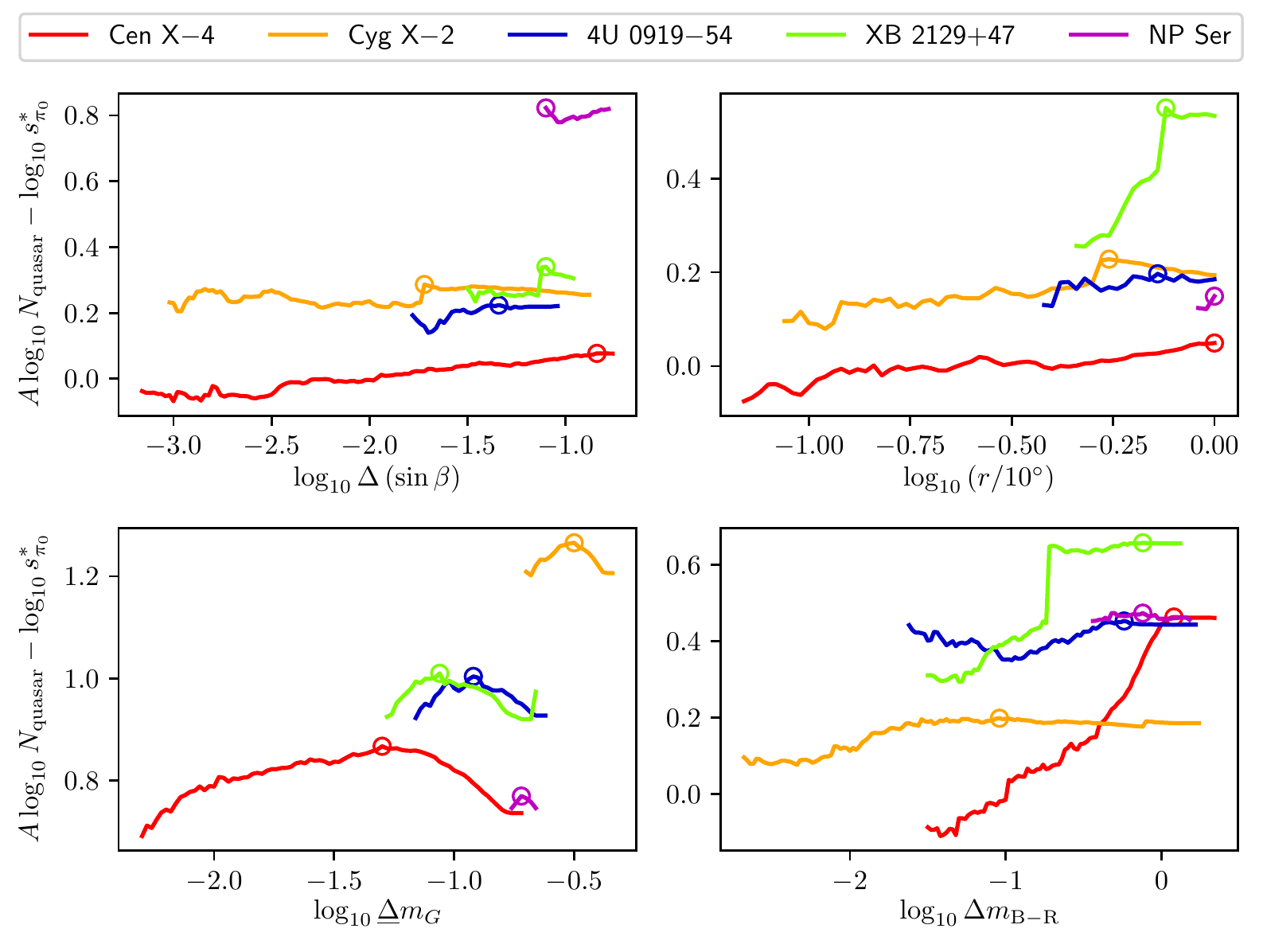}
    \caption{\small {\bf Top 4-panel block:} The solid lines present the marginalized relations between $s_{\pi_0}^{*}$ of background quasars and 4 filter parameters: $\Delta \sin{\beta}$ ($\beta$ denotes ecliptic latitude), search radius $r$ around the target, $\underline{\Delta} m_\mathrm{G}$ and $\Delta m_\mathrm{B-R}$.
    Here, $s_{\pi_0}^{*}=s_{\pi_0}/s_{\pi_0}^{\rm glb}$, where $s_{\pi_0}$ and $s_{\pi_0}^{\rm glb}$ represent the weighted standard deviation of quasar parallaxes and its global value (of the 1592629 quasars in the quasar catalog), respectively; 
    $\Delta m_\mathrm{B-R}$|$\Delta \sin{\beta}$ stands for the half width of the $m_\mathrm{B-R}$|$\sin{\beta}$ filter centred around the $m_\mathrm{B-R}^\mathrm{target}$|$\sin{\beta^\mathrm{target}}$ (used to pick like-$m_\mathrm{B-R}$|like-$\beta$ background quasars); $\underline{\Delta} m_\mathrm{G}$ defines the half relative width of the $m_\mathrm{G}$ filter around $m_\mathrm{G}^\mathrm{target}$ used to select like-$m_\mathrm{G}$ background quasars.
    The dashed curves lay out the marginalized relations between $N_\mathrm{quasar}$ and the 4 filter parameters, where $N_\mathrm{quasar}$ stands for number of remaining background quasars after being filtered. 
    To avoid small-sample effect, the calculations start from $N_{\rm quasar}=50$, or $N_{\rm quasar}^{\rm start}=50$.
    The dependence of $\pi_0$ on a variable ($r$, $\beta$, $m_\mathrm{G}$ or $m_\mathrm{B-R}$) is suggested if $s_{\pi_0}^{*}$ grows with a larger filter (such as the $m_\mathrm{G}$ dependence), and vice versa.
    {\bf Bottom 4-panel block:} 
    The marginalized relation between the index $w=A \log_{10}N_{\rm quasar}-\log_{10}s_{\pi_0}^{*}$ and each filter parameter, where $A=(\log_{10}s_{\pi_0}^{\rm *,max}-\log_{10}s_{\pi_0}^{\rm *,min})/(\log_{10}N_{\rm quasar}^{\rm max}-\log_{10}N_{\rm quasar}^{\rm start})$ is a scaling factor. The maximum of $w$ is chosen (circled out in both bottom and top blocks) as the ``optimal'' filter parameter.
    }
    \label{fig:S__DM_G__relation}
\end{figure}

Hence, the problem of $\pi_0$ determination using background quasars can be reduced to searching for optimal filter parameters (for each target), i.e. $r$, $\Delta \sin{\beta}$, $\underline{\Delta} m_\mathrm{G}$ and $\Delta m_\mathrm{B-R}$. 
For this search, we investigated the marginalized relation between $s_{\pi_0}^{*}$ and each of the 4 parameters (see the top 4-panel block of \ref{fig:S__DM_G__relation}), where $s_{\pi_0}^{*}=s_{\pi_0}/s_{\pi_0}^\mathrm{glb}$ stands for the weighted standard deviation of quasar parallaxes divided by its global value of 0.32\,mas. 
To avoid small-sample fluctuations, \ref{fig:S__DM_G__relation} starts at $N_\mathrm{quasar}=50$, or $N_\mathrm{quasar}^\mathrm{start}=50$, which can also prevent overly small-sized sub-samples.
To start with, we first introduce the $s_{\pi_0}^{*}$-to-$\underline{\Delta} m_\mathrm{G}$ relation and explain its implications. Relations between $s_{\pi_0}^{*}$ and other 3 parameters can be interpreted in the same way.

Unlike $\tilde{\sigma}_{\pi_0}$, if a sufficiently large ($\gtrsim50$) background-quasar sample had no $m_\mathrm{G}$ dependence (as opposed to \citealp{Lindegren18,Lindegren21}), $\log_{10} s_{\pi_0}^{*}$ would be expected to be largely independent to $\underline{\Delta} m_\mathrm{G}$, and to fluctuate around the global value 0.
This is clearly not the case for all of the 5 targets. From \ref{fig:S__DM_G__relation}, we found an obvious upward trend of $s_{\pi_0}^{*}$ with growing $\underline{\Delta} m_\mathrm{G}$. This upward trend shows that background quasars with similar $m_\mathrm{G}$ around $m_\mathrm{G}^\mathrm{target}$ tend to have similar parallaxes, which supports the $m_\mathrm{G}$-dependence of $\pi_0$ stated in \citet{Lindegren18,Lindegren21}.

Using the same interpretation, we also investigated marginalized relations of $s_{\pi_0}^{*}$ with respect to each of the other 3 filter parameters. As can be seen in \ref{fig:S__DM_G__relation}, most fields show a (weak) color dependence of $\pi_0$. For the two sky-position-related filter parameters, $\beta$ dependence of $\pi_0$ is clearly found in the target fields of \cyg\ and NP~Ser, whereas the $r$ dependence of $\pi_0$ is hardly noticeable in any of the 5 target fields. This contrast reinforces the belief that the $\beta$ dependence of $\pi_0$ contributes considerably to the position dependence of $\pi_0$ \citep{Lindegren21}. Additionally, $\log_{10} s_{\pi_0}^{*}$ converges to $0.0\pm0.1$ at $r=10\degr$ in all target fields, which roughly agrees with our assumption that the $r$-dependence of $\pi_0$ becomes negligible beyond 10\degr\ search radius; in other words, quasar parallaxes at $r>10\degr$ are almost not representative of the target $\pi_0$.

Knowing the way to interpret the relation between $s_{\pi_0}^{*}$ and each of the filter parameters, we can proceed to choose the optimal filter parameters. In choosing filter parameters, one wants the filtered sub-sample to have a consistently low level of $s_{\pi_0}^{*}$; meanwhile, one prefers a filtered sub-sample as large as possible to achieve reliable and precise $\pi_0$. To meet both demands is difficult when $s_{\pi_0}^{*}$ rises with a larger filter (as is the case for the $m_\mathrm{G}$ filter). In this regard, we created an index $w=A \log_{10} N_\mathrm{quasar} - \log_{10} s_{\pi_0}^{*}$, where $A=(\log_{10}s_{\pi_0}^{\rm *,max}-\log_{10}s_{\pi_0}^{\rm *,min})/(\log_{10}N_{\rm quasar}^{\rm max}-\log_{10}N_{\rm quasar}^{\rm start})$ is a scaling factor. Instead of the minimum of $s_{\pi_0}^{*}$ or the maximum of $N_{\rm quasar}$, we adopted the filter parameter at the peak of the index $w$ (see the bottom 4-panel block of \ref{fig:S__DM_G__relation}), so that an optimized trade-off between $s_{\pi_0}^{*}$ and $N_{\rm quasar}$ can be reached. Using this consistent standard, we reached the parameter of each filter for each target (see \ref{tab:zero_parallax_points}).

After applying the 4 filters, we acquired a quasar sub-sample of $N_\mathrm{quasar}$ and $s_{\pi_0}^{*}$ for each target, and obtained $\pi_0$ of the target from this sub-sample (see \ref{tab:zero_parallax_points}).
Due to the relative paucity of quasars identified at low Galactic latitudes $b$ (for the reasons discussed in  \citealp{Lindegren21}), NP~Ser, \uua\ and \xb\ (all of which have $|b|<4$\degr) receive relatively low values for $N_\mathrm{quasar}$ compared to \cen\ ($b=24$\degr), which partly lead to the relatively large $\tilde{\sigma}_{\pi_0}$. 
Nonetheless, we have shown that $\pi_0$ for targets at $|b|<6$\degr\ can be determined with nearby (on the sky) quasars.
On the other hand, \cyg, despite being located at $b=-11$\degr, has the smallest $N_\mathrm{quasar}$ and the largest $\tilde{\sigma}_{\pi_0}$ among the 5 targets.  This is mainly due to the relative rarity of bright quasars (\cyg\ has $m_\mathrm{G}=14.7$\,mag).

The negative $\log_{10} s_{\pi_0}^{*}$ of the quasar sub-sample for each target indicates that the sub-sample of (closely located, like-$\beta$, like-$m_\mathrm{G}$ and like-$m_\mathrm{B-R}$) quasars are representative of the target source.
Despite this good representativeness, the $\tilde{\sigma}_{\pi_0}$ estimated with \ref{eq:parallax_zero_point_uncertainty} is still an under-estimate for the uncertainty of $\pi_0$, as the sub-sample has a spread in each parameter and does not represent the target perfectly. We estimated the weighted average $m_\mathrm{G}$, $m_\mathrm{B-R}$ and $\sin{\beta}$ of the quasar sub-sample (where the weighting is $1/{\sigma_i}^2$, see \ref{eq:parallax_zero_point_uncertainty} for the definition of $\sigma_i$), which are presented in \ref{tab:zero_parallax_points} as $\overline{m_\mathrm{G}}$, $\overline{m_\mathrm{B-R}}$ and $\overline{\sin{\beta}}$, respectively.
According to \ref{tab:zero_parallax_points}, there are residual offsets of $\sin{\beta}$, $m_\mathrm{G}$ and $m_\mathrm{B-R}$ between the target and the average level of the sub-sample, which would result in extra systematic error for the $\pi_0$ estimates.
In order to estimate this systematic error of $\pi_0$, we calculated the empirical parallax zero points for the targets (see \ref{tab:zero_parallax_points}) using {\tt zero\_point.zpt} (\url{https://gitlab.com/icc-ub/public/gaiadr3_zeropoint}, \citealp{Lindegren21}), noted as $\pi_0^\mathrm{emp}$.
In the same way, we also estimated the empirical parallax zero points for each quasar of the sub-sample, then derived the weighted average empirical parallax zero point of the sub-sample, noted as $\overline{\pi_0^\mathrm{emp}}$ (see \ref{tab:zero_parallax_points}). The difference between $\pi_0^\mathrm{emp}$ and $\overline{\pi_0^\mathrm{emp}}$ is taken as an estimate for the systematic error of $\pi_0$, which is added in quadrature to the $\tilde{\sigma}_{\pi_0}$ calculated with \ref{eq:parallax_zero_point_uncertainty}.

No uncertainties are yet available for $\pi_0^\mathrm{emp}$. Regardless, all of our $\pi_0$ are consistent with the empirical counterparts at the $2\,\sigma$ confidence level.
We note that $\pi_0^\mathrm{emp}$ is only used to estimate the systematic error of $\pi_0$, and is not adopted in the discussions that follow.
The calibrated parallaxes $\pi_1-\pi_0$ are summarized in \ref{tab:zero_parallax_points}.

Despite the overall consistency between $\pi_0$ and $\pi_0^\mathrm{emp}$, it is noteworthy that all $\pi_0$ are larger than the $\pi_0^\mathrm{emp}$ counterparts, which can be a coincidence of small-number statistics. 
Alternatively, knowing that most of the targets are situated at low $b$, it might indicate a systematic offset between quasar-sub-sample-based $\pi_0$ and $\pi_0^\mathrm{emp}$ at low $b$.  
Interestingly, a recent parallax zero-point study based on $\sim$110,000 W Ursae Majoris variables showed that the W-Ursae-Majoris-variable-based $\pi_0$ are systematically larger than the $\pi_0^\mathrm{emp}$ counterparts at low $b$ (see Figure~3 of \citealp{Ren21}). 
Hence, a future quasar-sub-sample-based study involving a large sample of targets (and other parallax zero-point studies by different approaches) will be essential for probing $\pi_0^\mathrm{emp}$ at low $b$.
More specifically, if quasar-sub-sample-based $\pi_0$ at low $b$ are confirmed to be systematically larger than the $\pi_0^\mathrm{emp}$ counterparts, then it is likely that $\pi_0^\mathrm{emp}$ is systematically under-estimated at low $b$.

\begin{table*}
\caption{The information of the 4 filters (including search radius $r$, $\beta$ filter, $m_\mathrm{G}$ filter and $m_\mathrm{B-R}$ filter, see \ref{subsec:PRE_s_pi0} for explanation) and the parallax zero point $\pi_0$ calculated from the respective sub-sample of background quasars after applying the 4 filters. $\pi_1$ and $\pi_1-\pi_0$ stand for uncalibrated parallaxes and calibrated parallaxes, respectively. 
$\overline{m_\mathrm{G}}$, $\overline{m_\mathrm{B-R}}$ and $\overline{\sin{\beta}}$ represent the respective weighted average value of the three filter parameters (see \ref{subsec:PRE_s_pi0}).
$N_\mathrm{quasar}$ and $s_\mathrm{\pi_0}^{*}$ (defined in \ref{subsec:PRE_s_pi0}) are reported for the filtered quasar sub-sample in each target field. 
The empirical parallax zero-point solutions for the targets calculated with {\tt zero\_point.zpt} (\url{https://gitlab.com/icc-ub/public/gaiadr3_zeropoint}) are provided as $\pi_0^\mathrm{emp}$. The weighted average empirical parallax zero-points of the sub-samples (see \ref{subsec:PRE_s_pi0}) are presented as $\overline{\pi_0^\mathrm{emp}}$.
}
\label{tab:zero_parallax_points}
\centering
\resizebox{\textwidth}{!}{
\begin{tabular}{@{}ccccccccccc@{}}
\hline\hline
field & $r$ & $\sin{\beta^\mathrm{target}}$ & $\Delta \sin{\beta}$ & $m_\mathrm{G}^\mathrm{target}$ & $\underline{\Delta} m_\mathrm{G}$ & $m_\mathrm{B-R}^\mathrm{target}$ & $\Delta m_\mathrm{B-R}$ & $\overline{m_\mathrm{G}}$ & $\overline{m_\mathrm{B-R}}$ & $\overline{\sin{\beta}}$\\
& (\degr) &  & & (mag) &  & (mag) & (mag) & & &\\
\hline%
\cen\ & 10 & $-0.24$ & 0.14 & 17.85 & 5.0\% & 1.59 & 1.20 & 18.00 & 0.74 & -0.25 \\
\cyg\ & 5.5 & 0.74 & 0.02 & 14.70 & 31.6\% & 0.71 & 0.09 & 18.19 & 0.74 & 0.74\\
\uua\ & 7.2 & $-0.90$ & 0.05 & 17.15 & 12.0\% & 1.19 & 0.58 & 18.23 & 0.94 & -0.93\\
\xb\ & 7.6 & 0.84 & 0.08 & 17.58 & 8.7\% & 1.29 & 0.76 & 18.49 & 1.01 & 0.79\\
NP~Ser & 10 & 0.16 & 0.08 & 17.01 & 19.1\% & 1.51 & 0.76 & 17.61 & 1.57 & 0.16\\
\hline\hline
\end{tabular}
}

\vspace{0.3cm}
\resizebox{\textwidth}{!}{
\begin{tabular}{@{}cccccccc@{}}
\hline\hline
field & $N_\mathrm{quasar}$  &  $\log_{10}s_{\pi_0}^{*}$ &  $\pi_0^\mathrm{emp}$ & $\overline{\pi_0^\mathrm{emp}}$ & $\pi_0$ & $\pi_1$ & $\pi_1-\pi_0$\\
& & & (mas) & (mas) & (mas) & (mas) \\
\hline%
\cen\ & 772 & $-0.27$ & $-0.027$ & $-0.026$ & $-0.022\pm0.006\pm0.002$ & $0.53\pm0.13$ & $0.55\pm0.13$ \\
\cyg\ & 32 & $-0.29$ & $-0.030$ & $-0.020$ & $0.019\pm0.024\pm0.010$ &  $0.068\pm0.019$ & $0.050\pm0.032$\\
\uua\ & 91 & $-0.33$ & $-0.028$ & $-0.027$ & $-0.009\pm0.013\pm0.001$ &  $0.24\pm0.06$ & $0.25\pm0.06$\\
\xb\ & 63 & $-0.37$ & $-0.028$ & $-0.019$ & $0.004\pm0.018\pm0.009$ &  $0.50\pm0.08$ & $0.50\pm0.08$\\
NP~Ser & 40 & $-0.17$ & $-0.032$ & $-0.026$ & $-0.008\pm0.033\pm0.006$ & $0.69\pm0.08$ & $0.70\pm0.09$\\
\hline\hline
\end{tabular}
}
\end{table*}

\section{Discussion}
\label{sec:PRE_discussions}
In this section, we discuss the implications of the 5 calibrated parallaxes $\pi_1-\pi_0$ provided in \ref{tab:zero_parallax_points}. 

\subsection{NP~Ser and \gx\ revisited}
\label{subsec:PRE_NpSer_neq_Gx17}
\gx\ is one of the brightest X-ray sources on the sky, from which 43 PRE bursts have been recorded (see Table~1 of \citealp{Galloway20} and references therein); its distance was estimated to be 7.3--12.6\,kpc by treating its PRE bursts as standard candles (as explained in \ref{sec:PRE_intro}), where the distance uncertainty is dominated by the uncertain composition of the nuclear fuel \citep{Galloway20}. 
The optical source NP~Ser, $\approx1\farcs0$ away from \gx, was initially considered the optical counterpart of \gx\ \citep{Tarenghi72}; but the association has been ruled out, based on its sky-position offset from \gx\ \citep{Deutsch99} and its lack of optical variability (as expected for the optical counterpart of \gx) \citep[e.g.][]{Davidsen76,Margon78}.
Despite this, NP~Ser was initially treated as a potential counterpart and the Gaia properties of NP~Ser were analysed.
Regardless of the non-association of NP~Ser and \gx, the complex sky region around \gx\ for optical/infrared observations (see Figure~2 of \citealp{Deutsch99} for example) merits high-precision Gaia astrometry of NP~Ser, which will facilitate future data analysis of optical/infrared observations of \gx.

Using the astrometric parameters of NP~Ser in \ref{tab:before_calibration} and \ref{tab:zero_parallax_points}, we extrapolated the reference position of NP~Ser (from the reference epoch 2016.0\,yr, or MJD~57388) to MJD~47496, following the method described in Section~3.2 of \citet{Ding20}. 
The projected Gaia position of NP~Ser is offset from the VLA position of \gx\ (see Table~2 of \citealp{Deutsch99}) by 0\farcs$95\pm0$\farcs07 at MJD~47496.
Considering the distances to \gx\ and NP~Ser, nuclear fuel with cosmic abundances (73\% hydrogen) corresponds to the smallest PRE distance of $8.5\pm1.2$\,kpc for \gx\ \citep{Galloway20};
in comparison, the calibrated Gaia parallax of NP~Ser is $0.70\pm0.09$\,mas, corresponding to a distance of $1.44^{+0.21}_{-0.16}$\,kpc, which establishes NP~Ser as a foreground source of \gx.
The current sky-position offset between NP~Ser and \gx\ depends on the proper motion of \gx. Assuming \gx\ rotates around the Galactic centre in the Galactic disk with negligible peculiar velocity (with respect to the neighbourhood of \gx), the apparent proper motion of \gx\ would be $\mu_{\alpha}=-3.5$\,\maspy\ and $\mu_{\delta}=-6.7$\,\maspy\ given a distance of 8.5\,kpc (from the Earth).
Based on this indicative proper motion of \gx\ and the Gaia ephemeris of NP~Ser, NP~Ser is currently (at MJD~59394) 1\farcs0 away from \gx, and continues to move away from \gx\ on the sky; accordingly, it would become easier with time to resolve \gx\ from the foreground NP~Ser in optical/infrared observations.

\subsection{Posterior distances and compositions of nuclear fuel for \cen, \cyg\ and \uua}
\label{subsec:PRE_posterior_distances}
Among the (calibrated) parallaxes $\pi_1-\pi_0$ of the 4 PRE bursters (i.e. \cen, \cyg, \uua\ and \xb), $\pi_1-\pi_0$ for \cen, \uua\ and \xb\ are detected, whereas $\pi_1-\pi_0$ of \cyg\ is weakly constrained.
All PRE bursters except \xb\ have nominal PRE distances published that are inferred from their respective PRE bursts (see \ref{tab:constrain_composition} for PRE distances and their references).
Using Bayesian inference, we can incorporate the parallax of a PRE burster with its nominal PRE distance to constrain the composition of nuclear fuel and refine the nominal PRE distance. 

In \ref{sec:PRE_intro}, we have mentioned that the Eddington luminosity (measured by an observer at infinity) $L_{\mathrm{Edd},\infty}$ varies with NS mass $M_\mathrm{NS}$ and photosphere radius $R_\mathrm{P}$ (see \ref{eq:Eddington_luminosity}). In practice, the bolometric flux at the ``touchdown'' (when the photosphere drops back to the NS surface) of a PRE burst is compared to the theoretical $L_{\mathrm{Edd},\infty}$ at $R_\mathrm{P}=R_\mathrm{NS}$ (where $R_\mathrm{NS}$ stands for NS radius) to calculate the nominal PRE distance \citep{Galloway20}. Therefore, the theoretical $L_{\mathrm{Edd},\infty}$ should change with $M_\mathrm{NS}$ and $R_\mathrm{NS}$, which is, however, not taken into account in this work.
We note that, following \citet{Galloway20}, we assume $M_\mathrm{NS}=1.4\,\msun$ and $R_\mathrm{NS}=11.2$\,km \citep{Steiner18} for all PRE bursters.

\subsubsection{Bayes factor between two ends of the composition of nuclear fuel}
\label{subsubsec:PRE_bayes_factor_about_X}
As mentioned in \ref{sec:PRE_intro}, the Eddington luminosity of PRE bursts depends on the composition of nuclear fuel (accreted onto the NS surface from its companion) at the time of the outburst. This composition is normally parameterized by $X$, the mass fraction of hydrogen in nuclear fuel at the time of a PRE burst, which generally ranges from 0 to $\approx73$\%, the cosmic mass fraction of hydrogen. 
As the result of the dynamic nucleosynthesis process during accretion, $X$ varies from one PRE burst to another, but is always smaller than the hydrogen mass fraction of the donor star. 
$X$ can be roughly predicted by the theoretical ignition models of X-ray bursts, and mainly depends on the local accretion rate $\dot{m}$ \citep[e.g.][]{Fujimoto81}.

Nominal PRE distances for $X=0$, denoted as $D_0$, are summarized in \ref{tab:constrain_composition} along with their references.
At any given $X$, the nominal PRE distance $D_X= D_0/\sqrt{X+1}$ \citep{Lewin93}. As the fractional uncertainty of a nominal PRE distance measurement is independent of $X$, $\sigma_X=\sigma_0/\sqrt{1+X}$, where $\sigma_0$ and $\sigma_X$ represent the uncertainty of the nominal PRE distance at $X=0$ and at a given $X$, respectively.

As our first attempt to constrain $X$, we used the calibrated Gaia parallax of each PRE burster to determine which value of $X$ is more likely for the burster -- 0.73 or 0.
We calculated the Bayes factor $K=K^{X=0.7}_{X=0}$ using

\begin{equation}
\label{eq:bayes_factor}
\begin{split}
K &= \frac{\mathrm{P}(\pi_1-\pi_0|X=0.7)}{\mathrm{P}(\pi_1-\pi_0|X=0)}\\
&= \frac{\int^{\infty}_{0} \frac{1}{\sigma_{0.7}}\exp\left[-\frac{1}{2}\left(\frac{1/D-\mu_{\pi}}{\sigma_{\pi}}\right)^2-\frac{1}{2}\left(\frac{D-D_{0.7}}{\sigma_{0.7}}\right)^2\right]\mathrm{d}D}
{\int^{\infty}_{0} \frac{1}{\sigma_{0}}\exp\left[-\frac{1}{2}\left(\frac{1/D-\mu_{\pi}}{\sigma_{\pi}}\right)^2-\frac{1}{2}\left(\frac{D-D_{0}}{\sigma_{0}}\right)^2\right] \mathrm{d}D},
\end{split}
\end{equation}
where $\pi_1-\pi_0=\mu_{\pi}\pm\sigma_{\pi}$, $D|_{X=0.7}=D_{0.7} \pm \sigma_{0.7}$ and $D|_{X=0}=D_{0} \pm \sigma_{0}$. 
For the calculation of $K^{X=0.7}_{X=0}$, we did not use any Galactic prior (see \ref{eq:posterior_distance_PDF} for explanation), as the extra constraint given by a Galactic prior is negligible when $X$ is fixed.
The results of $K^{X=0.7}_{X=0}$ are presented in \ref{tab:constrain_composition}.

Among the three $\log_{10} K^{X=0.7}_{X=0}$ values, the one for \cen\ is the most offset from 0, which suggests the calibrated parallax of \cen\ substantially (when $0.5<|\log_{10}K^{X=0.7}_{X=0}|<1$, \citealp{Kass95}) favors $X=0$ over $X=0.7$. 
Merely 2 PRE bursts have ever been observed from \cen\ in 1969 \citep{Belian72} and 1979 \citep{Matsuoka80}. The long intervals between PRE bursts of \cen\ imply small accretion rates and low $X$ at the time of PRE bursts, despite hydrogen signatures observed from its donor star \citep[e.g.][]{Shahbaz14}. This expectation is confirmed by the $\log_{10} K^{X=0.7}_{X=0}$ of \cen.

It is noteworthy that hydrogen-poor nuclear fuel at the time of PRE bursts is also favored for \cyg\ and \uua. 
In fact, the previous independent test of the simplistic PRE model using GC PRE bursters also suggests hydrogen-poor nuclear fuel at the time of PRE bursts for the majority of the 12 GC PRE bursters \citep{Kuulkers03}.
The statistical inclination towards low $X$, if confirmed with a larger sample of PRE bursters, might be partly attributed to a selection effect explained as follows.

For each PRE burster, only the brightest PRE bursts are selected to calculate its nominal PRE distance \citep{Galloway08,Galloway20,Chevalier89}. This selection criterion results in a nominal PRE distance that tends to decrease with time as brighter bursts are discovered. 
This tendency is clearly shown by the fact that almost all the nominal PRE distances registered in Table~9 of \citet{Galloway08} are larger than their counterparts in Table~8 of \citet{Galloway20}. 
If we assume the variability of observed fluxes of PRE bursts is mainly caused by the fluctuation of $X$ (at the time of different PRE bursts), the selection criterion of PRE bursters would lead to an increasingly biased PRE-burst sample that is more likely low-$X$.
This selection effect would have a big impact on the $X$ of the frequent PRE burster \cyg\ (see $n_\mathrm{burst}$ in Table~1 of \citealp{Galloway20}), but less so for \uua\ and \cen. 
A low $X$ of \uua\ (if confirmed in the future) can be explained by the suggestion that the donor star of \uua\ is likely a helium white dwarf \citep{int-Zand05}.
On the other hand, the revised nominal PRE distances of PRE bursters \citep{Galloway20} relieve the tension between the pre-2008 nominal PRE distances (see the references in Table~1 of A21) and the Gaia DR2 distances (A21), as mentioned in \ref{sec:PRE_intro}.

\begin{table}
\caption{Bayes factor $K=K^{X=0.7}_{X=0}=P(\pi_1-\pi_0|X=0.7)/P(\pi_1-\pi_0|X=0)$ (ratio of two conditional probabilities), where $X$ refers to the mass fraction of hydrogen of the nuclear fuel.}
\label{tab:constrain_composition}
\centering
\begin{tabular}{@{}ccccc@{}}
\hline\hline
PRE & $D(X\!=\!0)$ & $\log_{10} K$ & $D^\mathrm{post}$ & $X^\mathrm{post}$ \\
burster & (kpc) & & (kpc) & \\
\hline\
%\vspace{1cm}
\cen\ & 1.2(3)\,$^\mathrm{a}$ & $-0.83$ & $1.4(2)$ & $0.2^{+0.3}_{-0.1}$ \\
\cyg\ & 11.6(9)\,$^\mathrm{b}$ &  $-0.51$ & $11.3^{+0.9}_{-0.8}$ & $0.2^{+0.2}_{-0.1}$\\
\uua\ & 3.9(2)\,$^\mathrm{b}$ & $-0.39$ & $3.5(3)$ & $0.3^{+0.3}_{-0.2}$ \\
\hline\hline
\end{tabular}
\tabnote{$^\mathrm{a}$ \citet{Chevalier89}; $^\mathrm{b}$ \citet{Galloway20}.}
\end{table}

\begin{figure*}
    \centering
	\begin{tabular}{cc}
    \includegraphics[width=75mm]{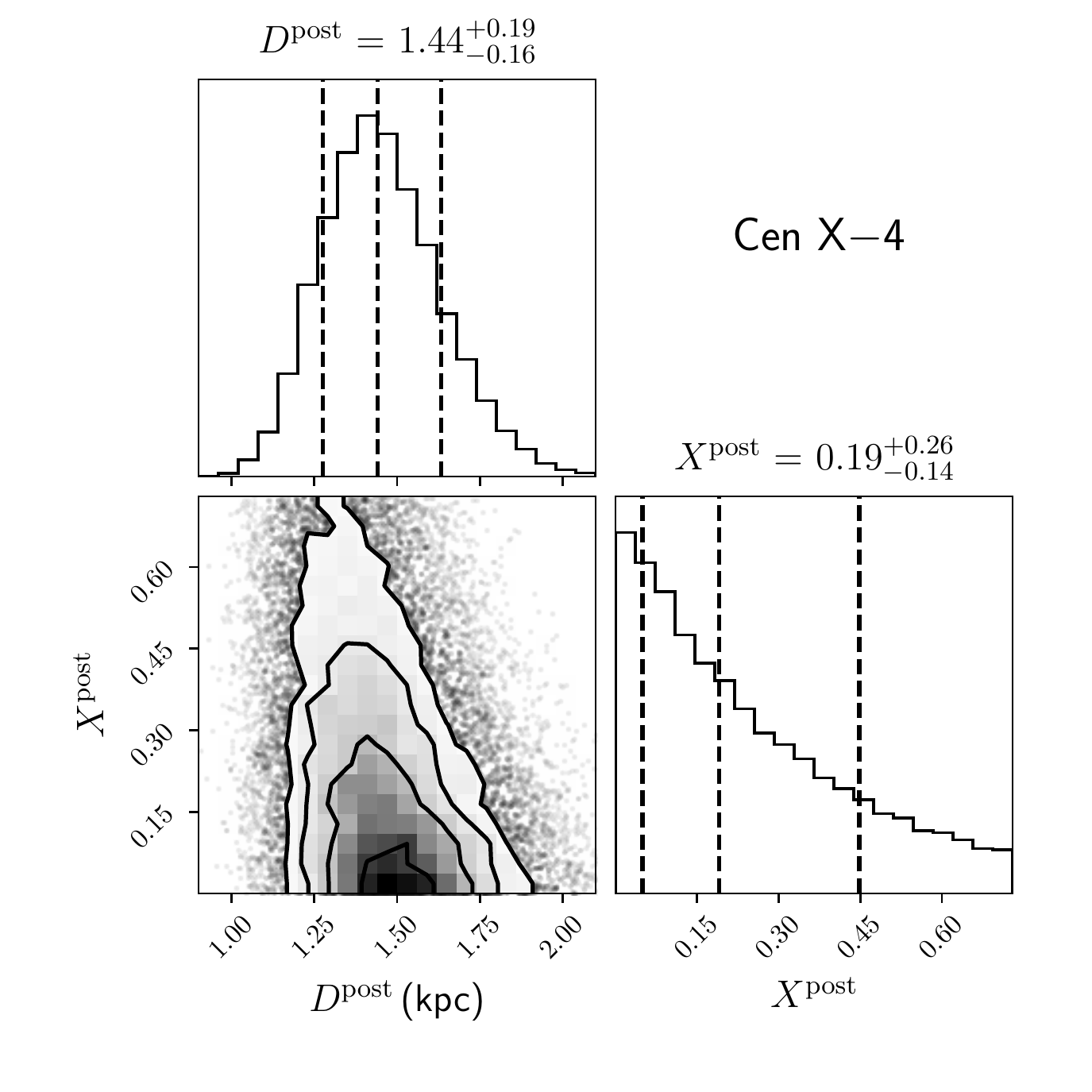} &   \includegraphics[width=75mm]{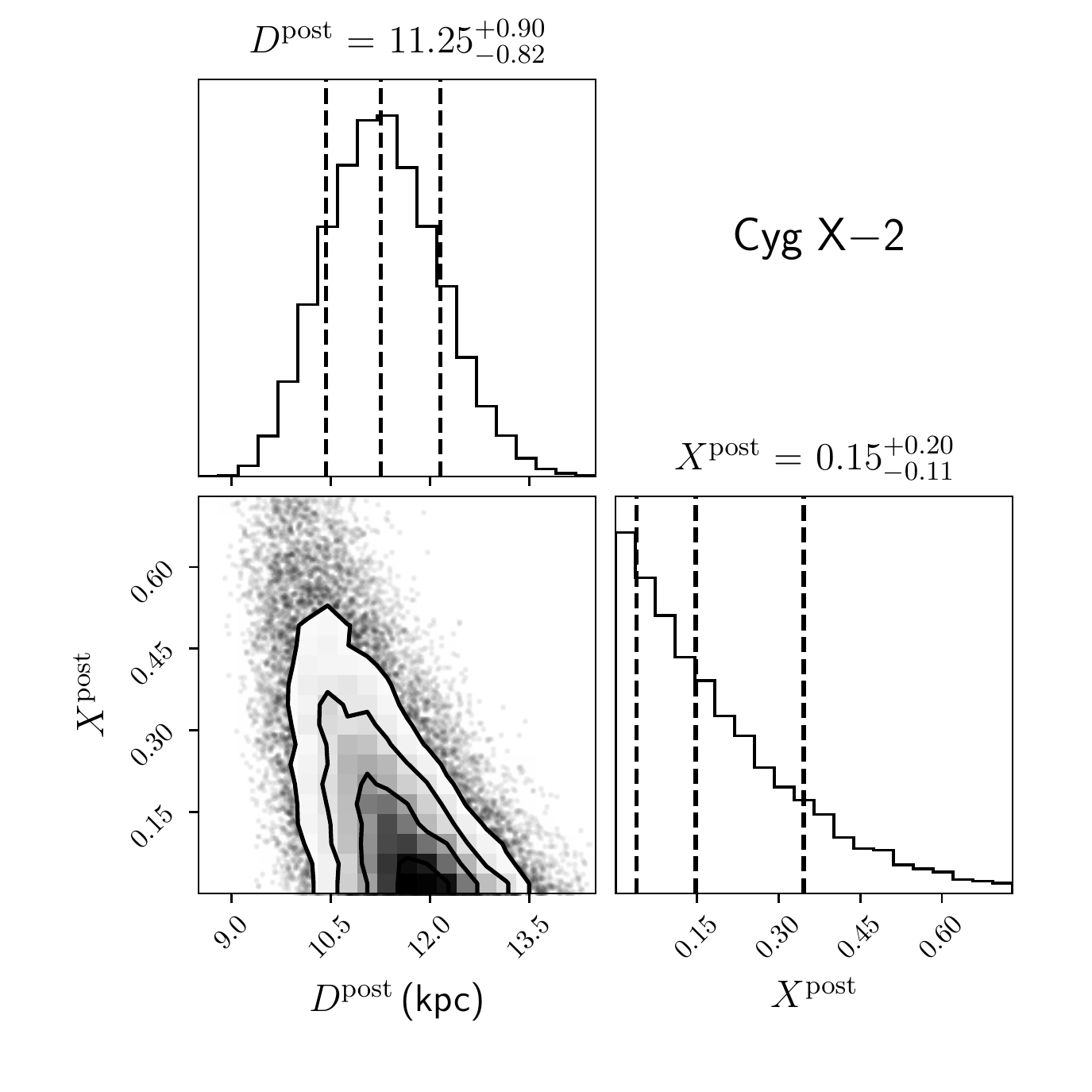} \\
    \multicolumn{2}{c}{\includegraphics[width=75mm]{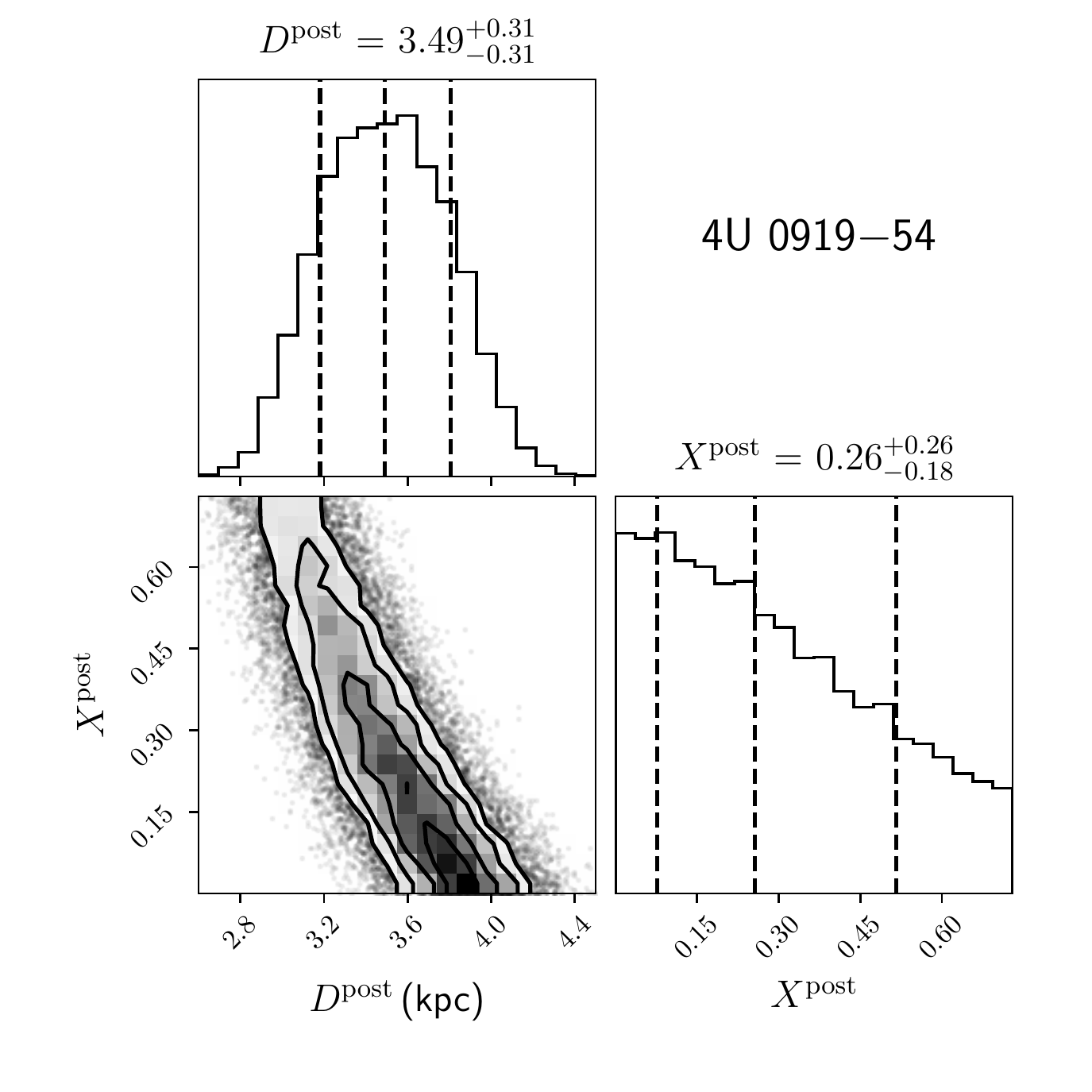} }\\
    \end{tabular}	

	\caption{2-D histograms and marginalized 1-D histograms for posterior distance $D^{\mathrm{post}}$ and posterior mass fraction of hydrogen of nuclear fuel $X^{\mathrm{post}}$  simulated with {\tt bilby} \citep{Ashton19} and plotted with {\tt corner.py} \citep{Foreman-Mackey16}. The $n$-th contour in each 2-D histogram contains $1-\exp\left(-n^2/2\right)$ of the simulated sample \citep{Foreman-Mackey16}.
	The vertical lines in the middle and two sides mark the median and central 68\% of the sample, respectively.}
    \label{fig:posterior_distances}
\end{figure*}

\subsubsection{Refining nominal PRE distances and compositions of nuclear fuel}
\label{subsubsec:PRE_post_D_and_X_assuming_PRE_model_is_right}
To better understand the statistical properties of $X$ (and to refine the PRE distance), we subsequently constrained both $X$ and the nominal PRE distance $D$ of each PRE burster with its calibrated parallax in a Bayesian way. The 2-D joint probability distribution (JPD) we applied is
\begin{equation}
\label{eq:posterior_distance_PDF}
\begin{split}
    \phi &= \phi(D,X\ |\ \pi_1-\pi_0,D_0) \\
    &\propto \phi(\pi_1-\pi_0,D_0\ |\ D,X) \\
    &\propto
    \rho(D) \exp\left[{-\frac{1}{2}\left( \frac{\frac{1}{D}-\mu_{\pi}}{\sigma_\pi}\right)^2-\frac{1}{2}\left( \frac{D-\frac{D_0}{\sqrt{X+1}}}{\frac{\sigma_0}{\sqrt{X+1}}}\right)^2}\right], \\
    & (D>0\ \text{and}\ X \in [0,0.73]),
\end{split}
\end{equation}
where $\rho(D)$ represents the
Galactic prior term, i.e. the prior probability of finding a PRE burster at a distance $D$ along a given line of sight. We approached $\rho$ by the Galactic mass distribution of LMXBs, which can be calculated with Equation~5 of \citet{Atri19} using parameters provided by \citet{Grimm02,Atri19}. 
Among the three components of the Galactic mass density (i.e. $\rho_\mathrm{bulge}$, $\rho_\mathrm{disk}$ and $\rho_\mathrm{sphere}$, see \citealp{Grimm02} for explanations), the Galactic-disk component $\rho_\mathrm{disk}$ dominates the Galactic mass density budget for LMXBs at the light of sight to any of the three PRE bursters. Therefore, for this work alone, we adopted $\rho_\mathrm{disk}$ (see Equation~5 of \citealp{Grimm02}) as the Galactic prior term $\rho$.

Following the JPD, we simulated posterior \{$D$, $X$\} pairs (hereafter referred to as $D^\mathrm{post}$ and $X^\mathrm{post}$), using {\tt bilby}\footnote{\url{https://lscsoft.docs.ligo.org/bilby/}} \citep{Ashton19}. 
For this simulation, we adopted a uniform distribution between 0 and 0.73 for $X$ and a uniform distribution between 0 and 50\,kpc for $D$ (where using a larger upper limit of $D$ does not change the result).
Based on the marginalized probability density functions (PDFs) of the simulated \{$D^\mathrm{post}$, $X^\mathrm{post}$\} chain, we estimated $D^\mathrm{post}$ and $X^\mathrm{post}$, which are presented in \ref{fig:posterior_distances} and \ref{tab:constrain_composition}.

In general, the conversion of parallax to distance depends on the adopted Galactic prior \citep{Bailer-Jones15}. As an example, A21 found that the application of different Galactic priors \citep[e.g.][]{Bailer-Jones21,Atri19} results in inconsistent distances (see Figure~3 of A21). 
Unlike A21, we applied an extra constraint on $D$ provided by the PRE distance (see \ref{eq:posterior_distance_PDF}), which gives a handle on $X$ and improves the constraint on $D$.
With this extra constraint, the dependence on the Galactic prior becomes negligible for all of the three PRE bursters: their $D^\mathrm{post}$ and $X^\mathrm{post}$ stay almost the same without the $\rho(D)$ term.

The PDFs of $X$ provide richer information on $X$ than $K^{X=0.7}_{X=0}$. As expected from $K^{X=0.7}_{X=0}$, the PDF of $X$ for \cen\ favors a hydrogen-poor nature of nuclear fuel.
In addition to the new constraint we put on $X$, the nominal PRE distance of each PRE burster incorporating the parallax information is supposed to be more reliable (though not necessarily more precise). We note that the refined nominal PRE distances are still model-dependent; their accuracies are subject to the validity of the simplistic PRE model.

\subsection{A roadmap for testing the simplistic PRE model with parallax measurements}
\label{subsec:PRE_roadmap_for_model_testing}
The discussion in \ref{subsec:PRE_posterior_distances} is based on the correctness of the simplistic PRE model. 
Now we discuss the potential feasibility of using parallaxes (of PRE bursters) to test the simplistic PRE model. 
We first define the PRE distance correction factor $\eta$ ($\eta>0$) using the relation 
\begin{equation}
\label{eq:eta}
D_X = \eta\ \frac{D_0}{\sqrt{X+1}},
\end{equation}
where $D_0$ stands for the nominal PRE distance at $X=0$ (derived from the simplistic PRE model), and $D_X$ refers to the ``true'' distance. If the simplistic PRE model is correct, then $\eta=1$. However, $\eta\ \neq\ 1$ is possible due to various causes including the stochasticity of PRE burst luminosity and anisotropy of PRE burst emission \citep[e.g.][]{Sztajno87}. The latter cause implies $\eta$ to be geometry-dependent (e.g. viewing-angle-related); if all PRE bursters have different geometry-dependent $\eta$, a search for a global deviation of $\eta$ from unity with a sample of PRE bursters would be less practical. Nevertheless, it remains practical to constrain $\eta$ on an individual basis. One way to make this constraint is using the generalized $X$, defined as
\begin{equation}
\label{eq:X_prime}
X'+1=\frac{X+1}{\eta^2}.
\end{equation}
Accordingly, $D_X=D_0/\sqrt{X'+1}$. Thanks to the same mathematical formalism, we can re-use \ref{eq:posterior_distance_PDF} as the JPD for statistical analysis, except that the domain of $X'$ is widened to $X'>-1$.

\begin{figure*}
    \centering
    \begin{tabular}{cc}
    \includegraphics[width=75mm]{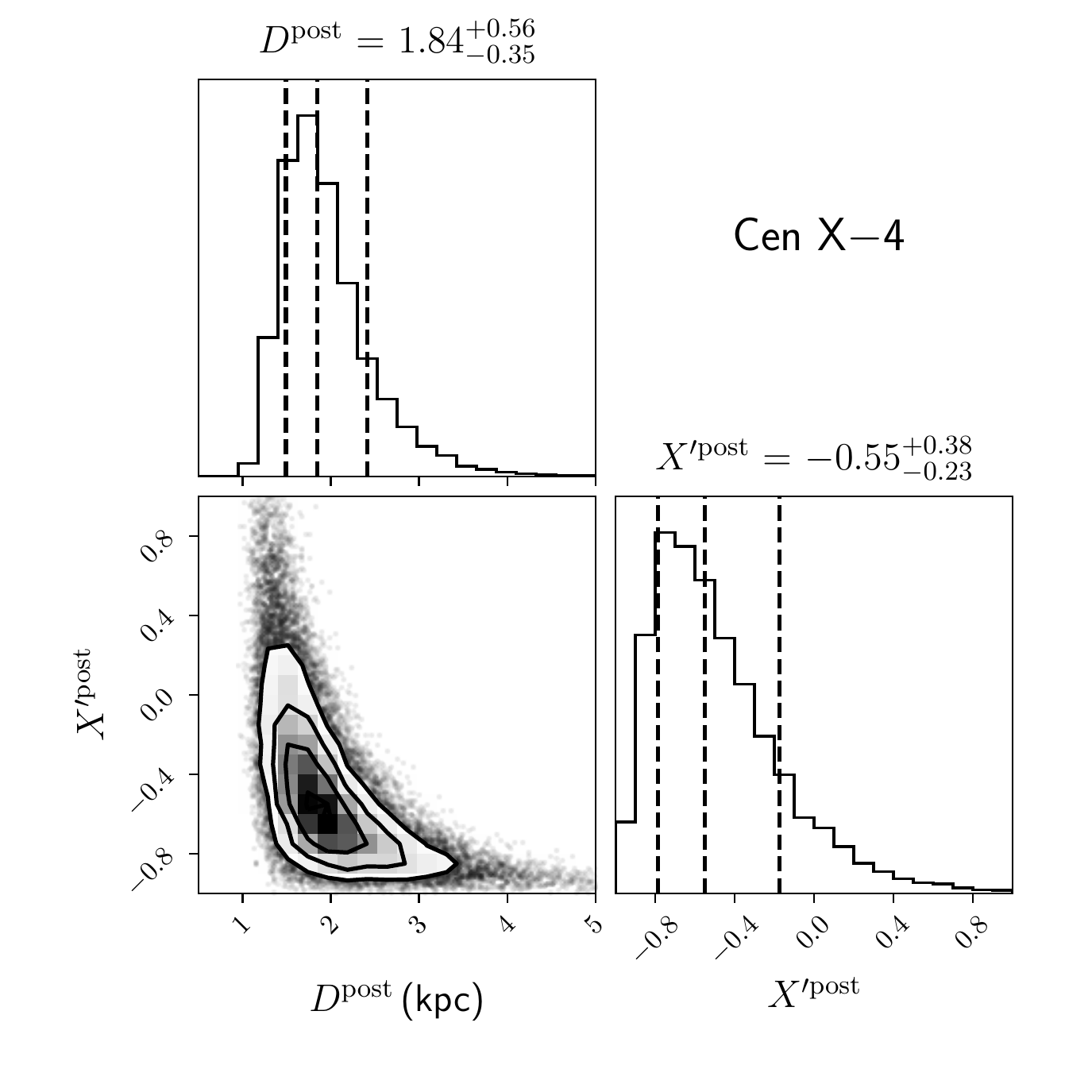} &   \includegraphics[width=75mm]{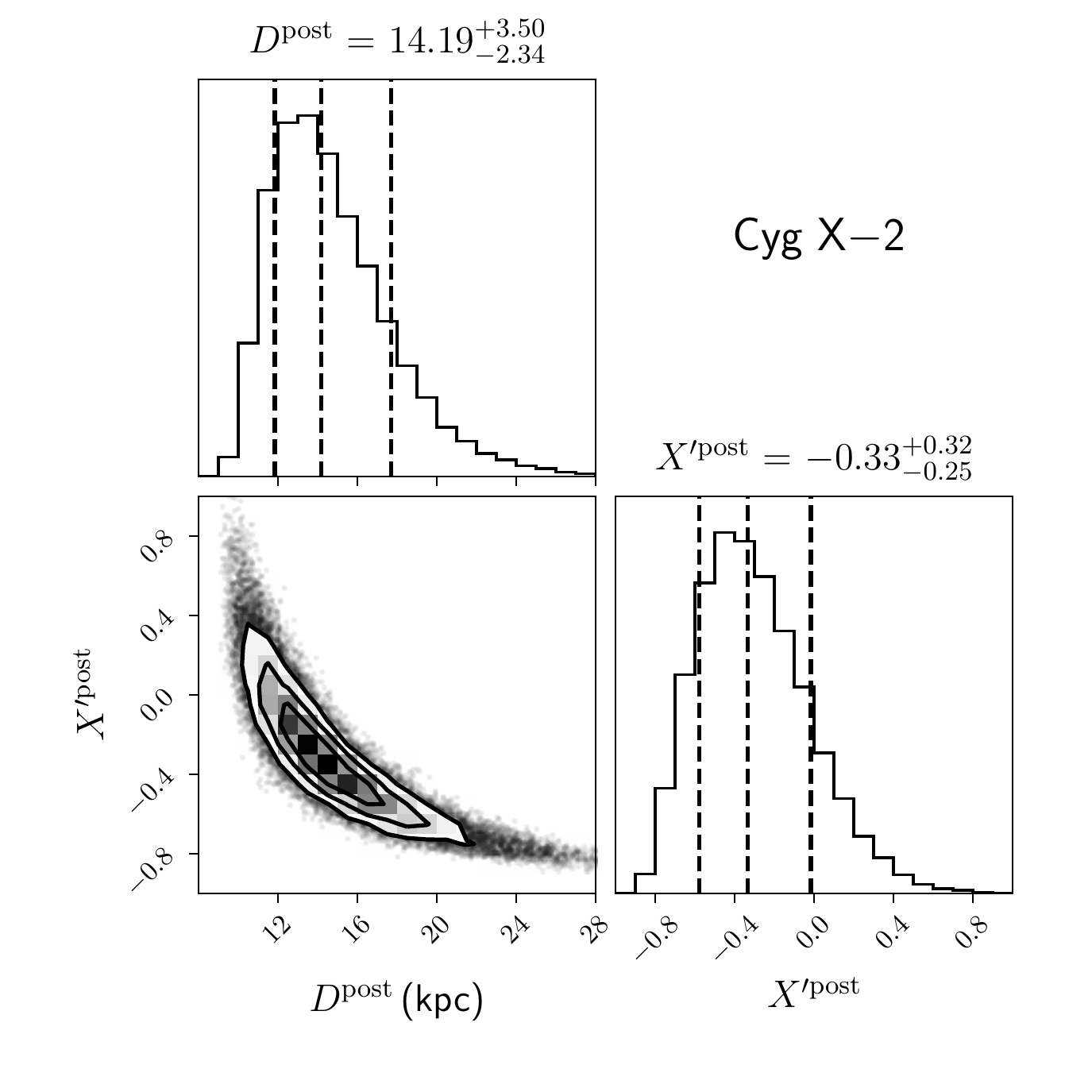} \\
    \multicolumn{2}{c}{\includegraphics[width=75mm]{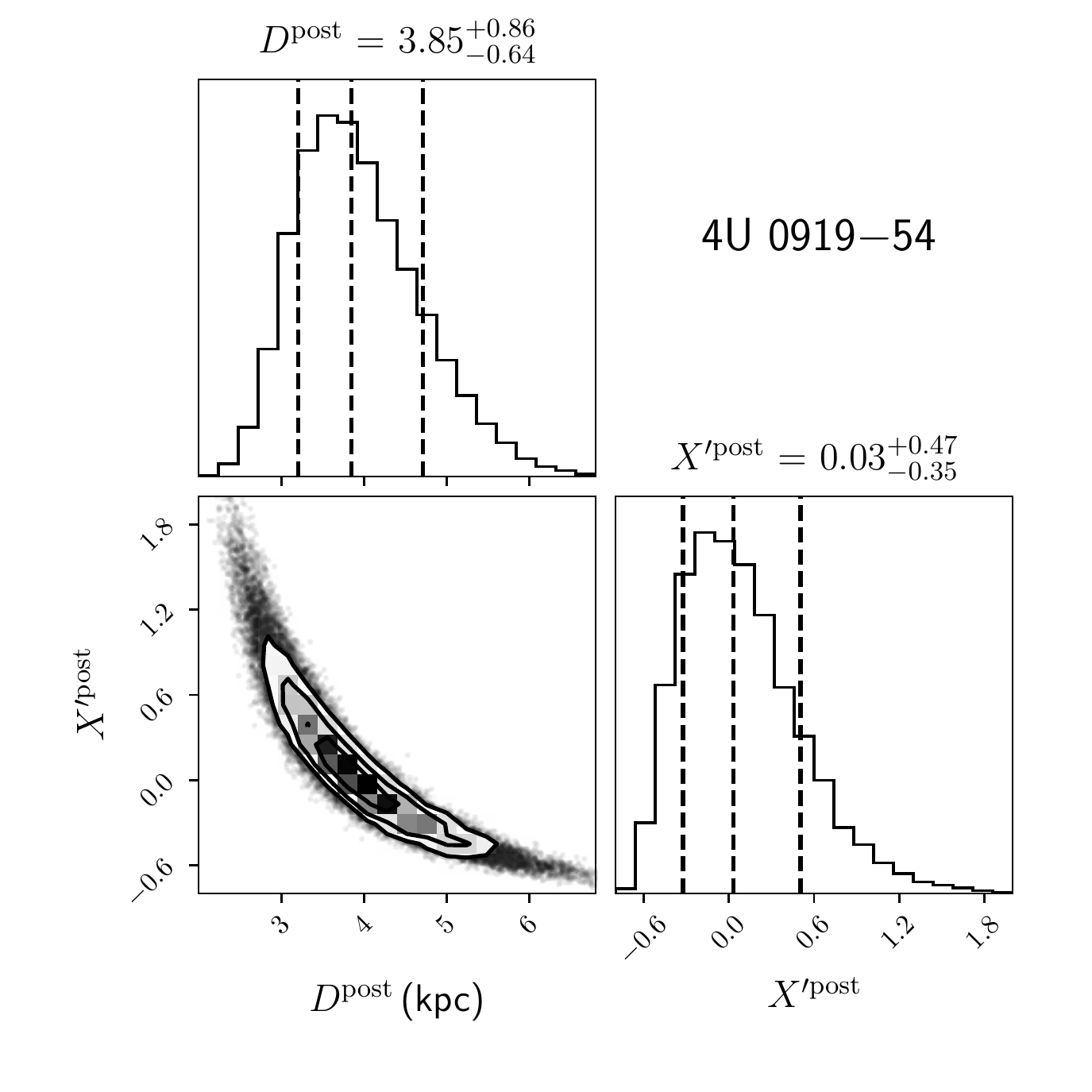} }\\
    \end{tabular}	

	\caption{2-D histograms and marginalized 1-D histograms of posterior distance $D^{\mathrm{post}}$ and $X'^{\mathrm{post}}$ (defined in \ref{subsec:PRE_roadmap_for_model_testing}) simulated with {\tt bilby} \citep{Ashton19} and plotted with {\tt corner.py} \citep{Foreman-Mackey16}. The $n$-th contour in each 2-D histogram contains $1-\exp\left(-n^2/2\right)$ of the simulated sample \citep{Foreman-Mackey16}.
	The vertical lines in the middle and two sides mark the median and central 68\% of the sample, respectively.}
    \label{fig:posterior_X1_D}
\end{figure*}

The simulated \{$D^{\mathrm{post}}$, $X'^{\mathrm{post}}$\} chain (posterior $D$ and $X'$) is depicted in \ref{fig:posterior_X1_D}. 
If the simplistic PRE model is correct, then $\eta=1$; as a result, $X'$ is supposed to fall into the ``standard'' domain of [0, 0.73]. None of the three PDFs of $X'$ decisively rules out $X'$ from this standard domain. However, we notice $X'<0$ at 90\% confidence for \cen, which moderately disfavors the simplistic PRE model.

As the domain of $X'$ is broadened from that of $X$, the fractional precision of $D^{\mathrm{post}}$ becomes predominantly reliant on that of the parallax.
In this regime, $D^{\mathrm{post}}$ and $X'^{\mathrm{post}}$ become more dependent on the adopted Galactic prior.
To understand the degree of this dependence, we re-estimated $D^{\mathrm{post}}$ and $X'^{\mathrm{post}}$ after resetting the Galactic prior to $\rho(D)=1$, which are summarized in \ref{tab:different_Galactic_prior}.
For \cyg\ with insignificant $\pi_1-\pi_0$, the dependence on the Galactic prior is evident, despite the consistency between the two sets of $D^{\mathrm{post}}$ and $X'^{\mathrm{post}}$.
For \cen\ and \uua\ both having relatively significant $\pi_1-\pi_0$, the degree of the dependence varies with the Galactic coordinates of the targets. In particular, we found the $D^{\mathrm{post}}$ and $X'^{\mathrm{post}}$ of \cen\ are robust against the selection of Galactic priors in \ref{eq:posterior_distance_PDF}.

\begin{table}
\caption{$D^{\mathrm{post}}$ and $X'^{\mathrm{post}}$ (see \ref{subsec:PRE_roadmap_for_model_testing}) inferred from the the Galactic prior $\rho(D)$ of \ref{eq:posterior_distance_PDF}, in comparison to the counterparts (noted as $\tilde{D}^{\mathrm{post}}$ and $\tilde{X'}^{\mathrm{post}}$) given $\rho(D)=1$.}
\label{tab:different_Galactic_prior}
\centering
\begin{tabular}{@{}ccccc@{}}
\hline\hline
PRE & $D^{\mathrm{post}}$ & $\tilde{D}^{\mathrm{post}}$ & $X'^{\mathrm{post}}$ & $\tilde{X'}^{\mathrm{post}}$ \\
burster & (kpc) & (kpc) &  &  \\
\hline\
%\vspace{1cm}
\cen\ & $1.8^{+0.6}_{-0.4}$ & $1.8^{+0.6}_{-0.3}$ & $-0.55^{+0.38}_{-0.23}$ & $-0.54^{+0.38}_{-0.24}$  \\
\cyg\ & $14^{+4}_{-2}$ & $18^{+13}_{-6}$ & $-0.3(3)$ & $-0.6^{+0.5}_{-0.3}$ \\
\uua\ & $3.9^{+0.9}_{-0.6}$ & $4.1^{+1.2}_{-0.8}$ & $0.0^{+0.5}_{-0.4}$ & $-0.1^{+0.5}_{-0.4}$\\
\hline\hline
\end{tabular}
\end{table}

Using $X'$ as a probe of the simplistic PRE model has the advantage of being independent of the PDF for $X$, which is usually not available. 
Future $D_0$ and $\pi_1-\pi_0$ measurements with higher precision will sharpen the above test of the simplistic PRE model. Lower fractional uncertainty of $\pi_1-\pi_0$ will not only reduce the strong correlation between $D^{\mathrm{post}}$ and $X'^{\mathrm{post}}$ for \cyg\ and \uua\ (shown in \ref{fig:posterior_X1_D}), but also lessen the dependence of inferred $D^{\mathrm{post}}$ and $X'^{\mathrm{post}}$ on the adopted Galactic prior.
On the other hand, constraining $X$ at PRE bursts independently (with, for example, ignition models of X-ray bursts) is the key to singling out $\eta$. Given that the simplistic PRE model is not significantly disfavored by any of the 3 PRE bursters, the refined distances in \ref{tab:constrain_composition} (based on the correctness of the simplistic PRE model) are still valid.

\subsection{Future prospects}
\label{subsec:PRE_future_prospects}
The potential of constraining $X$ and testing the simplistic PRE model using Gaia parallaxes of PRE bursters hinges on the achievable parallax uncertainties. 
Future Gaia data releases include Gaia Data Release 4 (DR4) (for the five-year data collection) and the data release after extended observations of up to 5.4 years (\url{https://www.cosmos.esa.int/web/gaia/release}).
We predict the achievable parallax ($\pi_1$) and proper motion uncertainties of \cen\ for future Gaia data releases using the simulation method detailed in the Appendix. 
In the same way, we estimated the predicted $\pi_1$ uncertainty $\hat{\sigma}_{\pi_1}$ of \cyg, \uua\ and \xb\ for future data releases and DR2, which are listed in \ref{tab:simulated_parallax_precision}. 
The predicted $\pi_1$ uncertainties for DR2, denoted as $\hat{\sigma}^\mathrm{DR2}_{\pi_1}$, generally agree with the real values $\sigma^\mathrm{DR2}_{\pi_1}$ (see \ref{tab:simulated_parallax_precision}).
In general, our simulation shows Gaia parallax and proper motion precision improves with observation time $t$ at roughly $t^{-1/2}$ and $t^{-3/2}$, respectively.

\begin{table}
\caption{Predicted parallax ($\pi_1$) uncertainties $\hat{\sigma}_{\pi_1}$ for \cen, \cyg\ and \uua\ obtained with simulations (see the Appendix for details). For comparison, real values of $\sigma^\mathrm{DR2}_{\pi_1}$ are provided as $\sigma^\mathrm{DR2}_{\pi_1}$.}
\label{tab:simulated_parallax_precision}
\centering
\begin{tabular}{@{}cccccc@{}}
\hline\hline
PRE & $\hat{\sigma}^\mathrm{DR2}_{\pi_1}$ & $\sigma^\mathrm{DR2}_{\pi_1}$ & $\sigma^\mathrm{EDR3}_{\pi_1}$ & $\hat{\sigma}^\mathrm{DR4}_{\pi_1}$ & $\hat{\sigma}^\mathrm{ext}_{\pi_1}$ $^*$\\
burster & (mas) & (mas) & (mas) & (mas ) & (mas)\\
\hline\
%\vspace{1cm}
\cen\ & 0.16 & 0.15 & 0.13 & 0.10 & 0.07 \\
\cyg\ & 0.024 & 0.024 & 0.019 & 0.014 & 0.010\\
\uua\ & 0.08 & 0.07 & 0.06 & 0.05 & 0.03\\
\xb\ & 0.10 & 0.09 & 0.08 & 0.06 & 0.04\\
\hline\hline
\end{tabular}
\tabnote{$^*$ we adopt an indicative 5-yr extension on top of the Gaia 5-yr mission.}
\end{table}

According to \ref{tab:zero_parallax_points}, the future parallax uncertainty of \cyg\ will be predominantly limited by the uncertainty of $\pi_0$, or $\sigma_{\pi_0}$. For this target, improvement in the estimation of the zero-point parallax correction and its uncertainty -- which is challenging due to the relative brightness of the source -- is essential.
In comparison, the $\sigma_{\pi_1}$ of \cen, \uua\ and \xb\ dominates the respective error budget of the calibrated parallax $\pi_1-\pi_0$. Therefore, one can expect $\pi_1-\pi_0$ of \cen, \uua\ and \xb\ to be potentially 1.9 times more precise (than EDR3) at the end of the extended Gaia mission (with 10-yr data), which would promise better constraints on $X$ and $X'$ for \cen\ and \uua.
On the other hand, to constrain $X$ and $X'$ for \xb\ with its already relatively precise Gaia parallax will require more PRE bursts to be detected from \xb.
Apart from Gaia astrometry, radio observations of PRE bursters during their outbursts using very long baseline interferometry technique can also potentially measure geometric parallaxes of PRE bursters, hence serving the same scientific goals outlined in this paper.

This work only deals with the simplistic PRE model, the hitherto most impactful PRE model. As is mentioned in \ref{sec:PRE_intro}, PRE models are not the only pathway to the distances of type I X-ray bursters. Bayesian inference based on a burst ignition model \citep{Cumming00} was recently realized to estimate parameters including $X$ and $D$ \citep{Goodwin19}. 
We believe such inferences will be made for more type I X-ray bursters in the near future.
In Bayesian analysis based on burst ignition models \citep[e.g.][]{Cumming00,Woosley04}, model-independent geometric parallaxes will serve as prior knowledge and help refine the inferred parameters including $D$ and $X$; the parallaxes can also judge which burst ignition model is more likely correct with a Bayes factor analysis.

Finally, we reiterate that we have assumed $M_\mathrm{NS}=1.4\,\msun$ and $R_\mathrm{NS}=11.2$\,km for all PRE bursters. Therefore, strictly speaking, $\eta$ should also reflect the deviation of $M_\mathrm{NS}$ and $R_\mathrm{NS}$ from their respective assumed values. 
Though a large deviation of $M_\mathrm{NS}$ from 1.4\,\msun\ is rare, it is not impossible \citep{Shao20}. 
In general, $L_{\mathrm{Edd},\infty}$ would increase with larger $M_\mathrm{NS}$ and $R_\mathrm{NS}$ (see \ref{eq:Eddington_luminosity}).
For example, $L_{\mathrm{Edd},\infty}(M_\mathrm{NS}=2, R_\mathrm{NS}=15)$ is 40\% greater than $L_{\mathrm{Edd},\infty}(M_\mathrm{NS}=1.4, R_\mathrm{NS}=11.2)$. 
Such a $L_{\mathrm{Edd},\infty}$ would correspond to $X'=-0.29$ assuming $X=0$ and the simplistic PRE model is correct. This value is consistent with $X'=-0.55^{+0.38}_{-0.23}$ for \cen. 
Given that the NS mass distribution is relatively well constrained compared to the NS radius distribution, it is possible to constrain the NS radius of a PRE burster in a Bayesian framework (slightly different from this work) provided prior knowledge of $X$, $D$ and $M_\mathrm{NS}$, assuming the simplistic PRE model is correct.

\section{Conclusion}
\label{sec:PRE_conclusion}
This work joins the few previous efforts to determine the Gaia parallax zero-point $\pi_0$ for a single source using nearby (on the sky) background quasars.
We provide a new template for $\pi_0$ determination (which also includes the $\pi_0$ uncertainty) approached by the weighted standard deviation of quasar parallaxes with respect to the filter parameters (see \ref{sec:PRE_parallax_calibration}). 
This work focuses on the small sample of PRE bursters with detected Gaia parallaxes $\pi_1$. A future work will apply the new template to a larger sample of targets, which can enable a comprehensive comparison with zero-parallax points estimated in other ways.

We also introduce a Bayesian framework to test the simplistic PRE model with parallaxes of PRE bursters on an individual basis. 
Though no parallax of the PRE bursters decisively disfavours the simplistic PRE model, the model is disfavored by the current \cen\ parallax at 90\% confidence.
At the end of the extended Gaia mission with 10-yr data, we expect the parallax precision to improve by a factor of 1.9, which will sharpen the test of the simplistic PRE model.
This test will also considerably benefit from an independent determination of $X$ (hydrogen mass fraction of nuclear fuel), which we highly encourage.

\section*{Acknowledgements}
The authors thank Ilya Mandel, Chris Flynn and Adelle Goodwin for useful discussions, and are grateful to the anonymous referee for the helpful comments on the manuscript.
H.D. is supported by the ACAMAR (Australia-ChinA ConsortiuM for Astrophysical Research) scholarship, which is partly funded by the China Scholarship Council.
A.T.D is the recipient of an ARC Future Fellowship (FT150100415).
Parts of this research were conducted by the Australian Research Council Centre of Excellence for Gravitational Wave Discovery (OzGrav), through project number CE170100004.
This work has made use of data from the European Space Agency
(ESA) mission Gaia (\url{https://www.cosmos.esa.int/gaia}), processed by
the Gaia Data Processing and Analysis Consortium (DPAC, \url{https://www.cosmos.esa.int/web/gaia/dpac/consortium}).
Data analysis was partly performed on OzSTAR, the Swinburne-based supercomputer.
%\end{acknowledgements}

\section*{Appendix: Simulating the parallax precision of \cen\ in future Gaia data releases}
\label{app:simulating_parallax_precision}
To simulate the uncalibrated parallax ($\pi_1$) precision of \cen\ for a future Gaia data release, we made two assumptions: {\bf 1)} \cen\ is observed by Gaia at equal intervals; 
{\bf 2)} $\sigma^i_k=\sigma^j_k$, ($k=\alpha, \delta$ and $i \neq j$), where $\sigma^i_\alpha$ and $\sigma^i_\delta$ stand for, respectively, the uncertainty of right ascension (RA) and declination at the $i$-th observation. 

The observation interval is determined with Table~1 of \citet{Gaia-Collaboration16}. At the ecliptic latitude of \cen\ ($-14\fdg1$), the number of observations is 63 over 5\,yr, corresponding to an interval of 29\,d.
It appears from \ref{tab:before_calibration} that, for a specific target, $\sigma_\alpha/\sigma_\delta \approx \sigma_{\mu_\alpha}/\sigma_{\mu_\delta}$ (where $\sigma_{\mu_\alpha}$ and $\sigma_{\mu_\delta}$ represent uncertainty of $\mu_\alpha$ and $\mu_\delta$ respectively), which implies the ratio of spatial resolution between RA and declination deviates slightly from unity and changes with pointing. 
Hence, we set $\sigma^i_\alpha / \sigma^i_\delta$ to 1.21, which is the $\sigma_{\mu_\alpha}/\sigma_{\mu_\delta}$ of \cen\ from EDR3.

Simulated positions of \cen\ were generated with {\tt pmpar} (available at \url{https://github.com/walterfb/pmpar}) using the ``predictor mode'', based on the astrometric parameters of \cen\ in \ref{tab:before_calibration}. 
We first simulated positions from MJD~56863 to MJD~57901 (the observing period of EDR3). The sole variable $\sigma^i_\alpha$ (and $\sigma^i_\delta$ accordingly) was tuned to make the parallax precision equal to 0.13\,mas, the current parallax precision of \cen. Subsequently, we stuck to the position uncertainties while extending the observing period to MJD~58689 (around the end day of the five-year period) and MJD~60516 (an indicative day 10 years after the first light of Gaia), which, respectively, yield parallax uncertainties of 0.095\,mas and 0.067\,mas.
Thus, we predict the parallax precision of \cen\ to improve by a factor of 1.3 and 1.9, respectively, with DR4 and 10-yr data.
Besides, we expect the proper motion precision of \cen\ to be enhanced by 2.7 and 6.7 times, respectively, with DR4 and 10-yr data.

\bibliographystyle{mnras}
\bibliography{haoding}
\chapter[Probing Alternative Theories of Gravity with VLBI Astrometry of Pulsar-white-dwarf Systems]{Probing Alternative Theories of Gravity with VLBI Astrometry of Pulsar-white-dwarf Systems}
\label{ch:J1012}

\newcommand{\psreaphsref}{J0958$+$5039}
\newcommand{\psreafrf}{J1118$+$1234}
\newcommand{\psreapibc}{J101307.3$+$531234}
\newcommand{\psreaibca}{J101307.4$+$530423}
\newcommand{\psreaibcb}{J101230.6$+$525826}
\newcommand{\psreaibcc}{J101204.0$+$531332}

This chapter is converted from \citet{Ding20} and an erratum \citep{Ding20d}. \citet{Ding20}, entitled ``Very Long Baseline Astrometry of PSR~J1012$+$5307 and its Implications on Alternative Theories of Gravity'', is the second publication (following \citealp{Vigeland18}) of the \mspsrpi\ catalogue.
%, and the first \mspsrpi\ paper focusing on gravitational tests. 
This chapter and \ref{ch:B1534} focus on tests of gravitational theories in different regimes. 
%This chapter focuses on constrains on dipole GW emission (where the most stringent constraints can be obtained from NS-WD systems due to the different binding energies), while Ch 7 focuses on strong-field GR tests that are best obtained from highly relativistic DNS systems.
\ref{ch:B1534} tests general relativity in the strong field regime using a double neutron star system, while this chapter constrains alternative theories of gravity with pulsar-white-dwarf systems that are potentially effective emitters of hypothetical dipole gravitational radiation \citep[e.g.][]{Eardley75}.
The astrometric inference of \psrea\ was revisited using Bayesian analysis in \ref{ch:mspsrpi}, as the Bayesian astrometric inference package {\tt sterne} had not been developed by the publication of \citet{Ding20}.
In response to a minor comment by the thesis examiners, $\dot{G}/G$ and $\kappa_D$ at 95\% confidence are added to \ref{subsec:J1012_alternative_gravity}.

\section{Abstract}

\psrea, a millisecond pulsar in orbit with a helium white dwarf (WD), has been timed with high precision for about 25 years. One of the main objectives of this long-term timing is to use the large asymmetry in gravitational binding energy between the neutron star and the WD to test gravitational theories.
Such tests, however, will be eventually limited by the accuracy of the distance to the pulsar. 
Here, we present VLBI (very long baseline interferometry) astrometry results spanning approximately 2.5 years for \psrea, obtained with the Very Long Baseline Array as part of the \mspsrpi\ project. These provide the first proper motion and absolute position for \psrea\ measured in a quasi-inertial reference frame.
From the VLBI results, we measure a distance of $0.83^{+0.06}_{-0.02}$\,kpc (all the estimates presented in the abstract are at 68\% confidence) for \psrea, which is the most precise obtained to date. Using the new distance, we improve the uncertainty of measurements of the unmodeled contributions to orbital period decay, which, combined with three other pulsars, places new constraints on the coupling constant for dipole gravitational radiation $\kappa_D=(-1.7\pm1.7)\times 10^{-4}$ and the fractional time derivative of Newton's gravitational constant $\dot{G}/G = -1.8^{\,+5.6}_{\,-4.7}\times 10^{-13}\,{\rm yr^{-1}}$ in the local universe. 
As the uncertainties of the observed decays of orbital period for the four leading pulsar-WD systems become negligible in $\approx10$ years, the uncertainties for $\dot{G}/G$ and $\kappa_D$ will be improved to $\leq1.5\times10^{-13}\,{\rm yr^{-1}}$ and $\leq1.0\times10^{-4}$, respectively, predominantly limited by the distance uncertainties.

%\end{abstract}

\section{Introduction} 
\label{sec:J1012_intro}

\subsection{Testing gravity theories with millisecond pulsars} \label{subsec:J1012_intro_test_gravity_with_psr}
When the beams of radiation emitted by rotating neutron stars (NSs) sweep across our line of sight, the result is a regular, lighthouse-like train of pulses. 
Thanks to their high rotational inertia, the spin period is extremely stable. The difference between observed pulse times of arrival (ToAs) and the model prediction for those ToAs is known as their residuals. The residuals are generally used to study unmodeled or imperfectly modeled physical process that would affect the ToAs \citep[e.g.][]{Lorimer12}.
Such processes include propagation through ionised material surrounding a binary companion \citep[e.g.][]{Lyutikov05} or through the interstellar medium \citep[ISM; e.g.][]{Lyne68,Bhat04} and gravitational phenomena.

Einstein's theory of general relativity (GR) is the simplest possible form among a class of candidate gravitational theories. Some alternatives to general relativity suggest a time dependence of Newton's gravitational constant $G$, which in most cases also necessitates dipolar \gw\ emission \citep{Will93}. There are several ways to test GR and constrain alternative theories of gravity with pulsars. 
Highly relativistic double neutron star systems \citep[e.g.][]{Damour92,Burgay03}, pulsar-white dwarf binaries \citep[e.g.][]{Lazaridis09,Freire12,Antoniadis13,Zhu15} and triple stellar systems hosting pulsars \citep{Archibald18} have probed different regions of phase space for deviations from the predictions of general relativity.
Taken collectively, an ensemble of pulsars can be used as a Pulsar Timing Array (PTA) to search for spatially correlated ToA variations that would betray the presence of nanohertz gravitational wave, such as those generated by supermassive binary black holes \citep{Detweiler79}.

The pulsars used in these tests belong to a sub-group of pulsars called “recycled” or ``millisecond” pulsars (MSPs).
In this work, we use these two terms interchangeably to refer to pulsars spun up via accretion from a companion donor star \citep{Alpar82}. 
One of the reasons why MSPs are important for these experiments is that they exhibit less intrinsic timing noise compared to non-recycled pulsars \citep{Shannon10}, thus providing much better timing stability \citep[e.g.][]{Perera19} and higher timing precision; this is important for the detection of small relativistic effects in their orbits.

In addition to studies based on pulsar timing, Very Long Baseline Interferometry (VLBI) astrometric experiments are also carried out on both MSPs and normal pulsars \citep[e.g.][]{Chatterjee09,Deller16,Vigeland18,Deller19}. 
By measuring annual geometric parallax and proper motion, VLBI astrometry can achieve model-independent estimates of distance, transverse velocity, and absolute positions for pulsars. It is significant not only in reducing the distance uncertainty and improving sensitivities of PTAs \citep{Madison13}, but also in various applications on a case by case basis.

\subsection{The \psrea\ binary system}
\label{subsec:J1012_intro_J1012}
\psrea\ is a millisecond pulsar (MSP) with a 5.3\,ms rotational period \citep{Nicastro95}. It has a helium white dwarf (WD) companion with mass $0.156\pm0.020$\,\msun\ (at 68\% confidence level, as is any other quoted uncertainty in this paper unless otherwise stated, \citealp{van-Kerkwijk96,Callanan98}) in a 0.6-day-long \citep{Lazaridis09} near-circular ($e<8\times10^{-7}$) orbit at a moderate inclination angle \citep{Driebe98,Lange01}. 
Spectroscopic observations of the WD were used to measure the mass ratio between \psrea\ and the WD companion to be $10.5\pm0.5$ by \citet{Callanan98} and $10.0\pm0.7$ by \citet{van-Kerkwijk04}. 
A new spectroscopic study of \psrea\ by \citet{Mata-Sanchez20} further refines the mass ratio to $10.44\pm0.11$.
\citet{Antoniadis16} re-visited the \citet{Callanan98} data with an updated model for the helium WD, which resulted in an updated mass estimate of $0.174\pm0.011$\,\msun. 
In this paper, we will use the new mass ratio $10.44\pm0.11$ and WD mass $0.174\pm0.011$\,\msun, that corresponds to a mass estimate of $\sim$1.8\,\msun\ for the NS in \psrea.

\subsection{Motivations for improving the distance to \psrea}
\label{subsec:J1012_intro_motivation}
The large difference in gravitational binding energy between the NS and WD in the \psrea\ system means that it would be an efficient emitter of dipolar gravitational waves in some alternate theories of gravity.
Timing observations of such binary pulsars are able to offer experimental tests for those theories.
Incorporating timing data from both \psrea\ and PSR J0437$-$4715, \citet{Lazaridis09} looked into the contributions to the time derivative of orbital period $\dot P_b$ and use the difference between the modeled and observed value $\dot P_b^{\,\mathrm{obs}}$ to constrain the coupling constant for  dipole gravitational radiation $\kappa_D$ to be $(0.3\pm2.5)\times10^{-3}$ and the fractional time derivative of Newton's gravitational constant $\dot{G}/G$ to be $(-0.7\pm3.3)\times10^{-12}~{\rm yr}^{-1}$, both at 95\% confidence. 
Using the same method but different pulsar-WD binaries, the best pulsar-based constraints, $\kappa_D=(-0.3\pm2.0)\times10^{-4}$ (at 68\% confidence) and $\dot{G}/G=(-1\pm9)\times10^{-13}\,{\rm yr^{-1}}$ (at 95\% confidence) are derived, respectively, by \citet{Freire12} and \citet{Zhu19}.

For \psrea\ the precision of the constraints on $\kappa_D$ and $\dot G$ is dominated by the uncertainty in $\dot P_b^{\,\mathrm{obs}}$,
and the distance to \psrea\ \citep{Lazaridis09}. 
Accordingly, improvements in the precision of distance estimates to pulsars such as \psrea\ have great potential to improve tests of alternate gravitational theories.
Furthermore, as the uncertainty of $\dot P_b^{\,\mathrm{obs}}$ decreases much faster than the uncertainty on the distance with pulsar-timing observations \citep{Bell96}, the latter will eventually dominate the error budget of $\kappa_D$ and $\dot G$.

Improving the distance to \psrea\ will also benefit some other studies. 
Additionally, as one of the pulsars having reliable independent distance measurements, \psrea\ was used by \citet{Yao17} to derive the latest model of the Galactic free electron density distribution. Therefore a more accurate distance to \psrea\ would further refine such a model.

\subsection{Measuring the distance to \psrea}
\label{subsec: distance2psr}
Several methods have been used in the past to estimate the distance to \psrea, the results of which are summarized in \ref{tab:J1012_astrometric_fits}.
A measure of $0.84\pm0.09$\,kpc was derived by \citet{Callanan98} using optical spectral-line observations of the WD companion. Pulsar timing is another way to measure the distance, as timing parallax is one of the outputs from parameter fits. \psrea\ is routinely timed by two PTAs, the European Pulsar Timing Array (EPTA) and the North American Nanohertz Observatory for Gravitational Waves (NANOGrav). To date, two timing parallaxes have been reported for \psrea\ utilizing solely EPTA data. 
\citet{Lazaridis09} reported a timing parallax of $1.22\pm0.26$\,mas using 15 years of multi-telescope data, corresponding to a distance of $0.82^{+0.22}_{-0.14}$\,kpc. A different timing parallax for \psrea\ $0.71\pm0.17$\,mas is reached more recently in \citet{Desvignes16}, showing $1.6\,\sigma$ tension with the previous timing result.
The EPTA and NANOGrav data for \psrea\ are also combined and analysed collectively under the International Pulsar Timing Array (IPTA) collaboration, leading to a distance estimate $0.7^{\,+0.2}_{\,-0.1}$\,kpc for \psrea\ \citep{Verbiest16}.

Apart from the above-mentioned methods, high-resolution trigonometric astrometry with VLBI or optical observations are able to provide model-independent distance estimation. A reliable distance to \psrea\ is essential for improving the uncertainty of $\dot G$ and dipole \gw\ emission (see \ref{subsec:J1012_alternative_gravity} for explanation). 
As well as GR tests, a distance based on trigonometric parallax reinforces the estimation of the bolometric luminosity of the companion WD, which reveals the WD radius. The WD radius can be translated to the WD mass when the mass-radius relation is worked out. Incorporating the known NS-WD mass ratio \citep{Mata-Sanchez20}, we can estimate the NS mass.

Prior to this work, an optical counterpart for \psrea\ has been identified by \citet{Jennings18} in the Gaia second data release (DR2) \citep{Gaia-Collaboration16,Gaia-Collaboration18}, carrying a tentative parallax of $1.3\pm0.4$\,mas. The parallax was then translated into a Gaia distance $0.79^{+0.73}_{-0.09}$\,kpc for \psrea\ incorporating other prior information \citep{Jennings18}.
In this work we focus on VLBI astrometry of \psrea\ as part of the \mspsrpi\ project, which is the extension of \psrpi\ project \citep{Deller16,Vigeland18,Deller19} focusing exclusively on MSPs. 
Throughout this paper, parameter uncertainties are quoted to 68\% confidence level unless stated otherwise.

\section{Observations and data reduction} \label{sec:J1012_observation}
\psrea\ was observed at L band (central frequency $\sim$1550 MHz) with the Very Long Baseline Array (VLBA) in eight epochs between July 2015 and November 2017 under the observation codes BD179 and BD192.  Each session was 1\,h long, and the observations are summarized in \ref{tab:J1012_eight_positions}. The observational setup is in general the same as other pulsars in \mspsrpi\ and \psrpi\ sample \citep[refer to][]{Deller19}, while using \psreaphsref\ as phase reference calibrator and \psreafrf\ to calibrate the instrumental bandpass. Four compact extragalactic radio sources within 9 arcminutes of \psrea\ were identified as suitable in-beam calibrators in early \mspsrpi\ observations, from which \psreapibc\ is chosen as the primary in-beam calibrator due to its relative brightness (\ref{tab:J1012_source_catalog}). 
The data were correlated using the DiFX software correlator \citep{Deller11a}
in two passes - gated and ungated. After gating, the S/N increases by $\approx40$\%.
The gated visibility datasets are subsequently processed in a python-based ParselTongue \citep{Kettenis06} pipeline calling {\tt AIPS} \citep{Greisen03} and {\tt DIFMAP} \citep{Shepherd94} functions, described in \citet{Deller19}. 
The reduction pipeline is publicly available now at \url{https://github.com/dingswin/psrvlbireduce}. It is to be released incorporating better readability, configurability, some new functions and extended diagnostic tools. The data for \psrea\ were reduced using the pipeline versioned a6b666e.
Multiple runs of the pipeline are made to iteratively flag bad visibility data (e.g., due to radio frequency interference), make uniform models for calibrators (including the phase calibrator, fringe finder and in-beam calibrators) and obtain reliable positions for \psrea.

\begin{table*}
	\centering
	\caption{Positions and uncertainties without $\mid$ with systematics}
	\label{tab:J1012_eight_positions}
	\begin{tabular}{ccccc} % four columns, alignment for each
		\hline
		\hline
		yyyy-mm-dd & Project & S/N & RA & Dec \\
		 & Code & & (J2000) & (J2000)\\ 
		\hline
		2015-07-16 & bd179e0 & 58.9 & $10^{\rm h}12^{\rm m}33\fs 439455(5|8)$ & 53\degr07'02\farcs14401(9$\mid$16)\\ 
		2015-11-15 & bd179e1 & 96.8 & $10^{\rm h}12^{\rm m}33\fs 439723(2|6)$ & 53\degr07'02\farcs13524(6$\mid$16) \\
		2016-11-11 & bd192e0 & 87.7 & $10^{\rm h}12^{\rm m}33\fs 440038(3|6)$ & 53\degr07'02\farcs11042(6$\mid$16) \\
		2016-11-19 & bd192e1 & 85.0 & $10^{\rm h}12^{\rm m}33\fs 440040(3|9)$ & 53\degr07'02\farcs10984(8$\mid$20) \\
		2017-05-10 & bd192e2 & 88.2 & $10^{\rm h}12^{\rm m}33\fs 439927(3|7)$ & 53\degr07'02\farcs09865(7$\mid$19) \\
		2017-05-29 & bd192e3 & 46.8 & $10^{\rm h}12^{\rm m}33\fs 439949(5|7)$ & 53\degr07'02\farcs09682(12$\mid$19) \\
		2017-06-11 & bd192e4 & 31.0 & $10^{\rm h}12^{\rm m}33\fs 439972(8|10)$ & 53\degr07'02\farcs09639(19$\mid$23) \\
		2017-11-26 & bd192e5 & 17.9 & $10^{\rm h}12^{\rm m}33\fs 440345(13|15)$ & 53\degr07'02\farcs08382(34$\mid$38)  \\
		\hline
	\end{tabular}
\end{table*}

\begin{table*}
	\centering
	\caption{Source catalog}
	\label{tab:J1012_source_catalog}
	\begin{tabular}{cccc} % four columns, alignment for each
		\hline
		\hline
		Source & Name in data & $\Delta_{\rm src-psr}$ & Purpose\\
		%& & (deg) & \\
		\hline
		\psrea\ & J1012$+$5307 & 0 & target \\
		\psreafrf\ & \psreafrf\ & 42\fdg6 & fringe finder/bandpass calibrator\\ 
		\psreaphsref\ & \psreaphsref\ & 3\fdg26 & phase calibrator \\
		\psreapibc\ & IBC00462 & 7\farcm51 & primary in-beam calibrator\\
		\psreaibca\ & IBC00412 & 5\farcm74 & in-beam calibrator\\
		\psreaibcb\ & IBC00421 & 8\farcm62 & in-beam calibrator\\
		\psreaibcc\ & IBC00460 & 7\farcm85 & in-beam calibrator\\
		\hline
	\end{tabular}
\end{table*}

\section{Systematic errors and parallax fits} \label{sec:J1012_result}
After data reduction, we determined positions of \psrea\ at eight epochs, which are summarized in \ref{tab:J1012_eight_positions}. The statistical positional uncertainties obtained from an image-plane fit are reported to the left of the "$\mid$" symbol in \ref{tab:J1012_eight_positions}.  However, we expect a significant contribution from systematic position shifts, and the uncertainties in \ref{tab:J1012_eight_positions} to the right of "$\mid$" symbol incorporate both the statistical uncertainty already mentioned and an empirical estimate of systematic uncertainty.  This estimate is made using the following empirical function rewritten from \citet{Deller19}:
\begin{equation}
\label{eq:J1012_empirical_sys_error}
\Delta_{sys}=A \times s \times \overline{\csc{\theta}} + B/S\,,
\end{equation}
where $\Delta_{sys}$ is the ratio of the systematic error to the synthesized beam size, $\theta$ stands for elevation angle, $\overline{\csc{\theta}}$ is the average $\csc{\theta}$ for a given observation (over time and antennas), $s$ is the angular separation in arcmin between \psreapibc\ and \psrea, $S$ represents the signal-to-noise ratio of \psreapibc, and $A=0.001$ and $B=0.6$ are empirically determined coefficients based on the \psrpi\ sample. The first term in \ref{eq:J1012_empirical_sys_error} represents propagation-related systematic errors, while the second term accounts for random errors resulting from the calibration solutions from the primary in-beam calibrator. In general, propagation-related systematic errors (proportional to the separation between calibrator and target) dominate the systematic (and indeed overall) error budget. This is still true in this work, even if the first term is significantly reduced by the usage of in-beam calibrators. 

For \psrea, the contribution from the first term is $\sim$3 times that of the second term. 
The full uncertainties are the addition in quadrature of statistical and systematic uncertainties. The inclusion of systematic uncertainties decreases the \rcs\ of least-squares astrometric fit from 7.9 to 1.9, which indicates that the uncertainty estimation is likely reasonable (although perhaps still modestly underestimated).

\subsection{Astrometric fitting} \label{subsec:J1012_pi&pm}
\begin{figure*}
\centering
\includegraphics[width=12cm]{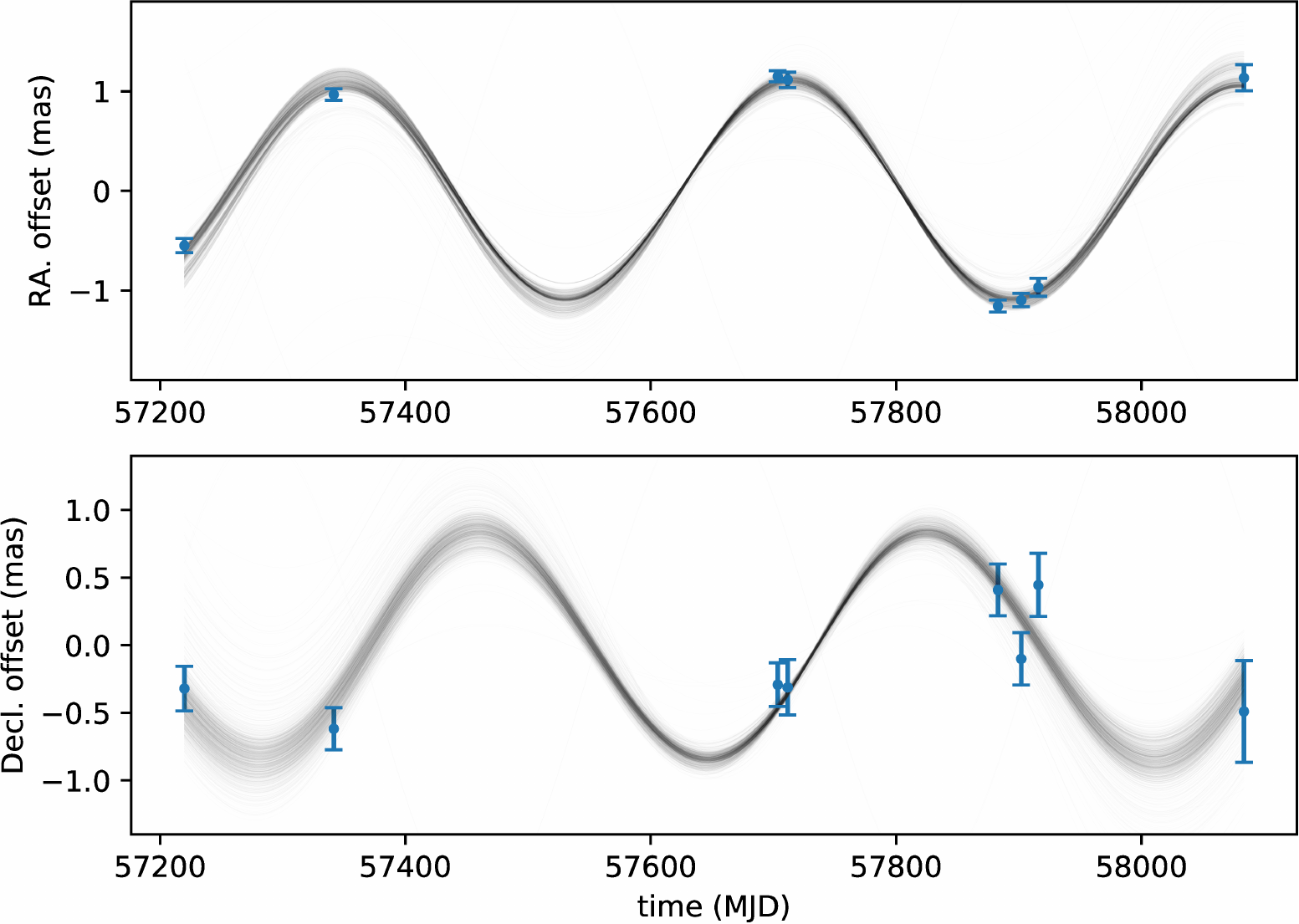}
\caption{Stacked sky position evolution removing best fit proper motion. Each line shows the fitted model out of a bootstrap run.}
\label{fig:J1012_bootstrap_parallax}
\end{figure*}

\begin{figure*}
\centering
\includegraphics[width=15cm]{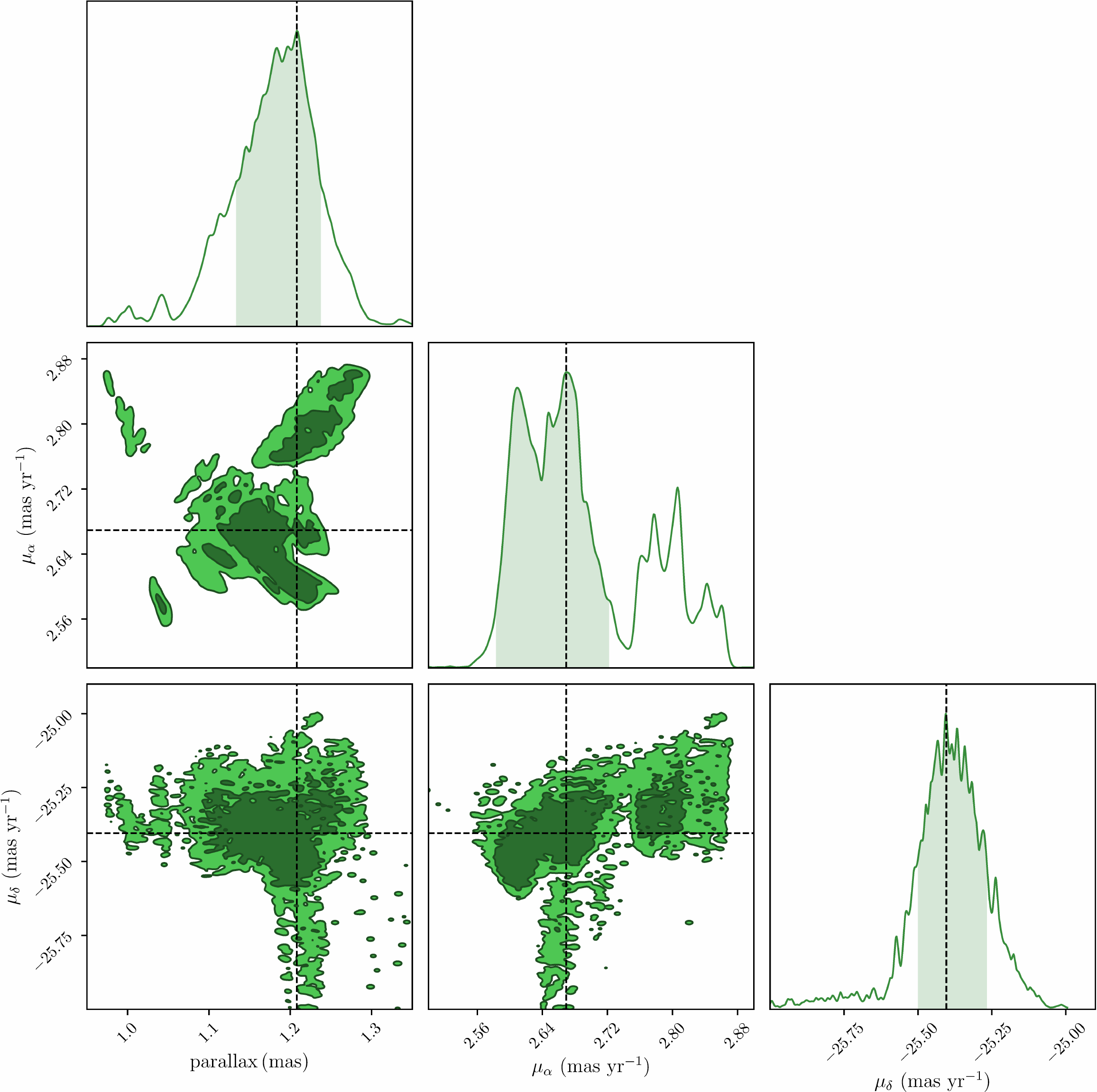}
\caption{
Error ``ellipses'' and marginalized histograms for parallax and proper motion.
In each histogram, the dashed line marks the measured value; the shade stands for the 68\% confidence interval.
In each error ``ellipse'', the dark and bright contour(s) enclose, respectively, 68\% and 95\% of the bootstrapped data points.
}
\label{fig:J1012_pdf_pyramid}
\end{figure*}

We used {\tt pmpar}\footnote{\url{https://github.com/walterfb/pmpar}} to perform astrometric fitting for parallax, proper motion and reference position. As found by \citet{Deller19} for PSR~B1913$+$16, using a bootstrap technique to estimate the astrometric parameters was consistent with but more conservative than a least-squares fit, and we followed this approach for \psrea. We bootstrapped 130000 times from the 8 positions with full uncertainties (i.e., in every run we drew positions eight times with replacement from the set of positions and performed astrometric fitting on the drawn sample) for estimation of the astrometric parameters. It is possible that the eight draws contain fewer than three effective epochs, which is the minimum required for astrometric fitting. In order to limit the number of severely biased fits yielded by overly short time baselines or negligible variation in the contribution of parallax between epochs,
we required at least 4 effective epochs for subsequent fitting in every bootstrap run.
\ref{fig:J1012_bootstrap_parallax} shows the stacked position evolution removing best fitted proper motion. Each line is the fitting result out of one bootstrap run. The densest part corresponds to the estimated parallax, position and proper motion. 
Using the probability density functions of parallax and proper motion obtained from the bootstrap runs (see \ref{fig:J1012_pdf_pyramid}), we determined the measured value and corresponding uncertainty, which are summarized in \ref{tab:J1012_astrometric_fits} along with the results of a simple linear least squares fit. 
For each astrometric observable (i.e. parallax, proper motion, or reference position), its measured value corresponds to the peak of its probability density function; its uncertainty is given by the narrowest interval that encloses 68.3\% of its 130000 bootstrapped results. 

As is seen in \ref{tab:J1012_astrometric_fits}, the uncertainties from bootstrap are much more conservative than direct fitting, and are possibly over-estimated since the \rcs\ for direct fitting is already close to unity. We use the bootstrap results in the following discussions. The histogram for $\mu_\alpha$ is bimodal in \ref{fig:J1012_pdf_pyramid}, which is discussed in \ref{subsec:J1012_timing_interplay}. Lutz-Kelker correction was not carried out since our significance of parallax $\varpi_0/\sigma>16$ is well over the critical value between 5.0 and 6.7, indicating a negligible Lutz-Kelker effect \citep{Lutz73}.

\begin{table*}
\caption{Parallax, proper motion, parallactic distance and transverse velocity for \psrea\ 
\label{tab:J1012_astrometric_fits}}
\resizebox{\textwidth}{!}{
\begin{tabular}{ccccccc}
\hline
\hline
method & $\varpi$ & $\mu_\alpha \equiv \dot{\alpha} \cos\delta$ & $\mu_{\delta}$ & $D$ & $v_t$ & references $^a$\\
& (mas) & ($\rm mas~{yr}^{-1}$) & ($\rm mas~{yr}^{-1}$) & (kpc) & ($\rm km~s^{-1}$) & \\
\hline

direct fitting & $1.17\pm 0.02$ & $2.68\pm 0.03$ & $-25.38\pm 0.06$ & $0.86\pm 0.02$ & $103.4\pm 1.9$ & This work\\
bootstrap & $1.21^{+0.03}_{-0.08}$ & $2.67^{+0.05}_{-0.09}$ & $-25.40^{+0.14}_{-0.09}$ & $0.83^{+0.06}_{-0.02}$ & $100.2^{+7.2}_{-2.7}$ & This work\\
\\
timing parallax (EPTA/2009) & {$1.22\pm 0.26$} & {$2.56\pm 0.01$} & {$-25.61\pm 0.02$} & $0.8\pm 0.2$ & $100.0\pm 21.3$ $^b$ & (1)\\
timing parallax (EPTA/2016) & {$0.71\pm 0.17$} & {$2.61\pm 0.01$} & {$-25.48\pm 0.01$} & $1.4^{+0.4}_{-0.3}$ $^c$ & $171.0\pm 41.0$ & (2) \\
orbital parallax (EPTA/2016) & $\cdots$ & $\cdots$ & $\cdots$ & $0.94\pm 0.03$ $^d$ & $\cdots$ & (2) \\
timing (IPTA) & $-$ & $-$ & $-$ & $0.7^{+0.2}_{-0.1}$ & $-$ & (3) \\
timing (NANOGrav) & $1.3\pm 0.4$ $^e$ & $2.66\pm 0.03$ & $-25.50\pm 0.04$ & $>0.5$ & $-$ & (4) \\
Gaia DR2 & $1.33\pm 0.41$ & $2.98\pm 0.52$ & $-26.94\pm 0.63$ & $0.79^{+0.73}_{-0.09}$ & $113^{+133}_{-12}$ & (5,~6,~7)\\
spectroscopy & $-$ & $-$ & $-$ & $0.84 \pm 0.09$ & $-$ & (8)\\
\hline
\end{tabular}
}
\tabnote{$^a$ (1)~\citet{Lazaridis09}, (2)~\citet{Desvignes16}, (3)~\citet{Verbiest16}, (4)~\citet{Arzoumanian18}, (5)~\citet{Gaia-Collaboration16}, (6)~\citet{Gaia-Collaboration18}, (7)~\citet{Jennings18}, (8)~\citet{Callanan98}.}

\tabnote{$^b$ Here, we have re-calculated $v_t$ using the parallax-based distance, rather than weighted distance for consistency \citep{Lazaridis09}.}

\tabnote{$^c$ $D_\varpi$: Lutz-Kelker correction not applied for consistency.}

\tabnote{$^d$ $D_{\dot P_b}$, distance derived from the $\dot P_b^{\,\mathrm{obs}}$ (time derivative of orbital period) budget in GR regime.}

\tabnote{$^e$ Classified by \citet{Arzoumanian18} as non-detection.}

\end{table*}

\subsection{Absolute position for \psrea} \label{subsec:J1012_abspos}
\begin{figure}
\centering
\includegraphics[width=14cm]{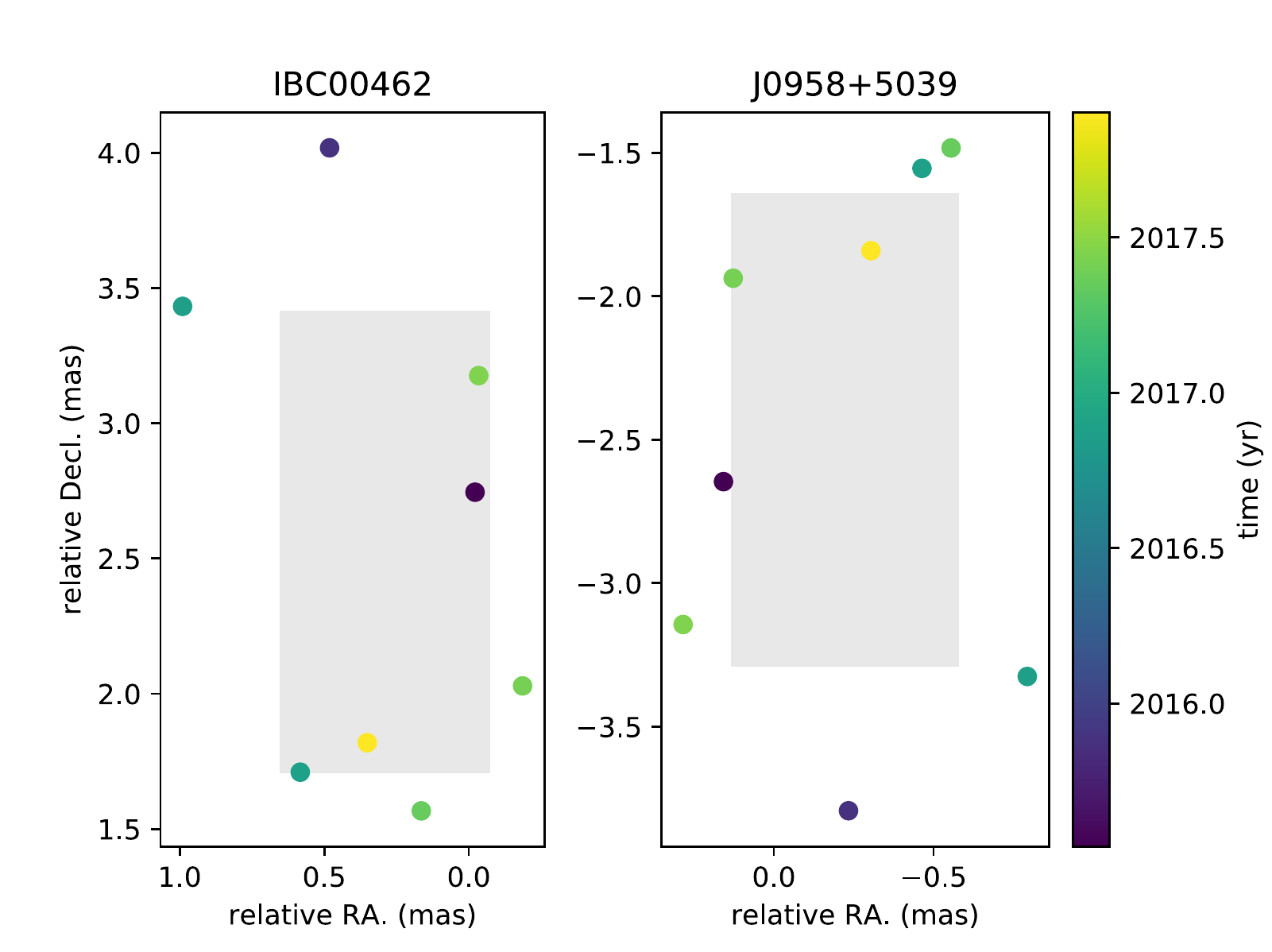}
\caption{Position scatter of the phase calibrator in reference to the primary in-beam calibrator (left) and primary in-beam calibrator referenced to the phase calibrator (right). The positions are relative to $09^{\rm h}58^{\rm m}37\fs 80944+$50\degr39'57\farcs4837 for J0958$+$5039 and $10^{\rm h}13^{\rm m}07\fs29548+$53\degr12'34\farcs3348 for IBC00462. The shaded rectangle in each panel shows the standard deviation of the position in RA and Dec.
\label{fig:J1012_position_scatter}}
\end{figure}

Absolute positions are of significance for comparing positions based on different reference frames.
The position for \psrea\ that we obtain from bootstrapping is anchored to \psreapibc, the primary in-beam calibrator. However, the absolute position of \psreapibc\ is not well determined, and we must estimate its position and uncertainty based on the \mspsrpi\ observations. In order to derive the absolute position for \psrea, we used \psreapibc\ to tie \psrea\ to \psreaphsref. We used two methods to make this connection. In the first approach the calibration solutions derived by \psreaphsref\ were transferred to \psreapibc, which was subsequently divided by the uniform \psreapibc\ model obtained from eight epochs; the centroid of the divided \psreapibc\ was located for each epoch; the average and scatter of the eight \psreapibc\ positions thus offer the information of the ``real" \psreapibc\ position relative to \psreaphsref\ and the systematic uncertainty of this position. The second way is in principle the same while in the reverse direction: the final solution derived by \psreapibc\ was applied to \psreaphsref. \ref{fig:J1012_position_scatter} shows the eight \psreapibc\ and \psreaphsref\ positions obtained in the two different ways. As is expected, from \ref{fig:J1012_position_scatter} we can conclude that 1) no time-dependence of position shifts is noticeable; 2) the scatter among the positions for the two objects, indicating the systematic errors around the mean position, is consistent in both RA and Dec. 
In each way we used the average position to tie \psreaphsref\ and \psreapibc, thus anchoring \psrea\ to \psreaphsref. The absolute positions derived from two ways are highly consistent; we proceed with their average position. Finally, we aligned \psrea\ to the latest \psreaphsref\ position\footnote{\url{http://astrogeo.org/vlbi/solutions/rfc_2019a/rfc_2019a_cat.html}}, which is measured at higher radio frequencies based primarily on dual-band 2.3/8.4 GHz observations \citep{Petrov08}. 

As the jet core of \psreaphsref\ is presumably the brightest spot in the \psreaphsref\ map, it is taken as the reference position for \psreaphsref\ after fringe fitting in {\tt AIPS}. Since the jet core moves upstream towards the central engine with increasing frequency \citep[e.g.][]{Bartel86,Lobanov98}, our presented absolute position is referenced to the jet core of \psreaphsref\ at L band, where its absolute position has not been determined.
\citet{Sokolovsky11} compiled multi-band observations on 20 AGNs and reported the median core shift between X band and L band is 1.15\,mas. A recent work by \citet{Plavin19} integrates long-term observations of 40 AGNs and concludes the core shift of AGNs between 8\,GHz and 2\,GHz is typically 0.5\,mas. They additionally found time variability of core shift at an average level of 0.3\,mas in 33 AGNs of the sample. With limited knowledge about the core shift of \psreaphsref, 
we split the median core shift 1.15\,mas evenly between the two axes and add them in quadrature to the errors of the absolute position of \psrea.

The absolute position we obtained for \psrea\ is shown in \ref{tab:J1012_absolute_position}.
We chose the midpoint of the eight VLBI epochs as the reference time for astrometric fitting to obtain the highest precision for the absolute position of \psrea, and note that extrapolating the position to earlier or later times will suffer progressively from the accumulation of proper motion uncertainty.

The uncertainty of the absolute position of \psrea\ comprises the bootstrap uncertainty of \psreaphsref\ position anchored to \psreapibc\ (i.e. the uncertainty derived from the normalized histogram of RAs/Decs as shown in \ref{fig:J1012_pdf_pyramid}), the systematic errors in \psreaphsref$-$\psreapibc\ connection (the scatter of eight positions) and the uncertainty of the absolute position of \psreaphsref. These components are added in quadrature. In order to make comparison to timing results, we also extrapolated the timing positions for \psrea\ to our reference epoch MJD~57700 using the ephemerides of \citet{Lazaridis09,Desvignes16,Arzoumanian18} (see \ref{tab:J1012_absolute_position}). Furthermore, we re-identified Gaia DR2 851610861391010944 as the optical counterpart for \psrea: its predicted position at MJD~57700 is $<5$\,mas from our VLBI position (as shown in \ref{tab:J1012_absolute_position}), while its proper motion and parallax are largely consistent with both the VLBI and timing results (\ref{tab:J1012_astrometric_fits}). 
The uncertainties of the timing and Gaia positions are estimated with Monte-Carlo simulation, assuming the astrometric parameters offered in literature (also reproduced in \ref{tab:J1012_astrometric_fits}) follow a gaussian distribution.

\begin{table*}
	\centering
	\caption{Absolute position for \psrea\ at MJD\,57700}
	\label{tab:J1012_absolute_position}
	\resizebox{\textwidth}{!}{
	\begin{tabular}{cccccc} 
		\hline
		\hline
		& This work & \citet{Lazaridis09} & \citet{Desvignes16} & \citet{Arzoumanian18} & Gaia DR2 \\
		\hline
	RA & $10^{\rm h}12^{\rm m}33\fs 
	4399(1)$ $^a$
	& $10^{\rm h}12^{\rm m}33\fs 43967(4)$ & $10^{\rm h}12^{\rm m}33\fs 43973(2)$ & $10^{\rm h}12^{\rm m}33\fs 439773(9)$ & $10^{\rm h}12^{\rm m}33\fs 43986(9)$ $^b$ \\
	Dec. & 
	53\degr07'02\farcs113(1) &
	53\degr07'02\farcs1094(4) & 53\degr07'02\farcs1113(1) & 53\degr07'02\farcs11090(9) & 53\degr07'02\farcs1098(9) \\
	\hline
	\end{tabular}
	}
	\tabnote{$^a$The uncertainty for both R.A. or declination includes an estimate of the systematic error introduced by core shift in the reference source between 1.5 GHz and 8.4 GHz, taken as 0.8\,mas in each axis as described in the text.}
	\tabnote{$^b$\citet{Gaia-Collaboration16,Gaia-Collaboration18,Jennings18}}
\end{table*}

\section{Discussion} 
\label{sec:J1012_discussion}

\subsection{Comparison to timing astrometry} 
\label{subsec:J1012_timing_interplay}
There are two published parallaxes and three proper motions for \psrea\ based on timing astrometry \citep{Lazaridis09,Desvignes16,Arzoumanian18}. The timing proper motions disagree significantly, as shown in \ref{tab:J1012_astrometric_fits}, indicating that the uncertainties have historically been somewhat underestimated. 
At the time of writing, the EPTA is the only PTA that detects a timing parallax for \psrea\ \citep{Lazaridis09,Desvignes16}.
Given the additional data available to the 2016 work, we would expect this to be the more accurate of the two EPTA results.
Our proper motion and reference position agree with both EPTA measurements, 
slightly favoring the 2016 measurement. 
Our measured parallax, on the other hand, agrees with the 2009 measurement but is in significant tension ($\approx2.6\,\sigma$) to the 2016 result. \ref{tab:J1012_astrometric_fits} shows that the \citet{Desvignes16} result is also inconsistent with other independently derived distance measurements, but the cause of the discrepancy is unknown.

When we consider the VLBI results for proper motion, a bimodality is
apparent in the probability density obtained for $\mu_\alpha$ in \ref{fig:J1012_pdf_pyramid}.
The sub-peak of $\mu_\alpha$ at $\approx$2.8\,\maspy\ is strongly disfavored by all timing results (cf. \ref{tab:J1012_astrometric_fits}). 
We diagnosed the origin of this bimodality by removing one epoch at a time from our bootstrap procedure and found that the inclusion of the second epoch (BD179E1) is responsible for the sub-peak of the $\mu_\alpha$ histogram.
However, since there is no clear evidence of a bad measurement at the second epoch (cf. \ref{fig:J1012_bootstrap_parallax}), we did not take any action such as removing this observation. The availability of pulsar timing proper motions does, however, offer the opportunity to study the effect of applying prior information when conducting the VLBI fitting.

VLBI astrometry is performed in a quasi-inertial reference frame, determined by numerous distant AGNs whose positions (as determined by VLBI observations) are assumed to be fixed.  Examples of realizations of such a reference frame include the International Celestial Reference Frame version 3 (ICRF3\footnote{\url{www.iers.org/IERS/EN/DataProducts/ICRF/ICRF3/icrf3.html}}) and the Radio Fundamental Catalog (RFC\footnote{\url{astrogeo.org/rfc/}}). Timing astrometry, on the other hand, is performed after referencing the pulse ToAs to the barycenter of the solar system, making use of a solar system ephemeris (SSE). Due to the different nature of the reference frames used by VLBI and timing astrometry, a small-scale 3-dimensional rotation between the two types of reference frames is possible, and this rotation could be time-dependent. As a result, small differences in the reference position might be seen between quantities measured using VLBI versus those measured using pulsar timing. Unless the time dependence of the frame misalignment was extremely large, however, the effect on proper motion and (especially) parallax would be extremely small compared to current levels of precision.

The 3-dimensional transformation between ICRF and barycentric frame can be decomposed into a 2-d translation and a 1-d rotation as the displacement is only $\sim1$\,mas level \citep{Wang17}. This can be visualized as the translation and rotation of a local 2-d frame in its surface. The proper motions obtained from VLBI and timing astrometry differ only when there is a noticeable 1-d frame rotation. Under the safe assumption that the angle of the 1-d frame rotation is smaller than 1\,arcmin, the effect of frame transformation on proper motion would be insignificant, and we can in principle make use of the timing proper motions as priors to the VLBI astrometric fitting. The longer time baseline of timing observations promises better precision of proper motion. Therefore if the assumption is met, the application of timing proper motion would potentially improve our parallax estimation.

\begin{figure*}
\centering
\includegraphics[width=14cm]{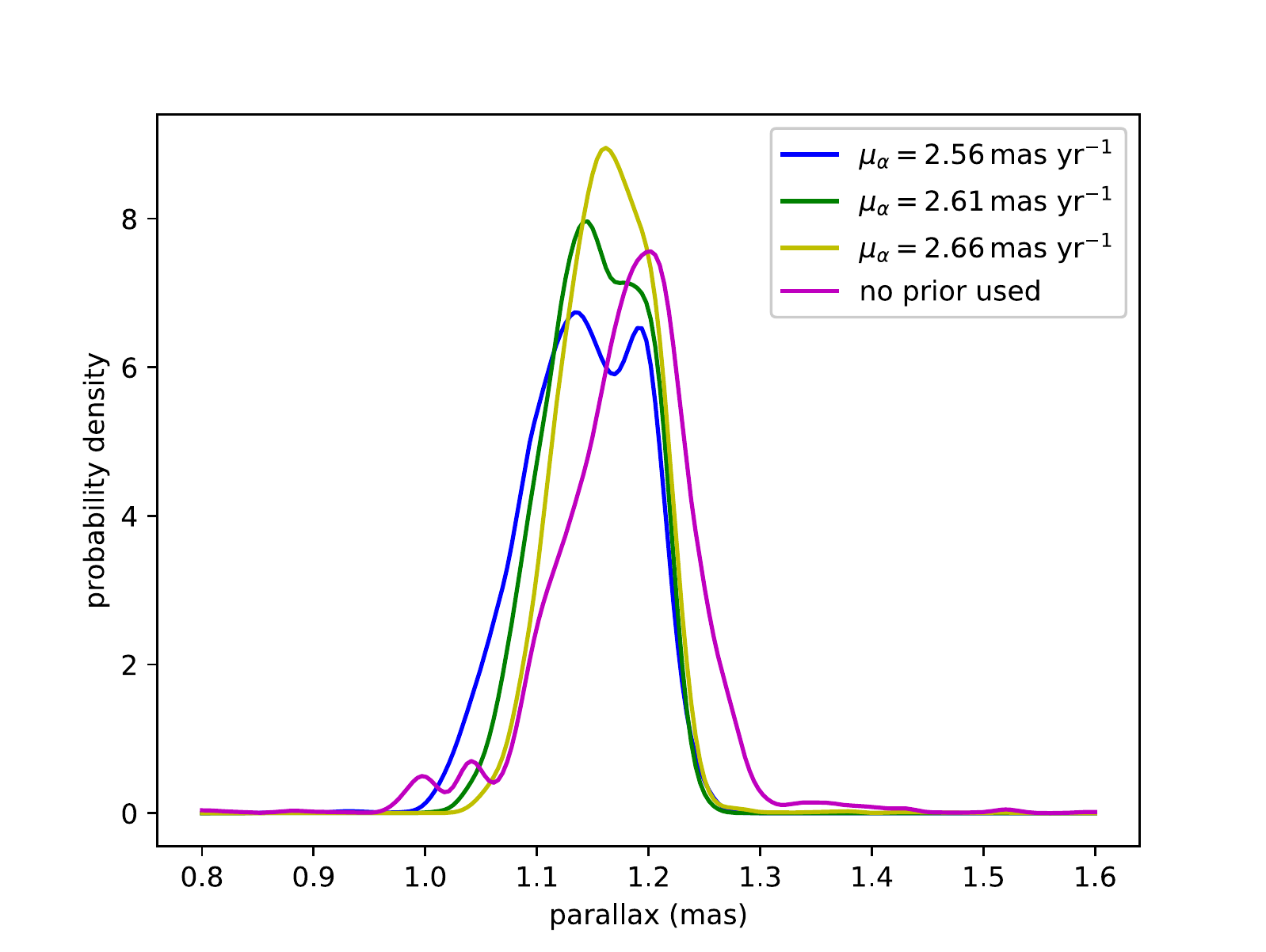}
\caption{Probability density functions of parallax (smoothed-out normalized histogram) out of bootstrapped astrometric fittings using $\mu_\alpha$ priors, in comparison with that of free astrometric fittings.}
\label{fig:J1012_timing_prior_pdf}
\end{figure*}

We fixed $\mu_\alpha$ to the 2009, 2016 and 2018 $\mu_\alpha$ respectively (see \ref{tab:J1012_astrometric_fits}) and ran bootstrapping again. The resultant parallax probability density functions are shown in \ref{fig:J1012_timing_prior_pdf}. The peak of the normalized histogram of parallax changes slightly with $\mu_\alpha$, as a result of the correlation between parallax and $\mu_\alpha$ (shown by the corresponding error ``ellipse'' in \ref{fig:J1012_pdf_pyramid}). In all cases, the effect of applying the timing proper motion prior is to reduce the most probable parallax value by a small fraction of a standard deviation, and for the most recent timing proper motion results \citep{Desvignes16,Arzoumanian18} the overall parallax uncertainty is reduced.

However, the timing measurements of $\mu_\alpha$ disagree by much more than their formal uncertainties, making it difficult to select the most accurate $\mu_\alpha$ to impose as a prior. Therefore, for simplicity and self-consistency, we use the VLBI results obtained with no priors from the timing proper motion as shown in \ref{tab:J1012_astrometric_fits} in following discussion, but note that 1) the application of a well-motivated prior based on timing could further improve the VLBI parallax and hence distance precision, and 2) whichever timing proper motion is chosen, the effect is to reduce the VLBI parallax and hence increase the estimated distance to the pulsar.

\subsection{Galactic path from updated 3D velocity}
\label{subsec:J1012_Gal_path}
The orbit of \psrea\ through the Galaxy was determined by
\citet{Lazaridis09}, who used their best-fit proper motion
and distance along with a radial velocity of $v_r=44\pm8\,{\rm
  km\,s}^{-1}$ estimated by \citet{Callanan98}. Tracing it back for 10\,Gyr in a model for the Galactic
potential they found that it is only rarely passes close to the Sun; rather
it spends more time out at Galacto-centric distances of 30\,kpc and
oscillating vertically up to 7\,kpc above/below the Galactic plane.  However, \citet{Freire11} used the same data to find different conclusions.  The Galacto-centric distances range from 4\,kpc to 7\,kpc, and the vertical oscillations go to 2\,kpc above/below the Galactic plane.

With our new distance as well as the improved radial velocity
$v_r\, =\, -21.3\, \pm \, 1.6\,{\rm km\,s}^{-1}$ \citep{Mata-Sanchez20}, we can once again repeat this
exercise. We use the \texttt{MWPotential2014} potential from
\texttt{galpy} \citep{Bovy15} for tracing the orbit of the pulsar.
In contrast to \citet{Lazaridis09} and similar to \citet{Freire11}, rather than spending little time near the Sun and orbiting out to 30\,kpc, we find that \psrea\ orbits
largely within the Solar circle, going from Galactocentric radii of
3.5--8.5\,kpc on a timescale of $\sim 125\,$Myr.  Similarly, rather
than oscillating to $\pm 7\,$kpc in the vertical direction it only moves
 to $\pm1\,$kpc (the difference between our results and those of \citealt{Freire11} are largely due to the different value of $v_r$ that we assumed).  These results are robust to the choice of
potential: using the same analytic potentials as \citet{Lazaridis09}
gives essentially the same overall orbits.  This is not surprising, as
the 3D space velocity of \psrea\ of $105\,{\rm km\,s}^{-1}$ relative to the
Local Standard of Rest (based on the Solar motion of \citealt{Schonrich10})
is largely consistent with the  MSP
distribution \citep{Matthews16}, or maybe a little high, and the
current vertical location in the Galaxy $D\sin b=663\,$pc is likewise
consistent with the MSP distribution \citep{Ng14}, suggesting a
vertical orbit that mimics the overall MSP population with a vertical
scale height of $<1\,$kpc.

\subsection{Constraints on $\dot G$ and dipole gravitational radiation} \label{subsec:J1012_alternative_gravity}
Since \citet{Lazaridis09} constrained alternative theories of gravity using timing analysis on \psrea, timing parameters including $\dot P_b^{\,\mathrm{obs}}$ (the observed time derivative of orbital period) have been updated by \citet{Desvignes16}. Despite the probable biased parallax presented by the 2016 work (as discussed in \ref{subsec:J1012_timing_interplay}), improved measurements for a number of other parameters are expected as a result of longer observation time, most notably (for our purposes) the orbital period derivative and proper motion. The $\dot P_b^{\,\mathrm{obs}}=(6.1\pm0.4)\times 10^{-14}$ reported by \citet{Desvignes16} is consistent with the 2009 counterpart $(5.0\pm1.4)\times 10^{-14}$, but is 3 times more precise. The accuracy of $D_{\dot P_b}$ reported by \citet{Desvignes16} is dominated by the precision of $\dot P_b^{\,\mathrm{obs}}$. 

Besides the improved timing precision, the companion mass $m_c$ and mass ratio $q$ of \psrea\ are also better constrained to $0.174\pm0.011$\,\msun\ \citep{Antoniadis16} and $10.44\pm0.11$ \citep{Mata-Sanchez20} respectively.
These, along with our new parallax, allows us to improve the constraints on the $\dot G$ (time derivative of Newton's gravitational constant) and dipole gravitational radiation. The method we adopt here has been developed by \citet{Damour88,Nordtvedt90,Damour91,Lazaridis09,Freire12,Zhu15}.

Contributions to the time variation of orbital period of \psrea\ can be summarized as 
\begin{equation}
\label{eq:J1012_Pbdot_budget}
{\dot P_b^{\,\mathrm{obs}}} = {\dot P_b^{\,\mathrm{Gal}}} + {\dot P_b^{\,\mathrm{Shk}}} + {\dot P_b^{\,\mathrm{GW}}} + {\dot P_b^{\,{\dot m}}} + {\dot P_b^{\,\mathrm{T}}} + {\dot P_b^{\,\mathrm{ex}}}\,,
\end{equation}
where $\dot P_b^{\,\mathrm{Gal}}$ and $\dot P_b^{\,\mathrm{Shk}}$ are not intrinsic to the \psrea\ binary, representing the effect of radial acceleration of \psrea\ induced by Galactic gravitational potential \citep{Damour91, Nice95} and transverse motion \citep{Shklovskii70}, respectively; $\dot P_b^{\,\mathrm{GW}}$, $\dot P_b^{\,{\dot m}}$ and $\dot P_b^{\,\mathrm{T}}$ are contributions intrinsic to the binary system resulting from gravitational-wave damping, mass loss of the binary \citep{Damour91,Freire12} and a deformed companion \citep{Smarr76,Freire12}, respectively; $\dot P_b^{\,\mathrm{ex}}$ stands for excess term of non-GR origins. As the non-intrinsic terms of $\dot P_b$ are dependent on distance and proper motion, we are able to refine $\dot P_b^{\,\mathrm{Gal}}$ and $\dot P_b^{\,\mathrm{Shk}}$ with our new astrometric results.

The Shklovskii term can be calculated with 
\begin{equation}
\label{eq:J1012_Shk}
\dot P_b^{\,\mathrm{Shk}} = \frac{(\mu_\alpha^2+\mu_\delta^2)D}{c} P_b\,,    
\end{equation}
where $\mu_{\alpha} \equiv \dot{\alpha} \cos\delta$, $D$ is the distance of \psrea\ from the Sun and $c$ is the speed of light. 
For the convenience of error propagation, the larger side of uncertainties for parallax and proper motion (\ref{tab:J1012_astrometric_fits}) are used as their symmetric uncertainties in the following calculation, i.e. $\varpi=1.21\pm0.08$\,mas, $\mu_{\alpha}=2.67\pm0.09$\,\maspy, $\mu_{\delta}=-25.40\pm0.14$\,\maspy.
Our new parallax and proper motion render $\dot P_b^{\,\mathrm{Shk}} = 68.6\pm4.4$\,\fsps, the uncertainty of which is 63\% of the counterpart in \citet{Lazaridis09}. In the same way as \citet{Zhu15} (and references therein), we updated $\dot P_b^{\,\mathrm{Gal}}=-5.5\pm0.2$\,\fsps\ with our new parallax-based distance to \psrea, taking the Sun-GC (Galactic Center) distance to be $R_0=8.122\pm0.031$\,kpc \citep{Gravity-Collaboration18} and circular speed of the local standard of rest to be $\Theta_0=233.3\pm1.4\,{\rm km~s^{-1}}$ \citep{McGaugh18}.
Combining the new mass ratio $q=10.44\pm0.11$ \citep{Mata-Sanchez20} into Equation~(21) from \citet{Lazaridis09}, we acquire $\dot P_b^{\,\mathrm{GW}} = -13\pm1$\,\fsps, where the uncertainty is dominated by the WD mass.

These updated contributions of $\dot P_b$, along with negligible $\dot P_b^{\,{\dot m}}$ and $\dot P_b^{\,\mathrm{T}}$ \citep{Lazaridis09}, give $\dot P_b^{\,\mathrm{ex}} = 10.6\pm6.1$\,\fsps, 2.6 times more precise than the counterpart in \citet{Lazaridis09}. The contributions to $\dot P_b^{\,\mathrm{ex}}$ as well as the derived $\dot P_b^{\,\mathrm{ex}}$ for our work and \citet{Lazaridis09} are summarized in \ref{tab:J1012_PbEx_contributions}.

\begin{table*}
	\centering
	\caption{Contributions to the excess orbital decay $\dot P_b^{\,\mathrm{ex}}$}
	\label{tab:J1012_PbEx_contributions}
	\resizebox{\textwidth}{!}{
	\begin{tabular}{cccccccc} 
		\hline
		\hline
		& $\dot P_b^{\,\mathrm{obs}}$ & $\dot P_b^{\,\mathrm{Shk}}$ & $\dot P_b^{\,\mathrm{Gal}}$ & $\dot P_b^{\,\mathrm{GW}}$ & $\dot P_b^{\,{\dot m}}$ & $\dot P_b^{\,\mathrm{T}}$ & $\dot P_b^{\,\mathrm{ex}}$ \\
		& \fsps\ & \fsps\ & \fsps\ & \fsps\ & \fsps\ & \fsps\ & \fsps\ \\
		\hline
	This work & 61(4) $^a$ & 68.6(4.4) & -5.5(2) & -13(1) & 0 & 0 & 10.6(6.1) \\
	\citet{Lazaridis09} & 50(14) & 70(7) & -5.6(2) & -11(2) & 0 & 0 & -4(16) \\
	\hline
	\end{tabular}
	}
	\tabnote{$^a$\citet{Desvignes16}}
\end{table*}

As already noted, some alternative theories of gravity demand dipole gravitational radiation and/or a time-dependence to Newton's gravitational constant $G$ in the local universe.
The new $\dot P_b^{\,\mathrm{ex}}$, consistent with zero at 1.7\,$\sigma$ confidence level, does not support alternative theories to GR. 
Nevertheless, we can make use of this measurement to set new limits to dipole gravitational radiation and $\dot G$ with
\begin{equation}
\label{eq:PbdotEx}
\dot P_b^{\,\mathrm{ex}} = \dot P_b^{\,\dot G} + \dot P_b^{\,\mathrm{dp}}\,,
\end{equation}
where $\dot P_b^{\,\dot G}$ and $\dot P_b^{\,\mathrm{dp}}$ represent orbital change caused by $\dot G$ and dipole gravitational radiation, respectively. The relation between $\dot G/G$ and $\dot P_b^{\,\dot G}$ is rewritten from \citet{Damour88,Nordtvedt90,Lazaridis09} as
\begin{equation}
\label{eq:J1012_PbdotGdot}
\frac{\dot P_b^{\,\dot G}}{P_b} = -2~\frac{\dot G}{G}\left[1-\left(1+\frac{1}{2}~\frac{1}{q+1}\right)s_p\right] \,,
\end{equation}
where $q=10.44\pm0.11$ \citep{Mata-Sanchez20} is the mass ratio between the pulsar and the companion, $s_p$ stands for the ``sensitivity" of the pulsar depending on its EoS, mass and the theory of gravity in concern \citep{Will93}.
The connection between $\dot P_b^{\,\mathrm{dp}}$ and $\kappa_D$, the putative coupling constant of dipole gravitational radiation, is reproduced from \citet{Lazaridis09} as
\begin{equation}
\label{eq:J1012_PbdotDipole}
P_b \dot P_b^{\,\mathrm{dp}} = -4\pi^2 T_\odot m_c~\frac{q}{q+1}~\kappa_D {s_p}^2 \,,
\end{equation}
where $T_\odot=G\msun/c^3=4.9255\,\mu$s, $m_c=0.174\pm0.011\,$\msun\ \citep{Antoniadis16} is the mass of the companion. 
Here we assume the higher-order terms of the ``sensitivities'' of the pulsar and the companion are negligible.

There are two ways to solve $\dot{G}/G$ and $\kappa_D$ from \ref{eq:PbdotEx}, \ref{eq:J1012_PbdotGdot}, \ref{eq:J1012_PbdotDipole}. They are both based on the universality of physical laws, i.e. $\dot G/G$ and $\kappa_D$ do not vary in the local universe. The first method is borrowing independent $\dot G/G$ or $\kappa_D$ from other measurements. The second one is using several well timed pulsars to solve or fit (when using more than 2 pulsars) $\dot G/G$ and $\kappa_D$ at the same time, introduced by \citet{Lazaridis09}.

To date, the most stringent limits on $\dot G/G$ are provided by lunar laser ranging (LLR), which yields $\dot G/G=(0.71\pm 1.52)\times 10^{-13}\,{\rm yr^{-1}}$ \citep[95\% confidence level][]{Hofmann18}, and modelling of the orbit of Mercury , which yields $|\dot G|/G=(4\pm 5)\times 10^{-14}\,{\rm yr^{-1}}$ \citep[95\% confidence level][]{Genova18}. The $\dot G/G$ from LLR can be translated into $\dot P_b^{\,\dot G}$, thus assisting us to gauge $\kappa_D$ separately. We use the LLR constraint in preference to that from \citet{Genova18} due to the ambiguity of the sign of the latter. In order to solve $\kappa_D$ in \ref{eq:J1012_PbdotDipole}, we assume $s_p=0.1(m_p/$\msun) (where $m_p=q m_c$), as proposed by \citet{Damour92a} and adopted by \citet{Lazaridis09,Zhu15}. We hence obtain $\dot P_b^{\,\dot G} = -0.19\pm0.20$\,\fsps\ and $\dot P_b^{\,\mathrm{dp}}=10.8\pm6.1$\,\fsps. The latter gives $\kappa_D=(-5.5\pm 6.6)\times 10^{-4}$ (95\% confidence level), which is 3.6 times as precise as the previous effort with \psrea\ made utilising the same approach \citep{Lazaridis09}. This estimate of $\kappa_D$ is, however, less precise than the $\kappa_D=(-0.8\pm1.6)\times10^{-4}$ (68\% confidence level) by \citet{Freire12} acquired with the same method while using PSR~J1738$+$0333.

As the second way to solve $\dot{G}/G$ and $\kappa_D$, we combined \psrea\ with PSR~J0437$-$4715, PSR~J1738$+$0333 and PSR~J1713$+$0747 to extract $\dot G/G$ and $\kappa_D$, following the method introduced by \citet{Lazaridis09}. The three other pulsars have been used to constrain $\dot G/G$ and $\kappa_D$ \citep{Verbiest08,Deller08,Freire12,Zhu19}. The parameters of the four pulsars we used to derive $\dot G/G$ and $\kappa_D$ are summarized in \ref{tab:J1012_4pulsar_parameters}. 
We approach $\dot{G}/G$ and $\kappa_D$ by least-square fitting, and their uncertainties by Monte-Carlo simulation. The marginalized $\dot G/G$ and $\kappa_D$ we obtain are
\begin{equation}
\label{eq:J1012_GdotG}
\dot G/G = -1.8^{\,+5.6}_{\,-4.7}\times 10^{-13}\,{\rm yr^{-1}}\,,
\end{equation}
\begin{equation}
\label{eq:J1012_kappa_D}
\kappa_D=(-1.7\pm1.7)\times 10^{-4}\
\end{equation}
at 68\% confidence, and 
\begin{equation}
\label{eq:J1012_GdotG_2sigma}
\dot G/G = -2^{\,+12}_{\,-21}\times 10^{-13}\,{\rm yr^{-1}}\,,
\end{equation}
\begin{equation}
\label{eq:J1012_kappa_D_2sigma}
\kappa_D=(-1.7^{\,+3.6}_{\,-3.4})\times 10^{-4}\
\end{equation}
at 95\% confidence.
These values are consistent with zero, and comparably precise to (and more conservative than) $\kappa_D=(-0.7\pm2.2)\times10^{-4}$ and $\dot{G}/G=(-1\pm9)\times10^{-13}\,{\rm yr^{-1}}$ measured at 95\% confidence by \citet{Zhu19}.
We note that we have adopted the relation $s_p=0.1(m_p/$\msun), and hence these two estimates are dependent on this assumed $s_p$ relation.
The mathematical formalism of $s_p$ hinges further on the EoS of NSs, which will be better constrained by NICER \citep{Bogdanov19,Bogdanov19a} and the gravitational-wave observatories \citep{Annala18} in the years to come.

\begin{table*}
	\centering
	\caption{Parameters of MSPs for estimating $\dot G/G$ and $\kappa_D$} 
	\label{tab:J1012_4pulsar_parameters}
	\resizebox{\textwidth}{!}{
	\begin{tabular}{ccccccc} % four columns, alignment for each
		\hline
		\hline
	pulsar & $P_b$ & $\dot P_b^{\,\mathrm{ex}}$ & $m_p$ & $m_c$ & $q$ & references  \\
		& d & \fsps\ & \msun\ & \msun\ &  & \\
		\hline
	PSR~J0437$-$4715 & 5.74 & 12(32) $^a$ & 1.44(7) & 0.224(7) & $-$ & \citet{Reardon16,Deller08} \\
	\psrea\ (this work) & 0.60 & 10.6(6.1) & $-$ & 0.174(11) & 10.44(11) & references in this paper \\
    PSR~J1713$+$0747 & 67.83 & 30(150) & 1.33(10) & 0.290(11) & $-$ & \citet{Zhu19} \\
    PSR~J1738$+$0333 & 0.35 & 2.0(3.7) & 1.46(6) & $-$ & 8.1(2) & \citet{Freire12} \\
	\hline
	\end{tabular}
	}
	\tabnote{$^a$We derived $\dot P_b^{\,\mathrm{ex}}$ for PSR~J0437$-$4715 with the results from \citet{Reardon16,Deller08}.}
\end{table*}

\section{Conclusions and future prospects}
\label{sec:J1012_conclusions}
\begin{enumerate}
\item This paper reports new VLBI astrometry of \psrea\ (\ref{tab:J1012_astrometric_fits}). Our new distance to \psrea, $0.83^{+0.06}_{-0.02}$\,kpc, is the most precise to date and consistent with major measurements. We present the first VLBI-based absolute position for \psrea, which paves the road for the frame link between the quasi-static International Celestial Reference Frame used by VLBI and the solar-system frame used by pulsar timing.

\item Using our new distance and proper motion, we reduce the uncertainty of the Shklovskii term in \ref{eq:J1012_Pbdot_budget}. On top of that, we set new constraints on the fractional time derivative of the Newton's gravitational constant $\dot G/G$ for the local universe and the coupling constant for dipolar gravitational radiation $\kappa_D$, combining three other millisecond pulsars, PSR~J0437$-$4715, PSR~J1738$+$0333 and PSR~J1713$+$0747. The new $\kappa_D$ is comparable to the most stringent constraint.

\item As is shown in \ref{tab:J1012_4pulsar_parameters}, among the four pulsars, the $\dot{P_b}^{\mathrm{ex}}$ of \psrea\ stands out with $>1\,\sigma$ offset from zero, which effectively brings the best-fit $\kappa_D$ away from zero. If we only use the other three pulsars and re-do the analysis, we obtain $\kappa_D=(-0.8^{\,+1.9}_{\,-1.7})\times10^{-4}$, where the uncertainty increases but the best-fit $\kappa_D$ becomes more consistent with zero. Therefore, whether $\dot{P_b}^{\mathrm{ex}}$ of \psrea\ will converge to zero with future timing and VLBI observations of \psrea\ is essential for the $\kappa_D$ test using pulsar-WD binaries. 
Assuming GR is correct, the 1.7\,$\sigma$ offset of $\dot P_b^{\,\mathrm{ex}}$ from zero implies
that the value of either our VLBI parallax and/or $\dot{P_b}^{\mathrm{obs}}$ is too high (given that the other contributing terms of $\dot{P_b}$ vary marginally).
This mild tension will be re-visited with a better $\dot{P_b}^{\mathrm{obs}}$ based on longer timing observations. 
As noted in \ref{subsec:J1012_timing_interplay}, we see that applying constraints to the proper motion of \psrea\ based on pulsar timing act to reduce the VLBI parallax, albeit within the current uncertainties, which would already mitigate the tension somewhat.
It is likely that the timing proper motions of \psrea\ acquired independently with EPTA and NANOGrav will converge to a value with negligible uncertainty in the future, allowing us to confidently apply this as a prior to our VLBI fit and improve our parallax estimate for \psrea. 
Furthermore, when the uncertainties of both $\dot P_b^{\,\mathrm{obs}}$ and the distance to \psrea\ are improved by a factor of 4 with new timing and VLBI observations, we are able to estimate the next uncertainty contributor to $\dot P_b^{\,\mathrm{ex}}$ - the WD mass (as well as the pulsar mass), assuming $\dot P_b^{\,\mathrm{ex}}=0$. This independent WD mass will help refine the relation between WD mass and WD atmospheric parameters in the helium-WD regime.

\item Looking into the future, the uncertainty of $\dot P_b^{\,\mathrm{obs}}$ for \psrea\ will quickly vanish \citep[as $t^{\,-2.5}$,][]{Bell96}, reducing the uncertainty of $\dot P_b^{\,\mathrm{ex}}$ to $\approx4.6$\,\fsps\ in $\approx10$ years. At that time, the Shklovskii term would become the leading error source of $\dot P_b^{\,\mathrm{ex}}$ for \psrea.
Inside the Shklovskii term, the distance (or parallax) dominates the error budget as the uncertainty of parallax improves as $t^{\,-0.5}$. That means distance uncertainty will eventually become the biggest barrier against better constraints on alternative theories of gravity. This will be the same for the analysis of most\footnote{For some systems, especially PSR~J1738$+$0333, the relatively low precision on the mass measurements - not the uncertainty on the distance - will be the main barrier towards improvement of the test.} other MSPs.
If we reduce the uncertainty of $\dot P_b^{\,\mathrm{obs}}$ to zero for each of the four above-mentioned pulsars (\psrea, PSR~J0437$-$4715, PSR~J1738$+$0333 and PSR~J1713$+$0747) and re-derive $\dot{G}/G$ and $\kappa_D$, we find the $1\,\sigma$ uncertainty reduces to $\leq1.5\times10^{-13}\,{\rm yr^{-1}}$ for $\dot{G}/G$ and $\leq1.0\times10^{-4}$ for $\kappa_D$. This simulation shows significantly better constraints on $\dot{G}/G$ and $\kappa_D$ can be made with continuous efforts on pulsar timing for $\approx10$ years. Beyond that, in order to further this study, we need to focus on improving the precision of distances to the pulsars of use.
\end{enumerate}

\section*{Acknowledgements}
HD is supported by the ACAMAR (Australia-ChinA ConsortiuM for Astrophysical Research) scholarship, which is partly funded by the China Scholarship Council (CSC). DLK and TJWL were supported by the NANOGrav Physics Frontiers Center, which is supported by the National Science Foundation award 1430284. Part of this research was carried out at the Jet Propulsion Laboratory, California Institute of Technology, under a contract with the National Aeronautics and Space Administration. 
The authors thank Norbert Wex and the anonymous referee for their important comments on this paper.
The data of this work come from the Very Long Baseline Array (VLBA), which is operated by the National Radio Astronomy Observatory (NRAO). The NRAO is a facility of the National Science Foundation operated under cooperative agreement by Associated Universities, Inc.
This work made use of the Swinburne University of Technology software correlator, developed as part of the Australian Major National Research Facilities Programme and operated under license.

\section*{Softwares}
DiFX \citep{Deller11a}, ParselTongue \citep{Kettenis06},
AIPS \citep{Greisen03},
DIFMAP \citep{Shepherd94},
psrvlbireduce (\url{https://github.com/dingswin/psrvlbireduce}),
pmpar (\url{https://github.com/walterfb/pmpar})

\newpage
\bibliographystyle{mnras}
\bibliography{haoding}



%\end{document}

% End of file `sample62.tex'.

\chapter[The Orbital-decay Test of General Relativity with VLBI
Astrometry of the Second Discovered Double Neutron Star]{The Orbital-decay Test of General Relativity \\with VLBI
Astrometry of the Second\\ Discovered Double Neutron Star}
\label{ch:B1534}.

This chapter is adapted from \citet{Ding21a} entitled ``The Orbital-decay Test of General Relativity to the 2\% Level with 6 yr VLBA
Astrometry of the Double Neutron Star PSR J1537+1155'', which is the third \mspsrpi\ paper (following \citealp{Vigeland18,Ding20}), and the only \mspsrpi\ paper that tests general relativity in the strong field regime.
This chapter marks the first time that the Bayesian astrometric inference package {\tt sterne} was used in this thesis. However, the Bayesian inference does not yet include the systematics-correcting parameter $\eta_\mathrm{EFAC}$ to be introduced in \ref{ch:mspsrpi}.

As is pointed out in \ref{ch:mspsrpi}, the Galactic-potential-induced orbital-decay term $\dot{P}_\mathrm{b}^\mathrm{Gal}$ (see \ref{sec:B1534_Pbdot_test}) is not correctly estimated due to a coding error. As a result, the ratio $\dot{P}_\mathrm{b}^\mathrm{int}/\dot{P}_\mathrm{b}^\mathrm{GW}$ between the observed orbital decay and the theoretical prediction is changed by a small but important fraction (1.8\%).
The corrected values of $\dot{P}_\mathrm{b}^\mathrm{Gal}$ and $\dot{P}_\mathrm{b}^\mathrm{int}/\dot{P}_\mathrm{b}^\mathrm{GW}$ are presented in \ref{ch:mspsrpi}.

\section{Abstract}

\psrfb, also known as \psrfbb, is the second discovered double neutron star (DNS) binary. More than 20 years of timing observations of \psrfb\ have offered some of the most precise tests of general relativity (GR) in the strong-field regime. 
As one of these tests, the gravitational-wave emission predicted by GR has been probed with the significant orbital decay ($\dot{P}_\mathrm{b}$) of \psrfb.
However, compared to most GR tests provided with the post-Keplerian parameters, the orbital-decay test was lagging behind in terms of both precision and consistency with GR, limited by the uncertain distance of \psrfb. 
With an astrometric campaign spanning 6 years using the Very Long Baseline Array, we measured an annual geometric parallax of $1.063\pm0.075$\,mas for \psrfb, corresponding to a distance of $0.94^{+0.07}_{-0.06}$\,kpc. This is the most tightly-constrained model-independent distance achieved for a DNS to date. After obtaining $\dot{P}_\mathrm{b}^\mathrm{Gal}$ (i.e., the orbital decay caused by Galactic gravitational potential) with a combination of 4 Galactic mass distribution models, we updated the ratio of the observed intrinsic orbital decay to the GR prediction to $0.977\pm0.020$, three times more precise than the previous orbital-decay test ($0.91\pm0.06$) made with \psrfb.

\section{Introduction}
\label{sec:B1534_intro}

\subsection{Pulsars in double neutron star systems}
\label{subsec:B1534_DNS_pulsars}
Double neutron stars (DNSs) are prized testbeds on which to evaluate theories of gravity and to probe the composition of neutron stars (NSs).
The DNS merger event GW170817 has been recorded both by gravitational-wave (GW) observatories and electromagnetically \citep[e.g.][]{Abbott17,Abbott17a,Goldstein17,Mooley18}, providing constraints on the interior composition of NSs \citep[e.g.][]{Annala18}. 
The same merger event also strengthens the belief that short Gamma Ray Bursts (SGRBs) are generated by DNS mergers \citep[e.g.][]{Coward12}, though most SGRBs are well beyond the horizon of the current ground-based GW detectors.
In addition, DNS mergers are considered the prime sources of \textit{r}-process elements \citep{Eichler89,Korobkin12,Drout17}. 
To test the connection between DNS mergers and the observed abundance of \textit{r}-process elements in the local universe, an estimate of the DNS merger rate is required, which can be constrained with observations of the Galactic DNS population \citep[e.g.][]{Kim15,Pol19}.

During their steady inspiral stage, DNS systems can be studied by measuring and modeling the pulse time-of-arrivals (ToAs) from a pulsar residing in a DNS system (hereafter referred to as a ``DNS pulsar"). 
So far, 16 known DNS pulsars and 3 suspected ones have been discovered from pulsar surveys (see Table~1 of \citealp{Andrews19}), including two found in globular clusters. 
Though in shallower gravitational potentials compared to DNS mergers, DNS pulsars provide some of the most precise tests on gravitational theories in the strong-field regime with long-term timing observations \citep[e.g.][]{Stairs03,Kramer06,Deller09}.
Gravitational theories are tested with DNS pulsars by comparing observed post-Keplerian (PK) parameters, which quantify effects beyond a simple Keplerian model of motion, to the predictions of a specific gravitational theory, e.g., the general theory of relativity (GR). However, the theory-dependent prediction of each PK parameters relies on the masses of the two DNS constituents. Therefore, one needs at least three PK measurements to test a gravitational theory, as two of them have to be used to determine the two DNS constituent masses (based on the theory to be tested).
The PK parameters include (but are not limited to) $\dot{P}_\mathrm{b}$, $\dot{\omega}$, $\gamma$, $r$ and $s$, which stand for, respectively, the orbital decay, the advance of periastron longitude, the Doppler coefficient, the ``range'' of the Shapiro delay effect and the ``shape'' of the Shapiro delay effect. 
To date, the best test of GR was provided with the double pulsar system PSR~J0737$-$3039A/B \citep{Kramer06}, thanks to the extra independent mass ratio constraint (unavailable for other DNSs).

\subsection{PSR~J1537$+$1155}
\label{subsec:B1534_J1537}
\psrfb\ (also known as \psrfbb, hereafter referred to as J1537) is the second DNS system discovered \citep{Wolszczan91} in a 10.1-hr orbit.
J1537 shows an exceptionally high proper motion among DNS pulsars (see Table~3 of \citealp{Tauris17}), which has been explained with an unusually large kick of 175--300\,\kmps\ received from the second supernova (in the evolution history of J1537) \citep{Tauris17}. 
Based on the timing observations of J1537, the combined $\dot{\omega}$ -- $\gamma$ -- $s$ test returned consistency with GR at the 0.17\% level \citep{Fonseca14}. However, its observed intrinsic $\dot{P}_\mathrm{b}$ deviated from GR prediction, a result which was thought to be due partly or mostly to the poorly constrained distance to the pulsar \citep{Stairs02,Fonseca14}. 
Furthermore, due to the exceptionally high proper motion, the large uncertainty in the distance to J1537 has become the primary limiting factor of the $\dot{P}_\mathrm{b}$ test (\citealp{Stairs98,Stairs02,Fonseca14}, also explained in \ref{sec:B1534_Pbdot_test}).

The hitherto most precise distance to J1537 is $1.051\pm0.005$\,kpc \citep{Fonseca14}, obtained by solving for the distance that matches the orbital period derivative observed with pulsar timing, assuming the correctness of GR \citep{Bell96}. However, such ``timing kinematic distances'' (which, in case of confusion, are conceptually different from the distances derived with radial velocities and a Galactic rotation model, e.g. \citealp{Kuchar94,Wenger18}) cannot be used to test theories of gravity, as GR has been assumed to be correct. 
To carry out the $\dot{P}_\mathrm{b}$ test of GR with J1537, one has to have an independent measurement of its distance \citep{Stairs02} in order to correct the distance-dependent terms from the observed orbital decay. 
Prior to this work, the best independent distance for J1537 has been based on its dispersion measurement (DM) along with a model of the distribution of Galactic free-electron density $n_e$, i.e., $0.7\pm0.2$\,kpc with the TC93 model \citep{Taylor93}. However, there are significant downsides with employing DM-based distances for this purpose.
While generally reliable for the population as a whole, DM-based distances can be inaccurate for individual sources \citep[e.g.][]{Deller09a}. This inaccuracy is more likely for sources at high Galactic latitudes $b$, such as J1537 at $b=48\degr$, due to sparser pulsars (that allow DM measurements) in those directions. Moreover, the two more recent $n_e$ models (NE2001 and YMW16, \citealp{Cordes02,Yao17}) have been built using timing-derived distance of J1537, meaning the DM distance of J1537 is no longer independent for these two $n_e$ models.

Compared to the aforementioned ways to measure the distance to J1537, geometric measurements of the distance to J1537 (based on the change in angle or relative distance to the source as the Earth orbits the Sun) offer the ability to measure the source distance to higher precision and free of model dependency.
Such geometric measurements can be realized with global fitting from pulsar timing or VLBI (very long baseline interferometry) observations in the radio band. Based on pulsar timing, the (timing) parallax of J1537 was measured to be $0.86\pm0.18$\,mas \citep{Fonseca14}.
However, as is pointed out in \citet{Fonseca14}, precise determination of timing parallax is often hampered by the covariance between parallax and DM; the stochastic variations in the latter introduced by the changing sightline between the pulsar and Earth can corrupt the timing parallax. Therefore, VLBI astrometry remains the best way to obtain the most precise model-independent geometric distance to J1537. 

In this letter, we present the astrometric results of J1537 obtained with VLBI observations spanning 6 years. Based on the new distance, we strengthen the $\dot{P}_\mathrm{b}$ test of GR with J1537.
Throughout this letter, the uncertainties are provided at 68\% confidence level unless otherwise stated.

\section{Observations and data reduction}
\label{sec:B1534_observations}
As part of the \mspsrpi\ program \citep[e.g.][]{Ding20}, J1537 was first observed with the \textit{Very Long Baseline Array} (VLBA) at around 1.5\,GHz from July 2015 to July 2017, which include 2 2-hr pilot observations under the project code BD179 and 9 1-hr observations under the project code BD192. The astrometric campaign was extended with 6 2-hr VLBA observations between August 2020 and July 2021 under the project code BD229. 
The observation and correlation strategy is identical to that of the \psrpi\ program \citep{Deller19}. ICRF~J150424.9$+$102939 and ICRF~J154049.4$+$144745 were observed as the band-pass calibrator and the primary phase calibrator, respectively.
FIRST~J153746.2$+$114215, 16\farcm3 away from J1537, has been identified and adopted as the secondary phase calibrator. 
At correlation of each observation, pulsar gating, based on pulse ephemerides of J1537 monitored with our timing observations, was applied to increase the S/N of detection.

All correlated data were reduced with the {\tt psrvlbireduce} (\url{https://github.com/dingswin/psrvlbireduce}) pipeline written in {\tt ParselTongue} \citep{Kettenis06}, which bridges python users to the two data-reduction packages {\tt AIPS} \citep{Greisen03} and {\tt DIFMAP} \citep{Shepherd94}. The final image-plane models of the two phase calibrators used for the data reduction can be found at \url{https://github.com/dingswin/calibrator_models_for_astrometry}.

The turbulent ionised interstellar medium between the Earth and J1537 leads to diffractive and refractive interstellar scintillation, which can increase or decrease the pulsar flux density \citep{Stairs02}. In four of the 17 VLBA epochs, scintillation reduced the brightness of J1537 below the detection threshold.
For each of the remaining 13 epochs of detection, the pulsar position and its statistical (or random) uncertainty was obtained by fitting an elliptical gaussian to the deconvolved pulsar image. The acquired pulsar positions are provided in \ref{tab:B1534_pulsar_positions}.

\begin{table*}
    \raggedright
    \caption{\psrfb\ positions in reference to  FIRST~J153746.2$+$114215}
     
        %\begin{tabular}{p{0.15\linewidth}
        %                p{0.2\linewidth}
        %                p{0.1\linewidth}
        %                p{0.30\linewidth}}
        \resizebox{\textwidth}{!}{
    	\begin{tabular}{cccccc} % four columns, alignment for each
		\hline
	project & obs. date & $\alpha_\mathrm{J2000}$ (RA.)  & $\delta_\mathrm{J2000}$ (Decl.)  & $(S/N)_\mathrm{J1537}$ $^\mathrm{b}$ & $(S/N)_\mathrm{SC}$ $^\mathrm{b}$ \\
	code & (yr) & & & &  \\
		\hline
	bd179f0	& 2015.5153 & $15^{\rm h}37^{\rm m}09\fs 96320(1)$ & $11\degr55'55\farcs0800(3)$ & 21.9 & 270.0 \\
	bd179f1 & 2015.5699 & $15^{\rm h}37^{\rm m}09\fs 96319(1)$ & $11\degr55'55\farcs0787(5)$ & 14.4 & 245.7 \\
	bd192f0 & 2016.5875 &  $15^{\rm h}37^{\rm m}09\fs 96327(2)$ & $11\degr55'55\farcs0516(7)$ & 9.6 & 125.9 \\
	bd192f3 & 2017.1111 & $15^{\rm h}37^{\rm m}09\fs 96349(4)$ & $11\degr55'55\farcs0397(13)$  & 7.5 & 100.0 \\
	bd192f4 & 2017.1820 & $15^{\rm h}37^{\rm m}09\fs 96353(3)$ & $11\degr55'55\farcs0365(10)$  & 6.8 & 199.4 \\
	bd192f5 & 2017.2421 & $15^{\rm h}37^{\rm m}09\fs 96348(3)$ & $11\degr55'55\farcs0372(10)$  & 5.5 & 276.7 \\
	bd192f8 & 2017.5755 & $15^{\rm h}37^{\rm m}09\fs 96341(2)$ & $11\degr55'55\farcs0296(9)$  & 5.3 & 293.3 \\
	bd229a & 2020.6611 & $15^{\rm h}37^{\rm m}09\fs 96370(3)$ & $11\degr55'54\farcs9495(9)$  & 5.3 & 239.6 \\
	bd229b & 2020.6693 & $15^{\rm h}37^{\rm m}09\fs 96373(2)$ & $11\degr55'54\farcs9489(7)$  & 7.5 & 230.1 \\
	bd229c & 2021.0564 & $15^{\rm h}37^{\rm m}09\fs 96388(1)$ & $11\degr55'54\farcs9391(4)$  & 19.2 & 160.1 \\
	bd229d & 2021.0646 & $15^{\rm h}37^{\rm m}09\fs 96390(1)$ & $11\degr55'54\farcs9388(5)$  & 25.1 & 174.5 \\
	bd229e & 2021.4935 & $15^{\rm h}37^{\rm m}09\fs 96381(2)$ & $11\degr55'54\farcs9279(8)$  & 7.3 & 132.3 \\
	bd229f & 2021.5019 & $15^{\rm h}37^{\rm m}09\fs 96383(2)$ & $11\degr55'54\farcs9282(7)$  & 8.1 & 123.7 \\
	\hline
	\end{tabular}
    }
    
    \tabnote{$^a$In this table, the positional uncertainties have included both random and systematic errors (see \ref{sec:B1534_results}). This table is available at \url{https://github.com/dingswin/publication_related_materials}, where the random errors for the positions can also be found.}
    \tabnote{$^b$ $(S/N)_\mathrm{J1537}$ and $(S/N)_\mathrm{SC}$ stand for the image S/N of (gated) J1537 and that of the secondary phase calibrator FIRST~J153746.2$+$114215, respectively.}
    
    \label{tab:B1534_pulsar_positions}
\end{table*}

\section{Astrometric results}
\label{sec:B1534_results}
Upon obtaining the 13 pulsar positions, we proceeded to estimate their systematic errors. 
This is because small residual calibration errors remain, even though direction-dependent calibration terms (of systematic errors) have been mitigated by the use of a close in-beam calibrator. 
We used the empirically derived expression from \citet{Deller19} to approach the systematic errors. For each epoch, the estimated systematic error was subsequently added in quadrature to the random error of the position. 
The positional uncertainties, including random and systematic errors, can be found in \ref{tab:B1534_pulsar_positions} alongside the pulsar positions. To make it easier for other researchers to reproduce the error budget, the image S/N for both J1537 and the secondary phase calibrator are also presented in \ref{tab:B1534_pulsar_positions}.
For the pulsar positions, the nominal systematic errors are around 0.14\,mas and 0.33\,mas in the right ascension (RA) and declination direction, respectively; in comparison, the median random errors are roughly twice the nominal systematic errors due to the faintness of J1537.

Based on the 13 pulsar positions and their
associated positional uncertainties (including the systematic errors described above),
we derived the astrometric results in three different methods: direct least-square fitting, bootstrap and Bayesian inference. 
Direct fitting was performed using {\tt pmpar} (\url{https://github.com/walterfb/pmpar}). A bootstrap was implemented as described in Section~3.1 of \citet{Ding20}. 
Compared to direct fitting and bootstrap, Bayesian analysis offers a simpler means to incorporate prior astrometric information (obtained elsewhere), and to infer extra orbital parameters (e.g., the longitude of ascending node and inclination angle) when positional precision allows \citep[e.g.][]{Deller13}.
We carried out Bayesian inference with {\tt sterne} (aStromeTry bayEsian infeReNcE, \url{https://github.com/dingswin/sterne}).
For the Bayesian inference, we assumed timing proper motion and parallax (reported in \citealp{Fonseca14}) follow Gaussian distributions, and used them as prior distributions for proper motion and parallax; the negligible (at the $5\,\mu$as level) reflex motion of J1537 (i.e., sky-position shifts due to the orbital motion) was not fitted.  
For both bootstrap and Bayesian analysis, we adopted the median value (of the marginalized sample) for an astrometric parameter as the estimate, and used the 16th and 84th percentiles to mark the $1\,\sigma$ uncertainty interval.

\begin{figure}
    \centering
	\includegraphics[width=13cm]{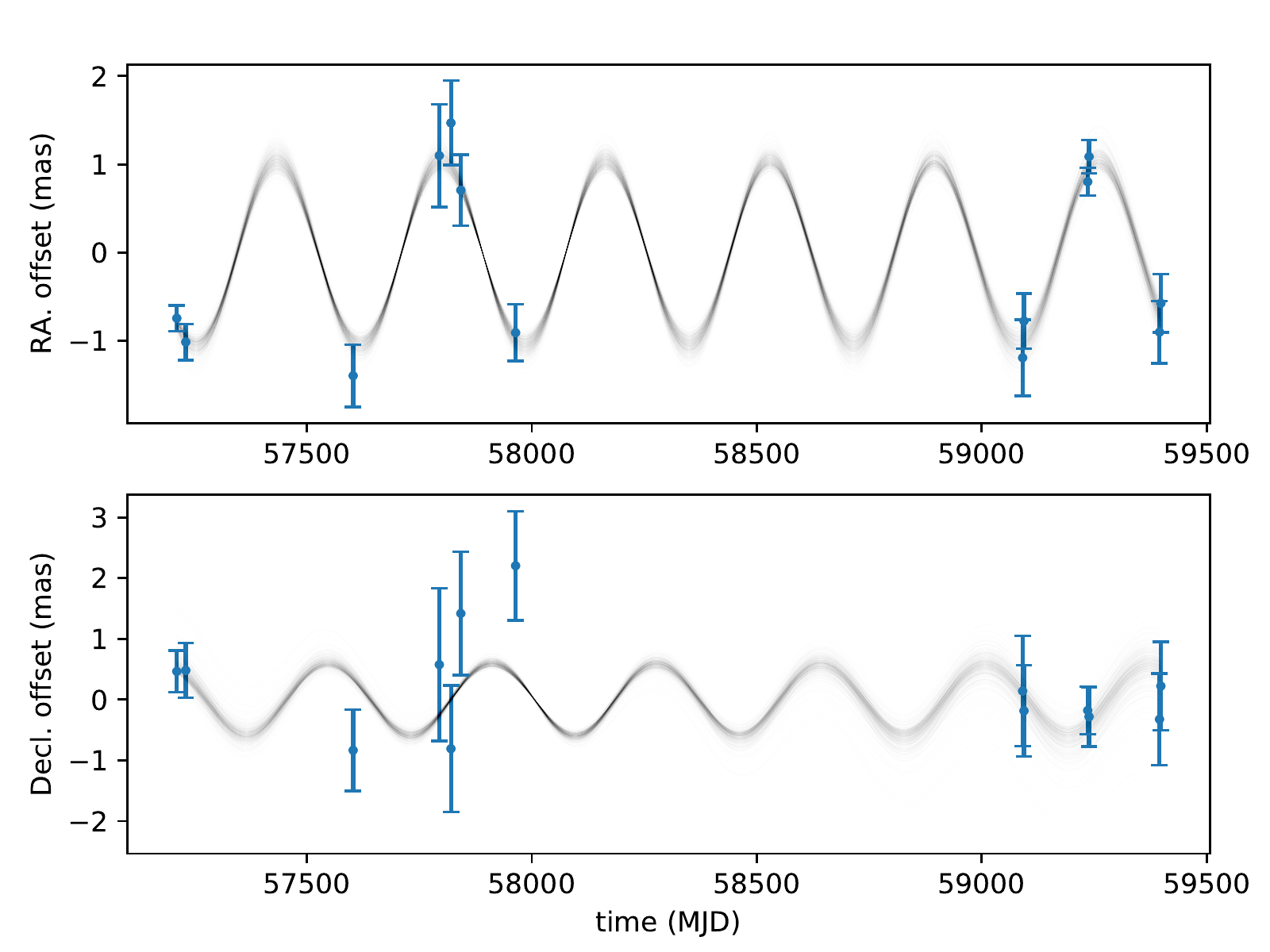}
    \caption{Parallax signature revealed by the \psrfb\ positions. Each quasi-sinusoidal curve represents the fitted model for a bootstrap run, after removing the best fit reference position and proper motion.}
    \label{fig:B1534_astrometric_model}
\end{figure}

\begin{figure*}
    \centering
	\includegraphics[width=14cm]{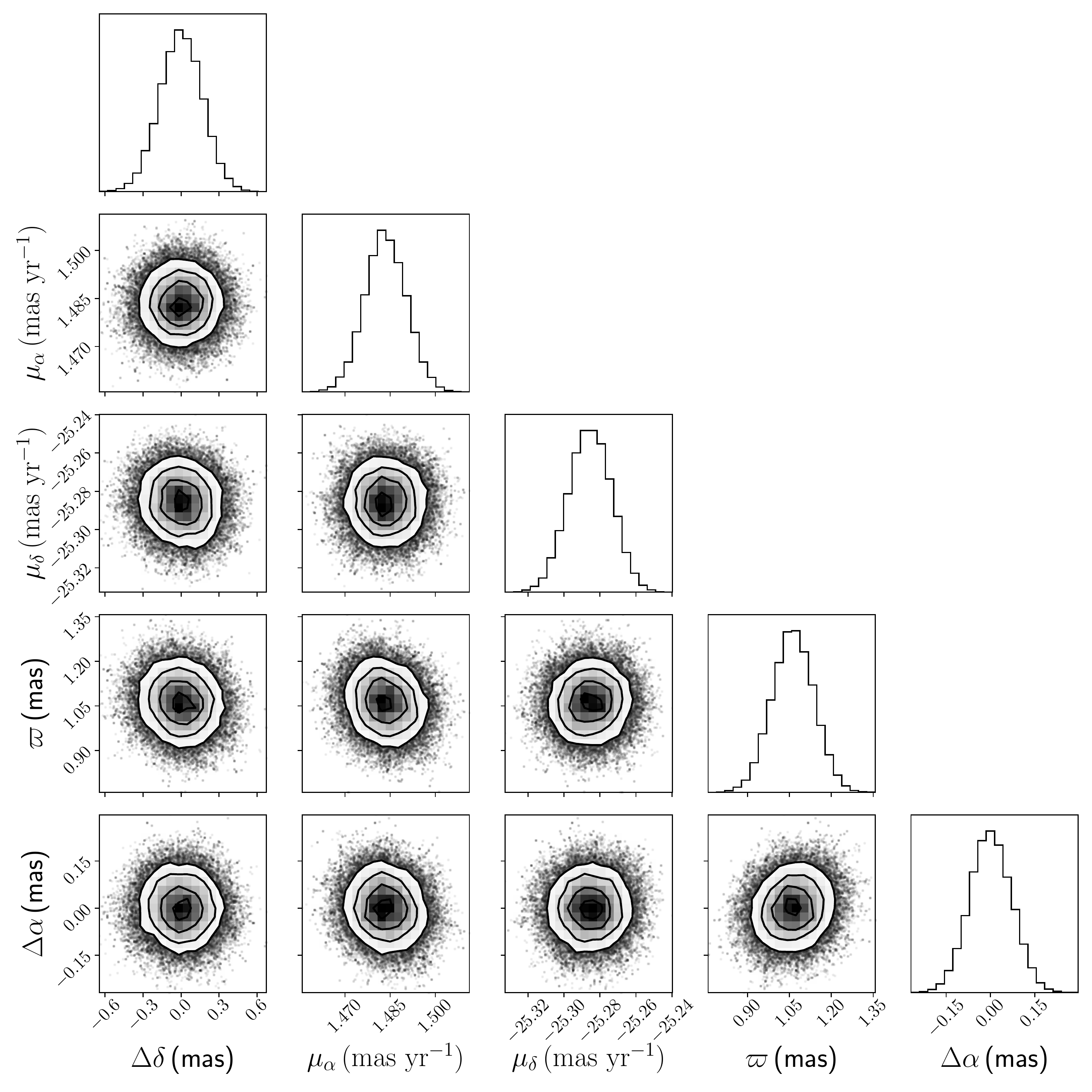}
    \caption{Error ``ellipses'' and the marginalized histograms for the posterior samples of the 5 astrometric parameters generated in the Bayesian analysis (see \ref{sec:B1534_results}). 
    The reference position offset is relative to the median reference position provided in \ref{tab:B1534_mu_and_pi}.}
    \label{fig:B1534_bayesian_cornerplot}
\end{figure*}

The astrometric results acquired with the three methods, as well as the three parallax-based distances, are summarized in \ref{tab:B1534_mu_and_pi}. For comparison, the distances based on dispersion measure (DM) and pulsar timing are reproduced in \ref{tab:B1534_mu_and_pi}. 
Here, the timing distance is quoted from the timing kinematic distance reported in \citet{Fonseca14}, which is derived from the orbital decay of J1537 by assuming GR is correct. In addition, the parallax signature, revealed by the 13 pulsar positions, is shown in \ref{fig:B1534_astrometric_model}. \ref{fig:B1534_bayesian_cornerplot} presents the posterior samples simulated with MCMC in the Bayesian analysis, which suggests negligible correlation in most (7 out of 10) pairs of astrometric parameters. However, small correlation is found between the three astrometric parameters having RA component, i.e., the reference RA $\alpha_\mathrm{J2000}$, the RA proper motion component $\mu_\alpha$ and the parallax $\varpi$. The largest correlation coefficient $|\rho|$ is 0.16 between $\varpi$ and $\alpha_\mathrm{J2000}$, while $|\rho|=0.14$ between $\mu_\alpha$ and $\varpi$.

According to \ref{tab:B1534_mu_and_pi}, the astrometric results obtained with the three methods agree with each other; the new model-independent distances are generally consistent with the DM-based distance and the timing kinematic distance. The consistency between the new model-independent distances and the timing kinematic distance will be further improved in \ref{sec:B1534_Pbdot_test} after updating the timing kinematic distance. The small reduced chi-square \rcs\ of 0.81 (for the method of direct fitting) implies that systematic errors for the 13 pulsar positions may have been slightly over-estimated. When applying the timing proper motion and parallax as prior information in the Bayesian analysis, \rcs\ only rises a little to 0.84, which indicates the timing proper motion and parallax \citep{Fonseca14} are consistent with the VLBI data.
Given a chi-square of $\sim17$ for 21 degrees of freedom, we did not see sufficient evidence to revise our estimated systematic uncertainties (reducing the estimated systematic uncertainty would bring the \rcs\ closer to unity and increase the parallax significance).

In the following discussion, we adopt the astrometric results derived with Bayesian analysis, which incorporates the VLBI and timing measurements. For those who want to use VLBI-only results (such as pulsar timers of J1537), we recommend the bootstrap results in \ref{tab:B1534_mu_and_pi}, as bootstrap can potentially correct improper error estimations to an appropriate level (see \citealp{Ding20c} as a good example), especially when the number of measurements is relatively large ($\gtrsim10$ for VLBI astrometry).

\begin{table*}
    \raggedright
    \caption{Reference position, proper motion and parallax measurements of \psrfb\ at the reference epoch MJD~57964}
     
        %\begin{tabular}{p{0.15\linewidth}
        %                p{0.2\linewidth}
        %                p{0.1\linewidth}
        %                p{0.30\linewidth}}
        \resizebox{\textwidth}{!}{
    	\begin{tabular}{cccccccc} % four columns, alignment for each
		\hline
	method & $\alpha_\mathrm{J2000}$ (RA.) $^\mathrm{a}$ & $\delta_\mathrm{J2000}$ (Decl.) $^\mathrm{a}$ & $\mu_\alpha \equiv \dot{\alpha}\cos{\delta}$ & $\mu_\delta$ & $\varpi$ & \rcs\ & $D$  \\
	   & & & (\maspy) & (\maspy) & (mas) & & (kpc)  \\
		\hline
	direct fitting	& $15^{\rm h}37^{\rm m}09\fs 963467(4)$ & $11\degr55'55\farcs0274(1)$ & $1.51\pm0.02$ & $-25.31\pm0.05$ & $1.06\pm0.07$ & 0.81 & $0.94^{\,+0.07}_{\,-0.06}$ \\
	bootstrap & $15^{\rm h}37^{\rm m}09\fs 963467(4)$ & $11\degr55'55\farcs0274(2)$ & $1.51\pm0.02$ & $-25.31^{\,+0.04}_{\,-0.05}$ & $1.07^{\,+0.09}_{\,-0.08}$ & --- & $0.93\pm0.07$  \\
	Bayesian inference $^\mathrm{b}$ & $15^{\rm h}37^{\rm m}09\fs 963469(5)$ & $11\degr55'55\farcs0274(2)$ & $1.483\pm0.007$ & $-25.29\pm0.01$ & $1.063\pm0.075$ & 0.84 & $0.94^{\,+0.07}_{\,-0.06}$  \\
	\\
	dispersion measure & --- & --- & --- & --- & --- & --- & $0.7\pm0.2$ $^\mathrm{c}$  \\
	pulsar timing & --- & --- & $1.482\pm0.007$  & $-25.29\pm0.01$ & $0.86\pm0.18$ & --- & $1.051\pm0.005$ $^\mathrm{d}$ \\
	\hline
	\end{tabular}
    }
    
    \tabnote{$^a$All reference positions in this table only indicate the relative positions with respect to the second phrase calibrator. Accordingly, the reference position errors do not take into account the position errors of the main and second phrase calibrators.}
    \tabnote{$^b$In the Bayesian analysis, we adopted the timing proper motion and parallax \citep{Fonseca14} as priors (assuming Gaussian distribution).}
    \tabnote{$^c$\citet{Taylor93}. For the two newer Galactic free-electron distribution models \citep{Cordes02,Yao17}, timing-derived distances of J1537 have been incorporated into their establishment, thus becoming correlated to the associated DM-based distances.}
    \tabnote{$^d$The timing results are reported in \citet{Fonseca14}; here, the quoted distance is the timing kinematic distance derived with the assumption that GR is correct (instead of with the timing parallax). We note that this timing kinematic distance is inferred with $\dot{P}_\mathrm{b}^{\,\mathrm{Gal}}=-3.5$\,\fsps\ (see \ref{tab:B1534_Pbdot_Gal} and \ref{sec:B1534_Pbdot_test} for more details) based on the Galactic mass distribution model by \citet{Nice95}.}
    
    \label{tab:B1534_mu_and_pi}
    \end{table*}

\section{Testing GR with the orbital decay of \psrfb}
\label{sec:B1534_Pbdot_test}
The observed orbital decay $\dot{P}_\mathrm{b}^\mathrm{obs}$ (or the observed time derivative of the orbital period) of J1537 has been estimated to be $-136.6\pm0.3$\,\fsps\ \citep{Fonseca14} from a global fit of the timing model, which can be attributed to
\begin{equation}
\label{eq:B1534_Pbdot_budget}
{\dot{P}_\mathrm{b}^{\,\mathrm{obs}}} = {\dot{P}_\mathrm{b}^{\,\mathrm{Gal}}} + {\dot{P}_\mathrm{b}^{\,\mathrm{Shk}}} + {\dot{P}_\mathrm{b}^{\,\mathrm{GW}}},
\end{equation}
where $\dot{P}_\mathrm{b}^{\,\mathrm{Gal}}$ and $\dot{P}_\mathrm{b}^{\,\mathrm{Shk}}$ stand for the extrinsic orbital decay due to the apparent effect of radial acceleration caused, respectively, by Galactic gravitational potential \citep{Damour91, Nice95} and by transverse motion \citep{Shklovskii70}; $\dot P_\mathrm{b}^{\,\mathrm{GW}}$ represents the intrinsic orbital decay as a result of the GW emissions from the inspiraling DNS. The estimation of the two extrinsic orbital-decay terms rely on the distance to J1537, while $\dot{P}_\mathrm{b}^{\,\mathrm{Shk}}$ also depends on the proper motion. 
On the other hand, the GR-based $\dot P_\mathrm{b}^{\,\mathrm{GW}}$ can be calculated, provided the orbital period $P_\mathrm{b}$, the orbital eccentricity $e$ and the masses of the two DNS constituents \citep{Peters63,Weisberg16}, all of which have been precisely determined with pulsar timing \citep{Fonseca14}. Hence, one can test GR by comparing the observed intrinsic orbital decay ${\dot{P}_\mathrm{b}^{\,\mathrm{int}}} =
{\dot{P}_\mathrm{b}^{\,\mathrm{obs}}} - {\dot{P}_\mathrm{b}^{\,\mathrm{Gal}}} - {\dot{P}_\mathrm{b}^{\,\mathrm{Shk}}}$ with $\dot P_\mathrm{b}^{\,\mathrm{GW}}$. For J1537, this test is the one (among the tests with PK parameters, see \ref{subsec:B1534_DNS_pulsars}) that showed the largest discrepancy with GR (see Figure~9 of \citealp{Fonseca14}), possibly due to the unreliable DM-based distance used for the test.

Using Equation~22 of \citet{Weisberg16}, we calculated $\dot P_\mathrm{b}^{\,\mathrm{GW}}=-192.45\pm0.06$\,\fsps.
Using the proper motion $\mu=\sqrt{\mu_\alpha^2+\mu_\delta^2}$ and the distance $D$ acquired with Bayesian inference (see \ref{tab:B1534_mu_and_pi}), we updated $\dot{P}_\mathrm{b}^{\,\mathrm{Shk}}=\mu^2 D/c \cdot P_\mathrm{b}=53\pm4$\,\fsps. 
The uncertainties for $\dot P_\mathrm{b}^{\,\mathrm{GW}}$ and $\dot{P}_\mathrm{b}^{\,\mathrm{Shk}}$ were derived with error propagation, which were subsequently confirmed by Monte-Carlo simulations.
In the $\dot{P}_\mathrm{b}^{\,\mathrm{Shk}}$ error estimation, we did not take into account the small correlation between $\mu_\alpha$ and $\varpi$ (mentioned in \ref{sec:B1534_results}). This is because the correlation between $\mu$ and $\varpi$ is still negligible, as the declination component dominates the proper motion (see \ref{tab:B1534_mu_and_pi}).

Following \citet{Zhu19}, we estimated $\dot{P}_\mathrm{b}^{\,\mathrm{Gal}}$ with different Galactic mass distribution models compiled in {\tt GalPot} (\url{https://github.com/PaulMcMillan-Astro/GalPot}, \citealp{McMillan17}).  
The $\dot{P}_\mathrm{b}^{\,\mathrm{Gal}}$ results are summarized in \ref{tab:B1534_Pbdot_Gal}. 
The errors on $\dot{P}_\mathrm{b}^{\,\mathrm{Gal}}$ can be attributed to two sources: the uncertainty in the measurements (such as distance and proper motion) and the inaccuracy of the Galactic mass distribution model.
The former $\dot{P}_\mathrm{b}^{\,\mathrm{Gal}}$ errors, at the $\leq0.1$\,\fsps\ level (see \ref{tab:B1534_Pbdot_Gal}), were derived with Monte-Carlo simulations.
We approached the latter $\dot{P}_\mathrm{b}^{\,\mathrm{Gal}}$ errors with the standard deviation of the $\dot{P}_\mathrm{b}^{\,\mathrm{Gal}}$ estimates listed in \ref{tab:B1534_Pbdot_Gal}. For this calculation of the standard deviation, we do not include the $\dot{P}_\mathrm{b}^{\,\mathrm{Gal}}$ based on the analytical model by \citet{Nice95}, because {\bf 1)} the analytical model is oversimplified (see the discussion in Appendix~A of \citealp{Zhu19}) and {\bf 2)} the resultant $\dot{P}_\mathrm{b}^{\,\mathrm{Gal}}$ is inconsistent with other models (see \ref{tab:B1534_Pbdot_Gal}). Accordingly, we adopted the average $\dot{P}_\mathrm{b}^{\,\mathrm{Gal}}$ of the four remaining Galactic mass distribution models \citep{Dehnen98,Binney11,Piffl14,McMillan17} as the $\dot{P}_\mathrm{b}^{\,\mathrm{Gal}}$ estimate. 
In this way, we obtained $\dot{P}_\mathrm{b}^{\,\mathrm{Gal}}=-1.9\pm0.2$\,\fsps, where the error budget has included the standard deviation (0.14\,\fsps) of $\dot{P}_\mathrm{b}^{\,\mathrm{Gal}}$. 
As {\tt Galpot} was not available in 2014, \citet{Fonseca14} adopted the $\dot{P}_\mathrm{b}^{\,\mathrm{Gal}}$ based on the Galactic mass distribution model of \citet{Nice95} (see \ref{tab:B1534_Pbdot_Gal}) for the calculation of the timing kinematic distance. Provided our new $\dot{P}_\mathrm{b}^{\,\mathrm{Gal}}$, the timing kinematic distance of $1.05$\,kpc (reported by \citealp{Fonseca14} and quoted in \ref{tab:B1534_mu_and_pi}) would decrease by 3\% to $1.02$\,kpc, thus becoming consistent with the new model-independent distance (see \ref{tab:B1534_mu_and_pi}).

\begin{table}
    \centering
    \caption{Galactic-potential-related orbital decay $\dot{P}_\mathrm{b}^{\,\mathrm{Gal}}$ as well as its two components (i.e., $\dot{P}_\mathrm{b,h}^{\,\mathrm{Gal}}$ and $\dot{P}_\mathrm{b,z}^{\,\mathrm{Gal}}$, respectively corresponding to the component horizontal and vertical to the Galactic plane) estimated with different models of Galactic mass distribution. The $\dot{P}_\mathrm{b}^{\,\mathrm{Gal}}$ uncertainties for the models are derived with Monte-Carlo simulations. For comparison, the $\dot{P}_\mathrm{b}^{\,\mathrm{Gal}}$ expected by GR (i.e., ${\dot{P}_\mathrm{b}^{\,\mathrm{obs}}} - {\dot{P}_\mathrm{b}^{\,\mathrm{Shk}}} - {\dot{P}_\mathrm{b}^{\,\mathrm{GW}}}$) is provided.}
    
    %\resizebox{\textwidth}{!}{
    \begin{tabular}{cccc} % four columns, 
	\hline
	\hline
	Galactic mass & $\dot{P}_\mathrm{b,h}^{\,\mathrm{Gal}}$ & $\dot{P}_\mathrm{b,z}^{\,\mathrm{Gal}}$ & $\dot{P}_\mathrm{b}^{\,\mathrm{Gal}}$  \\
	distribution model & (\fsps) & (\fsps) & (\fsps) \\
		\hline
	\citet{Nice95} & $1.1(1)$ & $-4.6(1)$ & $-3.51(6)^\mathrm{a}$ \\
	\citet{Dehnen98} & $0.120(1)$ & $-2.04(6)$ & $-1.92(6)^\mathrm{b}$  \\
	\citet{Binney11} & $0.131(3)$ & $-1.89(9)$ & $-1.76(9)$ \\
	\citet{Piffl14} & $0.132(1)$ & $-2.09(8)$ & $-1.96(8)$ \\
	\citet{McMillan17} & $0.141(2)$ & $-2.2(1)$ & $-2.1(1)$\\
	\hline
	${\dot{P}_\mathrm{b}^{\,\mathrm{obs}}} - {\dot{P}_\mathrm{b}^{\,\mathrm{Shk}}} - {\dot{P}_\mathrm{b}^{\,\mathrm{GW}}}$ & --- & --- & $2.5(3.8)$ \\
	
	\hline
	\end{tabular}
	%}
	
	\tabnote{$^a$For the calculation, we adopted $R_0=8.12\pm0.03$\,kpc (the distance from the Sun to the Galactic center) provided by \citet{Gravity-Collaboration18} and $\Theta_0=234.6\pm1.1\,{\rm km~s^{-1}}$ (the circular speed of the local standard of rest). We derived the $\Theta_0$ with the proper motion of Sgr~A* \citep{Reid20}, the aforementioned $R_0$ \citep{Gravity-Collaboration18} and the velocity of the Sun with respect to the local standard of rest \citep{Schonrich10}.}
    \tabnote{$^b$There are 4 models discussed in \citet{Dehnen98}. Here, we used the ``model 3'', which falls into the middle of the models 1 to 4, and is generally consistent with the other 3 models.}
    \label{tab:B1534_Pbdot_Gal}
\end{table}

\begin{figure}
    \centering
    \hspace*{-0.5cm}
	\includegraphics[width=13cm]{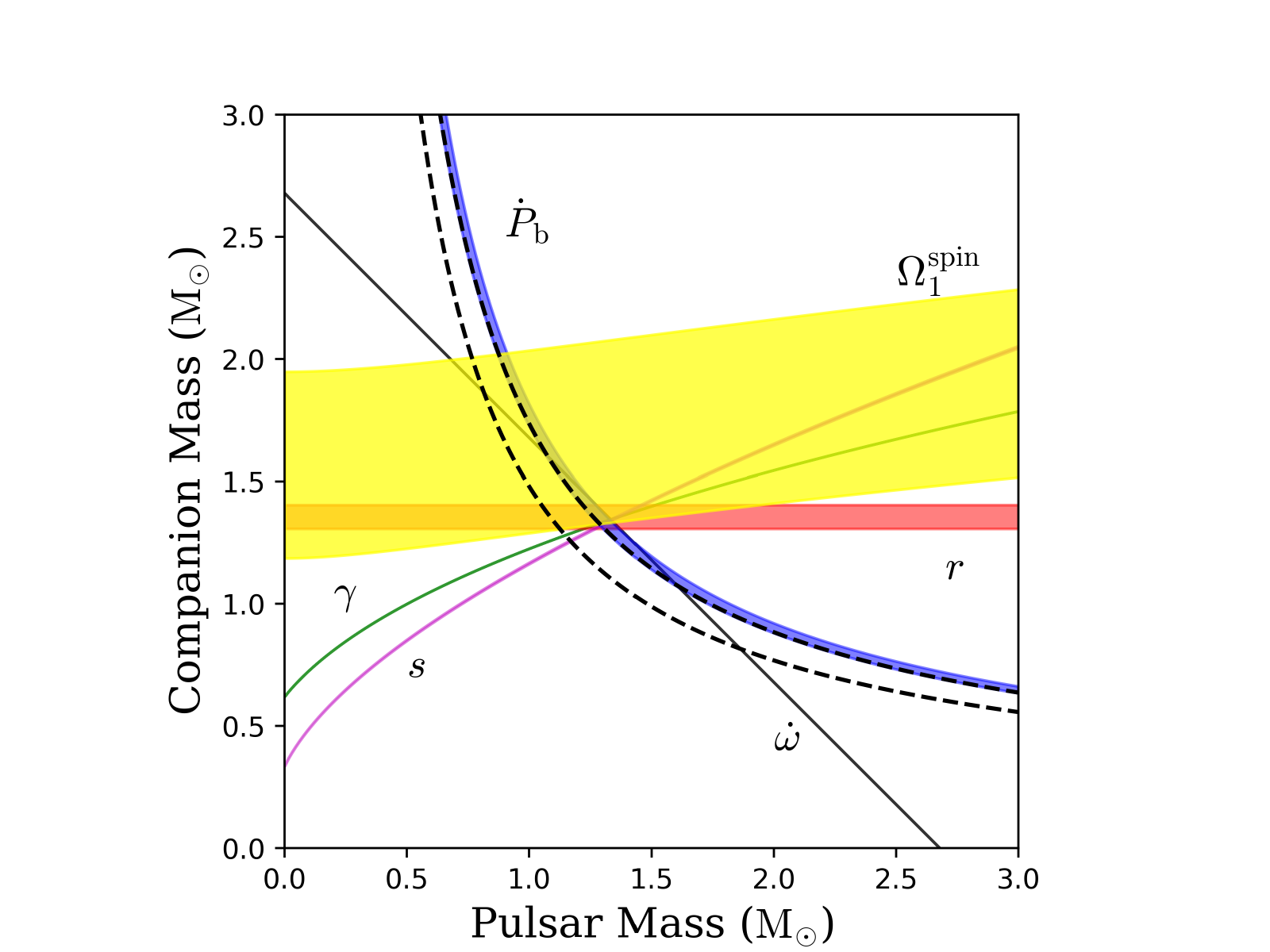}
    \caption{Mass-mass diagram of J1537. Its only difference from Figure~9 of \citet{Fonseca14} is the updated mass-mass constraint offered by the new observed intrinsic orbital decay ${\dot{P}_\mathrm{b}^{\,\mathrm{int}}}$, shown with the blue strip. For comparison, the mass-mass constraint given by the previous ${\dot{P}_\mathrm{b}^{\,\mathrm{int}}}=-0.17(1)\,\mathrm{ps~s^{-1}}$ inferred from the DM-based distance (see \ref{tab:B1534_mu_and_pi}) is provided with two dashed curves.
    For the other PK parameters, the green, pink, red, black and yellow strips stand for the mass-mass constraints posed by the time-averaged gravitational redshift $\gamma=2.0708(5)$\,ms, the Shapiro delay ``shape'' $s=0.977(2)$, the Shapiro delay ``range'' $r=6.6(2)\,\mu$s, the periastron advance rate $\dot{\omega}=1.755795(2)\,\mathrm{deg~{yr}^{-1}}$ and the pulsar precession rate $\Omega_1^\mathrm{spin}=0.59^{+0.12}_{-0.08}\,\mathrm{deg~{yr}^{-1}}$, respectively \citep{Fonseca14}.}
    \label{fig:B1534_mass_mass_diagram}
\end{figure}

Collectively, we reached ${\dot{P}_\mathrm{b}^{\,\mathrm{int}}}=-188.0\pm3.8$\,\fsps, corresponding to 
\begin{equation}
\label{eq:B1534_Pbdot_test_result}
\frac{\dot{P}_\mathrm{b}^{\,\mathrm{int}}}{\dot{P}_\mathrm{b}^{\,\mathrm{GW}}}=0.977\pm0.020 \,,
\end{equation}
which is the third most precise orbital-decay test of GR in the strong-field regime according to Table~3 of \citet{Weisberg16}. 
At the 2\% precision level, the new observed intrinsic orbital decay is within $1.2\,\sigma$ of the GR prediction (see \ref{fig:B1534_mass_mass_diagram}), which relieves the mild tension of the previous $\dot{P}_\mathrm{b}$ test (${\dot{P}_\mathrm{b}^{\,\mathrm{int}}}/{\dot{P}_\mathrm{b}^{\,\mathrm{GW}}}=0.91\pm0.06$ at 1.7\,$\sigma$ agreement, \citealp{Stairs02}). 

For visualisation, the mass-mass diagram of J1537 (updated from Figure~9 of \citealp{Fonseca14}) is presented in \ref{fig:B1534_mass_mass_diagram}, which involves 6 PK parameters. 
Apart from the 5 PK parameters already mentioned in \ref{sec:B1534_intro}, $\Omega_1^\mathrm{spin}$ stands for the precession rate of the pulsar. 
Each PK parameter is a function of the two DNS constituent masses. Therefore, each observed PK parameter (and its uncertainty) offers a constraint on the two masses. If GR is correct, all mass-mass constraints should converge at the ``true'' masses of the pulsar and its companion. In \ref{fig:B1534_mass_mass_diagram}, this convergence is visible with the new ${\dot{P}_\mathrm{b}^{\,\mathrm{int}}}$.

Looking into the future, the bottleneck of the orbital-decay test with J1537 will continue to be its parallax uncertainty, which would decrease with $t^{-1/2}$ (e.g. \citealp{Ding21}; here, $t$ stands for the on-source time instead of the time span) in the long term despite fluctuations of J1537 brightness. This process of precision enhancement would be accelerated with high-sensitivity VLBI observations, as the VLBA observations of J1537 are generally sensitivity-limited, especially when J1537 is down-scintillated.

\section*{Acknowledgements}
We are grateful to the anonymous referee for the swift and valuable comments, and appreciate the helpful discussions with Leonid Petrov, David Kaplan and Joseph Lazio. B.S. and A.L. thank Mitch Mickaliger for help with timing data acquisition and preparation.
H.D. is supported by the ACAMAR (Australia-ChinA ConsortiuM for Astrophysical Research) scholarship, which is partly funded by the China Scholarship Council (CSC).
A.T.D is the recipient of an ARC Future Fellowship (FT150100415).
Pulsar research at UBC is supported by an NSERC Discovery Grant and by the Canadian Institute of Advanced Research.
Pulsar research at JBCA is supported by a Consolidated Grant from STFC. 
Parts of this research were conducted by the Australian Research Council Centre of Excellence for Gravitational Wave Discovery (OzGrav), through project number CE170100004.
This work is based on observations with the Very Long Baseline Array (VLBA), which is operated by the National Radio Astronomy Observatory (NRAO). The NRAO is a facility of the National Science Foundation operated under cooperative agreement by Associated Universities, Inc.
Data reduction and analysis was performed on OzSTAR, the Swinburne-based supercomputer.
This work made use of the Swinburne University of Technology software correlator, developed as part of the Australian Major National Research Facilities Programme and operated under license. 


\bibliography{haoding}{}
\bibliographystyle{mnras}

\chapter[VLBI Astrometry of 18 Millisecond Pulsars]{VLBI Astrometry of 18 Millisecond Pulsars}
\label{ch:mspsrpi}

This chapter is converted from \citet{Ding23} entitled ``The  \mspsrpi\ catalogue: VLBA astrometry of 18 millisecond pulsars'', which  includes the re-analysis of three previously published \mspsrpi\ pulsars (see \ref{ch:J1012}, \ref{ch:B1534} and \citealp{Vigeland18}). 
The chapter presents the results for the entire \mspsrpi\ program, although additional observations (now underway) are necessary to achieve the sufficient absolute astrometric precision required for the study of solar system ephemerides (see \ref{subsec:SSE_VLBI}).

This is the culmination of the thesis, serving a wide range of scientific motivations (see \ref{sec:NS_formation}, \ref{sec:IISM}, \ref{sec:pulsar_astrometry},  \ref{sec:XRB_astrometry}). 
Methodologically, the chapter uses all the advanced phase referencing techniques mentioned in \ref{sec:ch2_VLBI_observing_strategies}, presents the addition of the systematics-correcting parameter $\eta_\mathrm{EFAC}$ (see \ref{sec:mspsrpi_parameter_inference}) to the Bayesian astrometric inference package {\tt sterne}, and applies multi-source astrometric inference for \psrkb\ (see \ref{subsubsec:mspsrpi_multi-source-inference}) with {\tt sterne}.
%$\eta_\mathrm{EFAC}$
For the sake of brevity, most figures and tables are submitted as supplementary materials, and meanwhile also made available online. 
%At the time of thesis submission, the paper was under review by {\it MNRAS}. As of 17 December 2022, it is in press.
A few minor corrections to this chapter, including the addition of \ref{fig:mspsrpi_VLBI_proper_motions__vs__timing_ones} and \ref{fig:mspsrpi_D_DM__vs__D_px}, are made in response to the comments by the thesis examiners.

\section{Abstract}
With unparalleled rotational stability, millisecond pulsars (MSPs)
serve as ideal laboratories for numerous astrophysical studies, many of which require precise knowledge of the distance and/or velocity of the MSP.
%While previous efforts have been made to refine our understanding of the spatial and velocity distributions of pulsars, to date none had focused specifically on a sample of MSPs.
Here, we present the astrometric results for 18 MSPs of the ``\mspsrpi" project focusing exclusively on astrometry of MSPs, which includes the re-analysis of 3 previously published MSPs.
%We have now addressed this gap with the ``\mspsrpi" project, and present here astrometric results for 15 previously unstudied MSPs alongside re-analyzed ones for 3 previously published sources.
On top of a standardized data reduction protocol, more complex strategies (i.e., normal and inverse-referenced 1D interpolation) were employed where possible to further improve astrometric precision.
We derived astrometric parameters 
using {\tt sterne}, a new Bayesian astrometry inference package that allows the incorporation of prior information based on pulsar timing where applicable.  We measured significant ($>3\,\sigma$) parallax-based distances for 15 MSPs, including $0.81\pm0.02$\,kpc for \psrfa\ --- the most significant model-independent distance ever measured for a double neutron star system. 
For each MSP with a well-constrained distance, we estimated its transverse space velocity and radial acceleration. Among the estimated radial accelerations, the updated ones of \psrea\ and \psri\ impose new constraints on dipole gravitational radiation and the time derivative of Newton's gravitational constant. Additionally, significant angular broadening was detected for \psrgb, which offers an independent check of the postulated association between the HII region Sh~2-27 and the main scattering screen of \psrgb.
Last but not least, the upper limit of the death line of $\gamma$-ray-emitting pulsars is refined with the new radial acceleration of the hitherto least energetic $\gamma$-ray pulsar \psrhb.

\section{Introduction}
\label{sec:mspsrpi_intro}

%\linenumbers

\subsection{Millisecond pulsars: a key for probing theories of gravity and detecting the gravitational-wave background}
\label{subsec:mspsrpi_gravity_test_intro}

Pulsars are an observational manifestation of neutron stars (NSs) that emit non-thermal electromagnetic radiation while spinning  \citep{Hewish69,Gold68,Pacini68}. 
Over 3000 radio pulsars have been discovered to date throughout the Galaxy and the nearest members of the Local Group \citep{Manchester05}.
Due to the large moment of inertia of pulsars, the pulses we receive on Earth from a pulsar exhibit highly stable periodicity. 
By measuring a train of pulse time-of-arrivals (ToAs) of a pulsar and comparing it against the model prediction, a long list of model parameters can be inferred \citep[e.g.][]{Detweiler79,Helfand80}.
This procedure to determine ToA-changing parameters is known as pulsar timing, hereafter referred to as timing.

In the pulsar family, recycled pulsars (commonly refereed to as millisecond pulsars, or MSPs), have the shortest rotational periods. 
They are believed to have been spun-up through the accretion from their donor stars during a previous evolutionary phase as a low-mass X-ray binary (LMXB) \citep{Alpar82}. 
As the duration of the recycling phase (and hence the degree to which the pulsar is spun-up) can vary depending on the nature of the binary, there is no clear cut spin period threshold that separates MSPs from canonical pulsars. In this paper, we define MSPs as pulsars with spin periods of $\lesssim 40$\,ms and magnetic fields $\lesssim10^{10}$~G. This range encompasses most partially recycled pulsars with NS companions, such as \psrfb\ (aka. PSR~B1534$+$12) and \psrfa.
Compared to non-recycled pulsars, ToAs from MSPs can be measured to higher precision due to both the narrower pulse profiles and larger number of pulses.  Additionally, MSPs exhibit more stable rotation \citep[e.g.][]{Hobbs10}; both factors promise a lower level of random timing noise. 
Consequently, MSPs outperform non-recycled pulsars in the achievable precision for
probing theories underlying ToA-changing astrophysical effects. In particular, MSPs provide the hitherto most precise tests for gravitational theories \citep[e.g.][]{Kramer21a,Zhu19,Freire12}.
Einstein's theory of general relativity (GR) is the simplest form among a group of possible candidate post-Newtonian gravitational theories. 
The discovery of highly relativistic double neutron star (DNS) systems \citep[e.g.][]{Hulse75,Wolszczan91,Burgay03,Lazarus16,Stovall18,Cameron18}
and their continued timing have resulted in many high-precision tests of GR and other gravity theories (\citealp{Fonseca14,Weisberg16,Ferdman20}, and especially \citealp{Kramer21a}). The precise timing, optical spectroscopy and VLBI observations of pulsar-white-dwarf (WD) systems have, in addition, achieved tight constraints on several classes of alternative theories of gravity \citep{Deller08,Lazaridis09,Freire12,Antoniadis13,Ding20,Guo21,Zhao22}.

Gravitational Waves (GWs) are changes in the curvature of spacetime (generated by accelerating masses), which propagate at the speed of light. Individual GW events in the Hz---kHz range have been detected directly with GW observatories (e.g. \citealp{Abbott16}; see the third Gravitational-Wave Transient Catalog\footnote{\url{https://www.ligo.org/science/Publication-O3aFinalCatalog/}}), and indirectly using the orbital decay of pulsar binaries \citep[e.g.][]{Taylor82,Weisberg16,Kramer21a,Ding21a}. 
Collectively, a gravitational wave background (GWB), formed with primordial GWs and GWs generated by later astrophysical events \citep{Carr80}, is widely predicted, but has not yet been confirmed by any observational means. In the range of $10^{-9}\,\mathrm{Hz}-0.1$\,Hz, supermassive blackhole binaries are postulated to be the primary sources of the GWB \citep{Sesana08}. 
In this nano-hertz regime, the most stringent constraints on the GWB are provided by pulsar timing \citep{Detweiler79}.

To enhance the sensitivity for the GWB hunt with pulsar timing, and to distinguish GWB-induced ToA signature from other sources of common timing ``noise'' (e.g., Solar-system planetary ephemeris error, clock error and interstellar medium, \citealp{Tiburzi16}), a pulsar timing array (PTA), composed of MSPs scattered across the sky (see \citealp{Roebber19} for spatial distribution requirement), is necessary \citep{Foster90}. 
After 2 decades of efforts, no GWB has yet been detected by a PTA, though common steep-spectrum timing noise (in which GWB signature should reside) has already been confirmed by several major PTA consortia \citep{Arzoumanian20,Goncharov21,Chen21,Antoniadis22}.

\subsection{Very long baseline astrometry of millisecond pulsars}
\label{subsec:mspsrpi_MSP_VLBI_astrometry}

In timing analysis, 
astrometric information for an MSP (reference position, proper motion, and annual geometric parallax) can form part of the global ensemble of parameters determined from ToAs. 
However, the astrometric signatures can be small compared to the ToA precision and/or covariant with other parameters in the model, especially for new MSPs that are timed for less than a couple of years \citep{Madison13}.
Continuing to add newly discovered MSPs into PTAs is considered the best pathway to rapid PTA sensitivity improvement \citep{Siemens13}, and is particularly important for PTAs based around newly commissioned high-sensitivity radio telescopes \citep[e.g.][]{Bailes20}.
Therefore, applying priors to the astrometric parameters can be highly beneficial for the timing of individual MSPs (especially the new ones) and for enhancing PTA sensitivities \citep{Madison13}.

Typically, the best approach to independently determine precise astrometric parameters for MSPs is the use of very long baseline interferometry (VLBI) observations, which can achieve sub-mas positional precision (relative to a reference source position) for MSPs in a single observation. By measuring the sky position of a Galactic MSP a number of times and modeling the position evolution,
VLBI astrometry can obtain astrometric parameters for the MSP.
Compared to pulsar timing, it normally takes much shorter time span 
%($\approx$2 years) 
to reach a given astrometric precision \citep[e.g.][]{Brisken02,Chatterjee09,Deller19}.

One of the limiting factors on searching for the GWB with PTAs is the uncertainties on the Solar-system planetary ephemerides (SSEs) \citep{Vallisneri20}, which are utilized to convert geocentric ToAs to ones measured in the (Solar-system) barycentric frame (i.e., the reference frame with respect to the barycentre of the Solar system). In pulsar timing analysis, adopting different SSEs (made by two main producers for various purposes) may lead to discrepant timing parameters \citep[e.g.][]{Wang17}. On the other hand, VLBI astrometry measures offsets with respect to a source whose position is measured in a quasi-inertial (reference) frame defined using remote quasars \citep[e.g.][]{Charlot20}. Although VLBI astrometry also relies on SSEs to derive annual parallax, it is robust against SSE uncertainties. In other words, for VLBI astrometry, using different SSEs in parameter inference would not lead to a noticeable difference (in the inferred parameters). Therefore, VLBI astrometry of MSPs can serve as an objective standard to be used to discriminate between various SSEs. Specifically, if an SSE is inaccurate, the barycentric frame based on the SSE would display rotation with respect to the quasar-based frame. This frame rotation can be potentially detectable by comparing VLBI positions of multiple MSPs against their timing positions (as measured using a common SSE) \citep{Chatterjee09,Wang17}. 
By eliminating inaccurate SSEs, VLBI astrometry of MSPs can suppress the SSE uncertainties, and hence enhance the PTA sensitivities.

Besides the GWB-related motivations, interferometer-based astrometric parameters (especially distances to MSPs) have been adopted to sharpen the tests of gravitational theories for individual MSPs \citep[e.g.][]{Deller09,Deller18,Guo21,Ding21a}.
Such tests are normally made by comparing the model-predicted and observed post-Keplerian (PK) parameters that quantify excessive gravitational effects beyond a Newtonian description of the orbital motion.
Among the PK parameters is the orbital decay $\dot{P}_\mathrm{b}$ (or the time derivative of orbital period). The intrinsic cause of $\dot{P}_\mathrm{b}$ in double neutron star systems is dominated by the emission of gravitational waves, which can be predicted using the binary constituent masses and orbital parameters \citep[e.g.][]{Lazaridis09,Weisberg16}.
To test this model prediction, however, requires any extrinsic orbital decay $\dot{P}_\mathrm{b}^\mathrm{ext}$ due to relative acceleration between the pulsar and the observer to be removed from the observed $\dot{P}_\mathrm{b}$. 
Such extrinsic terms depend crucially on the proper motion and the distance of the pulsar, however these (especially the distance) can be difficult to estimate from pulsar timing. Precise VLBI determination of proper motions and distances can yield precise estimates of these extrinsic terms and therefore play an important role in orbital-decay tests of gravitational theories.

Last but not least, pulsar astrometry is crucial for understanding the Galactic free-electron distribution, or the Galactic free-electron number density $n_\mathrm{e}(\vec{x})$ as a function of position. An $n_\mathrm{e}(\vec{x})$ model is normally established by using pulsars with well determined distances as benchmarks. As the pulsations from a pulsar allow precise measurement of its dispersion measure (DM), the average $n_\mathrm{e}$ between the pulsar and the Earth can be estimated given the pulsar distance. 
Accordingly, a large group of such benchmark pulsars across the sky would enable the establishment of an $n_\mathrm{e}(\vec{x})$ model. 
In a relevant research field, extragalactic fast radio bursts (FRBs) have been used to probe intergalactic medium distribution on a cosmological scale \citep[e.g.][]{Macquart20,Mannings21}, which, however, demands the removal of the DMs of both the Galaxy and the FRB host galaxy. The Galactic DM cannot be determined without a reliable $n_\mathrm{e}(\vec{x})$ model, which, again, calls for precise astrometry of pulsars across the Galaxy.

\subsection{The \mspsrpi\ project}
\label{subsec:mspsrpi_mspsrpi}

Using the Very Long Baseline Array (VLBA), the \psrpi\ project tripled the sample of pulsars with precisely measured astrometric parameters \citep{Deller19}, but included just three MSPs. 
The successor project, \mspsrpi, is a similarly designed VLBA astrometric program targeting exclusively MSPs. 
Compared to canonical pulsars, MSPs are generally fainter. To identify MSPs feasible for VLBA astrometry, a pilot program was conducted, which found 31 suitable MSPs. 
Given observational time constraints, we selected 18 MSPs with the foremost scientific importance as the targets of the \mspsrpi\ project, focusing primarily on sources observed by pulsar timing arrays. The 18 MSPs are listed in \ref{tab:mspsrpi_MSPs} along with their spin periods $P_\mathrm{s}$ and orbital periods $P_\mathrm{b}$ (if available) obtained with the ATNF Pulsar Catalogue\footnote{\label{footnote:PSRCAT}\url{https://www.atnf.csiro.au/research/pulsar/psrcat/}} \citep{Manchester05}. 
The results for 3 sources (\psrea, \psrfb, \psrga) involved in the project have been published \citep{Vigeland18,Ding20,Ding21a}. 
In this paper, we present the astrometric results of the remaining 15 MSPs studied in the \mspsrpi\ project. We also re-derived the results for the 3 published MSPs, in order to ensure consistent and systematic astrometric studies.

\begin{sidewaystable}
%\raggedright
\caption{List of pulsars and phase calibrators}
\label{tab:mspsrpi_MSPs}
%\begin{tabular}{@{}l@{\:}l@{\:}l@{}} % manual @ spacing to prevent this being too wide for a page
\resizebox{\textwidth}{!}{
%\rotatebox{270}{
\begin{tabular}{lcccccllllr}
\hline
\hline
PSR & $P_\mathrm{s}$ & $P_\mathrm{b}$ & Gating & Project Codes & Primary phase calibrator & $\Delta_\mathrm{PC-IBC}$ $\,^{f}$& Secondary phase calibrator & IBC $\,^{a}$ & $\Delta_\mathrm{psr-IBC}$ $\,^{b}$ & $S_\mathrm{unres}^\mathrm{IBC}$ $\,^{*}$\\
 & (ms) & (d) & gain & & & (deg) & & code & (arcmin) & (mJy)\\
\hline
\Psrb\ & 4.87 & --- & 1.75 & BD179B, BD192B & ICRF~J002945.8$+$055440 & 0.97 & FIRST~J003054.6$+$045908 & 00027 & 10.1 & 20.6\\
\Psrc\ & 3.86 & 0.29 & 1.85 & BD179C, BD192C & --- & --- & NVSS~J061002$-$211538$\,^{c}$ & 00238 & 15.4 & 112.8 \\
\Psrd\ & 28.85 & 8.3 & 2.15 & BD179D, BD192D & ICRF~J061909.9$+$073641 & 2.84 & NVSS~J062153$+$102206 & 00303 & $21.0 \! \Rightarrow \! 2.9$ $\,^{d}$ & 18.1\\
\Psrea\ $^{**}$ & 5.26 & 0.60 & 1.43 & BD179E, BD192E & ICRF~J095837.8$+$503957  & 3.38 & NVSS~J101307$+$531233 & 00462 & 7.51 & 20.8\\
\Psreb\ & 5.16 & --- & 1.69 & BD179E, BD192E & ICRF~J102838.7$-$084438 & 1.59 & FIRST~J102526.3$-$072216 & 00529 & 12.2 & 11.7\\
%\Psre\ & BD179B, BD192B & J0029$+$0554 & \\
\Psrfa\ & 40.93 & 8.6 & 2.27 & BD179F, BD192F & ICRF~J150644.1$+$493355 & 1.78 & NVSS~J151733$+$491626 & 00691 & $13.8 \! \Rightarrow \! 6.5$ $\,^g$ & 35.6 \\
\Psrfb\ $^{**}$ & 37.90 & 0.42 & 1.62 & BD179F, BD192F, BD229 & ICRF~J154049.4$+$144745 & 3.18 & FIRST~J153746.2$+$114215 & 00840 & 16.3 & 19.2\\
\Psrga\ $^{**}$ & 3.16 & 175 & 1.90 & BD179G, BD192F & ICRF~J164125.2$+$225704 & 0.79 & NVSS~J164018$+$221203 & 00920 & 12.1 & 98.0 \\
\Psrgb\ & 4.62 & 147 & 1.56 & BD179G, BD192G & ICRF~J163845.2$-$141550 & 2.49 & NVSS~J164515$-$122013   & 01120 & 24.3 & 6.0\\
\Psrha\ & 3.50 & --- & 1.78 & BD179H, BD192H & ICRF~J172658.9$-$225801 & 2.47 & NVSS~J172129$-$250538 & 01188 & 10.1 & 7.4\\
\Psrhb\ & 8.12 & --- & 1.86 & BD179H, BD192H & ICRF~J172658.9$-$225801 & 0.76 & NVSS~J172932$-$232722 & 01220 & 25.5 & 37.9 \\
\Psri\ & 5.85 & 0.35 & 1.98 & BD179I, BD192I & ICRF~J174037.1$+$031147 & 0.66 & NVSS~J173823$+$033305 & 01313 & 7.5 & 6.5 \\
\Psro\ & 3.05 & --- & 1.53 & BD179O, BD192O & ICRF~J182057.8$-$252812 & 0.61 & NVSS~J182301$-$250438 & 01433 & $23.8 \! \Rightarrow \! 3.7$ $\,^{d}$ & 3.3 \\
\Psrka\ & 4.09 & 116 & 1.64 & BD179K, BD192K & ICRF~J185250.5+142639 & 1.51 & NVSS~J185456$+$130110 & 01535 & 14.6 & 5.9 \\
\Psrl\ & 4.98 & 58.5 & 2.65 & BD179L, BD192L & ICRF~J191158.2$+$161146 & 3.16 & NVSS~J190957$+$130434 & 01769 & 8.7 & 4.9\\
\Psrma\ & 3.63 & 2.7 & 1.67 & BD179M, BD192M & ICRF~J190528.5$-$115332 & 1.76 & NVSS~J191233$-$113327 & 01816 & 21.9 & 23.4\\
\Psrmb\ & 7.65 & 10.9 & 2.12 & BD179M, BD192M & ICRF~J191207.1$-$080421 & 2.14 & NVSS~J191731$-$062435 & 01846 & 26.1 & 50.7\\
\Psrkb\ & 1.56 & --- & 1.35 & BD179K, BD192K & ICRF~J193510.4$+$203154  & 1.88 & NVSS~J194104$+$214913$\,^{e}$ & 01647 & $24.5 \! \Rightarrow \! 4.1$ & 10.6\\
 & & & & &   & 1.94 & NVSS~J194106$+$215304$\,^{e}$ & 01648 & $27.4 \! \Rightarrow \! 2.7$ & 1.8\\

\hline
\multicolumn{11}{l}{$\bullet$ The image models for the primary and secondary calibrators listed here are publicly available\textsuperscript{\ref{footnote:calibrator_models}}.}\\
\multicolumn{11}{l}{$\bullet$ $P_\mathrm{s}$ and $P_\mathrm{b}$ stand for spin period and orbital period, respectively.}\\
\multicolumn{11}{l}{$^*$ Unresolved flux intensity of the secondary phase calibrator at 1.55\,GHz.}\\
\multicolumn{11}{l}{$^{**}$ Published in \citet{Ding20,Ding21a,Vigeland18}.}\\
\multicolumn{11}{l}{$^a$ Secondary phase calibrators are named IBC\textit{XXXXX} in the BD179 and BD192 observing files, where ``\textit{XXXXX}'' represents a unique 5-digit IBC code.}\\
\multicolumn{11}{l}{$^b$ Angular separation between target and secondary calibrator.}\\
\multicolumn{11}{l}{$^c$ NVSS~J061002$-$211538, close to the pulsar on the sky, is bright enough to serve as primary phase calibrator.}\\
\multicolumn{11}{l}{$^d$ As 1D interpolation is applied, the pulsar-to-virtual-calibrator separation is also provided after ``$\Rightarrow$'' (see \ref{subsec:mspsrpi_dualphscal}).}\\
\multicolumn{11}{l}{$^e$ Here, inverse phase referencing is adopted, where the ``secondary phase calibrators'' are ultimately the targets (see \ref{subsec:mspsrpi_sophisticated_data_reduction}).}\\
\multicolumn{11}{l}{$^f$ Angular separation between primary and secondary calibrator.}\\
\multicolumn{11}{l}{$^g$ The NVSS~J151815$+$491105, a 4.5-mJy-bright source 6\farcm5 away from the pulsar, is used as the final reference source (see \ref{sec:mspsrpi_data_reduction}).}\\

\end{tabular}
}
\end{sidewaystable}

Along with the release of the catalogue results, this paper covers several scientific and technical perspectives. 
First, this paper explores novel data reduction strategies such as inverse-referenced 1D phase interpolation (see \ref{subsec:mspsrpi_sophisticated_data_reduction}). Second, a new Bayesian astrometry inference package is presented (see \ref{sec:mspsrpi_parameter_inference}).
Third, with new parallax-based distances and proper motions, we discriminate between the two prevailing $n_\mathrm{e}(\vec{x})$ models (see \ref{subsubsec:mspsrpi_DM_distances}), and investigate the kinematics of MSPs in \ref{subsec:mspsrpi_v_t}. 
Fourth, with new parallax-based distances of two MSPs, we re-visit the constraints on alternative theories of gravity (see \ref{sec:mspsrpi_orbital_decay_tests}).
Finally, discussions on individual pulsars are given in \ref{sec:mspsrpi_individual_pulsars}, which includes a refined ``death line'' upper limit of $\gamma$-ray pulsars (see \ref{subsec:mspsrpi_J1730}).
The study of SSE-dependent frame rotation, which depends on an accurate estimation of the reference points of our calibrator sources in the quasi-inertial VLBI frame, requires additional multi-frequency observations and will be presented in a follow-up paper.

Throughout this paper, we abide by the following norms unless otherwise stated. {\bfseries 1)} The uncertainties are provided at 68\% confidence level. {\bfseries 2)} Any mention of flux density refers to unresolved flux density $S_\mathrm{unres}$ in our observing configuration (e.g., a 10-mJy source means $S_\mathrm{unres}=10$\,mJy). {\bfseries 3)}
All bootstrap and Bayesian results adopt the 50th, 16th and 84th percentile of the marginalized (and sorted) value chain as, respectively, the estimate and its 1-$\sigma$ error lower and upper bound.
{\bfseries 4)} Where an error of an estimate is required for a specific calculation but an asymmetric error is reported for the estimate, the mean of upper and lower errors is adopted for the calculation.
{\bfseries 5)} VLBI positional uncertainties will be broken down into the uncertainty of the offset from a chosen calibrator reference point, and the uncertainty in the location of that chosen reference point. This paper focuses on the relative offsets which are relevant for the measurement of proper motion and parallax, and the uncertainty in the location of the reference source is presented separately.

\section{Observations and Correlation}
\label{sec:mspsrpi_observations}

All MSPs in the \mspsrpi\ catalogue
(see Table\ref{tab:mspsrpi_MSPs}) were observed at around 1.55\,GHz with the VLBA at 2\,Gbps data rate from mid-2015 to no later than early 2018.
The observing and correlation tactics for the \mspsrpi\ project are identical to those of the \psrpi\ project \citep{Deller19}. 
The primary phase calibrators were selected from the Radio Fundamental Catalogue\footnote{\url{astrogeo.org/rfc/}}.
The secondary phase calibrators were identified from 
the FIRST (Faint Images of the Radio Sky at Twenty-cm) catalogue \citep{Becker95} or the NVSS (NRAO VLA sky survey) catalogue \citep{Condon98} (for sky regions not covered by the FIRST survey) using a short multi-field observation.
Normally, more than one secondary phase calibrators are identified and observed together with the target. Among them, a main one that is preferably the brightest and the closest to the target is selected to carry out self-calibration; the rest secondary phase calibrators are hereafter referred to as redundant secondary phase calibrators. 
The primary and the main secondary phase calibrators for the astrometry of the 18 MSPs are summarized in \ref{tab:mspsrpi_MSPs}, alongside the project codes. 
At correlation time, pulsar gating was applied \citep{Deller11a} to improve the S/N on the target pulsars. The median values of the gating gain, defined as ${(S/N)_\mathrm{gated}}/{(S/N)_\mathrm{ungated}}$, are provided in \ref{tab:mspsrpi_MSPs}.

\section{Data Reduction and fiducial systematic errors}
\label{sec:mspsrpi_data_reduction}

We reduced all data with the {\tt psrvlbireduce} pipeline\footnote{\label{footnote:parseltongue}available at \url{https://github.com/dingswin/psrvlbireduce}} written in {\tt parseltongue} \citep{Kettenis06}, a {\tt python}-based interface for running functions provided by {\tt AIPS} \citep{Greisen03} and {\tt DIFMAP} \citep{Shepherd94}.
The procedure of data reduction is identical to that outlined in \citet{Ding20}, except for four MSPs --- \psrfa, \psrd, \psro\ and \psrkb.
For \psrfa, the self-calibration solutions acquired with NVSS~J151733$+$491626, a 36-mJy secondary calibrator 13\farcm8 away from the pulsar, are given to both the pulsar and NVSS~J151815$+$491105 --- a 4.5-mJy source about a factor of two closer to \psrfa\ than NVSS~J151733$+$491626. The  positions relative to NVSS~J151815$+$491105 are used to derive the astrometric parameters of \psrfa.
For the other exceptions, the data reduction procedures as well as fiducial systematics estimation are described in \ref{subsec:mspsrpi_dualphscal} and \ref{subsec:mspsrpi_sophisticated_data_reduction}.

At the end of the data reduction, a series of positions as well as their random errors $\sigma_i^\mathcal{R}$ (where $i\!=\!1, 2, 3,...$ refers to right ascension or declination at different epochs) are acquired for each pulsar. 
For each observation, on top of the random errors due to image noise, ionospheric fluctuations would introduce systematic errors that distort and translate the source, the magnitude of which generally increases with the angular separation between a target and its (secondary) phase calibrator \citep[e.g.][]{Chatterjee04,Kirsten15,Deller19}.
%With the exception of \psrd, \psro\ and \psrkb, 
We estimate fiducial values for these systematic errors $\sigma_i^\mathcal{S}$ of pulsar positions using the empirical relation (i.e., Equation~1 of \citealp{Deller19}) derived from the whole \psrpi\ sample.
While this empirical relation has proven a reasonable approximation to the actual systematic errors for a large sample of sources, for an individual observational setup $\sigma_i^\mathcal{S}$ may overstate or underestimate the true systematic error (see \ref{sec:mspsrpi_parameter_inference}). %Accordingly, it is more proper to evaluate the whole position errors with
We can account for our uncertainty in this empirical estimator by re-formulating the total positional uncertainty as
\begin{equation}
\label{eq:mspsrpi_EFAC}
\begin{split}
    \sigma_{i}\left(\eta_\mathrm{EFAC}\right)=\sqrt{(\sigma_i^\mathcal{R})^2+(\eta_\mathrm{EFAC} \cdot \sigma_i^\mathcal{S})^2} \,,
\end{split}
\end{equation}
where $\eta_\mathrm{EFAC}$ is a positive correction factor on the fiducial systematic errors. In this work, we assume $\eta_\mathrm{EFAC}$ stays the same for each pulsar throughout its astrometric campaign.
The inference of $\eta_\mathrm{EFAC}$ is described in \ref{sec:mspsrpi_parameter_inference}. 
We reiterate that the target image frames have been determined by the positions assumed for our reference sources (or virtual calibrators, see \ref{subsec:mspsrpi_dualphscal}), and that any change in the assumed reference source position would transfer directly into a change in the recovered position for the target pulsar.  Accordingly, the uncertainty in the reference source position must be accounted for in the pulsar's reference position error budget, after fitting the pulsar's astrometric parameters.

All pulsar positions and their error budgets are provided in the online\footnote{\label{footnote:pulsar_positions}\url{https://github.com/dingswin/publication_related_materials}} ``pmpar.in.preliminary'' and ``pmpar.in'' files. 
The only difference between ``pmpar.in.preliminary'' and ``pmpar.in'' (for each pulsar) files are the position uncertainties: ``pmpar.in.preliminary'' and ``pmpar.in'' offer, respectively, position uncertainties $\sigma_i(0)=\sigma_i^\mathcal{R}$ and $\sigma_i(1)=\sqrt{(\sigma_i^\mathcal{R})^2+(\sigma_i^\mathcal{S})^2}$. 
%is whether systematic errors are included.
%In the ``pmpar.in'' files, the positions errors $\sigma_i=\sqrt{(\sigma_i^\mathcal{R})^2+(\sigma_i^\mathcal{S})^2}$.
As an example, the pulsar positions for \psri\ are presented in \ref{tab:mspsrpi_pulsar_positions}, where the values on the left and right side of the `` | '' sign stand for, respectively, $\sigma_i(0)$ and $\sigma_i(1)$. 
Additionally, to facilitate reproducibility, the image models for all primary and secondary phase calibrators listed in \ref{tab:mspsrpi_MSPs} are released\footnote{\label{footnote:calibrator_models}\url{https://github.com/dingswin/calibrator_models_for_astrometry}} along with this paper.

\begin{table}
    \centering
    %\caption{Example pulsar positions}
    \caption{An example set of astrometric results for J1738+0333, where the presented uncertainty excludes the calibrator reference point uncertainty as described in the text.}
     
        %\begin{tabular}{p{0.15\linewidth}
        %                p{0.2\linewidth}
        %                p{0.1\linewidth}
        %                p{0.30\linewidth}}
       % \resizebox{\columnwidth}{!}{
    	\begin{tabular}{ccc} % four columns, alignment for each
		\hline
		\hline
	obs. date & $\alpha_\mathrm{J2000}$ (RA.)  & $\delta_\mathrm{J2000}$ (Decl.)   \\
	(yr) & &  \\
		\hline
	2015.6166 & $17^{\rm h}38^{\rm m}53\fs 969242(3|5)$ & $03\degr33'10\farcs90430(9|17)$ \\
	2015.8106 & $17^{\rm h}38^{\rm m}53\fs 969329(3|6)$ & $03\degr33'10\farcs90491(9|18)$ \\
	2016.6939 & $17^{\rm h}38^{\rm m}53\fs 969726(5|6)$ & $03\degr33'10\farcs90981(16|21)$ \\
	2017.1304 & $17^{\rm h}38^{\rm m}53\fs 970000(6|7)$ & $03\degr33'10\farcs91262(21|25)$ \\
	2017.2068 & $17^{\rm h}38^{\rm m}53\fs 970040(2|4)$ & $03\degr33'10\farcs91217(7|15)$ \\
	2017.2860 & $17^{\rm h}38^{\rm m}53\fs 970078(3|5)$ & $03\degr33'10\farcs91307(11|17)$ \\
	2017.2997 & $17^{\rm h}38^{\rm m}53\fs 970062(17|17)$ & $03\degr33'10\farcs91272(59|61)$ \\
	2017.7232 & $17^{\rm h}38^{\rm m}53\fs 970208(15|16)$ & $03\degr33'10\farcs91484(64|74)$ \\
	2017.7669 & $17^{\rm h}38^{\rm m}53\fs 970248(7|8)$ & $03\degr33'10\farcs91466(27|33)$ \\
	\hline 
	\multicolumn{3}{l}{$\bullet$ This table is compiled for \psri.}\\

	\multicolumn{3}{l}{$\bullet$ The values on the left and the right side of `` | '' are, respectively,}\\ 
	\multicolumn{3}{l}{\ \ \ statistical errors given in J1738+0333.pmpar.in.preliminary\textsuperscript{\ref{footnote:pulsar_positions}}, and}\\ 
	\multicolumn{3}{l}{\ \ \ systematics-included errors provided in J1738+0333.pmpar.in\textsuperscript{\ref{footnote:pulsar_positions}}.}
	\end{tabular}
	
    \label{tab:mspsrpi_pulsar_positions}
    
\end{table}

\subsection{1D interpolation on \psrd\ and \psro}
\label{subsec:mspsrpi_dualphscal}

One can substantially reduce propagation-related systematic errors using 1D interpolation with two phase calibrators quasi-colinear with a target \citep[e.g.][]{Fomalont03,Ding20c}. After 1D interpolation is applied, the target should in effect be referenced to a ``virtual calibrator'' much closer (on the sky) than either of the two physical phase calibrators, assuming the phase screen can be approximated by a linear gradient with sky position \citep{Ding20c}.

According to \ref{tab:mspsrpi_MSPs}, 7 secondary phase calibrators (or the final reference sources) are more than 20' away from their targets,
which would generally lead to relatively large systematic errors \citep[e.g.][]{Chatterjee04,Kirsten15,Deller19}.
Fortunately, there are 3 MSPs --- \psrd, \psro\ and \psrkb, for which the pulsar and its primary and secondary phase calibrators are near-colinear (see online\textsuperscript{\ref{footnote:pulsar_positions}} calibrator plans as well as \ref{fig:mspsrpi_J1939_calibration_plan}). 
Hence, by applying 1D interpolation, each of the 3 ``1D-interpolation-capable'' MSPs can be referenced to a virtual calibrator much closer than the physical secondary phase calibrator (see \ref{tab:mspsrpi_MSPs}). 

We implemented 1D interpolation on \psrd\ and \psro\ in the same way as the astrometry of the radio magnetar XTE~J1810$-$197 carried out at 5.7\,GHz \citep{Ding20c}.
Nonetheless, due to our different observing frequency (i.e., 1.55\,GHz), we estimated $\sigma^\mathcal{S}_i$ differently.
The post-1D-interpolation systematic errors should consist of {\bf 1)} first-order residual systematic errors related to the target-to-virtual-calibrator offset $\Delta_\mathrm{psr-VC}$ and {\bf 2)} higher-order terms.
Assuming negligible higher-order terms, we approached post-1D-interpolation $\sigma^\mathcal{S}_i$ with Equation~1 of \citet{Deller19}, using $\Delta_\mathrm{psr-VC}$ as the calibrator-to-target separation. 
The assumption of negligible higher-order terms will be tested later and discussed in \ref{subsubsec:mspsrpi_implications_for_1D_interpolation}.

\subsection{Inverse-referenced 1D interpolation on \psrkb}
\label{subsec:mspsrpi_sophisticated_data_reduction}
%For the sake of brevity while not overusing abbreviations in this catalogue paper, only in \ref{subsec:mspsrpi_sophisticated_data_reduction} we refer to \psrkb\ as J1939.
For \psrkb, normal 1D interpolation \citep{Fomalont03,Ding20c} is still not the optimal calibration strategy. The $\approx$10-mJy (at 1.55\,GHz) \psrkb\ is the brightest MSP in the northern hemisphere and only second to PSR~J0437$-$4715 on the whole sky.
After pulsar gating, \psrkb\ is actually brighter than the brightest secondary reference source NVSS~J194104$+$214913 (hereafter referred to as J194104).
%(see \ref{tab:mspsrpi_MSPs} for the gating gain). 
\psrkb\ is unresolved on VLBI scales, and does not show
%The lack of spatial structure in the \psrkb\ emission precludes 
long-term radio feature variations (frequently seen in quasars), which makes it an ideal secondary phase calibrator.
Both factors encouraged us to implement the inverse-referenced 1D interpolation (or simply inverse 1D interpolation) on \psrkb, where \psrkb\ is the de-facto secondary phase calibrator and the two ``secondary phase calibrators'' serve as the targets.
To avoid confusion, we refer to the two ``secondary phase calibrators'' for \psrkb\ (see \ref{tab:mspsrpi_MSPs}) as secondary reference sources or simply reference sources.

Though inverse phase referencing (without interpolation) has been an observing/calibration strategy broadly used in VLBI astrometry \citep[e.g.][]{Imai12,Yang16,Li18,Deller19}, inverse  interpolation is new, with the 2D approach of \citet{Hyland22} at 8.3 GHz being a recent and independent development.
We implemented inverse 1D interpolation at 1.55\,GHz on \psrkb\ in three steps (in addition to the standard procedure) detailed as follows.

\subsubsection{Tying \psrkb\ to the primary-calibrator reference frame}
\label{subsubsec:mspsrpi_tying_J1939_to_J1935}

Inverse 1D interpolation relies on the residual phase solutions $\Delta\phi_{n}(\vec{x},t)$ of self-calibration on \psrkb\ (where $\vec{x}$, $t$ and $n$ refers to, respectively, sky position, time and the $n$-th station in a VLBI array), which, however, change with $\Delta\vec{x}_\mathrm{psr}$ --- the displacement from the ``true'' pulsar position 
to its model position. 
When $|\Delta\vec{x}_\mathrm{psr}|$ is much smaller than the synthesized beam size $\theta_\mathrm{syn}$, the changes in $\Delta\phi_{n}(\vec{x},t)$ would be equal across all epochs, hence not biasing the resultant parallax and proper motion. 
However, if $|\Delta\vec{x}_\mathrm{psr}| \gtrsim \theta_\mathrm{syn}$, then the phase wraps of $\Delta\phi_{n}(\vec{x},t)$ would likely become hard to uncover.
The main contributor of considerable $|\Delta\vec{x}_\mathrm{psr}|$ is an inaccurate pulsar position. The proper motion of the pulsar would also increase $|\Delta\vec{x}_\mathrm{psr}|$ with time, if it is poorly constrained (or neglected). For \psrkb, the effect of proper motion across our observing duration is small compared to $\theta_\mathrm{syn}\sim10$\,mas.

In order to minimize $|\Delta\vec{x}_\mathrm{psr}|$, we shifted the pulsar model position, on an epoch-to-epoch basis, by $\Delta\vec{x}_\mathrm{cor}$ (which ideally should approximate $-\Delta\vec{x}_\mathrm{psr}$), to the position measured in the J1935 reference frame (see Section~4.1 of \citealp{Ding20c} for explanation of ``reference frame'').
This J1935-frame position was derived with the method for determining pulsar absolute position \citep{Ding20} (where J194104 was used temporarily as the secondary phase calibrator) except that there is no need to quantify the position uncertainty.
We typically found $|\Delta\vec{x}_\mathrm{psr}|\sim50$\,mas, which is well above $\theta_\mathrm{syn}\sim10$\,mas.
After the map centre shift, \psrkb\ becomes tied to the J1935 frame.

\subsubsection{1D interpolation on the tied \psrkb}
\label{subsubsec:mspsrpi_1D_inter_on_tied_J1939}
The second step of inverse 1D interpolation is simply the normal 1D interpolation on \psrkb\ that has been tied to the J1935 frame as described above (in \ref{subsubsec:mspsrpi_tying_J1939_to_J1935}).
When there is only one secondary reference source, optimal 1D interpolation should see the virtual calibrator moved along the interpolation line (that passes through both J1935 and \psrkb) to the closest position to the secondary reference source \citep[e.g.][]{Ding20c}. 
However, there are two reference sources for \psrkb\ (see \ref{tab:mspsrpi_MSPs}), 
and the virtual calibrator point can be set at a point that will enable both of them to be used.

\begin{figure*}
    \centering
	\includegraphics[width=16cm]{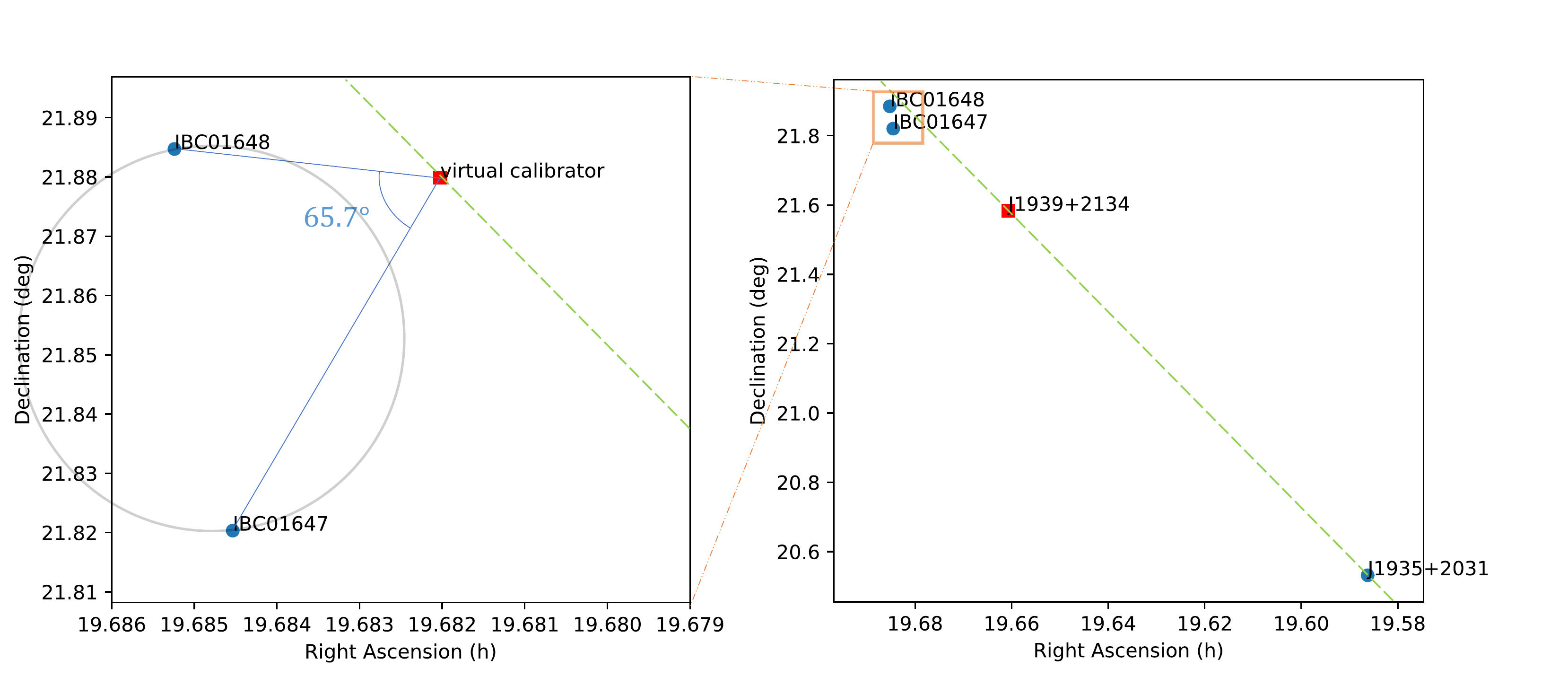}
    \caption{{\bf Right:} The calibrator plan for VLBI astrometry of \psrkb\ (see \ref{tab:mspsrpi_MSPs} for full source names), where \psrkb\ serves as the secondary phase calibrator and J1935 is the primary phase calibrator. {\bf Left:} The zoomed-in field for reference sources as well as the virtual calibrator (VC) along the J1935-to-pulsar line. For the inverse 1D interpolation on \psrkb, we used the VC location that forms the largest included angle ($65\fdg7$) with the two reference sources (see \ref{subsec:mspsrpi_sophisticated_data_reduction} for explanation), which corresponds to $\Delta_{VC-PC}/\Delta_{PC-psr}$=1.2836 (i.e., the VC-to-J1935 separation is 1.2836 times the J1935-to-pulsar separation). 
    %We find that the angular separation between VC and J1935 is 1.2836 times of the separation between J1935 and \psrkb.  %$\Delta_{VC-PC}/\Delta_{PC-psr}$=1.2836, where  %$\angle_\mathrm{J194104-VC-J194106}$ 
    }    
    \label{fig:mspsrpi_J1939_calibration_plan}
\end{figure*}

After calibration, a separate position series can be produced for each reference source.
%, after data reduction involving 1D interpolation. 
While we used each reference-source position series to infer astrometric parameters separately, we can also directly infer astrometric parameters with the combined knowledge of the two position series (which can be realized with $\tt sterne$\footnote{\label{footnote:sterne}\url{https://github.com/dingswin/sterne}}). 
If the errors in the two position series are (largely) uncorrelated, this can provide superior astrometric precision. 
Since position residuals should be spatially correlated, we would ideally set the virtual calibrator at a location such that the included angle between the two reference sources is $90\degr$.
While achieving this ideal is not possible, we chose a virtual calibrator location that forms the largest possible included angle (65\fdg7) with the two reference sources to minimise spatially correlated errors (see \ref{fig:mspsrpi_J1939_calibration_plan}).
This virtual calibrator is 1.2836 times further away from J1935 than \psrkb. 
Accordingly, the $\Delta\phi_{n}(\vec{x},t)$ solutions (obtained from the self-calibration on the tied \psrkb) were multiplied by 1.2836, and applied to the two reference sources.

\subsubsection{De-shifting reference source positions}
After data reduction involving the two steps outlined in \ref{subsubsec:mspsrpi_tying_J1939_to_J1935} and \ref{subsubsec:mspsrpi_1D_inter_on_tied_J1939}, one position series was acquired for each reference source. At this point, however, the two position series are not yet ready for astrometric inference, mainly because both proper motion and parallax signatures have been removed in the first step (see \ref{subsubsec:mspsrpi_tying_J1939_to_J1935}) when \psrkb\ was shifted to its J1935-frame position. Therefore, the third step of inverse 1D interpolation is to cancel out the \psrkb\ shift (made in the first step) by moving reference source positions by $-1.2836 \cdot \Delta\vec{x}_\mathrm{cor}$, where the multiplication can be understood by considering Figure~1 of \citealp{Ding20c}.
This de-shifting operation was carried out separately outside the data reduction pipeline\textsuperscript{\ref{footnote:parseltongue}}. After the operation, we estimated $\sigma_{ij}^\mathcal{S}$ of the reference sources (where $j\!=\!1, 2$ refers to an individual reference source) following the method described in \ref{subsec:mspsrpi_dualphscal}.
The final position series of the reference sources are available online\textsuperscript{\ref{footnote:pulsar_positions}}. The astrometric parameter inference based on these position series is 
outlined in \ref{sec:mspsrpi_parameter_inference}.

\section{Astrometric inference methods and quasi-VLBI-only astrometric results}
\label{sec:mspsrpi_parameter_inference}

After gathering the position series\textsuperscript{\ref{footnote:pulsar_positions}} with basic uncertainty estimation (see \ref{sec:mspsrpi_data_reduction}), we proceed to infer the astrometric parameters. 
The inference is made by three different methods: {\bf a)} direct fitting of the position series with {\tt pmpar}\footnote{\label{footnote:pmpar}\url{https://github.com/walterfb/pmpar}}, {\bf b)} bootstrapping (see \citealp{Ding20}) and {\bf c)} Bayesian analysis using {\tt sterne}\textsuperscript{\ref{footnote:sterne}} (see \citealp{Ding21a}).
The two former methods directly adopt $\sigma_i(1)=\sqrt{(\sigma_i^\mathcal{R})^2+(\sigma_i^\mathcal{S})^2}$ as the position errors. 
In Bayesian analysis, however, we followed \citet{Lentati14} and inferred $\eta_\mathrm{EFAC}$ along with other model parameters using the likelihood terms
\begin{equation}
\label{eq:mspsrpi_probability}
\begin{split}
    P_1 \propto \left(\prod_{i} \sigma_i\right)^{-1} \exp{\left[-\frac{1}{2} \sum_i \left(\frac{\Delta \epsilon_i}{\sigma_i}\right)^2\right]} \,,
\end{split}
\end{equation}
where $\sigma_i=\sigma_i(\eta_\mathrm{EFAC})$ obeys \ref{eq:mspsrpi_EFAC}; $\Delta \epsilon_i$ refers to the model offsets from the measured positions. 
To prevent possible confusion, we note here that $\eta_\mathrm{EFAC}$ is defined differently from the ``EFAC'' specified by Equation~10 of \citet{Lentati14}.
As is discussed in \ref{subsec:mspsrpi_Bayesian_as_major}, Bayesian inference outperforms the other two methods, and is hence consistently used to present final results in this work.  In all cases, the uncertainty in the reference source position should be added in quadrature to the uncertainty in the pulsar's reference position acquired with any method (of the three), in order to obtain a final estimate of the absolute positional uncertainty of the pulsar.

To serve different scientific purposes, we present two sets of astrometric results in two sections (i.e., \ref{sec:mspsrpi_parameter_inference} and \ref{sec:mspsrpi_inference_with_priors}), which differ in whether timing proper motions and parallaxes are used as prior information in the inference.

\subsubsection{Priors of canonical model parameters used in Bayesian analysis}
\label{subsubsec:mspsrpi_parameter_priors}
To facilitate reproduction of our Bayesian results, the priors (of Bayesian inference) we use for canonical model parameter are detailed as follows. Priors for the two orbital parameters can be found in \ref{subsec:mspsrpi_reflex_motion_inference}. We universally adopt the prior uniform distribution $\mathcal{U}$(0, 15) (i.e., uniformly distributed between 0 and 15) for $\eta_\mathrm{EFAC}$. This prior distribution can be refined for future work with an ensemble of results across many pulsars.
With regard to the canonical astrometric parameters (7 parameters for \psrkb\ and 5 for the other pulsars), we adopt $\mathcal{U}\left(X_0^\mathrm{(DF)}-20~\tilde{\sigma}_X^\mathrm{(DF)},~~ X_0^\mathrm{(DF)}+20~ \tilde{\sigma}_X^\mathrm{(DF)}\right)$ for each $X$, where $X$ refers to one of $\alpha_\mathrm{ref}$, $\delta_\mathrm{ref}$, $\mu_\alpha$, $\mu_\delta$ and $\varpi$. Here, $X_0^\mathrm{(DF)}$ stands for the direct-fitting estimate of $X$; $\tilde{\sigma}_X^\mathrm{(DF)}$ represents the
direct-fitting error corrected by the reduced chi-square $\chi^2_\nu$ (see \ref{tab:mspsrpi_astrometric_parameters}) with $\tilde{\sigma}_X^\mathrm{(DF)} \equiv \sigma_{X}^\mathrm{(DF)} \cdot \sqrt{\chi^2_\nu}$.
The calculation of prior range of $X$ is made with the function $\tt sterne.priors.generate\_initsfile$\textsuperscript{\ref{footnote:sterne}}.

\subsection{Astrometric inference disregarding orbital motion}
\label{subsec:mspsrpi_non_reflex_motion_inference}

\subsubsection{Single-reference-source astrometric inferences}

All MSPs (in this work) excepting \psrkb\ have only one reference source. For each of these single-reference-source MSPs, we fit for the five canonical astrometric parameters, i.e., reference position ($\alpha_\mathrm{ref}$ and $\delta_\mathrm{ref}$), proper motion ($\mu_\alpha \equiv \dot{\alpha} \cos{\delta}$ and $\mu_\delta$) and parallax ($\varpi$). In the Bayesian analysis alone, $\eta_\mathrm{EFAC}$ is also inferred alongside the astrometric parameters.  At this stage, we neglect any orbital reflex motion for binary pulsars -- the effects of orbital reflex motion are addressed in \ref{subsec:mspsrpi_reflex_motion_inference}.
The proper motions and parallaxes derived with single-reference-source astrometry and disregarding orbital motion are summarized in \ref{tab:mspsrpi_astrometric_parameters}. 
The reference positions are presented in \ref{subsec:mspsrpi_astrometric_results_non_PM_priors}.

\begingroup
\renewcommand{\arraystretch}{1.5} % Default value: 1

\begin{table*}
\raggedright
\caption{Proper motion $\left( \mu_\alpha,\mu_\delta \right)$ and parallax $\varpi$ from astrometry inferences disregarding orbital motion.}
\label{tab:mspsrpi_astrometric_parameters}
%\begin{tabular}{@{}l@{\:}l@{\:}l@{}} % manual @ spacing to prevent this being too wide for a page
\resizebox{\textwidth}{!}{
\begin{tabular}{lcccllcccccccc}
\hline
\hline
PSR & $\mu_\alpha^\mathrm{(DF)}$ & $\mu_\delta^\mathrm{(DF)}$ & $\varpi^\mathrm{(DF)}$ & \rcs\ & $\mu_\alpha^\mathrm{(Bo)}$ & $\mu_\delta^\mathrm{(Bo)}$ & $\varpi^\mathrm{(Bo)}$ & $\mu_\alpha^\mathrm{(Ba)}$ & $\mu_\delta^\mathrm{(Ba)}$ & $\varpi^\mathrm{(Ba)}$ & $\eta_\mathrm{EFAC}$ & $\eta_\mathrm{orb}$ & $P_\mathrm{b}$\\
 & (\maspy) & (\maspy) & (mas) & & (\maspy) & (\maspy) & (mas) & (\maspy) & (\maspy) & (mas) & & & (d)\\

\hline
\multicolumn{14}{c}{Non-1D-interpolated results}\\
\hline

{\Psrb}  & -6.13(4) & 0.33(9) & 3.02(4) & 1.4 & -6.1(1) & $0.34^{+0.09}_{-0.08}$ & $3.02^{+0.09}_{-0.08}$ &   -6.13(7) & $0.34^{+0.15}_{-0.16}$ & 3.02(7) & $1.35^{+0.45}_{-0.32}$ & --- & ---\\
\Psrc & 9.11(8) & 15.9(2) & 0.74(7) & 1.0 & $9.10^{+0.06}_{-0.05}$ & 15.9(2) & $0.73^{+0.05}_{-0.04}$ & 9.1(1) &  $15.96^{+0.25}_{-0.24}$ &  0.73(10) & $1.1^{+0.4}_{-0.2}$ & $3\!\times\!10^{-3}$ & 0.29\\ 
\Psrd & 3.51(9) & -1.32(16) & 0.88(7) & 3.7 & $3.5^{+0.3}_{-0.4}$ & -1.3(2) & $0.9^{+0.2}_{-0.3}$ & 3.4(2) & -1.3(5) & 0.85(21) & $2.33^{+0.56}_{-0.44}$ & 0.4 & 8.3\\ 
\Psrea $\,^*$ &  2.68(3) & -25.38(6) & 1.17(2) & 1.9 & $2.67^{+0.13}_{-0.06}$ & -25.39(12) & $1.18^{+0.05}_{-0.06}$ & 2.67(5) & $-25.39^{+0.14}_{-0.15}$ & $1.17^{+0.04}_{-0.05}$ & $1.7^{+0.6}_{-0.4}$ & 0.1 & 0.60 \\

\Psreb & -35.32(4) & -48.2(1) & 0.94(3) & 1.1 & $-35.32^{+0.05}_{-0.04}$ & -48.2(1) & $0.94^{+0.07}_{-0.06}$ &  -35.32(7) & -48.1(2) & 0.94(6) & $1.2^{+0.4}_{-0.3}$ & --- & ---\\

{\Psrfa} & -0.69(2) & -8.54(4) & 1.24(2) & 1.2 & -0.69(3) & $-8.53^{+0.10}_{-0.07}$ & 1.25(3) & -0.69(3) & -8.52(7) & 1.24(3) & $1.2^{+0.4}_{-0.3}$ & 4.9 & 8.6 \\

\Psrfb $\,^*$ & 1.51(2) & -25.31(5) & 1.06(7) & 0.81 & 1.51(2) & $-25.31^{+0.04}_{-0.05}$ & $1.07^{+0.09}_{-0.08}$ & 1.51(3) & $-25.30^{+0.05}_{-0.06}$ & $1.06^{+0.11}_{-0.10}$ & $0.54^{+0.57}_{-0.38}$ & 0.24 & 0.42 \\

\Psrga $\,^*$ & 2.199(56) & -11.29(9) & 0.676(46) & 1.1 & $2.20^{+0.06}_{-0.07}$ & -$11.29^{+0.15}_{-0.17}$ & $0.68^{+0.07}_{-0.06}$ & 2.20(9) & $-11.29^{+0.16}_{-0.14}$ &  0.68(7) & $1.29^{+0.66}_{-0.54}$ & 3.1 & 175 \\

{\Psrgb} & 6.2(2) & 3.3(5) & 1.3(1) & 0.8  & $6.1^{+0.7}_{-0.1}$ & 3.3(4) & $1.3^{+0.2}_{-0.5}$ & 6.2(2) &  3.3(6) & 1.33(18) & $1.0^{+0.3}_{-0.2}$ & 1.1 & 147\\
\Psrha & 2.5(1) & -1.9(3) & 0.02(7) & 4.4  & 2.5(6) & $-1.7^{+0.4}_{-0.7}$ & $0.2^{+0.2}_{-0.3}$ & 2.5(3) &  -1.9(9) & 0.0(2) & $3.1^{+0.8}_{-0.6}$ & --- & ---\\
{\Psrhb} & 20.3(1) & -4.79(26) & 1.56(9) & 1.7 & $20.32^{+0.18}_{-0.15}$ & $-4.80^{+0.33}_{-0.35}$ & $1.56^{+0.15}_{-0.17}$ & 20.3(2) & -4.8(5) & $1.57(18)$ & $1.4^{+0.3}_{-0.2}$ & --- & ---\\

\Psri & 6.97(4) & 5.18(7) & 0.51(3) & 1.8 & $7.00^{+0.06}_{-0.11}$ & 5.2(1) & $0.50^{+0.07}_{-0.06}$ & 6.98(8) & 5.18(16) & 0.50(6) & $1.9^{+0.7}_{-0.6}$ & 0.02 & 0.35\\ 

\Psro & -0.03(49) & -6.6(1.2) & -0.25(37) & 0.8 & $-0.03^{+0.28}_{-0.79}$ & $-6.8^{+1.3}_{-1.5}$ & $-0.20^{+0.54}_{-0.39}$ & -0.1(6) & -6.6(1.5) & -0.22(48) & $0.9^{+0.4}_{-0.3}$ & ---& ---\\

\Psrka & -1.37(9) & -2.8(2) & 0.49(6) & 1.0  & $-1.39^{+0.13}_{-0.28}$ & -2.8(2) & $0.48^{+0.07}_{-0.13}$ &  -1.36(10) & -2.8(2) & 0.50(6) & $0.45^{+0.46}_{-0.30}$ & 1.4 & 116\\

\Psrl & 0.49(8) & -6.85(15) & 0.26(7) & 0.15 & $0.49^{+0.07}_{-0.08}$ & -6.85(6) & $0.26^{+0.06}_{-0.02}$ & 0.50(4) & -6.85(9) & 0.25(3) & $0.19^{+0.15}_{-0.12}$ & 0.87 & 58.5\\
\Psrma & -13.8(1) & -10.3(2) & 0.38(9) & 0.9 & $-13.76^{+0.06}_{-0.08}$ & $-10.3^{+1.1}_{-0.3}$ & 0.36(8) &  -13.8(2) & -10.3(4) & $0.38^{+0.13}_{-0.14}$ & $1.1^{+0.4}_{-0.3}$ & $3\!\times\!10^{-4}$ & 2.7\\
\Psrmb & -7.12(8) & -5.7(2) & 0.60(7) & 1.1 & -7.1(1) & $-5.8^{+0.3}_{-0.2}$ & $0.60^{+0.12}_{-0.08}$ & -7.1(1) & -5.7(3) & 0.60(12) & $1.2^{+0.3}_{-0.2}$ & 0.27 & 10.9\\ 
\Psrkb $^{\,{(i)}}$ & 0.07(14) & -0.24(24) & 0.35(10) & 0.2 &  $0.06^{+0.36}_{-0.45}$ & $-0.24^{+0.10}_{-0.06}$ & $0.34^{+0.08}_{-0.25}$ & 0.08(10) & $-0.23^{+0.17}_{-0.16}$ & 0.34(7) & $0.45^{+0.15}_{-0.11}$ & --- & ---\\

\hline
\multicolumn{14}{c}{Single-reference-source 1D-interpolated results}\\
\hline

\Psrd & 3.68(6) & -1.33(9) & 0.94(4) & 6.3 & $3.53^{+0.36}_{-0.35}$ & -1.3(2) & $0.9^{+0.2}_{-0.3}$ & 3.5(2) &  -1.37(35) & 0.86(15) & $3.7^{+1.0}_{-0.7}$ & 0.4 & 8.3\\
\Psro & 0.3(3) & -3.7(6) & 0.1(3) & 1.5 & $0.4^{+0.5}_{-1.2}$ & $-4^{+1}_{-2}$ & $0.1^{+0.7}_{-0.4}$ & 0.3(6) & $-3.9^{+1.2}_{-1.3}$ & 0.1(5) & $1.6^{+0.8}_{-0.7}$ & ---& ---\\
\Psrkb $^{\,{(ii)}}$ & 0.08(4) & -0.45(6) & 0.38(3) & 1.3 &  $0.08^{+0.34}_{-0.07}$ & $-0.44^{+0.09}_{-0.12}$ & $0.36^{+0.05}_{-0.22}$ & 0.08(7) & -0.44(11) & $0.380^{+0.048}_{-0.049}$ & $1.4^{+0.7}_{-0.5}$ & --- & ---\\
\Psrkb $^{\,{(iii)}}$ & 0.3(3) & -0.3(2) & 0.36(19) & 1.3 & $0.2^{+1.0}_{-0.2}$ & -0.3(4) & $0.36^{+0.35}_{-0.58}$ & 0.3(4) & -0.3(4) & $0.38^{+0.29}_{-0.28}$ & $3.7^{+3.4}_{-2.6}$ & --- & ---\\ 

\hline
\multicolumn{14}{c}{Multi-reference-source 1D-interpolated results}\\
\hline

\Psrkb $^{\,{(iv)}}$ & --- & ---- & --- & --- &  --- & --- & --- & 0.08(7) & -0.43(11) & $0.384^{+0.048}_{-0.046}$ & $1.5^{+0.7}_{-0.6}$ & --- & ---\\

\hline
\multicolumn{14}{l}{$\bullet$ ``DF'', ``Bo'' and ``Ba'' stands for, respectively, direct fitting, bootstrap and Bayesian inference. \rcs\ is the reduced chi-square of direct fitting using {\tt pmpar}\textsuperscript{\ref{footnote:pmpar}}.}\\
\multicolumn{14}{l}{$\bullet$ The middle and top block presents, respectively, 1D-interpolated (see \ref{subsec:mspsrpi_dualphscal} and \ref{subsec:mspsrpi_sophisticated_data_reduction}) and non-1D-interpolated results.}\\
%\multicolumn{14}{l}{$\bullet$ All bootstrap and Bayesian results adopt the 50th, 16th and 84th percentile of the value chain as, respectively, the estimate and its 1-$\sigma$ error lower and upper bound.}\\
\multicolumn{14}{l}{$\bullet$ The bottom entry for \psrkb\ shows the result of multi-reference-source astrometry inference (see \ref{subsubsec:mspsrpi_multi-source-inference}).}\\
%\multicolumn{14}{l}{$\bullet$ For each pulsar, the set of astrometric results we adopt are boldfaced.}\\
\multicolumn{14}{l}{$\bullet$ $P_\mathrm{b}$ represents orbital period (see Table~4 for their references). $\eta_\mathrm{orb}$ is defined in \ref{eq:mspsrpi_reflex_motion_measurability}.}\\
%\multicolumn{14}{l}{$^{*}$ As 1D interpolation is applied, systematic errors are estimated in the way described in \ref{subsec:mspsrpi_dualphscal}.}\\
\multicolumn{14}{l}{$^*$ Already published in \citet{Vigeland18,Ding20,Ding21a}.}\\
\multicolumn{14}{l}{$^{(i)}$ Based on (non-1D-interpolated) J194104 positions inverse-referenced to \psrkb.}\\
\multicolumn{14}{l}{$^{(ii)}$ Using 1D-interpolated J194104 positions inverse-referenced to \psrkb.}\\
\multicolumn{14}{l}{$^{(iii)}$ Using 1D-interpolated J194106 positions inverse-referenced to \psrkb.}\\ 
\multicolumn{14}{l}{$^{(iv)}$ Based on 1D-interpolated J194104 and J194106 positions inverse-referenced to \psrkb\ (see \ref{subsec:mspsrpi_sophisticated_data_reduction}).}\\

\end{tabular}
}
\end{table*}
\endgroup

\subsubsection{Multi-source astrometry inferences}
\label{subsubsec:mspsrpi_multi-source-inference}
When multiple sources share proper motion and/or parallax (while each source having its own reference position), a joint multi-source astrometry inference can increase the degrees of freedom of inference (i.e., the number of measurements reduced by the number of parameters to infer), and tighten constraints on the astrometric parameters.
Multi-source astrometry inference has been widely used in maser astrometry (where maser spots with different proper motions scatter around a region of high-mass star formation, \citealp{Reid09a}), but has not yet been used for any pulsar, despite the availability of several bright pulsars with multiple in-beam calibrators (e.g., PSR~J0332$+$5434, PSR~J1136$+$1551) in the \psrpi\ project \citep{Deller19}. 
%despite the technique to use a combined reference source  multiple faint ($\sim3$\,mJy) reference sources have been combined to serve as 

\psrkb\ is the only source (in this work) that has multiple (i.e., two) reference sources, which provides a rare opportunity to test multi-reference-source astrometry.
%In multi-reference-source astrometry inference, all reference sources are expected to share the same proper motion and parallax, . That is to say, for 
%Unlike the single-reference-source inference, each reference position
We assumed that the position series of J194104 is uncorrelated with that of NVSS~J194106$+$215304 (hereafter J194106), and utilized {\tt sterne}\textsuperscript{\ref{footnote:sterne}} to infer the common parallax and proper motion, alongside two reference positions (one for each reference source). The acquired proper motion and parallax are listed in \ref{tab:mspsrpi_astrometric_parameters}. 
As inverse phase referencing is applied for \psrkb, the parallax and proper motion of \psrkb\ are the inverse of the direct astrometric measurements.
For comparison, the proper motion and parallax inferred solely with one reference source are also reported in \ref{tab:mspsrpi_astrometric_parameters}. 
Due to the relative faintness of J194106 (see \ref{tab:mspsrpi_MSPs}), the inclusion of J194106 only marginally improves the astrometric results (e.g., $\varpi$) over those inferred with J194104 alone. 

\begin{figure}
    \centering
	%\raggedright
	\includegraphics[width=14cm]{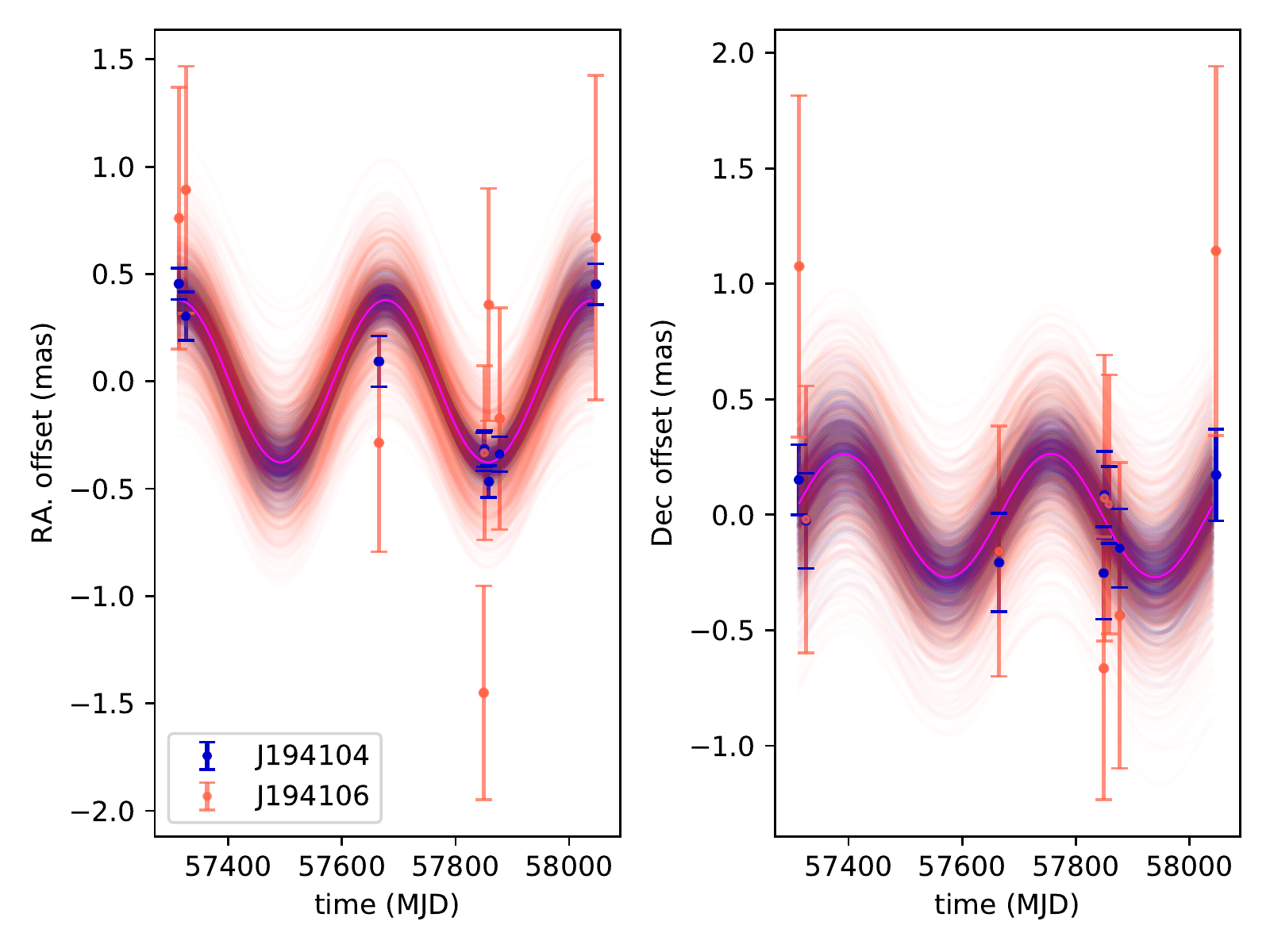}
    \caption{The common parallax signature of \psrkb\ revealed by the position measurements of both reference sources (see \ref{tab:mspsrpi_MSPs}). In both panels, the best-fit proper motion has been subtracted. The magenta curve in each panel represents the best-inferred astrometric model. The fuzzy region around the curve consists of various Bayesian simulations, the scatter of which can visualize the uncertainty level of the underlying model parameters (see \ref{subsubsec:mspsrpi_multi-source-inference}). As a result of the inverse referencing, the common parallax revealed here is actually the negative of the \psrkb\ parallax presented in \ref{tab:mspsrpi_astrometric_parameters}. 
    }    
    \label{fig:mspsrpi_J1939_parallax}
\end{figure}

The constraints on the parallax (as well as the proper motion) are visualized in \ref{fig:mspsrpi_J1939_parallax}. The best-inferred model (derived from the J194104 and J194106 positions) is illustrated with a bright magenta curve, amidst two sets of Bayesian simulations --- each set for a reference source.
Each simulated curve is a time series of simulated positions, with the best-inferred reference position ($\alpha_{\mathrm{ref},j}$ and $\delta_{\mathrm{ref},j}$, where $j$ refers to either J194104 or J194106) and proper-motion-related displacements (i.e., $\mu_\alpha \Delta t$ and $\mu_\delta \Delta t$, where $\Delta t$ is the time delay from the reference epoch) subtracted. As the simulated curve depends on the underlying model parameters, the degree of scatter of simulated curves would increase with larger uncertainties of model parameters. 
%In other words, broader scatter of simulated curves would indicate less constraints a series of position measurements have imposed on the model parameters.
Though sharing simulated parallaxes and proper motions with J194104, the simulated curves for J194106 exhibits broader scatter (than the J194104 ones) owing to more uncertain reference position (see \ref{subsec:mspsrpi_astrometric_results_non_PM_priors} for $\alpha_{\mathrm{ref,J194106}}$ and $\delta_{\mathrm{ref,J194106}}$).
The large scatter implies that the J194106 position measurements impose  relatively limited constraints on the common model parameter (i.e., parallax and proper motion), which is consistent with the findings from \ref{tab:mspsrpi_astrometric_parameters}.

\subsubsection{Implications for 1D/2D interpolation}
\label{subsubsec:mspsrpi_implications_for_1D_interpolation}
On the three 1D-interpolation-capable MSPs, we compared astrometric inference with both the 1D-interpolated and non-1D-interpolated position series (one at a time).
For \psrkb, the $\eta_\mathrm{EFAC}$ of the three 1D-interpolated realizations are consistent with each other, but larger than the non-1D-interpolated counterpart. This post-1D-interpolation inflation of $\eta_\mathrm{EFAC}$ also occurs to the other two 1D-interpolation-capable pulsars (see \ref{tab:mspsrpi_astrometric_parameters}), which suggests the post-1D-interpolation fiducial systematic errors $\sigma^\mathcal{S}_i$ might be systematically under-estimated. One obvious explanation for this under-estimation is that the higher-order terms of systematic errors are non-negligible (as opposed to the assumption we started with in \ref{subsec:mspsrpi_dualphscal}): they might be actually comparable to the first-order residual systematic errors (that are related to $\Delta_\mathrm{psr-VC}$) at the $\sim1.55$\,GHz observing frequencies.

On the other hand, the astrometric results based on the non-1D-interpolated J194104 positions inverse-referenced to \psrkb\ are less precise than the 1D-interpolated counterpart by $\approx40$\% , as is also the case for \psrd\ (see \ref{tab:mspsrpi_astrometric_parameters}). Moreover, the post-1D-interpolation parallax of \psro\ becomes relatively more accurate than the non-1D-interpolation negative counterpart. All of these demonstrate the utility of 1D/2D interpolation, even in the scenario of in-beam astrometry that is already precise. %Both post-1D-interpolation and non-1D-interpolation astrometric results will be further tested in \ref{sec:mspsrpi_inference_with_priors}.
In the remainder of this paper, we only focus on the 1D-interpolated astrometric results for the three 1D-interpolation-capable MSPs.

\subsection{Bayesian inference as the major method for \mspsrpi}
\label{subsec:mspsrpi_Bayesian_as_major}
We now compare the three sets of astrometric parameters (in \ref{tab:mspsrpi_astrometric_parameters}) obtained with different inference methods, and seek to proceed with only one set in order to simplify the structure of this paper.
Among the three inference methods we use in this work, direct least-square fitting is the most time-efficient, but is also the least robust against improperly estimated positional uncertainties. Conversely, the other two methods (i.e., bootstrap and Bayesian methods) do not rely solely on the input positional uncertainties, and can still estimate the model parameters and their uncertainties $\sigma_X^{(Y)}$ ($X\!=\!\mu_{\alpha}, \mu_{\delta}$ or $\varpi$; $Y\!=$\,``Bo'' or ``Ba'') more robustly in the presence of incorrectly estimated positional errors. 

\begin{figure*}
    \centering
	%\raggedright
	\includegraphics[width=15cm]{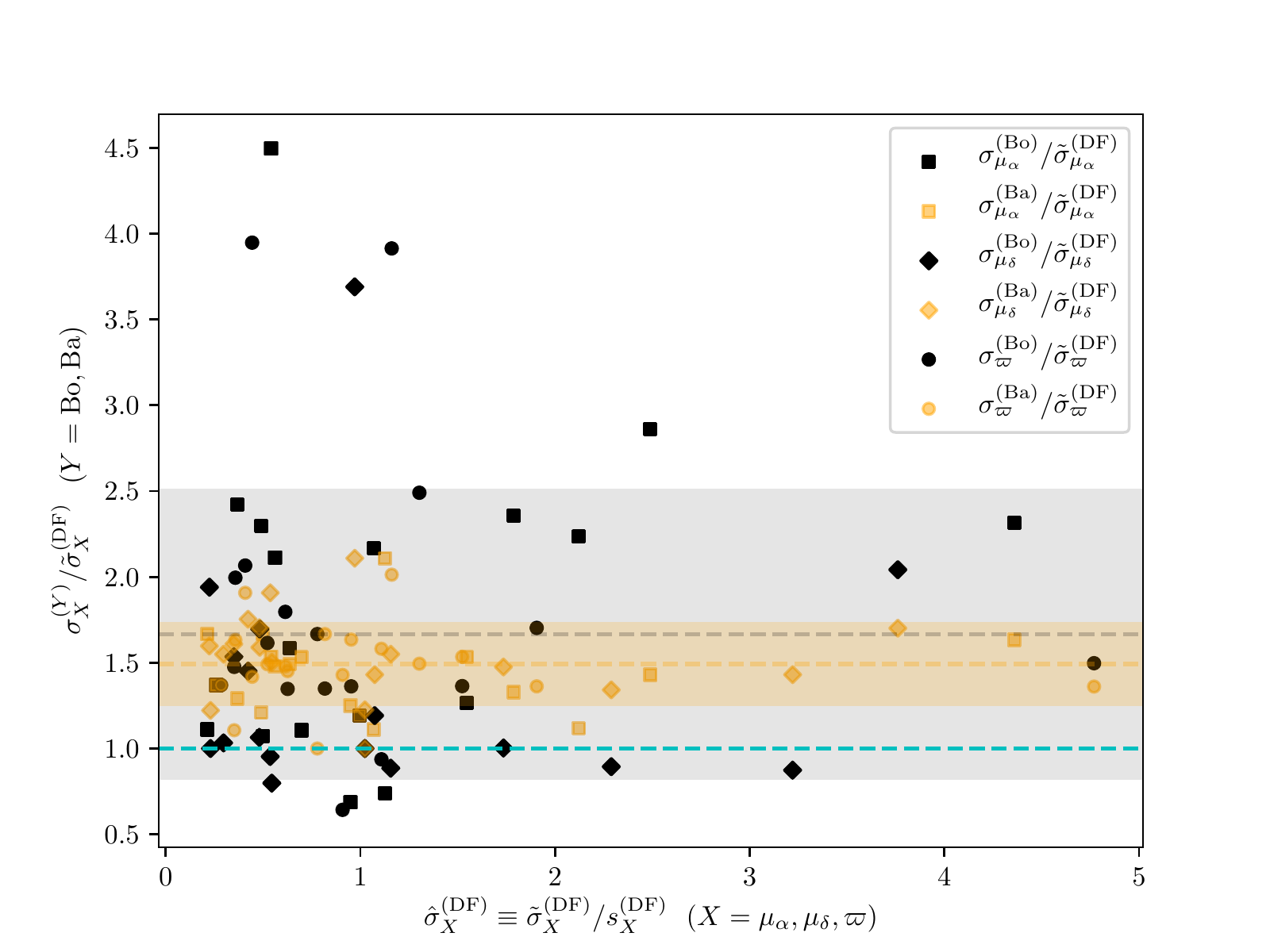}
    \caption{Bootstrap (denoted as ``Bo'') and Bayesian (``Ba'') errors (of three inferred parameters) divided by the corresponding \rcs-corrected direct-fitting errors. Here, $\tilde{\sigma}_X^{(DF)} \equiv \sigma_X^{(DF)} \cdot \sqrt{\chi^2_\nu}$ represents the $\chi^2_\nu$-corrected errors of direct fitting, where $X$ stands for one of the $\mu_\alpha$, $\mu_\delta$ and $\varpi$ groups. The dimensionless $\hat{\sigma}_X^{(DF)}$ is defined as an individual $\tilde{\sigma}_X^{(DF)}$ divided by the standard deviation $s_X^{(DF)}$ for all $\tilde{\sigma}_X^{(DF)}$ of the group $X$. The grey and orange shaded regions show, respectively, the standard deviation of $\sigma_X^\mathrm{(Bo)}/\tilde{\sigma}_X^\mathrm{(DF)}$ and $\sigma_X^\mathrm{(Ba)}/\tilde{\sigma}_X^\mathrm{(DF)}$ across all of the three groups (i.e., $\mu_\alpha$, $\mu_\delta$ and $\varpi$) around the respective mean value outlined with the grey and orange dashed lines.
    Both bootstrap and Bayesian errors are generally slightly higher than the level of direct-fitting errors illustrated with the cyan dashed line, and are well consistent with each other as anticipated. Despite the consistency, bootstrap errors show larger scatter than Bayesian ones. %Therefore, an individual bootstrap error estimation is less reliable.
    }    
    \label{fig:mspsrpi_Bo_errs_vs_Ba_errs}
\end{figure*}

Generally speaking, $\sigma_X^{(Y)}$ inferred from a pulsar position series are expected to change with the corresponding \rcs-corrected direct-fitting error $\tilde{\sigma}_X^\mathrm{(DF)} \equiv \sigma_{X}^\mathrm{(DF)} \cdot \sqrt{\chi^2_\nu}$.
%$\sigma_X^{(Y)}$ grows with larger $\tilde{\sigma}_X^\mathrm{(DF)}$.
In order to investigate the relation between $\sigma_X^{(Y)}$ and $\tilde{\sigma}_{X}^\mathrm{(DF)}$, we divided $\sigma_X^{(Y)}$ by $\tilde{\sigma}_{X}^\mathrm{(DF)}$ for each pulsar entry in the top block of \ref{tab:mspsrpi_astrometric_parameters}. 
The results are displayed in \ref{fig:mspsrpi_Bo_errs_vs_Ba_errs}. 
For the convenience of illustration, we calculated the dimensionless $\tilde{\sigma}_{X}^\mathrm{(DF)}$ defined as $\tilde{\sigma}_{X}^\mathrm{(DF)}/s_X^{(DF)}$ (where $s_X^{(DF)}$ represents  the standard deviation of $\tilde{\sigma}_{X}^\mathrm{(DF)}$ over the group $X$), which allows all the three sets (i.e., $\mu_{\alpha}$, $\mu_{\delta}$ and $\varpi$) of dimensionless $\tilde{\sigma}_{X}^\mathrm{(DF)}$ to be horizontally more evenly plotted in \ref{fig:mspsrpi_Bo_errs_vs_Ba_errs}. 

%Based on the whole \mspsrpi\ sample, we find out that $\sigma_X^{(Y)}$ is quasi-linear to $\tilde{\sigma}_X^\mathrm{(DF)}$. In other words, $\sigma_X^{(Y)}$ nearly scales with $\tilde{\sigma}_X^\mathrm{(DF)}$.
Across the entire \mspsrpi\ sample, we see that $\sigma_X^{(Y)}$ scales with $\tilde{\sigma}_X^\mathrm{(DF)}$ in a near-linear fashion.
%, as expected.
The mean scaling factors across all of the three parameter groups (i.e., $\mu_{\alpha}$, $\mu_{\delta}$ and $\varpi$) are $\left<\sigma^\mathrm{(Bo)}_X / \tilde{\sigma}_X^\mathrm{(DF)}\right>=1.67\pm 0.85$ and $\left<\sigma^\mathrm{(Ba)}_X / \tilde{\sigma}_X^\mathrm{(DF)}\right>=1.49\pm0.24$ (see \ref{fig:mspsrpi_Bo_errs_vs_Ba_errs}). The two mean scaling factors show that parameter uncertainties inferred using either a bootstrap or Bayesian approach will be slightly higher (and on average, consistent between the two approaches) than would be obtained utilising direct-fitting (illustrated with the cyan dashed line in \ref{fig:mspsrpi_Bo_errs_vs_Ba_errs}).

The more optimistic uncertainty predictions of $\tilde{\sigma}_X^\mathrm{(DF)}$ can be understood as resulting from two causes: first, it neglects both the finite width and the skewness of the $\chi^2$ distribution, and second, to achieve the expected $\chi^2$ it scales the {\em total} uncertainty contribution at each epoch, rather than the systematic uncertainty contribution alone.  When (as is typical for pulsar observations) the S/N and hence statistical positional precision can vary substantially between observing epochs, this simplified approach preserves the relative weighting between epochs, whereas increasing the estimated systematic uncertainty contribution acts to equalise the weighting between epochs (by reducing the position precision more for epochs where the pulsar was bright and the statistical precision high, than for epochs where the pulsar was faint and the statistical precision is already low).  

While the consistency between $\left<\sigma^\mathrm{(Bo)}_X / \tilde{\sigma}_X^\mathrm{(DF)}\right>$ and $\left<\sigma^\mathrm{(Ba)}_X / \tilde{\sigma}_X^\mathrm{(DF)}\right>$ suggests that both approaches can overcome this shortcoming in the direct fitting method, $\sigma^\mathrm{(Bo)}_X / \tilde{\sigma}_X^\mathrm{(DF)}$ shows a much larger scatter (3.5 times) compared to $\sigma^\mathrm{(Ba)}_X / \tilde{\sigma}_X^\mathrm{(DF)}$ (see \ref{fig:mspsrpi_Bo_errs_vs_Ba_errs}). To determine which approach best represents the true (and unknown) parameter uncertainties, it is instructive to consider the outliers in the bootstrap distribution results.  

First, consider cases where the bootstrap results in a lower uncertainty than $\tilde{\sigma}_X^\mathrm{(DF)}$.  
For the reasons noted above, we expect $\tilde{\sigma}_X^\mathrm{(DF)}$ to yield the most optimistic final parameter uncertainty estimates, and yet the bootstrap returns a lower uncertainty than $\tilde{\sigma}_X^\mathrm{(DF)}$ in a number of cases.
Second, the cases with the highest values of $\sigma^\mathrm{(Bo)}_X / \tilde{\sigma}_X^\mathrm{(DF)}$ reach $\gtrsim$3 on a number of occasions, which imply an extremely large (or very non-Gaussian) systematic uncertainty contribution, which would lead (in those cases) to a surprisingly low reduced $\chi^2$ for the best-fitting model.
Given the frequency with which these outliers arise, we regard it likely that bootstrap approach mis-estimates parameter uncertainties at least occasionally, likely due to the small number of observations available.
Therefore, we consider the Bayesian method described in this paper as the preferred inference method for the \mspsrpi\ sample, and consistently use the Bayesian results in the following discussions.
We note that as continued VLBI observing campaigns add more results, the systematic uncertainty estimation scheme applied to Bayesian inference can be further refined in the future.
%The recommended astrometric results of each MSP are boldfaced in \ref{tab:mspsrpi_astrometric_parameters}.

\subsection{Astrometric inference accounting for orbital motion}
\label{subsec:mspsrpi_reflex_motion_inference}
For some binary pulsars, VLBI astrometry can also refine parameters related to the binary orbit, on top of the canonical astrometric parameters. The orbital inclination $i$ and the orbital ascending node longitude $\Omega_\mathrm{asc}$ have been previously constrained for a few nearby pulsars, such as PSR~J1022$+$1001, PSR~J2145$-$0750 and PSR~J2222$-$0137 \citep{Deller13,Deller16,Guo21}.
To assess the feasibility of detecting orbital reflex motion with VLBI, we computed
\begin{equation}
\label{eq:mspsrpi_reflex_motion_measurability}
    \eta_\mathrm{orb}  \equiv \frac{2a_1}{1\,\mathrm{AU}} \cdot \frac{\varpi}{\sigma_{\varpi}} 
    = 2a_1 \cdot \left( \frac{1\,\mathrm{AU}}{\varpi}\right)^{-1} \cdot \frac{1}{\sigma_{\varpi}} = \frac{2a_1}{D} \cdot \frac{1}{\sigma_\varpi} \,,
\end{equation}
where $D$ and $a_1 \equiv a \sin{i}$ stands for, respectively, the distance (to the pulsar) and the orbital semi-major axis projected onto the sightline. 
On the other hand, $\tilde{\theta}_\mathrm{orb} \equiv 2a/D$ reflects the apparent angular size of orbit. Provided the parallax uncertainty $\sigma_{\varpi}$, $\tilde{\theta}_\mathrm{orb} / \sigma_{\varpi}$ quantifies the detectability of orbital parameters using VLBI astrometry. Hence,
\begin{equation}
\label{eq:mspsrpi_theta_orb}
   \frac{\tilde{\theta}_\mathrm{orb}}{\sigma_{\varpi}} \equiv \frac{2a}{D}\cdot \frac{1}{\sigma_{\varpi}} \geq  \eta_\mathrm{orb}
    \,.
\end{equation}
Since $i$ is usually unknown, the $\eta_\mathrm{orb}$ defined in \ref{eq:mspsrpi_reflex_motion_measurability} serves as a lower limit for $\tilde{\theta}_\mathrm{orb} / \sigma_{\varpi}$, and is used in this work to find out pulsar systems with $i$ and $\Omega_\mathrm{asc}$ potentially measurable with VLBI observations. 
In general, the orbital reflex motion should be negligible when $\eta_\mathrm{orb} \ll 1$, easily measurable when $\eta_\mathrm{orb} \gg 1$, and difficult to constrain (but non-negligible) when $\eta_\mathrm{orb} \sim 1$. By way of comparison, \citet{Guo21} were able to firmly constrain $\Omega_\mathrm{asc}$ and $i$ for PSR~J2222$-$0137 ($\eta_\mathrm{orb} = 10.2$), while \citet{Deller16} could place weak constraints for PSR~J1022$+$1001 and PSR~J2145$-$0750 ($\eta_\mathrm{orb} = 3.2$ and 1.6, respectively)
%For example, the $\eta_\mathrm{orb}$ for PSR~J1022$+$1001, PSR~J2145$-$0750 and PSR~J2222$-$0137 is, respectively, 3.2, 1.6 and 10.2. 

%For PSR~J1022$+$1001, PSR~J2145$-$0750 and PSR~J2222$-$0137, all of which have their $i$ and $\Omega_\mathrm{asc}$ refined with VLBI astrometry \citep{Deller16,Guo21}, $\eta_\mathrm{orb}$ is 3.2, 1.6 and 10.2, respectively.
%Accordingly, in this work, we consider the two aforementioned orbital parameters (of a binary pulsar) potentially detectable, if all the following conditions are met:
Accordingly, in this work, we fit for orbital reflex motion if all the following conditions are met:
\begin{enumerate}[label=(\roman*), leftmargin=*,align=left]
    \item $a_1$ is well determined with pulsar timing;
    \item $\eta_\mathrm{orb} > 1$;
    \item the orbital period $P_\mathrm{b} < 2$\,yr, where 2\,yr is the nominal time span of an \mspsrpi\ astrometric campaign.
\end{enumerate}
For the calculation of $\eta_\mathrm{orb}$, we simply use the direct-fitting parallax $\varpi^{\mathrm{(DF)}}$ for $\varpi$, and its \rcs-corrected uncertainty $\sigma_{\varpi}^\mathrm{(DF)} \cdot \sqrt{\chi^2_\nu}$ for $\sigma_{\varpi}$ (see \ref{tab:mspsrpi_astrometric_parameters}). The calculated $\eta_\mathrm{orb}$ as well as $P_\mathrm{b}$ are summarized in \ref{tab:mspsrpi_astrometric_parameters}. 
Among the 18 \mspsrpi\ pulsars, \psrfa, \psrga, \psrgb\ and \psrka\ meet our criteria (see \ref{tab:mspsrpi_astrometric_parameters}). 
%Hence, on the 3 pulsar systems, we also performed 7-parameter (i.e., the 5 canonical astrometric parameters plus $i$ and $\Omega_\mathrm{asc}$) astrometry inference. 
Hereafter, the 4 pulsars are referred to as the ``8P'' pulsars for the sake of brevity, as we would perform 8-parameter (i.e., the 5 canonical astrometric parameters and $\eta_\mathrm{EFAC}$ plus $i$ and $\Omega_\mathrm{asc}$) inference on them. %(besides the non-orbital-parameter inferences already summarized in \ref{tab:mspsrpi_astrometric_parameters}).
%We inferred the two orbital parameters (i.e., $i$ and $\Omega_\mathrm{asc}$) alongside canonical astrometric parameters for \psrgb, \psrka\ and \psrl, following the method described in \citet{Guo21}.
%However, due to the absence of prior constraint on $i$, we failed to refine the two orbital parameters for any of the 3 MSPs.

%In addition to the potential VLBI constraints on $i$ and $\Omega_\mathrm{asc}$, the observed time derivative of $a_1$, denoted as $\dot{a}_1$, may further tighten $i$ and $\Omega_\mathrm{asc}$ \citep[e.g.][]{Nice01,Deller16,Reardon21}, especially when all of the 8P pulsars already have substantial $\dot{a}_1$ measurements (see \ref{tab:mspsrpi_7_parameter_inference}). 

%, and $\mathcal{U}$(0, 360\degr) for $\Omega_\mathrm{asc}$ (of the 8P pulsars). 
For the 8-parameter inference, prior probability distributions of the canonical parameters and $\eta_\mathrm{EFAC}$ are described in \ref{subsubsec:mspsrpi_parameter_priors}. 
%The priors of $i$ and $\Omega_\mathrm{asc}$ are as follows. 
Both $i$ and $\Omega_\mathrm{asc}$ are defined in the TEMPO2 \citep{Edwards06} convention.
The prior probability distribution of $\Omega_\mathrm{asc}$ follows $\mathcal{U}$(0, 360\degr).
%At the basis, 
Sine distribution $\mathcal{S}$(0, 180\degr) is used for $i$ of the four 8P pulsars (i.e., the probability density $p(i) \propto \sin{i}$, $i \in \left[0, 180\degr\right]$). 
Where available, tighter constraints are applied to $i$ in accordance with \ref{tab:mspsrpi_7_parameter_inference} (also see the descriptions in \ref{sec:mspsrpi_individual_pulsars}).

Moreover, extra prior constraints can be applied to $i$ and $\Omega_\mathrm{asc}$ based on $\dot{a}_1$, the time derivative of $a_1$ \citep[e.g.][]{Nice01,Deller16,Reardon21}.
%under the premise that $\dot{a}_1$ is predominantly caused by apparent $i$ change due to sky position shift (of a binary system) \citep{Kopeikin96}. 
%This conduct becomes possible , which 
As $a_1 \equiv a\sin{i}$, 
\begin{equation}
\label{eq:mspsrpi_a1dot}
\begin{split}
    \frac{\dot{a}_1}{a_1} = \frac{\dot{a}}{a} + \frac{\partial i}{\partial t} \cot{i}  \approx \frac{\partial i}{\partial t} \cot{i}\,.
\end{split}
\end{equation}
Here, the $\dot{a}/a$ term reflects the intrinsic variation of the semi-major axis $a$ due to GR effects \citep{Peters64}, which is however $\sim$8 and $\sim$5 orders smaller than $\dot{a}_1/a_1$ for the 8P WD-pulsar systems and the DNS system \psrfa, respectively (see \citealp{Nice01} for an analogy).
%For the DNS system \psrfa, its $\dot{a}/a$ is roughly $\sim$5 orders smaller than its $\dot{a}_1/a_1$ (see \ref{tab:mspsrpi_7_parameter_inference}), assuming typical NS masses of 1.4\,\msun\ and $a_1$ of 40\,lt-s. 
Accordingly, the apparent $\dot{a}_1/a_1$ is predominantly caused by apparent $i$ change as a result of the sightline shift \citep{Kopeikin96}. When proper motion contributes predominantly to the sky position shift (as is the case for the 8P pulsars), 
\begin{equation}
\label{eq:mspsrpi_i_dot}
\begin{split}
    \frac{\partial i}{\partial t} = \mu \sin{\left(\theta_\mu-\Omega_\mathrm{asc}\right)}\,,
\end{split}
\end{equation}
where $\theta_\mu$ refers to the position angle (east of north) of the proper motion $\mu$ \citep{Kopeikin96,Nice01}.
%The first term has been analytically calculated by \citet{Peters64} to be
%Therefore, \ref{eq:mspsrpi_a1dot} can be used to restrict $i$ and $\Omega_\mathrm{asc}$ in their parameter space.
We incorporated the $\dot{a}_1/a_1$ measurements (with \ref{eq:mspsrpi_a1dot} and \ref{eq:mspsrpi_i_dot}) on top of other prior constraints,
%on $i$ and $\Omega_\mathrm{asc}$ (see \ref{tab:mspsrpi_7_parameter_inference}), 
and inferred $i$, $\Omega_\mathrm{asc}$, $\eta_\mathrm{EFAC}$ and the canonical five astrometric parameters for the 8P pulsars with {\tt sterne}\textsuperscript{\ref{footnote:sterne}}, following similar approaches taken by \citet{Deller16,Guo21}.

While we ultimately did not significantly constrain $i$ or $\Omega_\mathrm{asc}$ for any pulsars, including their non-negligible reflex motion in the inference is still necessary for correctly inferring the uncertainties of the non-orbital model parameters. 
The non-orbital inferred parameters are provided in \ref{subsec:mspsrpi_astrometric_results_non_PM_priors} below, along with all the non-8P pulsars.
%Furthermore, adopting timing proper motions as priors (of Bayesian inference) is expected to refine the $i$ and $\Omega_\mathrm{asc}$ estimations. However, we found this refinement merely marginal. Hence, the posterior constraints on orbital inclinations and ascending node longitudes (of the 8P pulsars) are not described until \ref{sec:mspsrpi_inference_with_priors},with an intention to avoid repetition.
As we found minimal differences between the constraints obtained on orbital parameters with or without the adoption of priors based on pulsar timing, we defer the presentation of the posterior constraints on orbital inclinations and ascending node longitudes (of the 8P pulsars) to \ref{sec:mspsrpi_inference_with_priors} in order to avoid repetition.

\begingroup
\renewcommand{\arraystretch}{1.4} % Default value: 1

\begin{table}
\raggedright
\caption{Prior constraints on $i$ and $\Omega_\mathrm{asc}$}
\label{tab:mspsrpi_7_parameter_inference}
%\begin{tabular}{@{}l@{\:}l@{\:}l@{}} % manual @ spacing to prevent this being too wide for a page
%\resizebox{\columnwidth}{!}{
\begin{tabular}{lcccc}
\hline
\hline
PSR & $\dot{a}_1$ & $\dot{a}_1/a_1$ & $i$ & $\Omega_\mathrm{asc}$ \\
 & ($10^{-15}~\text{lt-s~s}^{-1}$) & ($10^{-15}~\mathrm{s}^{-1}$) &  & (deg)  \\
\hline
%\Psrd & --- & --- & --- & 3.69(8) & -1.3(1) & 0.94(6) &  & \\

\Psrfa & -11(3) $^{a_1}$ & -0.55(15) & $\sin{i}\leq0.73$ $^{a_2}$ & ---   \\

\Psrga & 12(1) $^d$ & 0.22(2) & $\sin{i}=0.973(9)$ $^d$  & ---   \\

\Psrgb & -49.7(7) $^c$ & -1.98(3) & ---  & ---   \\
%& $-5.1(1)\!\times\!10^{-14}$ $^d$ & &  & 6.2(2) & 3.3(7) & $1.32^{+0.18}_{-0.19}$ &  & \\
\Psrka & 14(2) $^d$ & 0.34(5) & 85(14)\degr\ $^b$ & ---  \\
%\Psrl & -17(1) $^d$ & -0.80(5) & --- & --- & 0.50(4) & -6.86(9) & $0.25^{+0.04}_{-0.03}$ &  & &\\
\hline
%\multicolumn{6}{l}{$^e$ The narrowest circular $\Omega'_\mathrm{asc}$ interval encompassing 68\% of the simulations marks the 1-$\sigma$ upper and lower bound. The $\Omega'_\mathrm{asc}$ estimate is the median $\Omega'_\mathrm{asc}$ in}\\
%\multicolumn{6}{l}{\ \ \ this interval.}\\ 
%This 1-$\sigma$ $\Omega'_\mathrm{asc}$ interval is allowed to extend to over 360\degr\ or below 0\degr\ thanks to the periodicity of $\Omega'_\mathrm{asc}$.

\multicolumn{5}{l}{
$^{a_1}$\citet{Janssen08};
$^{a_2}$ inferred from the non-detection of Shapiro delay effects.}\\
\multicolumn{5}{l}{
$^b$ based on Shapiro delay measurements \citep{Faisal-Alam20}.}\\
\multicolumn{5}{l}{
$^c$\citet{Reardon21};
$^d$\citet{Perera19}.}\\

\end{tabular}

\end{table}
\endgroup

\subsection{The quasi-VLBI-only astrometric results}
\label{subsec:mspsrpi_astrometric_results_non_PM_priors}
%We have covered a lot of ground in \ref{sec:mspsrpi_parameter_inference}. 
To wrap up this section, we summarize in \ref{tab:mspsrpi_models_no_pm_prior} the full (including $\alpha_\mathrm{ref}$ and $\delta_\mathrm{ref}$) final astrometric results obtained with no exterior prior proper motion or parallax constraints, which we simply refer to as quasi-VLBI-only astrometric results (we add ``quasi'' because timing constraints on two orbital parameters, i.e., $i$ and $\dot{a}_1$, have already been used for the 8P pulsars).
These quasi-VLBI-only results are mainly meant for independent checks of timing results (which would enable the frame connection mentioned in \ref{subsec:mspsrpi_MSP_VLBI_astrometry}), or as priors for future timing analyses.
For the most precise possible pulsar parallaxes and hence distances, we recommend the use of the ``VLBI~$+$~timing'' results presented in \ref{sec:mspsrpi_inference_with_priors}.

\begingroup
\renewcommand{\arraystretch}{1.4} % Default value: 1
\begin{sidewaystable}
    \raggedright
    \caption{Final astrometric models inferred without using timing proper motions as priors.}
     
        %\begin{tabular}{p{0.15\linewidth}
        %                p{0.2\linewidth}
        %                p{0.1\linewidth}
        %                p{0.30\linewidth}}
        \resizebox{\textwidth}{!}{
    	\begin{tabular}{lccccccccccc} % four columns, alignment for each
		\hline
		\hline
	PSR & $t_\mathrm{ref}$ & $\alpha_\mathrm{ref}$ (J2000) $^*$  & $\sigma_\mathrm{\alpha_\mathrm{ref}}$ & $\delta_\mathrm{ref}$ (J2000) $^*$ & $\sigma_\mathrm{\delta_\mathrm{ref}}$ & $\mu_\alpha$ & $\mu_\delta$ & $\varpi$ & $\eta_\mathrm{EFAC}$ &   $\rho_{\mu_{\alpha},\varpi}$  & $\rho_{\mu_{\delta},\varpi}$ \\
	 & (MJD) & & (mas) & & (mas) & (\maspy) & (\maspy) & (mas) &   & & \\
		\hline
	\Psrb\ & 57849 & $00^{\rm h}30^{\rm m}27\fs 42502$ & $0.06[\pm0.3\pm0.2\pm0.8]$ & $04\degr51'39\farcs7159$ & $0.2[\pm0.8\pm0.3\pm0.8]$ & -6.13(7) & $0.34^{+0.15}_{-0.16}$ &  3.02(7)  & $1.35^{+0.45}_{-0.32}$  & 0.39 & 0.08 \\
	\Psrc & 57757& $06^{\rm h}10^{\rm m}13\fs 60053$ & $0.08[\pm0\pm3.5\pm0.8]$ & $-21\degr00'27\farcs7923$ & $0.2[\pm0\pm15.2\pm0.8]$ & 9.1(1) &  $15.96^{+0.25}_{-0.24}$ &  0.73(10)   & $1.1^{+0.4}_{-0.2}$  & 0.51 & 0.05\\
	\Psrd & 57685& $06^{\rm h}21^{\rm m}22\fs11617$ & $0.12[\pm2.1\pm0.5\pm0.8]$ & $10\degr02'38\farcs7261$ & $0.3[\pm2.8\pm0.7\pm0.8]$ & 3.5(2) &  -1.37(35) &  0.86(15)   & $3.7^{+1.0}_{-0.7}$ & 0.60 & -0.01 \\
	\Psrea & 57700 & $10^{\rm h}12^{\rm m}33\fs 43991$ & $0.04[\pm1.2\pm0.3\pm0.8]$ & $53\degr07'02\farcs1110$ & $0.1[\pm2.8\pm0.3\pm0.8]$ & 2.67(5) & $-25.39^{+0.14}_{-0.15}$ & $1.17^{+0.04}_{-0.05}$   & $1.7^{+0.6}_{-0.4}$  & 0.37 & -0.06 \\
	\Psreb & 57797& $10^{\rm h}24^{\rm m}38\fs65725$ & $0.06[\pm0.6\pm0.8\pm0.8]$ & $-07\degr19'19\farcs8014$ & $0.2[\pm1.4\pm1.5\pm0.8]$ & -35.32(7) & -48.1(2) &  0.94(6)  & $1.2^{+0.4}_{-0.3}$ & 0.23 & $2\!\times\!10^{-3}$ \\
	\Psrfa\ $^b$ & 57795 & $15^{\rm h}18^{\rm m}16\fs79817$ & $0.04[\pm0.5\pm0.4\pm0.8]$ & $49\degr04'34\farcs1132$ & $0.1[\pm1.4\pm0.4\pm0.8]$ & $-0.69^{+0.04}_{-0.03}$ & $-8.53^{+0.07}_{-0.09}$ & 1.238(36) & $1.4^{+0.6}_{-0.4}$ & -0.19 & -0.68 \\
	\Psrfb & 57964 & $15^{\rm h}37^{\rm m}09\fs96347$ & $0.06[\pm1.2\pm0.1\pm0.8]$ & $11\degr55'55\farcs0274$ & $0.1[\pm2.7\pm0.1\pm0.8]$ & 1.51(3) & $-25.30^{+0.05}_{-0.06}$ & $1.06^{+0.11}_{-0.10}$ & $0.5^{+0.6}_{-0.4}$ & -0.17 & 0.04\\
	\Psrga\ $^b$ & 57500 & $16^{\rm h}40^{\rm m}16\fs74587$ & $0.07[\pm0.4\pm0.3\pm0.8]$ & $22\degr24'08\farcs7642$ & $0.1[\pm0.3\pm0.6\pm0.8]$ & 2.19(9) & $-11.30^{+0.16}_{-0.13}$ & 0.68(8) & $1.3^{+0.7}_{-0.5}$ & -0.57 & -0.02\\
	\Psrgb\ $^b$ & 57700 & $16^{\rm h}43^{\rm m}38\fs16407$ & $0.1[\pm2.0\pm0.1\pm0.8]$ & $-12\degr24'58\farcs6531$ & $0.4[\pm6.2\pm0.1\pm0.8]$ & 6.2(2) & 3.3(6) & $1.31^{+0.17}_{-0.18}$ & $1.0^{+0.3}_{-0.2}$ & -0.43 & 0.04\\
	\Psrha & 57820 & $17^{\rm h}21^{\rm m}05\fs49936$ & $0.2[\pm3.0\pm0.3\pm0.8]$ & $-24\degr57'06\farcs2210$ & $0.6[\pm7.3\pm0.6\pm0.8]$ & 2.5(3) & -1.9(9) & 0.0(2) & $3.1^{+0.8}_{-0.6}$ & -0.13 & -0.01\\
	\Psrhb & 57821 & $17^{\rm h}30^{\rm m}21\fs 67969$ & $0.2[\pm0.5\pm0.3\pm0.8]$ & $-23\degr04'31\farcs1749$ & $0.5[\pm1.1\pm0.6\pm0.8]$ & 20.3(2) & -4.8(5) & $1.57(18)$ & $1.4^{+0.3}_{-0.2}$ & -0.38 & 0.01 \\
	\Psri & 57829 & $17^{\rm h}38^{\rm m}53\fs 97001$ & $0.06[\pm0.2\pm0.3\pm0.8]$ & $03\degr33'10\farcs9124$ & $0.1[\pm0.6\pm0.6\pm0.8]$ & 6.98(8) & 5.18(16) & 0.50(6) & $1.9^{+0.7}_{-0.6}$ & -0.53 & $4\!\times\!10^{-4}$ \\
	\Psro & 57836 & $18^{\rm h}24^{\rm m}32\fs00791$ & $0.4[\pm2.0\pm0.1\pm0.8]$ & $-24\degr52'10\farcs912$ & $1[\pm5.2\pm0.2\pm0.8]$ & 0.3(6) & $-3.9^{+1.2}_{-1.3}$ &  0.1(5) & $1.6^{+0.8}_{-0.7}$ & -0.65 & $-3\!\times\!10^{-3}$ \\
	\Psrka\ $^b$ & 57846 & $18^{\rm h}53^{\rm m}57\fs31785$ & $0.06[\pm0.8\pm0.2\pm0.8]$ & $13\degr03'44\farcs0471$ & $0.1[\pm1.6\pm0.4\pm0.8]$ & -1.4(1) & -2.8(2) & 0.49(7) & $0.5^{+0.6}_{-0.3}$ & -0.37 & 0.26 \\
	\Psrl & 57847 & $19^{\rm h}10^{\rm m}09\fs 70165$ & $0.03[\pm1.0\pm0.1\pm0.8]$ & $12\degr56'25\farcs4316$ & $0.06[\pm2.5\pm0.1\pm0.8]$ & 0.50(4) & -6.85(9) &  0.254(35) & $0.19^{+0.15}_{-0.12}$ & -0.48 & 0.03 \\
	\Psrma & 57768 & $19^{\rm h}11^{\rm m}49\fs 27544$ & $0.1[\pm0.9\pm0.3\pm0.8]$ & $-11\degr14'22\farcs5547$ & $0.3[\pm2.1\pm0.5\pm0.8]$ & -13.8(2) & -10.3(4) & $0.38^{+0.13}_{-0.14}$ & $1.1^{+0.4}_{-0.3}$ & -0.39 & -0.02\\
	\Psrmb & 57768 & $19^{\rm h}18^{\rm m}48\fs 02959$ & $0.1[\pm0.9\pm0.2\pm0.8]$ & $-06\degr42'34\farcs9335$ & $0.2[\pm2.2\pm0.4\pm0.8]$ & -7.1(1) & -5.7(3) & 0.60(12) & $1.2^{+0.3}_{-0.2}$ & -0.39 & -0.01 \\
	\Psrkb\ $^a$ & 57850 & $19^{\rm h}39^{\rm m}38\fs56134$  
	& $[0.07[\pm0.8\pm0.1\pm0.8]$ & $21\degr34'59\farcs1233$ & $0.2[\pm1.9\pm0.1\pm0.8]$ & 0.08(7) & -0.43(11) & $0.384^{+0.048}_{-0.046}$ & $1.5^{+0.7}_{-0.6}$ & -0.62 & -0.06 \\
	\hline 
	\multicolumn{12}{l}{$^*$ $\alpha_\mathrm{ref}$ and $\delta_\mathrm{ref}$ refer to the reference position at reference epoch $t_\mathrm{ref}$. The error budgets of the reference positions are provided in the adjacent columns, which include, from left to right, the error of relative}\\
	%This position should be regarded as a relative position, instead of an absolute one. The absolute pulsar positions can be inferred in }\\
	\multicolumn{12}{l}{\ \ \   reference position with respect to the reference point, the uncertainty of the reference point with regard to the main phase calibrator (estimated with Equation~1 of \citealp{Deller19}), the position uncertainty}\\
	\multicolumn{12}{l}{\ \ \  of the main phase calibrator, and the typical (0.8\,mas in each direction, \citealp{Sokolovsky11}) frequency-dependent core shift \citep[e.g.][]{Bartel86,Lobanov98} between 1.55\,GHz and $\sim8$\,GHz.}\\
	\multicolumn{12}{l}{\ \ \   We note that the errors outside ``[ ]'' are obtained with Bayesian inference, while the errors inside ``[ ]'' are only indicative. To properly determine the absolute pulsar position and its uncertainty requires}\\
	\multicolumn{12}{l}{\ \ \   the procedure described in Section~3.2 of \citet{Ding20}. This analysis will be made and presented in an upcoming paper.}\\
	\multicolumn{12}{l}{$\bullet$  $\rho_{\mu_{\alpha},\varpi}$ and $\rho_{\mu_{\delta},\varpi}$ stand for correlation coefficients between $\varpi$ and the two proper motion components.}\\
	\multicolumn{12}{l}{$\bullet$ The special parameter $\eta_\mathrm{EFAC}$ (that has been provided in \ref{tab:mspsrpi_astrometric_parameters} and \ref{tab:mspsrpi_7_parameter_inference}) is not reiterated in this table.}\\
	\multicolumn{12}{l}{$^a$ 
	Since inverse referencing is applied for \psrkb, the two reference sources are the de-facto targets. Accordingly, the proper motion and parallax are the negative values of the direct measurements out }\\
	\multicolumn{12}{l}{\ \ \ of inverse referencing. For the original astrometric model, the reference positions for the two reference sources J194104 and J194106 are, respectively, $19^{\rm h}41^{\rm m}04\fs319769(2)+21\degr49'13\farcs19731(7)$ and}\\
	\multicolumn{12}{l}{\ \ \  $19^{\rm h}41^{\rm m}06\fs86774(1)+21\degr53'04\farcs9594(2)$, where the uncertainties do not contain those of the reference point (i.e. the inside-the-bracket terms of $\sigma_\mathrm{\alpha_\mathrm{ref}}$ and $\sigma_\mathrm{\delta_\mathrm{ref}}$). The reference position of \psrkb}\\
	\multicolumn{12}{l}{\ \ \ presented in the table is estimated using normal  phase referencing with respect to J194104.}\\
	\multicolumn{12}{l}{$^b$ Results of the 8-parameter Bayesian inference are reported here; the constraints on $i$ and $\Omega_\mathrm{asc}$ are described in \ref{sec:mspsrpi_inference_with_priors} (see \ref{subsec:mspsrpi_reflex_motion_inference} for explanations).}\\

	\end{tabular}
	}
    \label{tab:mspsrpi_models_no_pm_prior}
    
\end{sidewaystable}
\endgroup

The reference positions $\alpha_\mathrm{ref}$ and $\delta_\mathrm{ref}$ we provide in \ref{tab:mspsrpi_models_no_pm_prior} are precisely measured, but only with respect to the assumed location of the in-beam calibrator source for each pulsar. In all cases, the uncertainties on the in-beam source locations (also shown in \ref{tab:mspsrpi_models_no_pm_prior}) dominate the total uncertainty in the pulsar's reference position.  A future work, incorporating additional multi-frequency observations of the in-beam calibrations, will enable significantly more precise pulsar reference positions to be obtained, as is discussed in \ref{subsec:mspsrpi_mspsrpi}.

\section{VLBI+timing astrometric results}
\label{sec:mspsrpi_inference_with_priors}
In Bayesian inference, the output of a model parameter $X_j$ (where $j$ refers to various model parameters) hinges on its prior probability distribution: generally speaking, tighter prior constraints (on $X_j$) that are consistent with data (in the sense of Bayesian analysis) would sharpen the output $X_j$. In cases where a strong correlation between $X_j$ and another model parameter $X_k$ is present, tighter prior $X_j$ constraints consistent with the data would potentially sharpen both the output $X_j$ and the output $X_k$.

\begin{figure}
    \centering
	%\raggedright
	\includegraphics[width=16cm]{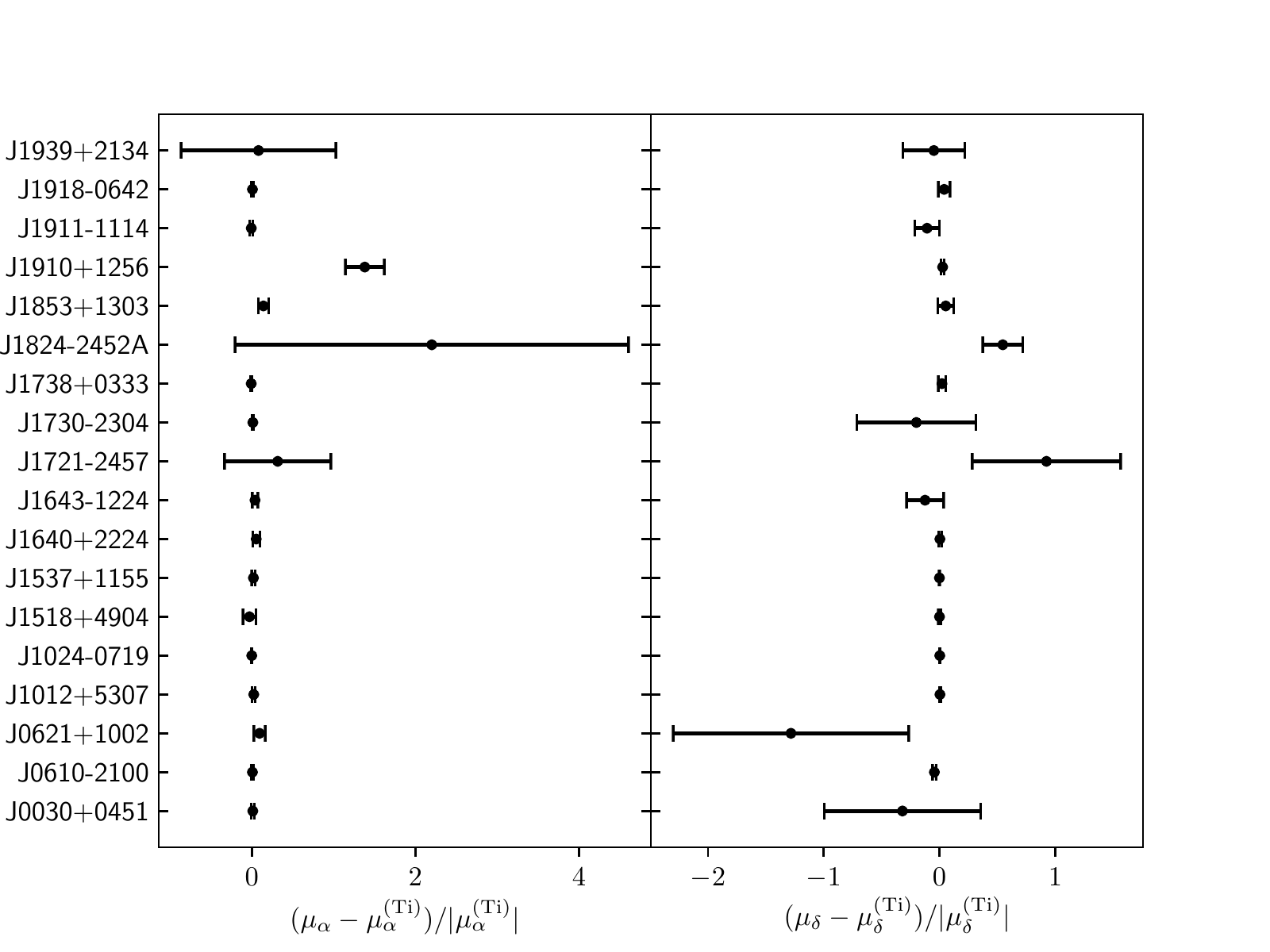}
    \caption{Fractional offsets of quasi-VLBI-only proper motions from timing ones. The denominators are only used as scaling factors (for the convenience of plotting), therefore their uncertainties do not contribute to the fractional offset uncertainties.
    }    
    \label{fig:mspsrpi_VLBI_proper_motions__vs__timing_ones}
\end{figure}

As noted in \ref{subsec:mspsrpi_MSP_VLBI_astrometry}, VLBI astrometry serves as the prime method to measure parallaxes of Galactic pulsars. A VLBI astrometric campaign (on a Galactic pulsar) normally spans $\sim2$ years, as a substantial parallax can likely be achieved in this timespan. On the other hand, most \mspsrpi\ pulsars have been timed routinely for $\gtrsim10$ years, which allows their proper motions to be precisely determined, as the precision on proper motion grows with $t^{3/2}$ for a regularly observed pulsar. In \ref{tab:mspsrpi_VLBI_timing_results}, we collect one timing proper motion (denoted as $\mu_\alpha^\mathrm{(Ti)}$ and $\mu_\delta^\mathrm{(Ti)}$) and one timing parallax ($\varpi^\mathrm{(Ti)}$) for each \mspsrpi\ pulsar. 
Among the published timing results, we select the timing proper motions measured over the longest timespan, and the $\varpi^\mathrm{(Ti)}$ having the smallest uncertainties.
\ref{fig:mspsrpi_VLBI_proper_motions__vs__timing_ones} is produced to visualize the difference between the quasi-VLBI-only proper motions and the timing ones.
According to \ref{tab:mspsrpi_models_no_pm_prior}, \ref{tab:mspsrpi_VLBI_timing_results} and \ref{fig:mspsrpi_VLBI_proper_motions__vs__timing_ones}, most timing proper motions are more precise than the quasi-VLBI-only counterparts. On the other hand, timing parallaxes are mostly less precise than the quasi-VLBI-only counterparts.
Nevertheless, adopting appropriate timing parallaxes as priors can still effectively lower parallax uncertainties.

The precisely measured $\mu_\alpha^\mathrm{(Ti)}$ and $\mu_\delta^\mathrm{(Ti)}$ provide the opportunity to significantly refine the quasi-VLBI-only proper motions. 
Furthermore, as shown with the Pearson correlation coefficients \citep{Pearson95} $\rho_{\mu_\alpha,\varpi}$ and $\rho_{\mu_\delta,\varpi}$ that we summarized in \ref{tab:mspsrpi_models_no_pm_prior}, large correlation between parallax and proper motion is not rare for VLBI astrometry. Therefore, using the $\mu_\alpha^\mathrm{(Ti)}$ and $\mu_\delta^\mathrm{(Ti)}$ measurements as the prior proper motion constraints in Bayesian inference can potentially refine both proper motion and parallax determination.

%Though employing tighter prior $X_j$ would generally tend to sharpen the posterior $X_j$, a $X_j$ prior inconsistent with the Bayesian data (i.e., the measured VLBI position series) may, however, widen the uncertainties of model parameters, as $\eta_\mathrm{EFAC}$ would be lifted to counter-balance rising \rcs. In this regard, we only adopt timing proper motions and parallaxes that are $\leq2\,\sigma$ discrepant. 
The astrometric results inferred with timing priors, hereafter referred to as VLBI+timing results, are reported in \ref{tab:mspsrpi_VLBI_timing_results}. To differentiate from the notation of quasi-VLBI-only astrometric parameter $Y$, we denote a VLBI+timing model parameter in the form of $Y'$.
%We note that $\varpi^\mathrm{(Ti)}$ are merely used for the comparison to $\varpi'$.
Comparing \ref{tab:mspsrpi_models_no_pm_prior} and \ref{tab:mspsrpi_VLBI_timing_results}, we find almost all VLBI+timing proper motions and parallaxes more precise than the quasi-VLBI-only counterparts; 
%except for the $\mu'_\alpha$ of \psrl\ and the $\mu'_\delta$ of \psrc. 
%Though both exceptions have merely marginally larger uncertainties than their quasi-VLBI-only counterparts
the most significant parallax precision enhancement occurs to \psrmb\ (by 42\%), followed by \psrkb\ (by 36\%) and \psrfb\ (by 33\%). 
% \psro\ (by 32\%), \psrb\ (by 29\%) and \psrga\ (by 25\%). 
Hence, we use the VLBI+timing results in the remainder of this paper.
%for the calculation of pulsar distances $d$ and space velocities $v_\perp$ (detailed in \ref{sec:mspsrpi_distances_and_velocities}).
%The difference between $\varpi'$ and $\varpi^\mathrm{(Ti)}$ are commented on in \ref{sec:mspsrpi_individual_pulsars}.

In 7 cases (i.e., \psrc, \psrgb, \psrhb, \psri, \psrka, \psro, \psrl), one of $\mu_\alpha^\mathrm{(Ti)}$, $\mu_\delta^\mathrm{(Ti)}$ or $\varpi^\mathrm{(Ti)}$ is more than 2\,$\sigma$ discrepant from the quasi-VLBI-only counterpart. Using such timing priors may widen the uncertainties of resultant model parameters, as $\eta_\mathrm{EFAC}$ would be lifted to counter-balance the increased \rcs. Without any indication that the discrepant timing values are less reliable, we use them as priors regardless. 
However, we caution the use of these 7 sets of VLBI+timing results, and would recommend the quasi-VLBI-only results to be considered if our adopted timing priors are proven inaccurate in future.

%\subsection{Caveats in }
We also now consider any  possible effects that could, despite our best efforts to characterise all sources of position noise, bias the fitted VLBI positions.  For any given VLBI calibrator source, evolution in the source structure can lead to a detectable position offset \citep[e.g.][]{Perger18,Zhang20c} that is then transferred to the target pulsar.  Due to the long timescales of AGN structure evolution, over the $\sim2$-year timescale of the \mspsrpi\ observations, this error may be quasi-linear in time and be absorbed into the pulsar proper motion \citep[e.g.][]{Deller13}.
Redundant secondary calibrators can be used to probe the astrometric effect of structure evolution. However, with small numbers of redundant calibrator sources, such probes are hardly conclusive, as the structure evolution of the redundant calibrators would also be involved.
Among the 7 pulsars showing $>2\,\sigma$ discrepancy  between quasi-VLBI-only and timing results (see \ref{tab:mspsrpi_VLBI_timing_results}), %More comments on the individual pulsars can be found in \ref{sec:mspsrpi_individual_pulsars}.
\psrb, \psrgb, \psrhb, \psri\ and \psro\ either display no relative motion between the redundant secondary calibrators and the main secondary calibrators or do not have any redundant calibrator (i.e. \psrgb), although the sub-optimal main secondary calibrators of \psrgb\ and \psro\ (see \ref{subsec:mspsrpi_J1643} and \ref{subsec:mspsrpi_J1824}) may likely affect the astrometric performance.
For \psrka, the main secondary calibrator has a clear jet aligned roughly with the right ascension (RA) direction, and thus source structure evolution is potentially significant.  
%The two redundant calibrators for \psrka\ do not display a clear common relative motion with respect to the main secondary calibrator, %s vary by up to 0.2 mas/yr with respect to the primary in-beam calibrator, non-negligible compared to our uncertainties.
%making it impossible to clearly ascribe the VLBI-timing discrepancy to source structure evolution. 
The two redundant calibrators for \psrka\ do display a relative proper motion of up to 0.2 mas/yr with respect to the main secondary calibrator, so while the mean relative motion seen between the two redundant secondary calibrators is small, calibrator structure evolution remains a possible explanation for the VLBI-timing discrepancy.
Finally, the main secondary calibrator of \psrl\ also exhibits a jet structure at a position angle of $\sim45$\degr. When using the only redundant calibrator of \psrl\ as the reference source, we obtained the VLBI-only result $\mu_\alpha=0.25\pm0.06$\,\maspy, $\mu_\delta=-7.3\pm0.1$\,\maspy\ and $\varpi=0.61\pm0.05$\,mas with Bayesian inference, where $\mu_\alpha$ becomes consistent with $\mu_\alpha^\mathrm{(Ti)}$ but $\mu_\delta$ and $\varpi$ are further away from the timing counterparts. 
The $\mu_\alpha$ consistency between VLBI and timing indicates that  structure evolution in our chosen calibrator is likely contributing to the VLBI-timing discrepancy. However, as the redundant calibrator is both fainter and further away from \psrl\ (compared to the main secondary calibrator), we do not use this source as the final reference source.

\begingroup
\renewcommand{\arraystretch}{1.4} % Default value: 1
\begin{sidewaystable}

%\raggedright
\caption{{\it Left of the dashed line:} proper motions $\{ \mu'_\alpha,\mu'_\delta \}$ and parallaxes $\varpi'$ inferred with timing proper motion and parallax priors. {\it Right:} distances $D$ and transverse space velocities $v_\perp$ based on $ \mu'_\alpha$, $\mu'_\delta$ and $\varpi'$.}
\label{tab:mspsrpi_VLBI_timing_results}
%\begin{tabular}{@{}l@{\:}l@{\:}l@{}} % manual @ spacing to prevent this being too wide for a page
\resizebox{\textwidth}{!}{
\begin{tabular}{lccccccc:cccccc}
\hline
\hline
PSR &  $\mu_\alpha^\mathrm{(Ti)}$ & $\mu_\delta^\mathrm{(Ti)}$ &
$\varpi^\mathrm{(Ti)}$ & $\mu_{\alpha}'$ & $\mu_\delta'$ & $\varpi'$  & $\eta'_\mathrm{EFAC}$  & DM  &
$d_\mathrm{DM}^\mathrm{(NE)}$ $^*$ & $d_\mathrm{DM}^\mathrm{(YMW)}$ $^*$ & $b$ & $D$ & $v_\perp$ $^{**}$ \\
 & (\maspy) & (\maspy) & (mas) & (\maspy) & (\maspy) & (mas) &   & ($\mathrm{pc~{cm}^{-3}}$) & (kpc) & (kpc) & (deg) & (kpc) & (\kmps) \\
\hline
\Psrb &  -6.2(1)  & 0.5(3) $^d$ & 3.08(8) $^d$ & -6.15(5) & 0.37(14) & 3.04(5) &  $1.3^{+0.4}_{-0.3}$  & 4.3 & 0.32(6) & 0.35(7) & -57.6 & $0.329^{+0.006}_{-0.005}$ & 15.4(2) \\
\Psrc  & 9.04(8)  & $^{!!}$16.7(1) $^c$ & --- & 9.06(7) & 16.6(1) & 0.72(11)  &  $1.4^{+0.4}_{-0.3}$  & 60.7 & 3.5(7) & 3.3(7) & -18.2 & $1.5^{+0.3}_{-0.2}$ & $120^{+25}_{-17}$ \\
\Psrd &  $^!3.2(1)$  & $^!-0.6(5)$ $^c$ & ---  & 3.27(9) & -1.1(3) & 0.74(14) &  $3.9^{+1.1}_{-0.8}$  & 36.5 & 1.4(3) & 0.42(8) & -2.0 & $1.6^{+0.5}_{-0.3}$ & $20^{+8}_{-5}$ \\ 
\Psrea & $^!2.61(1)$  & $-25.49(1)$ $^c$ & $^!$0.9(2) $^c$ & 2.61(1) & -25.49(1) & 1.14(4) &  $1.7^{+0.5}_{-0.4}$  & 9.02 & 0.41(8) & 0.8(2) & 50.9 & 0.877(35) & 95(4) \\
\Psreb & -35.270(17)  & -48.22(3) $^b$ & 0.83(13) $^b$ & -35.27(2) & -48.22(3) & 0.93(5) &  $1.2^{+0.4}_{-0.3}$  & 6.5 & 0.39(8) & 0.38(8) & 40.5 & 1.08(6) & 300(20) \\
%\Psre\ & BD179B, BD192B & J0029$+$0554 & \\
\Psrfa & $-0.67(4)$  & $-8.53(4)$ $^h$ & --- & -0.683(26) & -8.528(36) & $1.237^{+0.035}_{-0.031}$ &  $1.2^{+0.6}_{-0.4}$  & 11.61 & 0.6(1) & 1.0(2) & 54.3 & 0.81(2) & 16.0(6) \\
\Psrfb &  1.482(7)  & -25.285(12) $^g$ & 0.86(18) $^g$  & 1.484(7) & -25.286(11)  & 1.07(7)  & $0.5^{+0.5}_{-0.3}$   & 11.62 & 1.0(2) & 0.9(2) & 48.3 & $0.94^{+0.07}_{-0.06}$ & $102^{+8}_{-7}$ \\
\Psrga &  2.08(1) & -11.34(2) $^c$ & 0.6(4) $^c$ & 2.08(1) & -11.34(2) & 0.73(6) &  $1.3^{+0.7}_{-0.6}$  & 18.43 & 1.2(2) & 1.5(3) & 38.3 & $1.39^{+0.13}_{-0.11}$ & $53^{+5}_{-4}$ \\
\Psrgb &  $^!5.970(18)$  & 3.77(8) $^b$ & $^{!!}$0.82(17) $^b$ & 5.97(2) & 3.76(8) & 1.1(1) &  $1.0^{+0.3}_{-0.2}$  & 62.3 & 2.4(5) & 0.8(2) & 21.2 & $0.95^{+0.15}_{-0.11}$ & $41^{+5}_{-4}$ \\
\Psrha &  1.9(1.2)  & $^!-25(16)$ $^f$ & --- & 2.5(3) & -1.9(9) & 0.0(2) &  $3.1^{+0.8}_{-0.6}$  & 48.3 & 1.3(3) & 1.4(3) & 6.8 & $>1.5$ $^{\blacktriangle}$ & --- \\
\Psrhb & $^!20.06(12)$ & -4(2) $^b$ & $^{!!}$2.11(11) $^b$ & 20.1(1) & -4.7(6) & 2.0(1) &  $1.6^{+0.4}_{-0.3}$  & 9.62 & 0.5(1) & 0.5(1) & 6.0 & 0.51(3) & $54^{+3}_{-2}$ \\
\Psri &  7.037(5)  & 5.073(12) $^e$ & $^{!!}$0.68(5) $^e$ & 7.036(5) & 5.07(1) & 0.589(46) & $2.3^{+0.8}_{-0.6}$  & 33.8 & 1.4(3) & 1.5(3) & 17.7 & $1.74^{+0.15}_{-0.13}$ & $90^{+7}_{-6}$\\
\Psro &  -0.25(4)  & $^{!!!}-8.6(8)$ $^b$ & --- & -0.25(4) & -7.8(8) & 0.4(5)  & $2.5^{+0.8}_{-0.6}$  & 119.9 & 3.1(6) & 3.7(7) & -5.6 & $>0.5$ $^{\blacktriangle}$ & --- \\
\Psrka &  $^{!!}-1.63(2)$  & -2.96(4) $^c$ & 0.48(14) $^a$ & -1.62(2) & -2.96(4) & 0.53(7) &  0.9(5)  & 30.6 & 2.1(4) & 1.3(3) & 5.4 & 2.0(3) & $14^{+3}_{-2}$ \\
\Psrl &  $^{!!!!!}0.21(3)$  & $^!-7.04(5)$ $^c$ & 0.1(3) $^c$ & 0.24(3) & -7.03(5) & 0.36(6) &  $0.8^{+0.3}_{-0.2}$  & 38.1 & 2.3(5) & 1.5(3) & 1.8 & $3.4^{+1.0}_{-0.6}$ & $55^{+16}_{-9}$ \\
\Psrma &  -13.7(2)  & $^!-9.3(9)$ $^c$ & --- & -13.7(1) & $-10.2^{+0.4}_{-0.3}$ & $0.37^{+0.13}_{-0.14}$  &  $1.1^{+0.4}_{-0.3}$  & 31.0 & 1.2(2) & 1.1(2) & -9.6 & $>1.3$ $^{\blacktriangle}$ & --- \\
\Psrmb &  -7.15(2)  & -5.94(5) $^c$ & $^{!}$0.8(1) $^c$ & -7.15(2) & -5.93(5) & 0.71(7) &  $1.1^{+0.3}_{-0.2}$  & 6.1 & 1.2(2) & 1.0(2) & -9.1 & $1.48^{+0.19}_{-0.14}$ & $44^{+6}_{-5}$ \\
\Psrkb &  0.074(2)  & -0.410(3) $^c$ & $^{!}$0.28(5) $^a$ & 0.074(2) & -0.410(3) & 0.35(3)  & 1.2(6)  & 71.1 & 3.6(7) & 2.9(6) & -0.3 & $2.9^{+0.3}_{-0.2}$ & $80^{+9}_{-8}$ \\

\hline
\multicolumn{14}{l}{$\bullet$ ``Ti'' denotes historical pulsar timing results. $\rho$ stand for correlation coefficients of Bayesian inference without using timing proper motion priors.}\\
\multicolumn{14}{l}{$\bullet$ The ``!''s before $\mu_\alpha^\mathrm{(Ti)}$, $\mu_\delta^\mathrm{(Ti)}$ and $\varpi^\mathrm{(Ti)}$ convey the significance of the offset from the quasi-VLBI-only counterparts. In specific, the ``!'' repetition number $N_!$ means the offset significance is between}\\
\multicolumn{14}{l}{\ \ \ $N_!\,\sigma$ and $(N_!+1)\,\sigma$. VLBI+timing results obtained with timing proper motion or parallax priors that are more than 2\,$\sigma$ away from the quasi-VLBI-only counterparts (i.e., $\mu_\alpha^\mathrm{(Ti)}$,  $\mu_\delta^\mathrm{(Ti)}$ and $\varpi^\mathrm{(Ti)}$ }\\
\multicolumn{14}{l}{\ \ \ marked with $\geq2$ ``!''s) should be used with caution.}\\
\multicolumn{14}{l}{$\bullet$ $\mu_\alpha''$, $\mu_\delta''$ and $\varpi''$ designate results of Bayesian inference using timing proper motion priors.}\\
\multicolumn{14}{l}{$\bullet$ For each pulsar, we present the most precise timing estimates published in $^a$\citet{Faisal-Alam20},  $^b$\citet{Reardon21}, $^c$\citet{Perera19}, $^d$\citet{Arzoumanian18a}, $^e$\citet{Freire12},}\\
\multicolumn{14}{l}{\ \ \ $^f$\citet{Desvignes16}, $^g$\citet{Fonseca14} and $^h$\citet{Janssen08}.}\\ 
\multicolumn{14}{l}{$\bullet$ For comparison, we list $d_\mathrm{DM}^\mathrm{(NE)}$ and $d_\mathrm{DM}^\mathrm{(YMW)}$ derived with {\tt pygedm}\textsuperscript{\ref{footnote:pygedm}} given the sky position and the DM of a pulsar, based on the two latest $n_\mathrm{e}(\vec{x})$ models \citep{Cordes02,Yao17}.}\\
\multicolumn{14}{l}{$^*$20\% relative uncertainties are assumed for all DM-based distances (i.e., $d_\mathrm{DM}^\mathrm{(NE)}$ and $d_\mathrm{DM}^\mathrm{(YMW)}$).}\\ 
\multicolumn{14}{l}{$^{**}$Tangential space velocities corrected for the differential rotation of the Galaxy (see \ref{subsec:mspsrpi_v_t}).}\\ 
\multicolumn{14}{l}{$^{\blacktriangle}$The reciprocal of the 3\,$\sigma$ upper limit of the parallax is adopted as the lower limit of the distance.}\\ 

\end{tabular}
}
\end{sidewaystable}
\endgroup

\subsection{The posterior orbital inclinations and ascending node longitudes}
\label{subsec:mspsrpi_i_and_Omega_asc}
For the four 8P pulsars, orbital inclinations $i'$ and ascending node longitudes $\Omega'_\mathrm{asc}$ are also inferred alongside the five canonical parameters and $\eta'_\mathrm{EFAC}$ (see \ref{subsec:mspsrpi_reflex_motion_inference}). 
The full 8D corner plots out of the 8-parameter inferences are available online\textsuperscript{\ref{footnote:pulsar_positions}}.
%For the sake of brevity, in this paper, we only present truncated $i'-\mathrm{to}-\Omega'_\mathrm{asc}$ (2D) corner plots, which are bundled together in \ref{fig:mspsrpi_orbital_constraints}.
%The crescent shapes of the 2D histograms are the results of imposing the $\dot{a}_1$ constraints (see \ref{subsec:mspsrpi_reflex_motion_inference}). 
Prior constraints on $i'$ and $\Omega'_\mathrm{asc}$ have been provided in \ref{subsubsec:mspsrpi_parameter_priors}.
Owing to bi-modal features of all 1D histograms of $i'$, no likelihood component is substantially favored over the other. Hence, no tight posterior constraint on $i'$ is achieved for any 8P pulsar. Likewise, all 1D histograms of $\Omega'_\mathrm{asc}$ show multi-modal features, which precludes stringent constraints on $\Omega'_\mathrm{asc}$. 
%As the posterior $\Omega'_\mathrm{asc}$ distributions are more complex (than those of $i'$), they are described in \ref{sec:mspsrpi_individual_pulsars}.

\subsection{Comparison with Gaia results}
\label{subsec:mspsrpi_Gaia_results}

From the Gaia Data Release 2 \citep{Gaia-Collaboration18a}, Gaia counterparts for pulsars with optically bright companions have been identified and studied by \citet{Jennings18,Mingarelli18,Antoniadis21}. In the \mspsrpi\ sample, \psrea\ and \psreb\ have secure Gaia counterparts, while \psrl\ has a proposed Gaia counterpart candidate \citep{Mingarelli18}. 
In \ref{tab:mspsrpi_Gaia_results}, we updated the Gaia results for these three Gaia sources to the Gaia Data Release 3 (DR3, Gaia Collaboration et~al. A\&A in press).

For \psreb, the Gaia  proper motion \{$\mu_\alpha^\mathrm{(G)}$, $\mu_\delta^\mathrm{(G)}$\} and parallax $\varpi_1^\mathrm{(G)}$ are highly consistent with the VLBI+timing ones, which further strengthens the proposal that \psreb\ is in an ultra-wide orbit with a companion star (\citealp{Bassa16,Kaplan16}, also see \ref{subsec:mspsrpi_v_t} and \ref{subsec:mspsrpi_A_Gal}).
The Gaia proper motion and parallax of \psrea\ is largely consistent with the VLBI+timing counterparts. The $>1\,\sigma$ discrepancy between $\mu_\delta^\mathrm{(G)}$ and $\varpi_1^\mathrm{(G)}$ and the respective VLBI+timing counterparts can be explained by non-optimal goodness of (Gaia astrometric) fitting (GoF) (see \ref{tab:mspsrpi_Gaia_results}).
%Though the goodness of fitting is relatively 
On the other hand, the Gaia counterpart candidate for \psrl\ (proposed by \citealp{Mingarelli18}) possesses a $\mu_\alpha^\mathrm{(G)}$ 4\,$\sigma$ discrepant from the VLBI+timing one. Though this discrepancy is discounted by the relatively bad GoF by roughly a factor of 1.9 (see \ref{tab:mspsrpi_Gaia_results}), the connection between the Gaia source and \psrl\ remains inconclusive.
We note that the parallax zero-points $\varpi_0^\mathrm{(G)}$ \citep{Lindegren21} of the three Gaia sources are negligible and hence not considered, as $\varpi_0^\mathrm{(G)}$ is small ($|\varpi_0^\mathrm{(G)}|\lesssim0.02$\,mas, \citealp{Ding21}) compared to the uncertainty of $\varpi_1^\mathrm{(G)}$ (see \ref{tab:mspsrpi_Gaia_results}).

\begingroup
\renewcommand{\arraystretch}{1.4} % Default value: 1

\begin{table}
\raggedright
\caption{Gaia astrometric results}
\label{tab:mspsrpi_Gaia_results}
%\begin{tabular}{@{}l@{\:}l@{\:}l@{}} % manual @ spacing to prevent this being too wide for a page
%\resizebox{\columnwidth}{!}{
\begin{tabular}{lccccc}
\hline
\hline
PSR & Gaia DR3 & $\mu_\alpha^\mathrm{(G)}$ & $\mu_\delta^\mathrm{(G)}$ & $\varpi_1^\mathrm{(G)}$ & GoF.$^*$\\
 & source ID & (\maspy) & (\maspy) & (mas) & \\
\hline

\Psrea & 851610861391010944 & 2.7(3) & $^!$-25.9(3) & $^!$1.7(3) & -1.5\\

\Psreb & 3775277872387310208 & -35.5(3) & -48.35(36) & 0.86(28) & 0.4\\

\Psrl & 4314046781982561920$^?$ & $^{!!!!}$-2.3(6) & $^!$-6.1(6) & -0.1(8) & 1.9\\
\hline

\multicolumn{6}{l}{$\bullet$ Sources marked with ``?'' are tentative Gaia counterpart candidates.}\\
\multicolumn{6}{l}{$\bullet$ Values marked with $N$ ``!''s are $N \sigma - (N+1) \sigma$ offset from the VLBI+timing counterparts}\\
%\multicolumn{6}{l}{\ \ \ counterparts.}\\
\multicolumn{6}{l}{$^*$ Goodness of fitting, a parameter (of Gaia data releases) approximately following $\mathcal{N}(0,1)$ }\\
\multicolumn{6}{l}{\ \ \ distribution. A GoF closer to zero indicates better fitting performance.}\\

\end{tabular}

\end{table}
\endgroup

\section{Distances and Space velocities}
\label{sec:mspsrpi_distances_and_velocities}
In this section, we derive pulsar distances $D$ from parallaxes $\varpi'$ (see \ref{sec:mspsrpi_inference_with_priors}), and compare them to the dispersion-measure-based distances.
Incorporating the proper motions $\{\mu'_\alpha, \mu'_\delta\}$ (see \ref{sec:mspsrpi_inference_with_priors}), we infer the transverse space velocity $v_\perp$ (i.e., the velocity with respect to the stellar neighbourhood) for each pulsar, in an effort to enrich the sample of $\sim$40 MSPs with precise $v_\perp$ \citep{Hobbs05,Gonzalez11} and refine the $v_\perp$ distributions of MSP subgroups such as binary MSPs and solitary MSPs.
%and subsequently discuss the implications provided by the $v_\perp$ sample. 

\subsection{Parallax-based distances}
\label{subsec:mspsrpi_px_based_D}
Inferring a source distance from a measured parallax requires assumptions about the source properties, for which a simple inversion implicitly makes unphysical assumptions \citep[e.g.][]{Bailer-Jones21}. 
%We derived the parallax-based distances $D$ from the $\varpi'$ (of \ref{tab:mspsrpi_VLBI_timing_results}) in roughly the same way as \citet{Jennings18}. 
%We briefly recapitulate the distance inference method as follows, in order to facilitate comprehension and ready the mathematical formalism for further discussion. 
Various works \citep[e.g.][]{Lutz73,Verbiest12,Bailer-Jones15,Igoshev16} have contributed to developing and consolidating the mathematical formalism of parallax-based distance inference, which we briefly recapitulate as follows, in order to facilitate comprehension and ready the mathematical formalism for further discussion.

A parallax-based distance $D$ can be approached from the conditional probability density function (PDF) 
\begin{equation}
\label{eq:mspsrpi_px_based_D}
\begin{split}
    p(D|\varpi', l, b) \propto p(\varpi'|D) p(D, l, b),
\end{split}
\end{equation}
where $l$ and $b$ stands for Galactic longitude and latitude, respectively; 
$\varpi'=\varpi'_0 \pm \sigma_{\varpi'}$.
The first term on the right takes the form of 
%$\rho(\varpi'|D) \propto \exp{ \left[-\left(1/D-\varpi'\right)^2/\sigma^2_{\varpi'}\right]}$
\begin{equation}
\label{eq:mspsrpi_gaussian_distribution}
\begin{split}
    p(\varpi'|D) \propto \exp\left[{-\frac{1}{2}\left(\frac{1/D-\varpi'_0}{\sigma_{\varpi'}}\right)^2}\right],
\end{split}
\end{equation}
assuming $\varpi_0'$ is Gaussian-distributed, or more specifically, $\varpi_0' \sim \mathcal{N}\left(1/D, \sigma_{\varpi'}^2\right)$. 
The second term on the right side of \ref{eq:mspsrpi_px_based_D} can be approximated as $p(D,l,b) \propto D^2$, when the parent population $\Psi$ of the target celestial body is uniformly distributed spatially \citep{Lutz73}. Given a postulated (Galactic) spatial distribution $\rho(D,l,b)$ of $\Psi$, $p(D,l,b) \propto D^2 \rho(D,l,b)$.
Hence, 
\begin{equation}
\label{eq:mspsrpi_p_D_extended}
\begin{split}
    p(D|\varpi',l,b) \propto D^2 \rho(D,l,b) \exp\left[{-\frac{1}{2}\left(\frac{1/D-\varpi'_0}{\sigma_{\varpi'}}\right)^2}\right] \,.
\end{split}
\end{equation}
We join \citet{Verbiest12} and \citet{Jennings18} to adopt the $\rho(D,l,b)$ (of the ``Model C'') determined by \citet{Lorimer06a} for Galactic pulsars. While calculating the $\rho(D,l,b)$ with Equations~10 and 11 of \citet{Lorimer06a}, we follow \citet{Verbiest12} and \citet{Jennings18} to increase the scale height (i.e., the parameter ``$E$'' of \citealp{Lorimer06a}) to 0.5\,kpc to accommodate the MSP population. In addition, the distance to the Galactic centre (GC) in Equation~10 of \citealp{Lorimer06a} is updated to $d_\odot = 8.12\pm0.03$\,kpc \citep{Gravity-Collaboration18}. 
We do not follow \citet{Verbiest12,Igoshev16} to use pulsar radio fluxes to constrain pulsar distances, as pulsar luminosity is relatively poorly constrained.

\begin{figure}
    \centering
	%\raggedright
	\includegraphics[width=14cm]{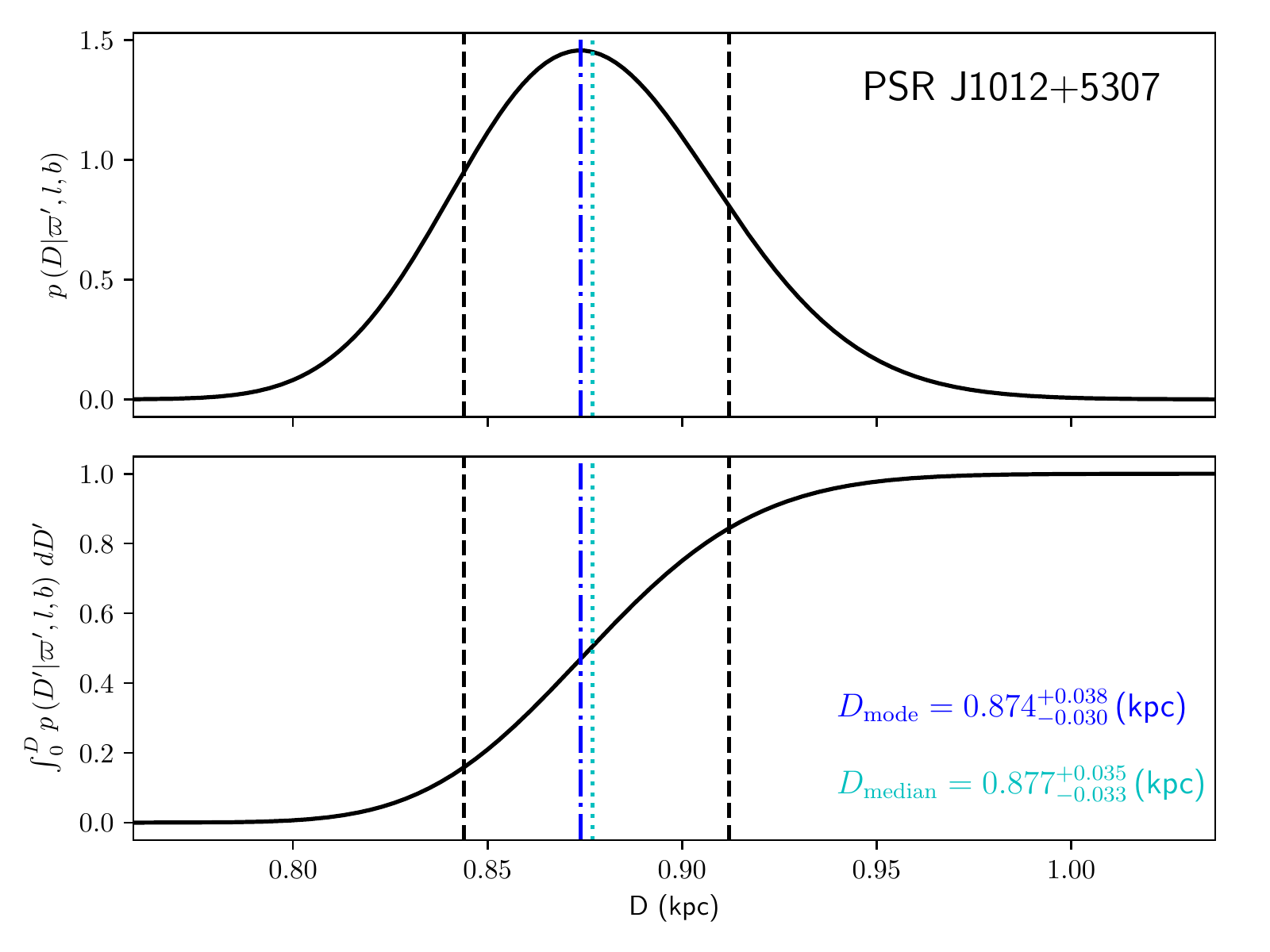}
    \caption{An example posterior probability density function $p(D|\varpi',l,b)$ (of distance) and its cumulative distribution function $\Phi(D|\varpi',l,b)=\int^{D}_{\,0} p(D'|\varpi',l,b)\,dD'$. The vertical dashed lines correspond to $\Phi(D|\varpi',l,b)\!=\!0.16$ and $\Phi(D|\varpi',l,b)\!=\!0.84$, which are respectively used as the lower and upper bound of the 1\,$\sigma$ uncertainty interval. The mode distance $D_\mathrm{mode}$ and median distance $D_\mathrm{median}$ are marked with dot-dashed blue line and dotted cyan line, respectively. Plots of this kind are also made for other \mspsrpi\ pulsars, and made available online\textsuperscript{\ref{footnote:pulsar_positions}}. Staying in line with the norm (see \ref{subsec:mspsrpi_MSP_VLBI_astrometry}) of this paper, we universally adopt $D_\mathrm{median}$ as the distances (i.e., $D$ in \ref{tab:mspsrpi_VLBI_timing_results}) for all \mspsrpi\ pulsars in this paper.
    }    
    \label{fig:mspsrpi_J1012_dist}
\end{figure}

Using the aforementioned mathematical formalism, we calculated $p(D|\varpi',l,b)$ for each \mspsrpi\ pulsar, and integrated it into the cumulative distribution function (CDF) $\Phi(D|\varpi',l,b)=\int^{D}_{\,0} p(D'|\varpi',l,b)\,dD'$. 
The $p(D|\varpi',l,b)$ and $\Phi(D|\varpi',l,b)$ is plotted for each pulsar and made available online\textsuperscript{\ref{footnote:pulsar_positions}}. An example of these plots are presented in \ref{fig:mspsrpi_J1012_dist}. The median distances $D_\mathrm{median}$ corresponding to $\Phi(D|\varpi',l,b)\!=\!0.5$ are taken as the pulsar distances, and summarized in \ref{tab:mspsrpi_VLBI_timing_results}. The distances matching $\Phi(D|\varpi',l,b)\!=\!0.16$ and $\Phi(D|\varpi',l,b)\!=\!0.84$ are respectively used as the lower and upper bound of the 1\,$\sigma$ uncertainty interval.

\subsubsection{Comparison with DM distances}
\label{subsubsec:mspsrpi_DM_distances}
As mentioned in \ref{subsec:mspsrpi_MSP_VLBI_astrometry}, the precise DM measured from a pulsar can be used to assess the pulsar distance, provided an $n_\mathrm{e}(\vec{x})$ model. Using {{\tt pygedm}\footnote{\label{footnote:pygedm}\url{https://github.com/FRBs/pygedm}}}, we compile into \ref{tab:mspsrpi_VLBI_timing_results} the DM distances (i.e., $d^\mathrm{(NE)}_\mathrm{DM}$ and $d^\mathrm{(YMW)}_\mathrm{DM}$) of each pulsar based on the two latest realisations of $n_\mathrm{e}(\vec{x})$ model --- the NE2001 model \citep{Cordes02} and the YMW16 model \citep{Yao17}. 
For all the DM distances, we adopt typical 20\% fractional uncertainties.
We have obtained significant ($\geq3\,\sigma$) parallax-based distances $D$ for 15 out of 18 \mspsrpi\ pulsars. These distances enable an independent quality check of both $n_\mathrm{e}(\vec{x})$ models.

\begin{figure}
    \centering
	%\raggedright
	\includegraphics[width=14cm]{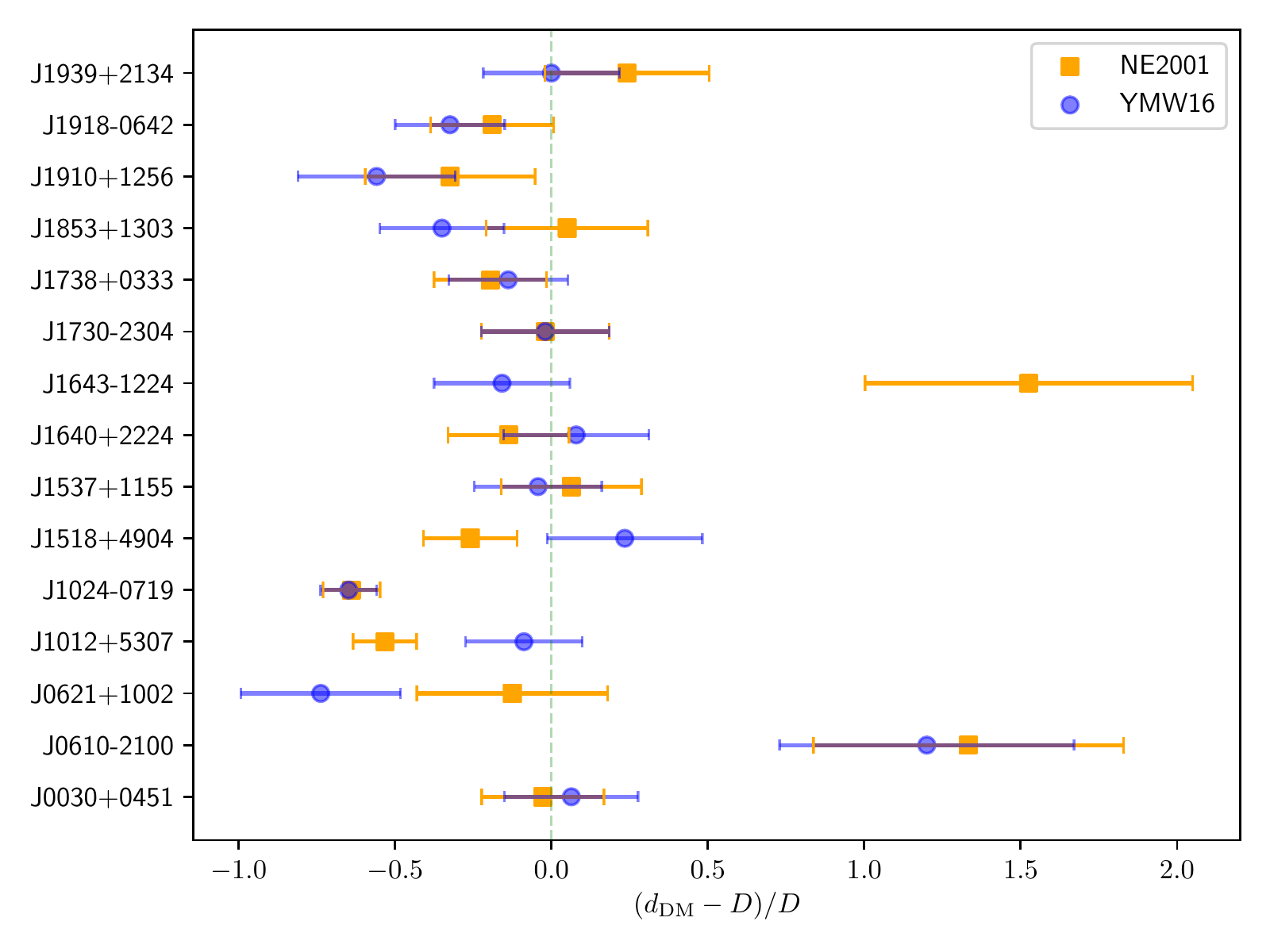}
    \caption{Fractional offset of the DM distances from the respective parallax-based distances, compiled for the 15 pulsars with significant parallax measurements. Here, the denominators of the fractional offset only serve as scaling factors; hence, their uncertainties are not propagated onto the uncertainties of the fractional offsets. The green dashed line represents perfect consistency between a DM distance and the corresponding parallax-based distance.
    }    
    \label{fig:mspsrpi_D_DM__vs__D_px}
\end{figure}

As is shown with \ref{tab:mspsrpi_VLBI_timing_results} and illustrated by \ref{fig:mspsrpi_D_DM__vs__D_px}, among the 15 pulsars with parallax-based distance measurements, YMW16 is more accurate than NE2001 in three cases (i.e., \psrea, \psrgb\ and \psrkb), but turns out to be the other way around in four cases (i.e. \psrd, \psrka, \psrl\ and \psrmb). In other 8 cases, the $D$ cannot discriminate between the two models.
The small sample of 15 $D$ measurements shows that NE2001 and YMW16 remain comparable in terms of outliers.
%remains an important reference of pulsar distances, and should not be simply categorized as an outdated $n_\mathrm{e}(\vec{x})$ model.
In 2 (out of the 15) cases (i.e., \psrc, \psreb), $D$ is about $2.6\,\sigma$ and $6.8\,\sigma$ away from either DM distance, which reveals the need to further refine the $n_\mathrm{e}(\vec{x})$ models. Such a refinement can be achieved with improved pulsar distances including the ones determined in this work.
 
\subsection{Transverse space velocities}
\label{subsec:mspsrpi_v_t}

Having determined the parallax-based distances $D$ and the proper motions $\{\mu'_\alpha, \mu'_\delta\}$, we proceed to calculate transverse space velocities $v_\perp$ for each pulsar, namely the transverse velocity with respect to the neighbouring star field of the pulsar. In estimating the transverse velocity of a pulsar neighbourhood, we assume the neighbourhood observes circular motion about the axis connecting the North and South Galactic Poles, which is roughly valid given that all \mspsrpi\ pulsars with significant ($>3\,\sigma$) $D$ share a median $|z|=D\sin{|b|}$ of 0.3\,kpc. Using the Galactic rotation curve from \citet{Reid19} and the full circular velocity of the Sun $247\pm1$\,\kmps, we derived the apparent transverse velocity of the neighbourhood $v_{\perp,N}$, thus obtaining $v_\perp$ by subtracting the apparent transverse velocity of the pulsar by $v_{\perp,N}$. Here, the full circular velocity (denoted as $\Theta_0+\Vsun$ in \citealp{Reid19}) is calculated with $d_\odot = 8.12\!\pm\!0.03$\,kpc \citep{Gravity-Collaboration18} and the proper motion of Sgr~$\mathrm{A^*}$ from \citet{Reid19}.

To estimate the uncertainty of $v_\perp$, we simulated a chain of 50,000 distances for each pulsar based on the $p(D|\varpi',l,b)$ that we have obtained in \ref{subsec:mspsrpi_px_based_D}. Besides, we also acquired chains of 50,000 $\mu'_\alpha$ and $\mu'_\delta$ given the VLBI+timing proper motions of \ref{tab:mspsrpi_VLBI_timing_results}, assuming $\mu'_\alpha$ and $\mu'_\delta$ follow Gaussian distributions. 
With these chains of $D$, $\mu'_\alpha$ and $\mu'_\delta$, we calculated 50,000 $v_\perp$ values, which form a PDF of $v_\perp$ for each pulsar. The $v_\perp$ inferred from the PDFs are summarized in \ref{tab:mspsrpi_VLBI_timing_results}.

\begin{figure*}
    \centering
	%\raggedright
	\includegraphics[width=15cm]{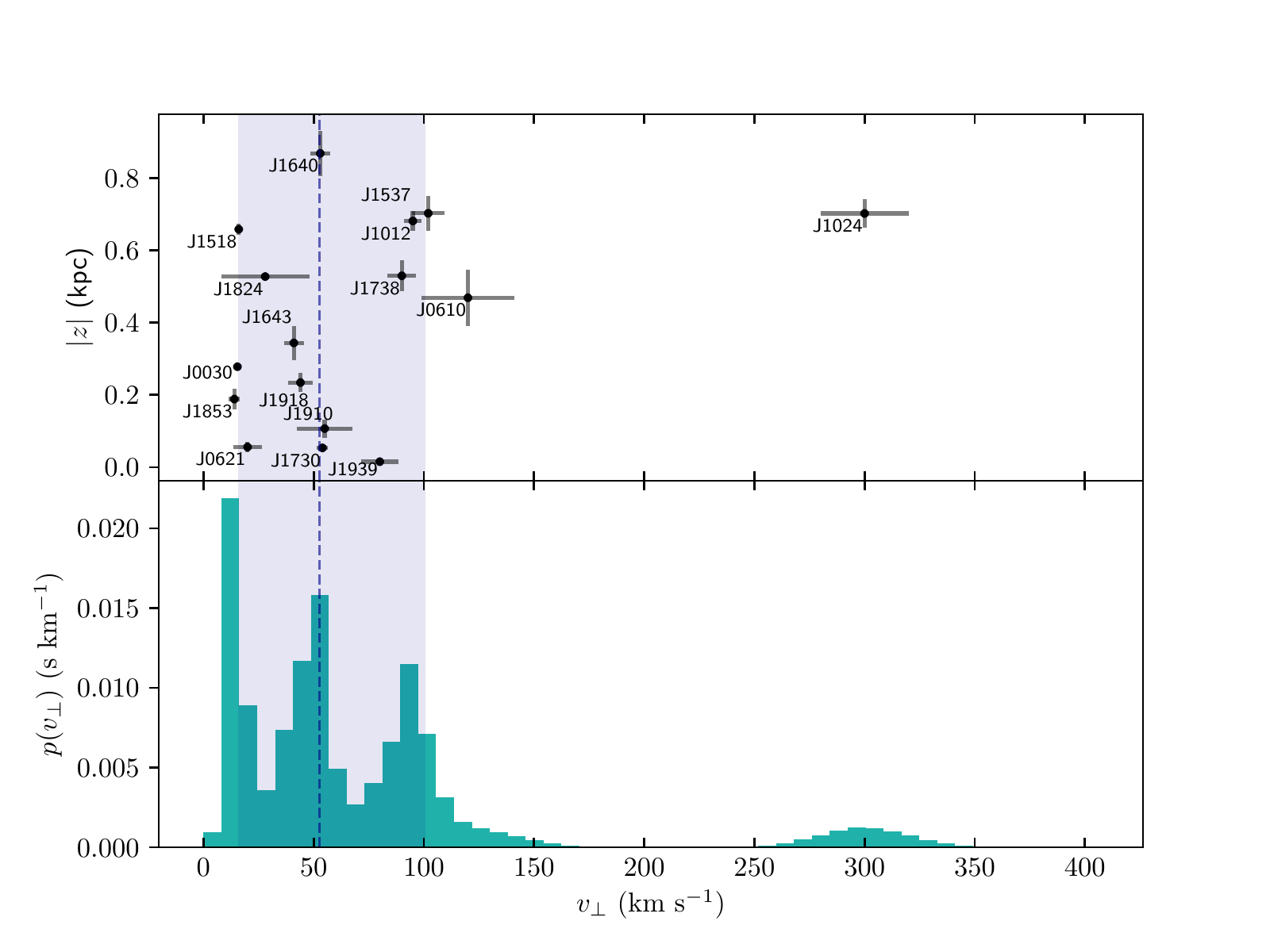}
    \caption{{\bfseries Upper:} The transverse space velocities $v_\perp$ versus the Galactic vertical heights $|z|=D\sin{|b|}$ of the 16 \mspsrpi\ pulsars with significant ($>3\,\sigma$) distance measurements (including 15 parallax-based distances and a globular cluster distance). {\bfseries Lower:} The probability density function (PDF) of $v_\perp$ for the 16 MSPs.
    The median of the $v_\perp$ PDF is marked with the dashed line, while the 1\,$\sigma$ error interval is shown with the shaded region.
    }    
    \label{fig:mspsrpi_v_t}
\end{figure*}

In \ref{fig:mspsrpi_v_t}, we illustrate the $v_\perp$ in relation to $|z|$ for 16 pulsars with precise distance estimates. Among the 16 pulsars, only \psro\ does not have a significant parallax-based distance. Nevertheless, its $v_\perp$ can be inferred by incorporating its proper motion with the astrometric information (i.e., distance and proper motion) of its host globular cluster (see \ref{subsec:mspsrpi_J1824}).
No clear correlation is revealed between $v_\perp$ and $|z|$, which reinforces our decision to treat all \mspsrpi\ pulsars across the $|z|\lesssim1$\,kpc regime equally. 
%The equally weighted mean transverse space velocity, denoted as $\bar{v}_\perp$, equals to $71\pm70$\,\kmps.
By concatenating the simulated $v_\perp$ chains, we acquired the PDF for the 16 MSPs (see \ref{fig:mspsrpi_v_t}), which gives $v^\mathrm{(MSP)}_\perp=53^{+48}_{-37}$\,\kmps.

Amongst the \mspsrpi\ sources, \psreb\ is an obvious outlier, with a velocity of $\sim$300 km s$^{-2}$ that is 3$\sigma$ above the mean. As proposed by \citet{Bassa16} and \citet{Kaplan16}, \psreb\ is theorized to have been ejected from a dense stellar region, thus possibly following a different $v_\perp$ distribution from typical field MSPs (isolated along with their respective companions throughout their lives). In this regard, we turn our attention to the binary sample of pulsars with well determined orbital periods $P_\mathrm{b}$ (see $P_\mathrm{b}$ of \ref{tab:mspsrpi_astrometric_parameters}), and obtain $v_\perp^\mathrm{(BI)}=50^{+49}_{-34}$\,\kmps\ for field binary MSPs.
Based on this small sample, we do not find the $v_\perp$ of the three solitary MSPs (i.e., \psrb, \psrhb\ and \psrkb) to be inconsistent with $v_\perp^\mathrm{(BI)}$.  
Neither are the two DNSs (i.e., \psrfa\ and \psrfb).
If we exclude the two DNSs from the binary sample, we would come to $v_\perp^\mathrm{(WD)}=50^{+46}_{-31}$\,\kmps\ for the \mspsrpi\ pulsars with WD companions, which is highly consistent with  $v_\perp^\mathrm{(BI)}$ and $v^\mathrm{(MSP)}_\perp$.
%We note we essentially assume that each of the 16 MSPs in the sample is equally representative of the that thes simple concatenation of the simulated $v_\perp$ chains assume

Compared to $113\pm17$\,\kmps\ previously estimated for a sample of $\sim40$ MSPs \citep{Gonzalez11}, our $v_\perp^\mathrm{(MSP)}$ is largely consistent but on the smaller side.
\citet{Boodram22} recently shows that MSP space velocities have to be near zero to explain the Galactic Centre $\gamma$-ray excess \citep[e.g.][]{Abazajian12}.
Interestingly, the $v_\perp$ PDF based on our small sample of 16 shows a multi-modal feature, with the lowest mode consistent with zero.
Specifically, the 7 \mspsrpi\ pulsars with the smallest $v_\perp$ share an equally weighted mean $v_\perp$ of only 25\,\kmps, which suggests MSPs with extremely low space velocities are not uncommon. 
Accordingly, we suspect the MSP origin of the GC $\gamma$-ray excess can still not be ruled out based on our sample of $v_\perp$.

%Compared to the previous work \citep{Gonzalez11} that compiles $v_\perp$ of MSPs, our $\bar{v}_\perp^\mathrm{(BI)}$ is consistent with this work adds new MSPs to the MSP sample with precise $v_\perp$

\section{Radial accelerations of pulsars and orbital-decay tests of gravitational theories}
\label{sec:mspsrpi_orbital_decay_tests}

As described in \ref{subsec:mspsrpi_MSP_VLBI_astrometry}, %millisecond pulsars serve as ideal probes of various theories of gravity. To quantify excessive gravitational effects beyond Newtonian orbital motions, multiple post-Keplerian (PK) parameters have been introduced. Among the PK parameters is the orbital decay $\dot{P}_\mathrm{b}$ (or the time derivative of orbital period). The intrinsic cause of $\dot{P}_\mathrm{b}$ is the emission of gravitational waves during the inspiraling phase. 
VLBI astrometry of pulsars, in conjunction with pulsar timing, can enhance the orbital-decay tests of gravitational theories.
%However, this is not the only cause of the observed orbital decay $\dot{P}_\mathrm{b}^\mathrm{obs}$. 
For binary systems involved in this work, the observed orbital decay has three significant components:
\begin{equation}
\label{eq:mspsrpi_Pb_budget}
\begin{split}
    \dot{P}_\mathrm{b}^\mathrm{obs} = \dot{P}_\mathrm{b}^\mathrm{GW} + \dot{P}_\mathrm{b}^\mathrm{Shk} + \dot{P}_\mathrm{b}^\mathrm{Gal}\,,
\end{split}
\end{equation}
where $\dot{P}_\mathrm{b}^\mathrm{GW}$ reflects the effect of gravitational-wave damping intrinsic to a binary system, while $\dot{P}_\mathrm{b}^\mathrm{Shk}$ and $\dot{P}_\mathrm{b}^\mathrm{Gal}$ are both extrinsic contributions caused, respectively, by relative line-of-sight accelerations (of pulsars) $\mathcal{A}_\mathrm{Shk}$ and $\mathcal{A}_\mathrm{Gal}$. 
Specifically, $\dot{P}_\mathrm{b}^\mathrm{Shk}=\mathcal{A}_\mathrm{Shk}/c \cdot P_\mathrm{b} =\mu^2 D/c \cdot P_\mathrm{b}$ (where $\mu^2={\mu'_\alpha}^2+{\mu'_\delta}^2$) is the radial acceleration caused by the tangential motion of pulsars \citep{Shklovskii70}, which becomes increasingly crucial for pulsars with larger $\mu$ (e.g. \psrfb, \citealp{Ding21a}), as $\mathcal{A}_\mathrm{Shk} \propto \mu^2$. 
On the other hand, 
\begin{equation}
\label{eq:mspsrpi_PbGal}
\begin{split}
    \dot{P}_\mathrm{b}^\mathrm{Gal} &= \frac{\mathcal{A}_\mathrm{Gal}}{c} P_\mathrm{b} 
    = \frac{\left[-\nabla \varphi\left(\vec{x}\right)\right]\big|^{\vec{x}_\mathrm{target}}_{\vec{x}_{\odot}} \cdot \vec{e}_{r}}{c} P_\mathrm{b}
\end{split}
\end{equation}
is a consequence of the gravitational pull (or push) exerted by the Galaxy.
%Galactic mass distribution. 
Here, $\varphi(\vec{x})$ and $\vec{e}_{r}$ are, respectively, the Galactic gravitational potential (as a function of Galactic position $\vec{x}$) and the unit vector in the Earth-to-pulsar direction. 
%Unlike the model-independent $\dot{P}_\mathrm{Shk}$, $\dot{P}_\mathrm{Gal}$ would vary with models of $\varphi(\vec{x})$, which should be taken into account as a systematic error of $\dot{P}_\mathrm{Gal}$ \citep[e.g.][]{Ding21a}.

In order to test any theoretical prediction of $\dot{P}_\mathrm{b}^\mathrm{GW}$, it is necessary to estimate $\mathcal{A}_\mathrm{Shk}$ and $\mathcal{A}_\mathrm{Gal}$ and remove their effect on $\dot{P}_\mathrm{b}^\mathrm{obs}$. Besides this impact, the radial accelerations $\mathcal{A}_\mathrm{Shk}$ and $\mathcal{A}_\mathrm{Gal}$ would, more generally, affect the time derivative of all periodicities intrinsic to a pulsar system, which include the pulsar spin period derivative $\dot{P}_\mathrm{s}$. 
Similar to $\dot{P}_\mathrm{b}^\mathrm{Shk}$ and $\dot{P}_\mathrm{b}^\mathrm{Gal}$, $\dot{P}_\mathrm{s}^\mathrm{Shk}=\mathcal{A}_\mathrm{Shk}/c \cdot P_\mathrm{s}$ and $\dot{P}_\mathrm{s}^\mathrm{Gal}=\mathcal{A}_\mathrm{Gal}/c \cdot P_\mathrm{s}$ (where ${P}_\mathrm{s}$ stands for the spin period of a pulsar).
As MSPs consist of nearly half of the $\gamma$-ray pulsar population, determining the extrinsic terms of $\dot{P}_\mathrm{s}$ and the intrinsic spin period derivative $\dot{P}_\mathrm{s}^\mathrm{int}=\dot{P}_\mathrm{s}^\mathrm{obs}-\dot{P}_\mathrm{s}^\mathrm{Shk}-\dot{P}_\mathrm{s}^\mathrm{Gal}$ is essential for exploring the ``death line'' (i.e., the lower limit) of high-energy emissions from pulsars \citep[e.g.][]{Guillemot16}. 
In \ref{subsec:mspsrpi_Shk} and \ref{subsec:mspsrpi_A_Gal}, we evaluate $\mathcal{A}_\mathrm{Shk}$ and $\mathcal{A}_\mathrm{Gal}$ one after another.
The evaluation only covers pulsars with significant $D$, as both $\mathcal{A}_\mathrm{Shk}$ and $\mathcal{A}_\mathrm{Gal}$ are distance-dependent.
%In particular, Shklovskii corrections would become increasingly crucial for pulsars with larger $\mu$ (e.g. \psrfb, \citealp{Ding21a}), as $\mathcal{A}_\mathrm{Shk} \propto \mu^2$. 
%Since the Shklovskii contribution to observable periodicities of a pulsar system is always positive, it is also referred to in this paper as Shklovskii dilation. 

\subsection{Shklovkii effects}
\label{subsec:mspsrpi_Shk}
We estimate the model-independent $\mathcal{A}_\mathrm{Shk}$ in a way similar to the estimation of $v_\perp$ (see \ref{subsec:mspsrpi_v_t}). Three chains of 50,000 $\mu'_\alpha$, $\mu'_\delta$ and $D$ were simulated from their respective PDFs. Using the relation $\mathcal{A}_\mathrm{Shk} = \left({\mu'_\alpha}^2+{\mu'_\delta}^2\right) D$, 50,000 $\mathcal{A}_\mathrm{Shk}$ were calculated to assemble the PDF of $\mathcal{A}_\mathrm{Shk}$ for each pulsar with significant $D$. The $\mathcal{A}_\mathrm{Shk}$ inferred from the PDFs are compiled in \ref{tab:mspsrpi_Shk} along with their resultant $\dot{P}_\mathrm{s}^\mathrm{Shk}$ and $\dot{P}_\mathrm{b}^\mathrm{Shk}$.

\begingroup
\renewcommand{\arraystretch}{1.4} % Default value: 1

\begin{sidewaystable}

\raggedright
\caption{Extrinsic terms of $\dot{P}_\mathrm{s}$ and $\dot{P}_\mathrm{b}$ for 15 pulsars with significant $D$.}
\label{tab:mspsrpi_Shk}
%\begin{tabular}{@{}l@{\:}l@{\:}l@{}} % manual @ spacing to prevent this being too wide for a page
\resizebox{\textwidth}{!}{
\begin{tabular}{lcc:ccccc:ccccc}
\hline
\hline
PSR & $\mathcal{A}_\mathrm{Shk}$ & $\mathcal{A}_\mathrm{Gal}$ & $P_\mathrm{s}$ & $\dot{P}_\mathrm{s}^\mathrm{Shk}$ & $\dot{P}_\mathrm{s}^\mathrm{Gal}$ & $\dot{P}_\mathrm{s}^\mathrm{obs}$ & $\dot{P}_\mathrm{s}^\mathrm{int}$ & $\dot{P}_\mathrm{b}^\mathrm{Shk}$ & $\dot{P}_\mathrm{b}^\mathrm{Gal}$ & $\dot{P}_\mathrm{b}^\mathrm{obs}$ & $\dot{P}_\mathrm{b}^\mathrm{int}$ & $\dot{P}_\mathrm{b}^\mathrm{GW}$ \\
 & ($\mathrm{pm~s^{-2}}$) $^*$ & ($\mathrm{pm~s^{-2}}$)  &  (ms) & ($\mathrm{zs~s^{-1}}$) $^*$ & ($\mathrm{zs~s^{-1}}$) & ($\mathrm{zs~s^{-1}}$) & ($\mathrm{zs~s^{-1}}$) & ($\mathrm{fs~s^{-1}}$) & ($\mathrm{fs~s^{-1}}$) $^*$ & ($\mathrm{fs~s^{-1}}$) & ($\mathrm{fs~s^{-1}}$) & ($\mathrm{fs~s^{-1}}$)\\
\hline
%\Psrd & --- & --- & --- & 3.69(8) & -1.3(1) & 0.94(6) &  & & \\
\Psrb & 9.1(2) & -33.0(3.7) & 4.87 & 0.148(3) & -0.54(6) & 10.2 & 10.59(6) &  --- & --- & --- & --- & ---\\
\Psrc & $3.9^{+0.8}_{-0.6}\!\times\!10^2$ & -9(2) & 3.86 & $5.0^{+1.0}_{-0.7}$ & -0.12(3) & 12.3 $^a$ & 7.4(9) &  $32^{+6}_{-5}$ & -0.72(17) & $-70(30)$ $^a$ & -101(31) & $\sim-4.6$ $^a$ \\
\Psrd & $14^{+4}_{-3}$ & 23.8(4.5) & 28.85 & $1.3^{+0.4}_{-0.3}$ & 2.3(4) & 47.3 & 43.7(6) &  $33^{+11}_{-7}$ & 57(11) & ---& ---&---\\
\Psrea & $419^{+17}_{-15}$ & -23.5(2.4) & 5.26 & 7.3(3) & -0.41(4) & 17.1 & 10.2(3) &  73(3) & -4.1(4) & 61(4) & -7.9(5.0) & -13(1) $^b$ \\
\Psreb & $2.8(2)\!\times\!10^3$ & -40(3) & 5.16 & 48(3) & -0.69(5) & 18.6 & -29(3) $^{**}$ & --- & ---& --- & ---&---\\
\Psrfa & 43(1) & -48.5(3.2) & 40.93 & 5.9(2) & -6.6(4) & 27.2 & 27.9(5) &  107(3) & -120(8) & $2.4(2.2)\!\times\!10^2$ & $2.6(2.2)\!\times\!10^2$ & $\sim-1.2$ $^e$ \\
\Psrfb & $4.4(3)\!\times\!10^2$ & -42(3) & 37.90 & $55.6^{+4.0}_{-3.5}$ & -5.3(4) & 2422.5 & 2372(4) & $53.3^{+3.8}_{-3.3}$ & -5.1(4) & -136.6(3) & -185(4) & -192.45(6) $^c$ \\
\Psrga & $1.3(1)\!\times\!10^2$ & -48.5(4.3) & 3.16 & 1.4(1) & -0.51(5) & 2.8 & 1.9(1) & $6.8^{+0.6}_{-0.5}\!\times\!10^3$ & $-2.45(22)\!\times\!10^3$ & ---& --- &---\\
\Psrgb & $35^{+5}_{-4}$ & 1.0(1.7) & 4.62 & $0.53^{+0.08}_{-0.06}$ & $2(3)\!\times\!10^{-2}$ & 18.5 & 17.95(7) & $1.5(2)\!\times\!10^3$ & 41(72) & ---& ---&---\\
%\Psrha & &  & & & & & & \\
\Psrhb & $158^{+9}_{-8}$ & 11.7(9) & 8.12 & 4.3(2) & 0.32(2) & 20.2 & 15.6(2) & --- & ---& --- & --- &---\\
\Psri & $95^{+8}_{-7}$ & -6.2(1.6) & 5.85 & $1.86^{+0.16}_{-0.14}$ & -0.12(3) & 24.1 & 22.4(2) & $9.7^{+0.9}_{-0.7}$ & -0.64(16) & -17(3) & -26.1(3.1) & $-27.7^{+1.5}_{-1.9}$ $^d$\\
%\Psro & & & & & & & & \\
\Psrka & $17^{+3}_{-2}$ & -16(5) & 4.09 & $0.23^{+0.04}_{-0.03}$ & -0.22(7) & 8.7 & 8.69(8) &  $5.6^{+0.9}_{-0.7}\!\times\!10^2$ & $-5(2)\times10^2$ & --- & --- &---\\
\Psrl & $1.2^{+0.4}_{-0.2}\!\times\!10^2$ & -35(17) & 4.98 & $2.0^{+0.6}_{-0.4}$ & -0.6(3) & 9.7 & 8.3(6) & $2.0^{+0.6}_{-0.4}\!\times\!10^3$ & $-6(3)\times10^2$ & ---& --- &---\\
%\Psrma & & & & & & & & --- \\
\Psrmb & $93^{+12}_{-9}$ & 6.8(1.5) & 7.65 & $2.4^{+0.3}_{-0.2}$ & 0.17(4) & 25.7 & 23.1(3) & $2.9^{+0.4}_{-0.3}\!\times\!10^2$ & 21(5) & ---& --- &---\\
\Psrkb & $0.37^{+0.04}_{-0.03}$ & -60(10) & 1.56 & $1.9(2)\!\times\!10^{-3}$ & -0.31(5) & 105.1 & 105.41(5) & --- & --- & --- & --- &---\\

\hline

\multicolumn{13}{l}{$\bullet$ Unless otherwise specified, we adopted $P_\mathrm{s}$, $\dot{P}^\mathrm{obs}_\mathrm{s}$, $P_\mathrm{b}$ and $\dot{P}^\mathrm{obs}_\mathrm{b}$ from the ATNF Pulsar Catalogue\textsuperscript{\ref{footnote:PSRCAT}}.}\\ %{\tt PSRCAT}\footnote{\label{ft:psrcat}\url{https://www.atnf.csiro.au/research/pulsar/psrcat/}}.}\\
\multicolumn{13}{l}{$\bullet$ Other references: $^a$\citet{van-der-Wateren22}; $^b$\citet{Ding20} and references therein; $^c$\citet{Ding21a} and references therein; $^d$\citet{Freire12}; $^e$\citet{Janssen08}.}\\
\multicolumn{13}{l}{$^{*}$ ``zs'' means zeptosecond, which is $10^{-21}$\,s. Besides, ``pm'' and ``fs'' represent $10^{-12}$\,m and $10^{-15}$\,s, respectively.}\\
\multicolumn{13}{l}{$^{**}$ The negative $\dot{P}_\mathrm{s}^\mathrm{int}$ is the result of the radial acceleration caused by a far-away companion \citep{Bassa16,Kaplan16}.}\\

\end{tabular}
}
\end{sidewaystable}
\endgroup

\subsection{Relative radial accelerations due to Galactic gravitational pull}
\label{subsec:mspsrpi_A_Gal}
We estimate $\mathcal{A}_\mathrm{Gal}$  in the same way as \citet{Ding21a}, following the pioneering work of \citet{Zhu19}. 
To briefly demonstrate this method, we present, in \ref{tab:mspsrpi_A_Gal}, the $\mathcal{A}_\mathrm{Gal}$ based on five different $\varphi(\vec{x})$ models for the 15 pulsars with significant $D$ measurements. 
The five $\varphi(\vec{x})$ models are denoted as NT95 \citep{Nice95}, DB98 \citep{Dehnen98}, BT08 \citep{Binney11}, P14 \citep{Piffl14} and M17 \citep{McMillan17} in this paper.
%The five $\varphi(\vec{x})$ models are, respectively, the ``NT95'' model analytically formulated by \citet{Nice95}
%Compared to the other $\varphi(\vec{x})$ models, the oversimplified analytical solution of NT95 more likely renders outlying $\mathcal{A}_\mathrm{Gal}$ (in respective groups of five $\mathcal{A}_\mathrm{Gal}$), such as the $\mathcal{A}_\mathrm{Gal}^\mathrm{(NT95)}$ of \psri, \psrfb, \psrea\ and \psrga. In response to the frequent outlying manner, we chose to not use the NT95 solutions.Since it is unclear which of the remaining four models is the most accurate, we derived $\mathcal{A}_\mathrm{Gal}$ with all of the four $\varphi(\vec{x})$ models (see \citealp{Ding21a}). We summarize $\mathcal{A}_\mathrm{Gal}$ of each pulsar in \ref{tab:mspsrpi_Shk} alongside the associated $\dot{P}_\mathrm{b}^\mathrm{Gal}$ (if $P_\mathrm{b}$ is well determined) and $\dot{P}_\mathrm{s}^\mathrm{Gal}$. 
The results obtained with NT95, which uses a simple analytical approach, are frequently discrepant compared to the other 4 $\varphi(\vec{x})$ models. Accordingly, and following \citet{Ding21a}, we exclude it and use the remaining 4 models to derive the estimate for $\mathcal{A}_\mathrm{Gal}$ and its uncertainty, which we present in \ref{tab:mspsrpi_Shk} (along with $\dot{P}_\mathrm{b}^\mathrm{Gal}$ and $\dot{P}_\mathrm{s}^\mathrm{Gal}$).

Incorporating the $\dot{P}_\mathrm{s}^\mathrm{Shk}$ derived in \ref{subsec:mspsrpi_Shk}, we calculated the intrinsic spin period derivative $\dot{P}_\mathrm{s}^\mathrm{int}=\dot{P}_\mathrm{s}^\mathrm{obs}-\dot{P}_\mathrm{s}^\mathrm{Shk}-\dot{P}_\mathrm{s}^\mathrm{Gal}$.
%\hao{comment on \psreb\ here.}
We note that the negative $\dot{P}_\mathrm{s}^\mathrm{int}$ of \psreb\ is probably the consequence of radial acceleration induced by a putative companion in an extremely wide orbit with \psreb\ (\citealp{Bassa16,Kaplan16}, also see \ref{subsec:mspsrpi_Gaia_results}).
In addition to $\dot{P}_\mathrm{s}^\mathrm{int}$, $\dot{P}_\mathrm{b}^\mathrm{int}=\dot{P}_\mathrm{b}^\mathrm{obs}-\dot{P}_\mathrm{b}^\mathrm{Shk}-\dot{P}_\mathrm{b}^\mathrm{Gal}$ are estimated for four pulsar systems with reported $\dot{P}_\mathrm{b}^\mathrm{obs}$. 
The improved \psri\ parallax as well as the re-assessed \psrea\ parallax calls for an update to the constraint on alternative theories of gravity \citep[e.g.][]{Freire12,Zhu19,Ding20}, which is discussed in \ref{subsec:mspsrpi_alternative_gravity}.

While performing the $\mathcal{A}_\mathrm{Gal}$ analysis, we found an error in the code that had been used to implement the calculation of \ref{eq:mspsrpi_PbGal} for the \citet{Ding21a} work (which, to be clear, is not an error in the {\tt GalPot}\footnote{\label{footnote:galpot}\url{https://github.com/PaulMcMillan-Astro/GalPot}} package that provides the $\varphi(\vec{x})$ models). Therefore, we note that the $\dot{P}_\mathrm{b}^\mathrm{Gal}$ of \psrfb\ in \ref{tab:mspsrpi_Shk} is a correction to the \citet{Ding21a} counterpart. 
%A dedicated erratum will follow the release of this paper.
Further discussions on \psrfb\ can be found in \ref{subsec:mspsrpi_J1537}.

Last but not least, assuming GR is correct, the approach taken above can be inverted to infer
$\mathcal{A}_\mathrm{Gal}^\mathrm{(GR)}\!=\!(\dot{P}_\mathrm{b}^\mathrm{obs} - \dot{P}_\mathrm{b}^\mathrm{GW} - \dot{P}_\mathrm{b}^\mathrm{Shk})c/P_\mathrm{b}$, which can be used to constrain Galactic parameters for the local environment (of the Solar system) \citep{Bovy20}, or probe the Galactic dark matter distribution in the long run \citep{Phillips21}. The  $\mathcal{A}_\mathrm{Gal}^\mathrm{(GR)}$ for the three viable pulsars are listed in \ref{tab:mspsrpi_A_Gal}. 
%In three out of the four cases (including \psrnb),  $\mathcal{A}_\mathrm{Gal}$ derived from the NT95 analytical solution is further away from the GR expectation compared to the later $\varphi(\vec{x})$ models. 

%In this work, we correct

\begingroup
\renewcommand{\arraystretch}{1.4} % Default value: 1

\begin{table}
\raggedright
\caption{Radial accelerations due to Galactic gravitational pull based on different models of Galactic gravitational potential}
\label{tab:mspsrpi_A_Gal}
%\begin{tabular}{@{}l@{\:}l@{\:}l@{}} % manual @ spacing to prevent this being too wide for a page
\resizebox{\columnwidth}{!}{
\begin{tabular}{lcccccc}
\hline
\hline
PSR   & $\mathcal{A}_\mathrm{Gal}^\mathrm{(NT95)}$ & $\mathcal{A}_\mathrm{Gal}^\mathrm{(DB98)}$ & $\mathcal{A}_\mathrm{Gal}^\mathrm{(BT08)}$ & $\mathcal{A}_\mathrm{Gal}^\mathrm{(P14)}$ & $\mathcal{A}_\mathrm{Gal}^\mathrm{(M17)}$ & $\mathcal{A}_\mathrm{Gal}^\mathrm{(GR)}$ $^{*}$ \\
 & ($\mathrm{pm~s^{-2}}$) &  ($\mathrm{pm~s^{-2}}$) & ($\mathrm{pm~s^{-2}}$) & ($\mathrm{pm~s^{-2}}$) & ($\mathrm{pm~s^{-2}}$) & ($\mathrm{pm~s^{-2}}$)\\
\hline
\Psrb & -29(3) & $^!$-37.0(2) & $^!$-27.3(2) & -35.3(2) & -32.5(3) & --- \\
\Psrc & $^!$-12(1) & $-8.6^{+1.2}_{-0.8}$ & $^!$ $-6.0^{+1.0}_{-0.5}$ & $-10.9^{+0.7}_{-0.3}$ & $-8.8^{+1.0}_{-0.4}$ & --- \\
\Psrd & 24(4) & 23(5) & 24(5) & 24(5) & 25(5) & --- \\
\Psrea & $^{!!!}$-32.0(6) & -24.0(2) & $^!$-19.44(9) & -24.80(9) & -25.80(6) & 05(29) \\
\Psreb & $^!$-45.1(9) & -38.5(6) & $^!$-35.6(9) & -42(1) & -43(1) & --- \\
\Psrfa & -47.5(5) & -48.1(5) & $^!$-44.8(7) & -50.4(7) & $^!$-51.9(7) & --- \\
\Psrfb & $^{!!!!}$-29(1) & -42(1) & -39(2) & -43(2) & -45(2) & $21^{+28}_{-31}$ \\
\Psrga & $^{!!!}$-33(1) & -46(3) & -45(3) & -50(3) & -52(4) & --- \\
\Psrgb & $^{!!}$10(3) & $^!$ $-1.2^{+0.8}_{-0.6}$ & $^!$ $3.2^{+0.7}_{-0.5}$ & $0.6^{+0.7}_{-0.6}$ & $1.3^{+0.6}_{-0.4}$ & --- \\
\Psrhb & 13.2(8) & 10.8(6) & 12.1(6) & 11.5(6) & 12.5(7) & --- \\
\Psri  & $^{!!!!!!}$10.1(1.8) & $-6.4^{+0.4}_{-0.6}$ & $-4.2^{+0.8}_{-1.1}$ & $-7.5^{+0.7}_{-1.0}$ & $-6.9^{+0.8}_{-1.2}$ & 9(35) \\
\Psrka & -13(3) & $-13^{+3}_{-4}$ & $-13^{+3}_{-5}$ & $-19^{+4}_{-5}$ & $-16^{+4}_{-5}$ & --- \\
\Psrl  & -35(13) & $-29^{+8}_{-16}$ & $-31^{+10}_{-18}$ & $-42^{+12}_{-21}$ & $-36^{+10}_{-20}$ & --- \\
\Psrmb & $^{!!}$14(2) & 5.9(5) & $^!$8.7(5) & $^!$5.0(2) & 7.4(3) & --- \\
\Psrkb & -64(8) & -53(8) & -56(8) & -67(9) & -63(9) & --- \\

\hline

\multicolumn{7}{l}{$\bullet$ NT95, DB98, BT08, P14 and M17 refer to five different $\varphi(\vec{x})$ models (see \ref{subsec:mspsrpi_A_Gal} for}\\
\multicolumn{7}{l}{\ \ \ the references).}\\
\multicolumn{7}{l}{$\bullet$ The ``!''s indicate, in the same way as \ref{tab:mspsrpi_VLBI_timing_results}, the significance of the offset between the $\mathcal{A}_\mathrm{Gal}$}\\
\multicolumn{7}{l}{\ \ \ in \ref{tab:mspsrpi_Shk} and that of each $\varphi(\vec{x})$ model.}\\
\multicolumn{7}{l}{$^*$ $\mathcal{A}_\mathrm{Gal}^\mathrm{(GR)}\!=\!(\dot{P}_\mathrm{b}^\mathrm{obs} - \dot{P}_\mathrm{b}^\mathrm{GW} - \dot{P}_\mathrm{b}^\mathrm{Shk})c/P_\mathrm{b}$ is based on the assumption that GR is correct.}\\

\end{tabular}
}
\end{table}
\endgroup

\subsection{New constraints on alternative theories of gravity}
%\subsection{DNS}
\label{subsec:mspsrpi_alternative_gravity}
In the GR framework, the excess orbital decay $\dot{P}_\mathrm{b}^\mathrm{ex}=\dot{P}_\mathrm{b}^\mathrm{int}-\dot{P}_\mathrm{b}^\mathrm{GW}$ is expected to agree with zero. However, some alternative theories of gravity expect otherwise due to their predictions of non-zero dipole gravitational radiation and time-varying Newton's gravitational constant $G$. Both phenomena are prohibited by GR. Namely, in GR, the dipole gravitational radiation coupling constant $\kappa_D=0$, and $\dot{G}/G=0$. The large asymmetry of gravitational binding energy of pulsar-WD systems makes them ideal testbeds for dipole gravitational emissions \citep[e.g.][]{Eardley75}. 
In an effort to test (and possibly eliminate) alternative theories of gravity, increasingly tight constraints on $\kappa_D$ and $\dot{G}/G$ have been placed using multiple pulsar-WD systems \citep{Deller08,Freire12,Zhu19,Ding20}.

With the reassessed astrometric results of \psrea, the $\dot{P}_\mathrm{b}^\mathrm{ex}$ of \psrea\ changes from $10.6\pm6.1\,\mathrm{fs~s^{-1}}$ in \citet{Ding20} to $5.1\pm5.1\,\mathrm{fs~s^{-1}}$. This change is mainly caused by three reasons: {\bf 1)} priors are placed on the proper motion during inference in this work (but not in \citealp{Ding20}); {\bf 2)} a Bayesian framework is applied in this work (while \citealp{Ding20} reported bootstrap results); {\bf 3)} this work adopts PDF medians as the estimates (while \citealp{Ding20} used PDF modes). Though barely affecting this work (see \ref{fig:mspsrpi_J1012_dist}), the choice between PDF mode and median makes a difference to \citet{Ding20} given that their parallax PDF is more skewed (see Figure~4 of \citealp{Ding20}).
After employing the new VLBI+timing distance, the $\dot{P}_\mathrm{b}^\mathrm{ex}$ of \psri\ has shifted from $2.0\pm3.7\,\mathrm{fs~s^{-1}}$ \citep{Freire12} to $1.6\pm3.5\,\mathrm{fs~s^{-1}}$. More discussions on \psri\ can be found in \ref{subsec:mspsrpi_J1738}.

With the new $\dot{P}_\mathrm{b}^\mathrm{ex}$ of \psrea\ and \psri, we updated the constraints on $\kappa_D$ and $\dot{G}/G$ in exactly the same way as \citet{Ding20}. The prerequisites of this inference are reproduced in \ref{tab:mspsrpi_PbdotEx}, where the two underlined $\dot{P}_\mathrm{b}^\mathrm{ex}$ are the only difference from the Table~6 of \citet{Ding20}.  
We obtained 
\begin{equation}
\label{eq:mspsrpi_gdot_kD}
\begin{split}
    \dot{G}/G &= -1.6^{\,+5.3}_{\,-4.8}\times10^{-13}\,\mathrm{yr^{-1}}, \\
    \kappa_D &= -1.1^{\,+2.4}_{\,-0.9}\times10^{-4}.
\end{split}
\end{equation}
Compared to \citet{Ding20}, $\kappa_D$ becomes more consistent with zero, while the new uncertainties of $\kappa_D$ and $\dot{G}/G$ remain at the same level.

\begingroup
\renewcommand{\arraystretch}{1.4} % Default value: 1

\begin{table}
\centering
\caption{Excess orbital decay $\dot{P}_\mathrm{b}^\mathrm{ex}=\dot{P}_\mathrm{b}^\mathrm{obs}-\dot{P}_\mathrm{b}^\mathrm{Shk}-\dot{P}_\mathrm{b}^\mathrm{Gal}$ and other prerequisites for constraining $\dot{G}/G$ and $\kappa_D$}
\label{tab:mspsrpi_PbdotEx}
%\begin{tabular}{@{}l@{\:}l@{\:}l@{}} % manual @ spacing to prevent this being too wide for a page
%\resizebox{\columnwidth}{!}{
\begin{tabular}{lccccc}
\hline
\hline
PSR   & $P_\mathrm{b}$ & $\dot{P}_\mathrm{b}^\mathrm{ex}$ & $m_\mathrm{p}$ & $m_\mathrm{c}$ & $q$  \\
 & (d) &  ($\mathrm{fs~s^{-1}}$) & (\msun) & (\msun) &   \\
\hline
\Psrna & 5.74 & 12(32) & 1.44(7) & 0.224(7) & --- \\
\Psrea & 0.60 & \underline{5.1(5.1)} & --- & 0.174(11) & 10.44(11) \\
\Psrnb & 67.83 & 30(150) & 1.33(10) & 0.290(11) & --- \\
\Psri & 0.35 & \underline{1.6(3.5)} & 1.46(6) & --- & 8.1(2) \\

\hline

\multicolumn{6}{l}{$\bullet$ $m_\mathrm{p}$, $m_\mathrm{c}$ and $q$ stand for, respectively, pulsar mass, companion mass and}\\
\multicolumn{6}{l}{\ \ \ mass ratio (i.e., $m_\mathrm{p}/m_\mathrm{c}$).
See \citet{Ding20} for their references.}\\

\end{tabular}

\end{table}
\endgroup

%\section{Rotation between ICRF and planetary-ephemeris-dependent barycentric frame}

\section{Individual pulsars}
\label{sec:mspsrpi_individual_pulsars}

In this section, we discuss the impacts of the new astrometric measurements (particularly the new distances) on the scientific studies around individual pulsars. Accordingly, special attention is paid to the cases where there is no published timing parallax $\varpi^\mathrm{(Ti)}$.
%(or the most precise $\varpi^\mathrm{(Ti)}$ compiled in \ref{tab:mspsrpi_VLBI_timing_results} remains discrepant from our $\varpi'$). 
In addition, we also look into the two pulsars (i.e. \psrha\ and \psro) that have $\varpi'$ consistent with zero, in an effort to understand the causes of parallax non-detection.
%Besides, the posterior distributions of $i'$ and $\Omega'_\mathrm{asc}$ (regarding the four 8P pulsars) are described in this section as well.
%The constraint on the scattering screens from angular broadening effects would be covered here, and summarized in a table.

%\subsection{\psrb}
%The sole pulsar well studied by NICER. The new astrometric results are very consistent with timing results.

\subsection{\psrc}
\label{subsec:mspsrpi_J0610}
\psrc\ is the third black widow pulsar discovered \citep{Burgay06}, which is in a 7-hr orbit with an extremely low-mass ($\approx0.02$\,\msun, \citealp{Pallanca12}) star.
Adopting a distance of around 2.2\,kpc, \citet{van-der-Wateren22} obtained a $\gamma$-ray emission efficiency $\eta_\gamma \equiv 4 \pi F_\mathrm{\gamma} D^2 / \dot{E}^\mathrm{int}$ in the range of 0.5--3.7, where $\dot{E}^\mathrm{int}$ and $F_\mathrm{\gamma}$ are, respectively, the intrinsic NS spin-down power and the $\gamma$-ray flux above 100\,MeV.

In addition, \citet{van-der-Wateren22} estimated a mass function
\begin{equation}
\label{eq:mspsrpi_mass_func}
f(m_\mathrm{p},q)=m_\mathrm{p} \frac{\sin^3{i}}{q(q+1)^2}=\frac{4\pi^2 a_1^3}{G P_\mathrm{b}^2}
\end{equation}
of $5.2\times10^{-6}$\,\msun\ for the \psrc\ system (where $q \equiv m_\mathrm{p}/m_\mathrm{c}$). Besides, they determined the irradiation temperature (of the companion) $T_\mathrm{irr}=2820\pm190$\,K as well as the projected orbital semi-major axis $a_1=7.3\times10^{-3}$\,lt-s. 
Combining these three estimates, we calculated the heating luminosity
\begin{equation}
\label{eq:mspsrpi_L_irr}
\begin{split}
   L_\mathrm{irr} &\equiv 4 \pi \left[\frac{a_1 (1+q)}{\sin{i}}\right]^2 \sigma_\mathrm{SB} T_\mathrm{irr}^4 \\
   &\approx 4 \pi a_1^2 \left[ \frac{m_\mathrm{p}}{f(m_\mathrm{p},q)} \right]^{2/3} \sigma_\mathrm{SB} T_\mathrm{irr}^4 \\
   &\sim  9.1\times10^{32} \left( \frac{m_\mathrm{p}}{1.4\,\msun}\right)^{2/3} \mathrm{erg~s^{-1}}, 
\end{split}
\end{equation}
%the heating luminosity $L_\mathrm{irr} \equiv 4 \pi A$
where $\sigma_\mathrm{SB}$ represents the Stefan-Boltzmann constant.

Our new distance $D=1.5^{+0.3}_{-0.2}$\,kpc to \psrc\ is less than half the DM-based distances (see \ref{tab:mspsrpi_VLBI_timing_results}), and significantly below that assumed by \citet{van-der-Wateren22}. 
Assuming a NS moment of inertia $I_\mathrm{NS}=10^{45}\,\mathrm{g~cm^2}$,
the $\dot{P}_\mathrm{s}^\mathrm{int}$ of \psrc\ (see \ref{tab:mspsrpi_Shk}) corresponds to an intrinsic spin-down power 
\begin{equation}
\label{eq:mspsrpi_Edot}
    \dot{E}^\mathrm{int} \equiv 4 \pi^2 I_\mathrm{NS} \dot{P}_\mathrm{s}^\mathrm{int}/P_\mathrm{s}^3
\end{equation}
of $(5.1\pm0.5)\times10^{33}\,\mathrm{erg~s^{-1}}$, which is roughly twice as large as the $\dot{E}^\mathrm{int}$ range calculated by \citet{van-der-Wateren22}.
In conjunction with a smaller $\gamma$-ray luminosity $L_\gamma=4 \pi F_\mathrm{\gamma} D^2$ (due to closer distance), the $\dot{E}^\mathrm{int}$ reduced $\eta_\gamma$ to around 0.37 (from $0.5 < \eta_\gamma < 3.7$ estimated by \citealp{van-der-Wateren22}), disfavoring unusually high $\gamma$-ray beaming towards us. %Assuming the irradiation of the companion is powered essentially by the spin-down energy of the NS, smaller amount of total $\gamma$-ray emissions implies an irradiation efficiency about twice as high as the $\epsilon/\eta_\gamma\sim0.25$ evaluated by \citet{van-der-Wateren22} (see the paper for the definition of $\epsilon$).
%Conversely, assuming isotropic $\gamma$-ray emissions, $4 \pi F_\mathrm{\gamma} D^2 < \dot{E}^\mathrm{int}$, thus obtaining $I_\mathrm{NS} > \eta_\gamma \cdot 10^{45}\,\mathrm{g~cm^2}=3.5\times10^{44}\,\mathrm{g~cm^2}$.
Moreover, the heating efficiency $\epsilon_T$ drops to $\sim0.17$ (from $0.15<\epsilon_T<0.77$ evaluated by \citealp{van-der-Wateren22}), disfavoring the scenario where the NS radiation is strongly beamed towards the companion.

%\subsection{\psrd}
%\label{subsec:mspsrpi_J0621}
%An IMBP (intermediate-mass binary pulsar) with 29\,ms spin period and 0.97\,\msun\ \citep{Splaver02}. After implementing 1D interpolation, the significance of $\varpi'$ just exceeded 5\,$\sigma$.

%\subsection{\psrea}
%\label{subsec:mspsrpi_J1012}
%\psrea\ is in a 0.6-d orbit with a 0.174-\msun\ \citep{Antoniadis16} helium WD companion. \citet{Mata-Sanchez20} estimated the most precise mass ratio $q=10.44\pm0.11$ for the system. 

%\subsection{\psreb}
%\label{subsec:mspsrpi_J1024}

%The 5.2-ms \psreb\ is the only MSP (in the \mspsrpi\ catalogue) that has an unphysical negative $\dot{P}_\mathrm{s}^\mathrm{int}$ (see \ref{tab:mspsrpi_Shk}), which can be explained by the existence of a far-away companion star on the near side of the NS \citep{Bassa16,Kaplan16}. More explicitly, \citet{Kaplan16} proposed that \psreb\ is in a 2--20\,kyr orbit with a low-mass ($\approx0.4$\,\msun) main-sequence star.Our astrometry shows a large transverse velocity of 300\,km/s, probably consistent with the ultra-wide orbit. \hao{there is an F2 for this; $\dot{P}_\mathrm{s}^\mathrm{int}<0$.}

\subsubsection{On the DM discrepancy}
\label{subsubsec:mspsrpi_J0610_DM}
In \ref{subsubsec:mspsrpi_DM_distances}, we noted that our VLBI parallax-derived distance and the DM model-inferred distance to this pulsar differed 
substantially.
%by more than $7\,\sigma$. 
Specifically, PSR~J0610$-$2100 has a measured $\mathrm{DM} = 60.7$~pc~cm${}^{-3}$ while the NE2001 model predicts 27.5~pc~cm${}^{-3}$ for a line of sight of length 1.5\,kpc.  We attribute this discrepancy to thermal plasma or ``free electrons’’ along the line of sight that is not captured fully by a ``clump’’ in the NE2001 model.  The NE2001 model includes this ``clump’’ to describe the effects due to the Mon~R2 \ion{H}{2} region, centered at a Galactic longitude and latitude of (214\degr, -12.6\degr), 
located at an approximate distance of $\sim0.9$\,kpc \citep{Herbst75}.  However, the WHAM survey shows considerable H$\alpha$ in this direction, extending over tens of degrees.  Lines of sight close to the pulsar show changes in the H$\alpha$ intensity by factors of two, but an approximate value toward the pulsar is roughly 13~Rayleighs, equivalent to an emission measure $\mathrm{EM} = 29$~pc~cm${}^{-6}$ (for a temperature $T = 8000$~K).  Using standard expressions, as provided in the NE2001 model, to convert EM to \hbox{DM}, there is sufficient H$\alpha$ intensity along the line of sight to account for the excess DM that we infer from the difference between our parallax-derived distance and the NE2001 model distance.

\subsection{\psrfa}
\label{subsec:mspsrpi_J1518}
The 41-ms \psrfa, discovered by \citet{Sayer97}, is one of the only two DNSs in the current sample.
According to \citet{Janssen08}, the non-detection of Shapiro delay effects suggests $\sin{i} \leq 0.73$ at 99\% confidence level. 
Accordingly, we adopted 0.73 as the upper limit of $\sin{i}$, and carried out 8-parameter Bayesian inference, which led to a bi-modal posterior PDF on $i'$ and a multi-modal PDF on $\Omega'_\mathrm{asc}$ (see the online corner plot\textsuperscript{\ref{footnote:pulsar_positions}}). The predominant constraints on both $i'$ and $\Omega'_\mathrm{asc}$ come from the $\dot{a}_1$ measurement (\citealp{Janssen08} or see \ref{tab:mspsrpi_7_parameter_inference}). %as is shown with the crescent-shaped 2D histogram of $i'$ and $\Omega'_\mathrm{asc}$ in \ref{fig:mspsrpi_orbital_constraints}.
Though there are 3 major likelihood peaks for the $\Omega'_\mathrm{asc}$, two of them gather around 171\degr, making the PDF relatively more concentrated.
When a much more precise $\dot{a}_1$ measurement is reached with new timing observations, the existing VLBI data will likely place useful constraints on $i'$ and $\Omega'_\mathrm{asc}$. So will additional VLBI observations.
%Assuming $\Omega'_\mathrm{asc}$, we estimated $\Omega'_\mathrm{asc}=195^{+25}_{-47}$\,deg, where 
%On the other hand, the $\sin{i} \leq0.73 $ constraint corresponds to $i\leq47\degr$ or $i\geq133\degr$. Our Bayesian inference does not favor one side of inclination over another.
%This slight preference is enforced by the relatively strong correlation between RA and $\Omega_\mathrm{asc}$ ($\rho_{\alpha,\Omega_\mathrm{asc}}=-0.63$).

In addition to $i'$ and $\Omega'_\mathrm{asc}$, the 8-parameter Bayesian inference also renders a $40\,\sigma$ parallax $\varpi'$, which becomes the most significant parallax achieved for a DNS. 
In contrast, to detect a timing parallax $\varpi^\mathrm{(Ti)}$ for \psrfa\ would take $\gtrsim600$ years \citep{Janssen08}, due to its relatively high ecliptic latitude of 63\degr.
%\hao{add a $\dot{P}_\mathrm{b}$ limit.}

\subsection{\psrfb}
\label{subsec:mspsrpi_J1537}
\psrfb, also known as PSR~B1534$+$12, is the second discovered DNS \citep{Wolszczan91}. The DNS displays an exceptionally high proper motion amongst all Galactic DNSs (see Table~3 of \citealp{Tauris17}), leading to an unusually large Shklovskii contribution to observed timing parameters. Therefore, precise astrometry of the DNS plays an essential role in its orbital decay test of GR.
The most precise astrometric parameters of \psrfb\ are provided by \citet{Ding21a} based on the same dataset used in this work, which result in $\dot{P}_\mathrm{b}^\mathrm{Shk}=53\pm4$\,\fsps. 
Subsequently, \citet{Ding21a} estimated $\dot{P}_\mathrm{b}^\mathrm{Gal}=-1.9\pm0.2$\,\fsps, and concluded $\dot{P}_\mathrm{b}^\mathrm{int}/\dot{P}_\mathrm{b}^\mathrm{GW}=0.977\pm0.020$.

In this work, we inferred $\eta_\mathrm{EFAC}$ on top of the canonical astrometric parameters, which is the only difference from the Bayesian method of \citet{Ding21a}. Despite this difference, our astrometric results of \psrfb\ remain almost the same as \citet{Ding21a}. So is our re-derived $\dot{P}_\mathrm{b}^\mathrm{Shk}=53.3^{+3.8}_{-3.3}$\,\fsps. 
However, as is mentioned in \ref{subsec:mspsrpi_A_Gal}, the $\dot{P}_\mathrm{b}^\mathrm{Gal}$ estimated by \citet{Ding20c} is incorrect due to a coding error. After correction, $\dot{P}_\mathrm{b}^\mathrm{Gal}$ drops to $-5.1\pm0.4$\,\fsps\ (see \ref{tab:mspsrpi_Shk}). Consequently, we obtained $\dot{P}_\mathrm{b}^\mathrm{int}/\dot{P}_\mathrm{b}^\mathrm{GW}=0.96\pm0.02$. %Namely, the observed intrinsic orbital decay becomes just within $2\,\sigma$ of the GR prediction, which is yet not gravely concerning.

As \citet{Ding21a} have pointed out, the limiting factor of the GR orbital decay test using \psrfb\ remains the distance precision, which generally improves as $t^{-1/2}$ with additional observations, but can be accelerated if more sensitive instrumentation can be deployed.
%In this work, we re-assessed proper motion and distance of \psrfb\ in a Bayesian method slightly different from \citet{Ding21a}, in that we infer
%Recently, \citet{Ding21a} estimated $\dot{P}_\mathrm{b}^\mathrm{Gal}=-1.9\pm0.2$\,\fsps.
On the other hand, the extremely high braking index of 157 (two orders higher than the normal level) calculated from the rotational frequency $\nu_\mathrm{s} \equiv 1/P_\mathrm{s}$, its first derivative $\dot{\nu}_\mathrm{s}$ and its second derivative $\ddot{\nu}_\mathrm{s}$ \citep{Fonseca14}  indicate likely timing noise contributions that may affect the observed orbital period derivative to some degree.
%, which may bias $\dot{P}_\mathrm{b}^\mathrm{obs}$. 
This will be clarified with continued timing observations and refined timing analysis.
%\hao{have a look at the F2 of \psrfb.}

\subsection{\psrga}
\label{subsec:mspsrpi_J1640}

\psrga\ is a 3.2-ms MSP \citep{Foster95} in a wide ($P_\mathrm{b}=175$\,d) orbit with a WD companion \citep{Lundgren96}. 
The \mspsrpi\ results for \psrga\ have been published in \citet{Vigeland18}, which are highly consistent with our re-assessed quasi-VLBI-only results (see Table~2 of \citealp{Vigeland18} and \ref{tab:mspsrpi_models_no_pm_prior}).
Our 8-parameter Bayesian inference renders 
a 1D histogram of $\Omega'_\mathrm{asc}$ with 4 likelihood components at 0\degr, 140\degr, 200\degr\ and 320\degr, which is predominantly shaped by the prior on $\dot{a}_1$ from pulsar timing (see \ref{tab:mspsrpi_7_parameter_inference}).
%According to \citet{Perera19}, $\sin{i}=0.973\pm0.009$, which means $i=77\pm2$\degr\ or $i=103\pm2$\degr. This ambiguity dominates the $i$ uncertainty. With the Bayes factor $P(i>90\degr)/P(i<90\degr)=2$, our 8-parameter Bayesian inference still does not favor one $i$ range over another. 

\subsection{\psrgb}
\label{subsec:mspsrpi_J1643}

\psrgb\ is a 4.6-ms pulsar in a 147-d orbit with a WD companion \citep{Lorimer95a}. 
%The pulsar has a $\dot{a}_1$ measurement at the 1.4\% precision level (see \ref{tab:mspsrpi_7_parameter_inference}), which dominates the constraint on $i'$ and $\Omega'_\mathrm{asc}$. Despite the precise $\dot{a}_1$, the 
As a result of multi-path propagation due to inhomogeneities in the ionised interstellar medium (IISM), the pulse profiles of \psrgb\ are temporally broadened \citep[e.g.][]{Lentati17}.   
As the Earth-to-pulsar sightline moves through inhomogeneous scattering ``screen(s)'' (in the IISM), the temporal broadening $\tau_\mathrm{sc}$ varies with time; at 1\,GHz, $\tau_\mathrm{sc}$ fluctuates up and down by $\lesssim 5\,\mu$s on a yearly time scale \citep{Lentati17}.
Meanwhile, the moving scattering ``screen(s)'' would also change the radio brightness of the pulsar. This effect, as known as pulsar scintillation, is used to constrain the properties of both \psrgb\ and the scattering screen(s) between the Earth and the pulsar \citep{Mall22}. The scintillation of \psrgb\ has previously been modelled with both isotropic and anisotropic screens \citep{Mall22}. The isotropic model renders a pulsar distance  $D=1.0\pm0.3$\,kpc and locates the main scattering screen at $D_\mathrm{sc}=0.21\pm0.02$\,kpc; in comparison, the anisotropic model yields a pulsar distance $D=1.2\pm0.3$\,kpc, and necessitates a secondary scattering screen $0.34\pm0.09$\,kpc away (from the Earth) in addition to a main scattering screen at $0.13\pm0.02$\,kpc distance \citep{Mall22}.
%The distance to the main scattering screen derived from the anisotropic model strengthens the association between is  
On the other hand, the HII region Sh~2-27 in front of \psrgb\ is suspected to be associated with the main scattering screen of the pulsar. 
This postulated association is strengthened by the agreement between the distance to the main scattering screen (based on the two-screen anisotropic model, \citealp{Mall22}) and the distance to the HII region (i.e., $112\pm3$\,pc, \citealp{Ocker20}).

\subsubsection{Independent check on the postulated association between the HII region Sh~2-27 and the main scattering screen}
\label{subsubsec:mspsrpi_HII_region}

Besides the pulse broadening effect, the scattering by the IISM would lead to apparent angular broadening of the pulsar, which has been detected with VLBI at $\gtrsim8$\,GHz \citep[e.g.][]{Bower14}.
By the method outlined in Appendix~A of \citet{Ding20c}, we measured a semi-angular-broadened size $\theta_\mathrm{sc}=3.65\pm0.43$\,mas for \psrgb, which is the only significant $\theta_\mathrm{sc}$ determination in the \mspsrpi\ catalogue. 
Likewise, the secondary in-beam calibrator of \psrgb\ is also scatter-broadened, which may likely introduce additional astrometric uncertainties (see more explanation in \ref{subsec:mspsrpi_J1721}).

As both pulse broadening and angular broadening are caused by the IISM deflection, $\theta_\mathrm{sc}$, $\tau_\mathrm{sc}$, the pulsar distance $D$, and the distance(s) $D_\mathrm{sc}$ to the scattering screen(s) are geometrically related. Assuming there is one dominant thin scattering screen, we make use of following approximate relation 
\begin{equation}
\label{eq:mspsrpi_tau_sc}
\begin{split}
   \frac{\theta_\mathrm{sc}^2}{2c \tau_\mathrm{sc}} = \frac{1}{D_\mathrm{sc}} - \frac{1}{D}\,\,\,\,\,\,\,\,\,\, (\mathrm{when}\,\, \theta_\mathrm{sc} \lesssim 1\degr),
\end{split}
\end{equation}
where $c$ stands for the speed of light in vacuum.

\begin{figure}
    \centering
	%\raggedright
	\includegraphics[width=12cm]{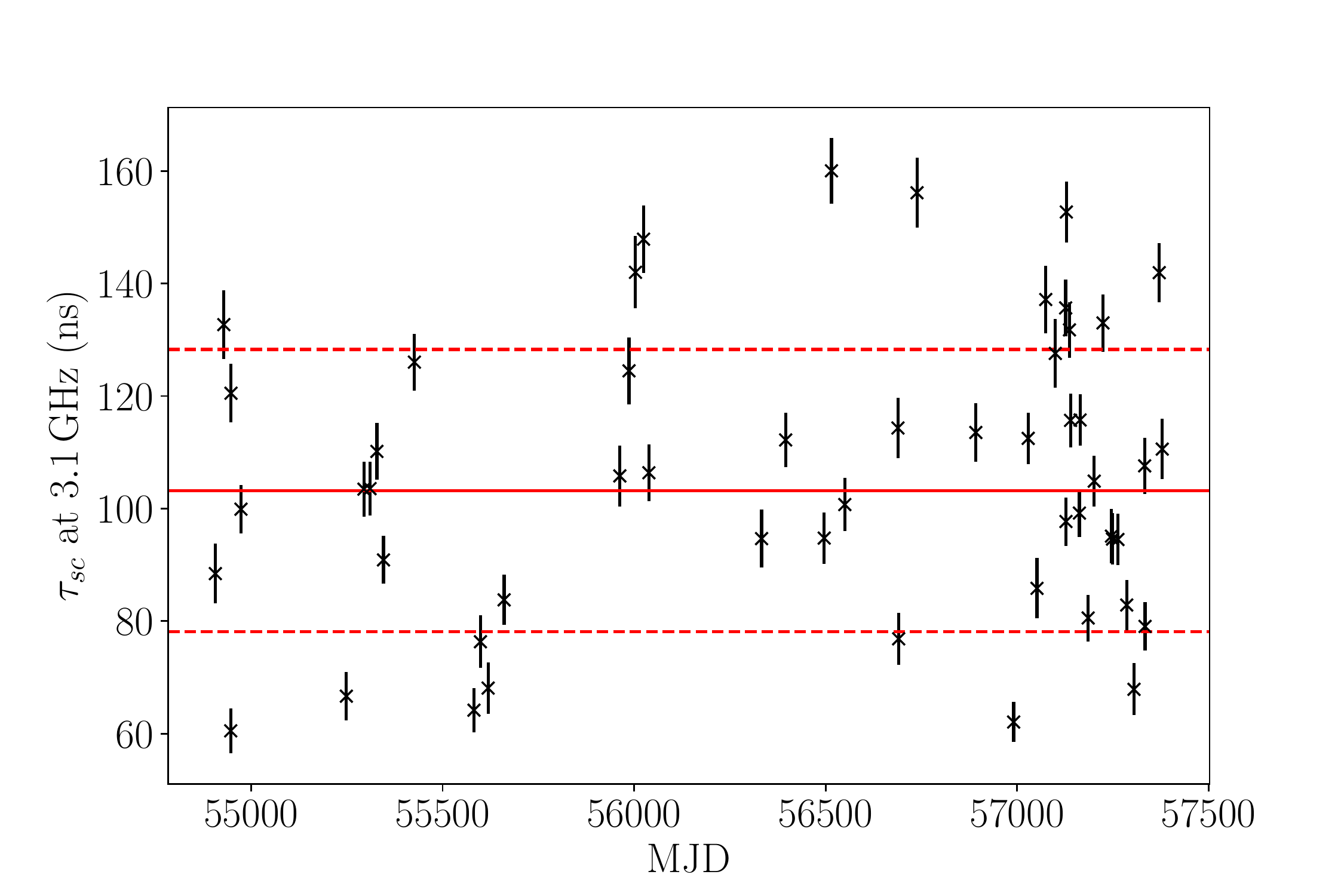}
    \caption{Temporal broadening $\tau_\mathrm{sc}$ of \psrgb\ at 3.1\,GHz. The solid red line and the dashed red line show the mean temporal broadening and a 68\% confidence interval, respectively.
    }    
    \label{fig:mspsrpi_tau_sc}
\end{figure}

To estimate the unknown $\tau_\mathrm{sc}$ at our observing frequency of $\sim$1.55\,GHz, we used the data spanning MJD~54900---57500 from the PPTA second data release \citep{Kerr20}. 
We analysed the dynamic spectra of observations centred around 3.1\,GHz and recorded with the PDFB4 processor, using the scintools\footnote{\url{https://github.com/danielreardon/scintools}} package \citep{Reardon20}. A model was fit to the auto-correlation function of each dynamic spectrum, which has an exponential decay with frequency \citep{Reardon19}. The characteristic scintillation scale (in frequency) $\Delta\nu_d$ is related to the scattering timescale with $\tau_{sc} = 1/(2\pi\Delta\nu_{d})$. 
We found the mean temporal broadening $\tau_\mathrm{sc}=103\pm25$\,ns at 3.1\,GHz, with fluctuations of $\lesssim60$\,ns (see \ref{fig:mspsrpi_tau_sc}).
%Assuming a frequency-scaling relation $\tau_\mathrm{sc} \propto \nu^{-11/3}$ (where $\nu$ stands for observing frequency) associated with the Kolmogorov turbulence \citep[e.g.][]{Armstrong95}, we obtained $\tau_\mathrm{sc}=1.3\pm0.3\,\mu$s at our observing frequency 1.55\,GHz.
To convert this $\tau_\mathrm{sc}$ to our observing frequency 1.55\,GHz, we compare the maximum degree (i.e., 60\,ns) of fluctuations at 3.1\,GHz to that (i.e., $5\,\mu$s, \citealp{Lentati17}) at 1\,GHz, and acquired an indicative scaling relation 
\begin{equation}
    \label{eq:mspsrpi_tau_sc_scaling}
    \tau_\mathrm{sc} \propto \nu^{-3.9},
\end{equation}
where $\nu$ is the observing frequency. This relation reasonably agrees with the scaling relation $\tau_\mathrm{sc} \propto \nu^{-11/3}$ associated with the Kolmogorov turbulence \citep[e.g.][]{Armstrong95}. 
With the indicative scaling relation, we calculated $\tau_\mathrm{sc}=1.54\pm0.37\,\mu$s.
It is timely to note that $\theta_\mathrm{sc}^2 / \tau_\mathrm{sc}$ (on the left side of \ref{eq:mspsrpi_tau_sc}) is frequency-independent. By combining \ref{eq:mspsrpi_tau_sc} and \ref{eq:mspsrpi_tau_sc_scaling}, we come to another equivalent indicative scaling relation
\begin{equation}
\label{eq:mspsrpi_theta_scaling}
    \theta_\mathrm{sc}\propto\nu^{-1.95}.
\end{equation}

Substituting $\tau_\mathrm{sc}=1.54\pm0.37\,\mu$s, $\theta_\mathrm{sc}=3.65\pm0.43$\,mas and $D=0.95^{+0.15}_{-0.11}$\,kpc into \ref{eq:mspsrpi_tau_sc}, we obtained $D_\mathrm{sc}=86^{+30}_{-24}$\,pc, where the uncertainty is derived with a Monte-Carlo simulation. This $D_\mathrm{sc}$ is consistent with the distance to the HII region Sh~2-27 \citep{Ocker20}, hence independently supporting the association between the HII region and the main scattering screen of \psrgb.

\subsubsection{Probing scintillation models}
\label{subsubsec:mspsrpi_scintillation_models}

Apart from the above check on the connection between the HII region Sh~2-27 and the main scattering screen, the angular broadening of \psrgb\ also promises a test of the aforementioned isotropic scintillation model proposed by \citet{Mall22}. 
To do so, we changed the pulsar distance to the one inferred with the model (i.e., $1.0\pm0.3$\,kpc). 
With this change, we derive $D_\mathrm{sc}=86^{+30}_{-24}$\,pc, which disagrees with $0.21\pm0.02$\,kpc based on the isotropic model. 
% 
%At 1.4\,GHz, we would expect $\theta_\mathrm{sc}\propto\nu^{-1.95}=4.45\pm0.52$\,mas.
%Though \citet{Mall22} is based on observations at 1.4\,GHz different from ours, the 
To investigate the impact of a different scaling relation $\tau_\mathrm{sc} \propto \nu^{-\alpha_\mathrm{sc}}$, we inferred $\tau_\mathrm{sc}=4.3\,\mu$s with both $D$ and $D_\mathrm{sc}$ based on the isotropic model \citep{Mall22}, which corresponds to an unreasonably large $\alpha_\mathrm{sc}=5.4$. 
Hence, we conclude that our $\theta_\mathrm{sc}$ and $\tau_\mathrm{sc}$ cannot easily reconcile with the one-screen isotropic model proposed by \citet{Mall22}.
%On the other hand, when the pulsar distance is changed to the one (i.e., $1.2\pm0.3$\,kpc, \citealp{Mall22}) provided by the anisotropic model, we came to $D_\mathrm{sc}=87^{+31}_{-24}$\,pc, which is roughly consistent with the main scattering screen distance inferred by the model.
%Therefore, our results favour the two-screen anisotropic model over the one-screen isotropic model.

Fundamentally, the irreconcilability implies a one-screen model might be incapable of describing both scintillation and angular broadening effects of \psrgb.
In principle, it is possible to analyse a multi-screen model with a $\theta_\mathrm{sc}(t)$ series (at various time $t$) and its associated $\tau_\mathrm{sc}(t)$, instead of only using their mean values. 
However, this analysis is not feasible for this work, as $\tau_\mathrm{sc}$ and $\theta_\mathrm{sc}$ were not measured on the same days.
Nonetheless, we can still investigate whether our observations of \psrgb\ can reconcile with the scintillation observations \citep{Mall22} in the context of a two-screen model.

In the scenario of two thin scattering screens, we derived the more complicated relation
\begin{equation}
\label{eq:mspsrpi_tau_sc2}
%\begin{split}
   \begin{cases}
   2c\tau_\mathrm{sc} &= k_1 \beta_\mathrm{sc}^2+k_2\beta_\mathrm{sc}\theta_\mathrm{sc}+k_3\theta_\mathrm{sc}^2\,\,\,\,\,\,\,\, (\theta_\mathrm{sc} \lesssim 1\degr \text{ and } \beta_\mathrm{sc} \lesssim 1\degr)\\[5pt]
   k_1 &= \frac{(D-D_\mathrm{sc2})(D-D_\mathrm{sc1})}{D_\mathrm{sc2}-D_\mathrm{sc1}} \\[5pt]
   k_2 &= -\frac{2D_\mathrm{sc1}(D-D_\mathrm{sc2})}{D_\mathrm{sc2}-D_\mathrm{sc1}}\\[5pt]
   k_3 &= \frac{D_\mathrm{sc1} D_\mathrm{sc2}}{D_\mathrm{sc2}-D_\mathrm{sc1}},
   \end{cases}
%\end{split}
\end{equation}
where $D_\mathrm{sc1}$ and $D_\mathrm{sc2}$ are the distance to the first and the second scattering screen, respectively; $\beta_\mathrm{sc}$ represents a half of the opening angle of the second scattering screen (closer to the pulsar) as seen from the pulsar.
As a side note, \ref{eq:mspsrpi_tau_sc} can be considered a special case (i.e., $D=D_\mathrm{sc2}$) of \ref{eq:mspsrpi_tau_sc2}.
In \ref{eq:mspsrpi_tau_sc2}, all parameters except $\beta_\mathrm{sc}$ are known, either determined with the anisotropic two-screen model or obtained in this work. Hence, it is feasible to constrain the geometric parameter $\beta_\mathrm{sc}$ with the known parameters.

However, \ref{eq:mspsrpi_tau_sc2} can yield unphysical solutions (i.e., $\beta_\mathrm{sc}>0$). We applied the simple condition 
\begin{equation}
\label{eq:mspsrpi_two_screen_condition1}
\begin{split}
   %\begin{cases}
    \frac{\theta_\mathrm{sc}^2}{2c \tau_\mathrm{sc}} \leq \frac{1}{D_\mathrm{sc1}} - \frac{1}{D} 
   %\end{cases}
\end{split}
\end{equation}
to ensure \ref{eq:mspsrpi_tau_sc2} gives physical solutions of $\beta_\mathrm{sc}$. 
This equation is equivalent to $D_\mathrm{sc1} \leq D_\mathrm{sc}$, 
where $D_\mathrm{sc}$ corresponds to the one-screen solution of \ref{eq:mspsrpi_tau_sc}. This is because $D_\mathrm{sc1} > D_\mathrm{sc}$ would always lead to longer routes, thus exceeding the $\tau_\mathrm{sc}$ budget. It is important to note that \ref{eq:mspsrpi_two_screen_condition1} is valid for a model with any number of scattering screens. Hence, we recommend to use \ref{eq:mspsrpi_two_screen_condition1} in scintillation model inferences as a prior condition, to cater for the constraints imposed by $\theta_\mathrm{sc}$ and $\tau_\mathrm{sc}$ (and thereby truncate the parameter space of a scintillation model).

%Therefore, a physical solution of $\beta_\mathrm{sc}>0$ would at least suggest the reconcilability between the scintillation and our VLBI observations of \psrgb.

%In \ref{eq:mspsrpi_tau_sc2}, frequency-dependent changes in $\theta_\mathrm{sc}$ and $\tau_\mathrm{sc}$ do not cancel out (as they did in the one-screen case). Hence, we converted $\theta_\mathrm{sc}$ and $\tau_\mathrm{sc}$ from 1.55\,GHz to the observing frequency 1.4\,GHz of \citet{Mall22}, using the respective scaling relations (i.e., \ref{eq:mspsrpi_tau_sc_scaling} and \ref{eq:mspsrpi_theta_scaling}). We substituted $D_\mathrm{sc1}$, $D_\mathrm{sc2}$ and $D$ based on the anisotropic two-screen model into \ref{eq:mspsrpi_tau_sc2} alongside the converted $\theta_\mathrm{sc}$ and $\tau_\mathrm{sc}$, and did not obtain a physical solution.

To test the anisotropic two-screen model \citep{Mall22} with our $\theta_\mathrm{sc}$ and $\tau_\mathrm{sc}$, we calculated $D_\mathrm{sc}=89^{+33}_{-26}$\,pc with the pulsar distance (i.e., $D=1.2\pm0.3$\,kpc) based on the anisotropic two-screen model. This $D_\mathrm{sc}$ is consistent with $D_\mathrm{sc1}=129\pm15$\,pc \citep{Mall22} (therefore not ruling out $D_\mathrm{sc1} < D_\mathrm{sc}$). That is to say, our $\theta_\mathrm{sc}$ and $\tau_\mathrm{sc}$ measurements can loosely reconcile with the anisotropic two-screen model \citep{Mall22}.
%Accordingly, $D_\mathrm{sc1}-D_\mathrm{sc}=40^{+30}_{-36}$\,pc
In comparison, we reiterate our finding that a one-screen model is difficult to describe both scintillation and angular broadening effects of \psrgb.

 %through which the pulsar emissions reach us. 
%A physical solution to $\beta_\mathrm{sc}$ indicates that the scintillation and VLBI observations can reconcile in the framework of a two-screen model. 

\subsection{\psrha}
\label{subsec:mspsrpi_J1721}
\psrha\ is a 3.5-ms solitary MSP discovered at intermediate Galactic latitudes \citet{Edwards01}.
The main secondary phase calibrator of \psrha\ (and indeed, all the sources near it on the plane of the sky) is heavily resolved due to IISM scattering, which leads to non-detections on the longest baselines and a lack of calibration solutions for some antennas, reducing the spatial resolution of the VLBI observations. The non-uniform IISM distribution also leads to refractive image wander as the line-of-sight to the pulsar changes \citep[e.g.,][]{Kramer21a}, which is most pronounced for heavily scatter-broadened sources such as the calibrator for \psrha. In conjunction with the lower spatial resolutions, which reduces positional precision to begin with, this additional noise term likely results in the parallax non-detection (see \ref{tab:mspsrpi_models_no_pm_prior}).

\subsection{\psrhb}
\label{subsec:mspsrpi_J1730}
\psrhb\ is a solitary pulsar spinning at $P_\mathrm{s}=8.1$\,ms \citep{Lorimer95a}. Being so far the least-energetic (in terms of $\dot{E}^\mathrm{int}$) $\gamma$-ray pulsar \citep{Guillemot16}, the pulsar plays a key role in exploring the death line of NS high-energy radiation. 
Substituting $\dot{P}_\mathrm{s}^\mathrm{int}$ and $P_\mathrm{s}$ of \ref{eq:mspsrpi_Edot} with values listed in \ref{tab:mspsrpi_Shk}, we substantially refined the $\dot{E}^\mathrm{int}$ death line (of all $\gamma$-ray-emitting pulsars) to 
\begin{equation}
    \dot{E}_\mathrm{death} \leq \dot{E}^\mathrm{int}_\mathrm{J1730} = (1.15\pm0.01)\times10^{33}\left(\frac{I_\mathrm{NS}}{10^{45}~\mathrm{g~cm^2}}\right)\,\mathrm{erg~s^{-1}},
\end{equation}
which is consistent with (but on the higher side of) the previous estimate $(8.4\pm2.2)\times10^{32}\,\mathrm{erg~s^{-1}}$ by \citet{Guillemot16} (assuming the same $I_\mathrm{NS}$).
%five times smaller than the $\dot{E}^\mathrm{int}$ of \psrc\ (see \ref{subsec:mspsrpi_J0610}).
On the other hand, we updated the $\gamma$-ray luminosity (above 100\,MeV) to $L_\gamma=4 \pi D^2 F_\mathrm{\gamma}=(3.1\pm1.6)\times10^{32}\,\mathrm{erg~s^{-1}}$, where the precision is limited by the less precise $F_\gamma$ \citep{Guillemot16}. Accordingly, we obtained $\eta_\gamma = 0.27\pm0.14$.

\subsection{\psri}
\label{subsec:mspsrpi_J1738}
\psri, discovered from a 1.4-GHz high-Galactic-latitude survey with the 64-m Parkes radio telescope \citep{Jacoby09}, is a 5.85-ms pulsar in a 8.5-hr orbit with a WD companion. Thanks to the relatively short $P_\mathrm{b}$, the WD-pulsar system plays a leading role in constraining alternative gravitational theories that predict dipole gravitational radiation \citep{Freire12,Zhu15}.

Our VLBI-only $\varpi$ is $2.3\,\sigma$ away from the most precise $\varpi^\mathrm{(Ti)}$ (see \ref{tab:mspsrpi_models_no_pm_prior} and \ref{tab:mspsrpi_VLBI_timing_results}). After adopting timing priors, $\varpi'=0.589\pm0.046$\,mas becomes closer to $\varpi^\mathrm{(Ti)}=0.68\pm0.05$\,mas \citep{Freire12}, meaning that $\dot{P}^\mathrm{Shk}_\mathrm{b}$ is only 1.2 times larger than the previous estimate.
On the other hand, our re-assessed $\dot{P}^\mathrm{Gal}_\mathrm{b}$, based on more realistic $\varphi(\vec{x})$ models, is smaller than that estimated with the NT95 $\varphi(\vec{x})$ model \citep{Freire12}.
Combining the unchanged  $\dot{P}^\mathrm{obs}_\mathrm{b}=-17\pm3$\,\fsps\, the re-derived  $\dot{P}^\mathrm{Int}_\mathrm{b}=-26.1\pm3.1$\,\fsps\ is almost the same as the previous estimate, as the change of $\dot{P}^\mathrm{Gal}_\mathrm{b}$ happens to nearly cancels out that of $\dot{P}^\mathrm{Shk}_\mathrm{b}$.
%shifts to the smaller side of the previous estimate, while becoming more consistent with the previously estimated $\dot{P}^\mathrm{GW}_\mathrm{b}=-27.7^{+1.5}_{-1.9}$\,\fsps\ \citep{Freire12}.

Future pulsar timing or VLBI investigation into the discrepancy between $\varpi^\mathrm{(Ti)}$ \citep{Freire12} and $\varpi$ is merited by the importance of the pulsar system. Specifically, if the true parallax turns out to be around 0.5\,mas, it would not only mean that $\dot{P}^\mathrm{Shk}_\mathrm{b}$ is 1.4 times higher than the estimate by \citet{Freire12}, but also suggest the WD radius $R_\mathrm{WD}$ to be 1.4 larger (as $R_\mathrm{WD} \propto D$ according to Equation~1 of \citealp{Antoniadis12}). A higher $R_\mathrm{WD}$ would lead to  lighter WD and NS (as the mass ratio is well determined), thus smaller $\dot{P}^\mathrm{GW}$.
%the new distance would also have an impact on the WD radius $R_\mathrm{WD}$, which feeds into estimates of both the WD and NS mass. According to Equation~1 of \citet{Antoniadis12}, $R_\mathrm{WD} \propto D$. 
%Scaling the previous $R_\mathrm{WD}=0.037\,R_\odot$ \citep{Antoniadis12} by the ratio of parallax-based distances (i.e, 2.2/1.47), we obtained $R_\mathrm{WD}=(5.5\pm0.6)\times10^{-2}\,R_\odot$, where we assume the new parallax uncertainty dominates the $R_\mathrm{WD}$ uncertainty. This larger $R_\mathrm{WD}$ would correspond to a smaller WD mass $m_\mathrm{c}$ (as well as a smaller NS mass $m_\mathrm{p}$, since the mass ratio $q$ is well determined by \citealp{Antoniadis12}). Deriving this $m_\mathrm{c}$ is, however, beyond the scope of this paper. 

%Therefore, in this paper, we stick to the $\dot{P}^\mathrm{GW}_\mathrm{b}=-27.7^{+1.5}_{-1.9}$\,\fsps\ \citep{Freire12} calculated with the previous $m_\mathrm{c}$ and $m_\mathrm{p}$ reported by \citet{Antoniadis12}.

\subsection{\psro}
\label{subsec:mspsrpi_J1824}
\psro\ is a 3-ms solitary pulsar discovered in the Globular cluster M28 (NGC~6626) \citep{Lyne87a}. 
The calibration configuration for this pulsar was sub-optimal, as the best in-beam phase calibrator for \psro\ was both resolved and faint (3.3\,mJy, see \ref{tab:mspsrpi_MSPs}), leading to noisy solutions, especially on the longest baselines.  This is likely responsible for the parallax non-detection (see \ref{tab:mspsrpi_models_no_pm_prior}), and indicates that higher sensitivity to enable improved calibration solutions would be advantageous in any future VLBI campaign.

The proper motion of M28 is estimated to be $\mu_\alpha^\mathrm{M28}=-0.28\pm0.03$\,\maspy\ and $\mu_\delta^\mathrm{M28}=-8.92\pm0.03$\,\maspy\ \citep{Vasiliev21} with Gaia Early Data Release 3 (EDR3). Hence, the relative proper motion of \psro\ with respect to M28 is $\Delta\mu_\alpha=0.03\pm0.05$\,\maspy\ and $\Delta\mu_\delta=1.1\pm0.8$\,\maspy.
Combining the M28 distance $D=5.4\pm0.1$\,kpc estimated by \citet{Baumgardt21}, we obtained the transverse space velocity $v_\perp=28\pm20$\,\kmps\ for \psro, which is smaller than the typical escape velocity ($\approx50$\,\kmps) of a globular cluster. Therefore, the pulsar is probably (as expected) bound to M28.

%, unless accelerated by other stars in its future course.
%which is unsurprisingly on the larger side of the $\bar{v}_\perp^\mathrm{(BI)}=63\pm39$\,\kmps\ of field MSPs (see \ref{subsec:mspsrpi_v_t}).
%Subtracting our proper motion of \psro\ (see \ref{tab:mspsrpi_VLBI_timing_results}) by that of NGC~6626, we acquired the relative proper motion

%\subsection{\psrka} \label{subsec:mspsrpi_J1853}

%\subsection{\psrl}
%\label{subsec:mspsrpi_J1910}

%\subsection{\psrmb} \label{subsec:mspsrpi_J1918}

%\subsection{\psrkb} \label{subsec:mspsrpi_J1939}

\section{Summary and Future prospects}
\label{sec:mspsrpi_summary}

%The \mspsrpi\ project is a large 1.55-GHz VLBA astrometry program focusing on 18 MSPs. This work presents astrometric results of all \mspsrpi\ pulsars, which includes the re-assessment of three previously published pulsars. 
In this \mspsrpi\ release paper, we have presented VLBI astrometric results for 18 MSPs, including a re-analysis of three previously published sources.  From the 18 sources, we detect significant parallaxes for all but three.
For each \mspsrpi\ pulsar, at least one self-calibratable in-beam calibrator was identified to serve as the reference source of relative astrometry. In three cases, 1D interpolation, a more complex observing and data reduction strategy, is adopted to further suppress propagation-related systematic errors. Among the three pulsars, \psrkb\ is the brightest MSP in the northern hemisphere. Hence, we took one step further to perform inverse-referenced 1D interpolation using \psrkb\ as the in-beam calibrator. 
Compared to the pioneering Multi-View study of \citet{Rioja17} at 1.6\,GHz, the larger number of observations and targets here provides more opportunities to characterise the interpolation performance, which is crucial for ultra-precise astrometric calibration schemes proposed for future VLBI arrays incorporating the Square Kilometre Array (SKA).
%These three cases are the first 1D interpolation realizations at $\lesssim4$\,GHz, which are important for investigating higher-order residual systematic errors of 1D/2D interpolation in the context of the Multi-View observing strategy \citep{Rioja17} proposed for future ultra-precise astrometry with the Square Kilometre Array (SKA). 
Based on a small sample of three, we found that $\eta_\mathrm{EFAC}$ has consistently inflated after applying 1D interpolation (see \ref{subsubsec:mspsrpi_implications_for_1D_interpolation}). This inflation implies that the higher-order terms of in the phase screen approximation may not be negligible, and could become the limiting factor of the ultra-precise SKA-based astrometry using the Multi-View strategy. Further investigations of the same nature, especially using temporally simultaneous (in-beam) calibrators, at low observing frequencies are merited and encouraged.

In this paper, we present two sets of astrometric results -- the quasi-VLBI-only results (see \ref{sec:mspsrpi_parameter_inference}) and the VLBI+timing results (see \ref{sec:mspsrpi_inference_with_priors}).
Both sets of astrometric results are inferred with the astrometry inference package {\tt sterne}\textsuperscript{\ref{footnote:sterne}}. 
The former set of results is largely independent of any input based on pulsar timing, making use only of orbital parameters as priors in the inference of orbital reflex motion, which affects only four pulsars from our sample and is near-negligible in any case.  The latter, however, additionally adopts the latest available timing parallaxes and proper motions as priors of inference wherever possible, affecting all pulsars in our sample. While the latter approach typically gives more precise results, we note that this is dependent on the accuracy of the timing priors, and identify seven pulsars (\psrc, \psrgb, \psrhb, \psri, \psro, \psrka\ and \psrl) for which disagreement between the quasi-VLBI-only and the timing priors mean that the VLBI+timing results should be treated with caution. 
From the VLBI perspective, we looked into possible causes of additional astrometric uncertainties, including non-optimal calibrator quality (see \ref{subsec:mspsrpi_J1721} and \ref{subsec:mspsrpi_J1824}) and calibrator structure evolution (see \ref{sec:mspsrpi_inference_with_priors}).
In future, proper motion uncertainties (including any unaccounted systematic error due to calibrator structure evolution) can be greatly reduced with only $\lesssim2$ extra observations per pulsar. For example, a 10-yr time baseline can improve the current VLBI-only proper motion precision by roughly a factor of 8.

From the VLBI+timing parallaxes $\varpi'$, we derived distances $D$  using \ref{eq:mspsrpi_p_D_extended}. Incorporating the PDFs of $D$ and proper motions \{$\mu'_\alpha$, $\mu'_\delta$\}, we estimated the transverse space velocities $v_\perp$ for 16 pulsars with significant distance determination, and found their $v_\perp$ to be generally on the smaller side of the previous estimates \citep{Hobbs05,Gonzalez11}.
%the field binary MSPs (that never left their original companions) generally show smaller $v_\perp$ compared to the MSPs born in dense stellar regions (see \ref{subsec:mspsrpi_v_t} and \ref{subsec:mspsrpi_J1824}). 
\citet{Boodram22} propose that MSPs must have near-zero space velocities in order to explain the Fermi Galactic centre excess.
Our relatively small space velocities inferred for 16 MSPs suggest that MSPs may not be ruled out as the source of the Galactic $\gamma$-ray centre excess.
Moreover, smaller space velocities of MSPs would also potentially address the globular cluster retention problem outlined in \ref{subsec:NS_kinematics}.
If the multi-modal feature of the $v_\perp$ is confirmed with a sample of $\sim50$ MSPs, it may serve as a kinematic evidence for alternative formation channels of MSPs (\citealp{Bailyn90,Gautam22}, also see \citealp{Ding23a} as an analogy). 

In addition, we estimated the radial accelerations of pulsars with their distances and proper motions (see \ref{sec:mspsrpi_orbital_decay_tests}), which allows us to constrain the intrinsic spin period derivative $\dot{P}_\mathrm{s}^\mathrm{int}$ and the intrinsic orbital decay $\dot{P}_\mathrm{b}^\mathrm{int}$ (see \ref{tab:mspsrpi_Shk}). We used the improved $\dot{P}_\mathrm{s}^\mathrm{int}$ of \psrhb\ to place a refined upper limit to the death line of $\gamma$-ray pulsars (see \ref{subsec:mspsrpi_J1730}), and the $\dot{P}_\mathrm{b}^\mathrm{int}$ (of \psrea\ and \psri) to constrain alternative theories of gravity (see \ref{subsec:mspsrpi_alternative_gravity}).
%By subtracting the effects of radial accelerations, we are able to estimate the intrinsic 
As already noted by \citet{Ding20}, the orbital decay tests (of gravitational theories) with the three viable \mspsrpi\ systems (i.e., \psrea, \psrfb\ and \psri) will be limited by the distance uncertainties, as parallax precision improves much slower than the $\dot{P}_\mathrm{b}^\mathrm{obs}$ precision \citep{Bell96}. 
%Though only contributing to a small proportion of the $\dot{P}_\mathrm{b}^\mathrm{int}$ uncertainty, an inaccurate $\mathcal{A}_\mathrm{Gal}$ estimate would likely bias the orbital decay test.

%1. Prospects: Gaia --> Galactic acceleration terms.

Moreover, we detected significant angular broadening of \psrgb, which we used to {\bfseries 1)} provide an independent check of the postulated connection between the HII region Sh~2-27 and the main scattering screen of \psrgb, and {\bfseries 2)} test the scintillation models proposed by \citet{Mall22}.
In future scintillation model inferences, angular broadening and temporal broadening measurements, where available, are suggested to be used as priors using \ref{eq:mspsrpi_two_screen_condition1}, in order to achieve more reliable (and potentially more precise) scintillation model parameters. 
Such an inference would also complete the geometric information of the deflection routes (using \ref{eq:mspsrpi_tau_sc2}, for example, in the two-screen case).

\section*{Acknowledgements}

The authors thank Y. Kovalev, L. Petrov and N. Wex for useful discussions.
HD is supported by the OzGrav scholarship through the Australian Research Council project number CE170100004.
%ATD is the recipient of an ARC Future Fellowship (FT150100415). 
PF thanks the continued support by the Max-Planck-Gesellschaft. SC, JMC, EF, DK and TJWL are members of the NANOGrav Physics Frontiers Center, which is supported by NSF award PHY-1430284.
%This research were conducted by the Australian Research Council Centre of Excellence for Gravitational Wave Discovery (OzGrav), .
This work is based on observations with the Very Long Baseline Array (VLBA), which is operated by the National Radio Astronomy Observatory (NRAO). The NRAO is a facility of the National Science Foundation operated under cooperative agreement by Associated Universities, Inc.
Pulsar research at Jodrell Bank Observatory is supported by a consolidated grant from STFC.
Pulsar research at UBC is supported by an NSERC Discovery Grant and by the Canadian Institute for Advanced Research.
The Nan\c{c}ay Radio Observatory is operated by the Paris Observatory, associated with the French Centre National de la Recherche Scientifique (CNRS). We acknowledge financial support from the ``Programme National de Cosmologie et Galaxies'' (PNCG) and ``Programme National Hautes Energies'' (PNHE) of CNRS/INSU, France.
The Parkes radio telescope (Murriyang) is part of the Australia Telescope, which is funded by the Commonwealth Government for operation as a National Facility managed by CSIRO.
The Wisconsin $\mathrm{H\alpha}$ Mapper and its $\mathrm{H\alpha}$ Sky Survey have been funded primarily by the National Science Foundation. The facility was designed and built with the help of the University of  Wisconsin Graduate School, Physical Sciences Lab, and Space Astronomy Lab. NOAO staff at Kitt Peak and Cerro Tololo provided on-site support for its remote operation.
Data reduction and analysis was performed on OzSTAR, the Swinburne-based supercomputer.
This work made use of the Swinburne University of Technology software correlator, developed as part of the Australian Major National Research Facilities Programme and operated under license. {\tt sterne} as well as other data analysis involved in this work made use of {\tt numpy} \citep{Harris20}, {\tt scipy} \citep{Virtanen20}, {\tt astropy} \citep{The-Astropy-Collaboration18} and the {\tt bilby} package \citep{Ashton19}.
%This work made use of the ATNF Pulsar Catalogue (v.~1.67) \citep{Manchester05}.

%%%%%%%%%%%%%%%%%%%%%%%%%%%%%%%%%%%%%%%%%%%%%%%%%%
\section*{Data and Code Availability}
\label{sec:mspsrpi_data_availability}
\begin{itemize}
    \item The VLBA data can be downloaded from the NRAO Archive Interface at \url{https://data.nrao.edu} with the project codes in \ref{tab:mspsrpi_MSPs}.
    \item Image models of phase calibrators are provided at \url{https://github.com/dingswin/calibrator_models_for_astrometry}.
    \item Supplementary materials supporting this paper can be found at \url{https://github.com/dingswin/publication_related_materials}.
    \item The data reduction pipeline {\tt psrvlbireduce} is available at \url{https://github.com/dingswin/psrvlbireduce} (version ID: b8ddafd).
    \item The astrometry inference package {\tt sterne} can be accessed at \url{https://github.com/dingswin/sterne}.
\end{itemize}



\bibliographystyle{mnras}
\bibliography{haoding} % if your bibtex file is called example.bib

\chapter{Conclusion}
\label{ch:conclusion}

\section{Thesis summary}
\label{sec:thesis summary}

This thesis presents precise astrometric results for three classes of Galactic neutron stars (NSs) --- NS X-ray binaries (XRBs), magnetars and recycled pulsars (including double neutron stars and other millisecond pulsars), using either the Very Long Baseline Array (VLBA) or the Gaia space telescope \citep{Gaia-Collaboration16}. 
Observationally, the thesis obtains 15 significant ($>3\,\sigma$) parallaxes of millisecond pulsars (MSPs) from the \mspsrpi\ project --- the largest ever astrometric survey of MSPs (see \ref{subsec:mspsrpi_mspsrpi}), and determined the first two magnetar parallaxes including
a 5\,$\sigma$ preliminary parallax for the fastest-spinning magnetar \swift\ (see \ref{sec:J1818_astrometry_progress}).

Methodologically, this thesis involves the development of two new packages --- the {\tt psrvlbireduce}\footnote{\url{https://github.com/dingswin/psrvlbireduce}} VLBI data reduction pipeline and the {\tt sterne}\footnote{\url{https://github.com/dingswin/sterne}} astrometric Bayesian inference package.
Using the two packages, the thesis made the first realizations of 1D interpolation (see \ref{subsec:J1810_dualphscal}, \ref{sec:J1818_astrometry_progress} and \ref{subsec:mspsrpi_dualphscal}), inverse-referenced 1D interpolation (see \ref{subsec:mspsrpi_sophisticated_data_reduction}) and multi-source astrometric inference (see \ref{subsubsec:mspsrpi_multi-source-inference}) on pulsars. Based on a small sample of 3 pulsars astrometrically measured using 1D interpolation, the thesis reveals that the higher-order term of interpolation might be non-negligible at 1.5\,GHz (see \ref{subsubsec:mspsrpi_implications_for_1D_interpolation}), which may become the limiting factor of the Multi-View observing strategy \citep{Rioja17} designed for the Square Kilometre Array.  Additionally, a generic quasar-based method to estimate parallax zero-points is proposed for Gaia astrometry (see \ref{subsec:PRE_s_pi0}).

Using the precise astrometric results, the thesis fulfills abundant scientific motivations (outlined in \ref{ch:Introduction}), which are categorized into 4 perspectives. 
Firstly, based on 16 \mspsrpi\ pulsars with significant distance estimation, a multi-modal transverse space velocity $v_\perp$ distribution was determined for MSPs (see \ref{subsec:mspsrpi_v_t}). One mode of this $v_\perp$ distribution can be approximated by a normal distribution that is consistent with zero, which may potentially observationally resolve the two retention problems described in \ref{subsec:NS_kinematics} --- the globular cluster retention problem \citep[e.g.][]{Pfahl02} and the observed Galactic Centre $\gamma$-ray excess \citep[e.g.][]{Abazajian12,Boodram22}. 
The low $v_\perp$ mode also agrees with the prediction of alternative MSP and NS formation channels mentioned in \ref{subsec:NS_kinematics}.
%related to white dwarfs (WDs), i.e., the accretion-induced formation channel \citep{Bailyn90,Gautam22} and the WD-WD merger channel \citep{Michel87}.
In a similar vein, this thesis proposes to use the $v_\perp$ distribution of magnetars as the probe of magnetar formation channels (see \ref{sec:J1818_intro}).
Previous study of magnetar $v_\perp$ distribution based on a sample of 8 objects has shown a mean $v_\perp$ of around 150\,\kmps\ \citep{Tendulkar13,Lyman22} that is consistent with normal pulsars \citep{Hobbs05}. This thesis determined the astrometry of 2 of the 5 constant radio magnetars, and hence increased the sample of magnetars with precise velocities (see \ref{subsec:probe_magnetar_formation}), and further confirms the relatively small $v_\perp$ as opposed to the earlier postulate of \citet{Duncan92} (which suggested magnetars might be very high-velocity objects).

Secondly, the significant proper motions and parallaxes measured for \mspsrpi\ pulsars enable precise determination of the extrinsic time derivatives of pulsar spin periods and orbital periods (where applicable) (see \ref{subsec:pulsar_timing} and \ref{sec:XRB_astrometry}). By deducting the extrinsic terms from the observed values, the intrinsic spin period derivatives $\dot{P}_\mathrm{s}^\mathrm{int}$ and intrinsic orbital decays $\dot{P}_\mathrm{b}^\mathrm{int}$ are estimated. The $\dot{P}_\mathrm{s}^\mathrm{int}$ allow precise estimation of the spin-down power (see \ref{subsec:NS_emission_sources}), which refines the death line of $\gamma$-ray pulsars using \psrhb\ (see \ref{sec:XRB_astrometry} and \ref{subsec:mspsrpi_J1730}) --- the $\gamma$-ray pulsar with the smallest spin-down power \citep{Lorimer95a,Guillemot16}.
The $\dot{P}_\mathrm{b}^\mathrm{int}$ improves the orbital-decay test of general relativity (GR) with the double neutron star system \psrfb\ (see \ref{sec:B1534_Pbdot_test} and \ref{subsec:mspsrpi_J1537}), and tightens the constraints on alternative gravitational theories (that predict time variation of the Newton's gravitational constant and dipolar gravitational radiation) using four pulsar-WD systems (see \ref{subsec:J1012_alternative_gravity} and \ref{subsec:mspsrpi_alternative_gravity}). 

Thirdly, 4 photospheric radius expansion (PRE) bursters (explained in \ref{subsec:PRE_bursts} and \ref{sec:PRE_intro}) with significant parallaxes from the Gaia Early Data Release 3 (EDR3) are used to refine the PRE burster distances, and constrain the composition $X$ of the nuclear fuel accumulated on the NS surface at the time of PRE bursts, assuming that the simplistic PRE model is correct. 
Furthermore, by relaxing the $X$ prior (of Bayesian inference) to include unphysical ranges, the thesis novelly probes the simplistic PRE burst model that has been used to estimate the distances to most PRE bursters (see \ref{subsec:PRE_roadmap_for_model_testing}). It is found that the simplistic PRE burst model largely stands the test by the PRE burster parallaxes, at the Gaia EDR3 parallax uncertainty level.

Last but not least, this thesis also constrains the distribution of ionised interstellar medium (IISM) in different ways.
The parallax-based distances of 15 \mspsrpi\ pulsars are used to test the two prevailing models \citep{Cordes02,Yao17} of Galactic free-electron distribution $n_\mathrm{e}(\Vec{x})$, which shows that the two models have comparable reliability.
In addition, the angular-broadened size $\theta_\mathrm{sc}$ (see \ref{subsec:scattering_screen}) of \psrgb\ is precisely determined with VLBA observations at 1.55\,GHz.
Incorporating a temporal broadening estimate $\tau_\mathrm{sc}$, the $\theta_\mathrm{sc}$ {\bfseries 1)} independently supports the association between the main scattering screen of \psrgb\ emission and the HII region Sh 2-27, and {\bfseries 2)} favors the two-screen scintillation model over the one-screen one (both proposed by \citealp{Mall22}). 
%It is found that the association between the main scattering screen (of \psrgb) and the HII region Sh 2-27 is independently supported by $\theta_\mathrm{sc}$, $\tau_\mathrm{sc}$ and the parallax-based distance of \psrgb.
Where $\tau_\mathrm{sc}$ and $\theta_\mathrm{sc}$ are known, the thesis proposes to use \ref{eq:mspsrpi_two_screen_condition1} to restrict the parameter space of viable scintillation models.

\section{Future prospects}
\label{sec:future_prospects}

%\subsection{Future works}
%\label{subsec:follow_up_works}

Although a good amount of scientific outputs have been derived from the \mspsrpi\ catalogue (see \ref{sec:thesis summary}), a few further studies promised by the \mspsrpi\ data are not covered by this thesis. Specifically, the solar system ephemerides (SSEs) can be examined using the absolute positions of MSPs (see \ref{subsec:SSE_VLBI}), which can potentially refine the prior knowledge of SSEs for the {\tt BAYESEPHEM} framework \citep{Arzoumanian18} or other frameworks of its kind, then improve the pulsar timing array (PTA) sensitivities. 
%The SSE examination and the impact on PTA sensitivities will be carr
This work of SSE examination is underway.
The absolute MSP positions will be estimated in the same way as \ref{subsec:J1012_abspos}, but based on refined phase calibrator positions determined at 1.5\,GHz with a parallel project.
This incorporated effort will likely achieve $\lesssim0.5$\,mas precision for each MSP position, enabling the most sensitive SSE examination.

On another line of research, the novel quasar-based method proposed to estimate Gaia parallax zero-points $\varpi_0$ has so far only been used on five sources (see \ref{subsec:PRE_s_pi0}). To increase the statistical importance of the $\varpi_0$ estimation method, future work will optimize the method, and apply it to a larger sample covering both low and high Galactic latitudes $b$, which may further examine the excess $\varpi_0$ (compared to the empirical $\varpi_0$ of \citealp{Lindegren21}) at low $b$  found by \citet{Ren21} (and supported by \citealp{Ding21} or \ref{ch:PRE}).

%\subsection{Future prospects}
%\label{subsec:future_prospects}

Meanwhile, many of the studies presented in this thesis are limited by their small sample sizes, hence prioritizing increases in the dimension of the astrometric sample. 
%While this thesis advances the astrometric studies of magnetars, NS XRBs and MSPs, more research opportunities are revealed
For example, the establishment of a statistically significant magnetar $v_\perp$ distribution requires dozens of magnetars precisely measured astrometrically. However, only about a dozen magnetars are sufficiently bright at optical/infrared or radio (see the McGill Online Magnetar Catalogue\footnote{\url{http://www.physics.mcgill.ca/~pulsar/magnetar/main.html}}, \citealp{Olausen14}), which limits the sample size of magnetars that can be precisely measured astrometrically.
Until very recently, magnetars were believed to be short-lived electromagnetic sources that quickly exhaust their energy sources and become invisible, which is consistent with the small ($\sim30$) number of magetars ever identified \citep{Olausen14}.
Nevertheless, the recent discovery of ultra-long period pulsars \citep{Hurley-Walker22,Caleb22} suggests a potentially previously overlooked radio magnetar population, which may supercharge the investigation of the magnetar $v_\perp$ distribution that would shed more light on the magnetar formation channels (see \ref{sec:J1818_intro}).

Likewise, the tentative multi-modal $v_\perp$ distribution of MSPs revealed by 16 \mspsrpi\ pulsars merits a larger astrometric survey of MSPs. The multi-modal feature, if confirmed with $\sim50$ MSPs carefully and systematically studied astrometrically, would strongly indicate alternative formation channels (\citealp{Michel87,Bailyn90,Tauris13,Gautam22,Miyaji80}, also see \ref{subsec:NS_kinematics}).
Furthermore, an extended astrometric survey focusing on newly discovered MSPs is essential for the PTA research. As mentioned in \ref{subsec:VLBI_PTA}, VLBI astrometry of newly discovered MSPs can potentially maximize the pace of PTA sensitivity improvement \citep{Siemens13,Madison13}, assuming that the new MSPs are added to the PTA.

As newly discovered MSPs tend to be fainter than historical MSPs (not to mention that MSPs are already generally fainter than canonical pulsars), MSP astrometry would become increasingly limited by the sensitivity of a VLBI array.
%Due to the relative faintness of MSPs, VLBI astrometry of the MSPs would probably require a VLBI network joined by a high-sensitivity telescope.
%Meanwhile, adding newly deployed high-sensitivity into VLBI array would substantially improve the sensitivity of the array, thus making VLBI astrometry for fainter radio sources plausible. 
A high-sensitivity VLBI array, e.g. FAST-VLBI \citep{Zhang17,Chen20}, SKA-VLBI \citep{Paragi14} and ngVLA \citep{McKinnon19}, would enable precise astrometry of further-away ($\gtrsim5$\,kpc) MSPs, 
%Among the potential targets of high-sensitivity astrometry, pulsars further away ($\gtrsim5$\,kpc) from the Earth 
which are often more important for refining the $n_\mathrm{e}(\Vec{x})$ model. 
%The flat-spectrum radio magnetars at large distances offer rare chances to push the boundary of precise astrometry further, while 
Alongside the sensitivity improvement is the demand for advanced calibration techniques (see \ref{sec:ch2_VLBI_observing_strategies}).
Specifically, in-beam phase referencing (see \ref{subsec:ch2_in_beam_astrometry}) would be the optimal strategy for FAST-VLBI due to its relatively long slewing time, while the interpolation strategy would be less effective (see \ref{subsec:ch2_interpolation}). In comparison, interpolation may potentially serve as the standard phase calibration strategy for SKA-VLBI \citep{Rioja17}, capitalizing on the beamforming capacity of the SKA (and ideally other components of the SKA-VLBI array).
As the parallaxes to be measured get smaller, the calibration techniques have to keep developing.

Apart from the use of high-sensitivity arrays, astrometry spanning a long time (e.g. \ref{sec:J1810_observations}) can also greatly improve astrometric precision, especially for the proper motion. This ``long-haul'' astrometry can be realized with two groups of VLBI observations many years apart. 
Specifically, the \mspsrpi\ pulsars were astrometrically measured roughly 6 years ago. Adding $\lesssim2$ long-haul VLBI observations per source can substantially reduce the proper motion uncertainties, which may resolve the discrepancy between the VLBI-only and the timing proper motions of some of the \mspsrpi\ pulsars (see \ref{sec:mspsrpi_inference_with_priors}).

To improve the astrometric precision of Gaia sources also requires continued observations. This is particularly important for probing the simplistic PRE model using the Gaia parallaxes of the few PRE bursters (see \ref{subsec:PRE_roadmap_for_model_testing}), but would be ultimately limited by the lifespan of the Gaia mission.
%Future Gaia release probes PRE model. 
In comparison, VLBI astrometry of PRE bursters with relatively frequent outbursts (e.g. \aql) would potentially achieve better astrometric precision in the long run, therefore being essential for the study of PRE models.

%Last but not least, the Bayesian astrometric inference framework, currently embodied as the {\tt sterne} package, can be further improved.

\bibliographystyle{mnras}
\bibliography{haoding}

% Appendices and glossary of terms

%\appendix

%\include{appendices/appendixA}

%\include{appendices/appB}
%\clearpage
%\input{glossary}

% REFERENCES
%\newpage
%\addcontentsline{toc}{chapter}{Bibliography}
%% Choose the bibliographic style file to use
%\bibliographystyle{bibstyle}
%\bibliographystyle{mnras}
%\bibliographystyle{aasjournal}
%\bibliography{mybibliography,haoding}

% List of publications
%\clearpage
%\addcontentsline{toc}{chapter}{Publications}
%\input{publications}

\end{document}